# Quantum information processing in modular cavity QED architectures

Zoé McIntyre

Department of Physics
McGill University
Montreal, Quebec, Canada

May 7, 2025

A thesis submitted to McGill University in partial fulfillment of the requirements of the degree of Doctor of Philosophy

# Abstract

This thesis contains a collection of articles exploring various aspects of quantum information processing with cavity quantum electrodynamics (QED), starting with qubit noise spectroscopy and building towards the longer-term goal of modular quantum-computing architectures equipped with protocols for controlling and correcting the states of distantly separated qubits. To render the results accessible to a general audience, the first chapter presents a self-contained introduction to the field of cavity QED. Subsequent chapters contain the original text of published articles or articles submitted for publication, all of which were prepared as part of PhD research conducted at McGill University.

Following the introductory material in Chapter 1, we show in Chapter 2 how measurements of the field emitted by a cavity can be leveraged for in-situ qubit noise spectroscopy in the presence of significant inhomogeneous broadening. We also identify a signature of genuinely quantum noise in the cavity output field originating from the non-commutation of bath operators acting on the qubit. In Chapter 3, we present a novel quantum-optical effect, based on destructive interference, whereby a suitable modulation of a longitudinal cavity-qubit coupling can be used to entangle the state of a qubit with the path taken by an incoming multiphoton wavepacket. Entanglement between a qubit and a which-path degree-of-freedom, generated in this manner, can in turn be used to generate entanglement between distant stationary qubits. As shown in Chapter 4, qubit–which-path entanglement can also be exploited for maximally sensitive estimation of a phase in a Mach-Zehnder interferometry setup, i.e., sensing at the quantum Cramér-Rao bound. The measurement protocols given for this purpose rely on homodyne detection rather than on the more experimentally challenging number-resolving measurements typically considered in the literature. In Chapter 5, we discuss strategies for realizing stabilizer measurements using qubit-conditioned phase shifts applied to propagating pulses of radiation. We find that in the context of a subsystem surface code, photon loss during such stabilizer measurements would not introduce any horizontal hook errors on the code qubits. We also present in Chapter 5 a six-qubit entangled "tetrahedron" state that could be prepared using these stabilizer checks, and which could serve as the resource state for controlled quantum teleportation of a two-qubit state. In the sixth and final chapter, we give protocols for performing entangling gates between distant stationary qubits using time-bin encoded photons.

# Abrégé

Cette thèse contient une collection d'articles explorant divers aspects du traitement de l'information quantique avec l'électrodynamique quantique en cavité, en commençant par la spectroscopie du bruit des qubits et en progressant vers l'objectif à plus long terme d'architectures modulaires d'informatique quantique équipées de protocoles pour contrôler et corriger les états de qubits distants. Afin de rendre les résultats plus accessibles à un grand public, le premier chapitre présente une introduction autonome au domaine de l'électrodynamique quantique en cavité. Les chapitres suivants contiennent les textes originaux d'articles publiés ou soumis pour publication, tous préparés dans le cadre de recherches doctorales menées à l'Université McGill.

Après l'introduction du premier chapitre, nous montrons au deuxième chapitre comment les mesures du champ émis par une cavité peuvent être exploitées pour la spectroscopie du bruit avec un qubit en présence d'un élargissement inhomogène considérable. Nous identifions également une signature de bruit véritablement quantique dans le champ de sortie de la cavité provenant de la non-commutation des opérateurs de bain agissant sur le qubit. Dans le troisième chapitre, nous présentons un nouvel effet d'optique quantique, basé sur l'interférence destructive, dans lequel une modulation appropriée d'un couplage longitudinal cavité-qubit peut être utilisée pour intriquer l'état d'un qubit avec le chemin emprunté par un paquet d'ondes de photons multiple. Une telle intrication pourrait à son tour être utilisée pour générer une intrication entre des qubits distants. Comme le montre le quatrième chapitre, l'intrication entre un qubit et un degré de liberté « quel chemin » peut également être exploitée pour une estimation extrêmement sensible d'une phase dans une configuration d'interférométrie de Mach-Zehnder, c'est-à-dire pour une détection à la limite quantique de Cramér-Rao. Les protocoles de mesure donnés à cette fin reposent sur la détection homodyne plutôt que sur les mesures de résolution des nombres généralement envisagées dans la littérature scientifique. Dans le cinquième chapitre, nous discutons des stratégies pour réaliser des mesures de stabilisation à base de décalages de phase appliqués à des impulsions de rayonnement. Nous constatons que dans le contexte d'un code de surface de sous-système, la perte de photons pendant ces mesures de stabilisation n'introduirait pas d'erreurs de crochet horizontal sur les qubits du code. Nous présentons également au cinquième chapitre un état « tétraèdre » intriqué à six qubits qui pourrait être préparé en utilisant ces vérifications de stabilisateurs, et qui pourrait servir d'état ressource pour la téléportation quantique contrôlée d'un état à deux qubits. Dans le sixième et dernier chapitre, nous donnons des protocoles pour effectuer des portes d'intrication entre des qubits distants en utilisant des photons encodés dans le temps.

# Preface

**Contributions of authors**
This thesis is written in a manuscript-based format. The work presented in Chapters 2-6 was completed by the author during PhD studies at McGill University under the supervision of Prof. W. A. Coish. In all cases, the theory and calculations were developed and performed by the author with help from WAC. All manuscripts were written by the author, again with help from WAC.

**Contributions to original knowledge**
Chapters 2-5 of this thesis contain the original text of manuscripts that have been published. Chapter 6 is a manuscript to be submitted. Each of these chapters includes its own introduction, conclusion, and in some cases, appendices. Below, we give the journal reference for each chapter and list the original contributions to knowledge found therein.

CHAPTER 2
**Non-Markovian transient spectroscopy in cavity QED**
Z. McIntyre and W. A. Coish
*Phys. Rev. Res.* **4**, L042039 (2022)

The contributions to original knowledge of Chapter 2 are:

- An expression for the output field of a cavity coupled to a qubit undergoing dynamical decoupling, together with a protocol for extracting the qubit echo envelope from this output field
- The identification of a signature of quantum noise (due to non-commuting bath operators) in the output field
- An expression for the stretched-exponential decay of the qubit echo envelope in the presence of inhomogeneously broadened Purcell decay

CHAPTER 3
**Photonic which-path entangler based on longitudinal cavity-qubit coupling**
Z. M. McIntyre and W. A. Coish
*Phys. Rev. Lett.* **132**, 093603 (2024)

The contributions to original knowledge of Chapter 3 are:

- The identification of a novel quantum-optical effect wherein a parametrically modulated longitudinal cavity-qubit coupling can be used to entangle the state of a qubit with the path taken by a multiphoton wavepacket

- A parameter regime where such an effect could be realized with a flopping-mode spin qubit
- A protocol for converting qubit–which-path entanglement into entanglement between distant stationary qubits

CHAPTER 4
**Homodyne detection is optimal for quantum interferometry with path-entangled coherent states**
Z. M. McIntyre and W. A. Coish
*Phys. Rev. A* **110**, L010602 (2024)

The contributions to original knowledge of Chapter 4 are:

- To our knowledge, the first measurement protocol for Heisenberg-limited, optimally sensitive quantum interferometry using only homodyne detection
- An analysis of how photon loss impacts the precision that can be achieved using this measurement protocol

CHAPTER 5
**Flying-cat parity checks for quantum error correction**
Z. M. McIntyre and W. A. Coish
*Phys. Rev. Res.* **6**, 023247 (2024)

The contributions to original knowledge of Chapter 5 are:

- An analysis of the errors introduced by photon loss during multiqubit stabilizer measurements realized using qubit-conditioned phase shifts on classical pulses of radiation
- A protocol for preparing genuine multipartite entanglement (in the form of a "tetrahedron" state) using such stabilizer measurements, as well as a concrete procedure for using this tetrahedron state as a resource for controlled quantum teleportation of a two-qubit state

CHAPTER 6
**Protocols for inter-module two-qubit gates mediated by time-bin encoded photons**
Z. M. McIntyre and W. A. Coish
*arXiv: 2503.03938*

The contributions to original knowledge of Chapter 6 are:

- Protocols for long-range two-qubit gates mediated by time-bin encoded photons
- An analysis of stationary-qubit dephasing due to the absorption of a photonic time-bin qubit by a dielectric environment

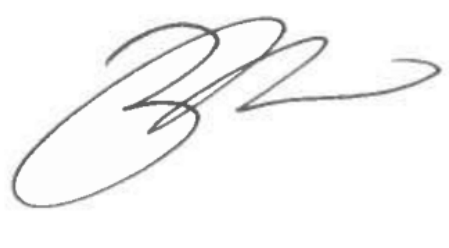

Zoé McIntyre

# Acknowledgments

First and foremost, I would like to thank my supervisor Bill for his unwavering support, optimism, patience, and for believing in me. I'd also like to thank Profs. Lily Childress, Jack Sankey, Kartiek Agarwal, and Kai Wang for serving as members of my PhD committee at various points over the years.

I'm fortunate to have enjoyed great company from past and current members of Bill's group, and from the rotating cast of 422. Thank you Melissa AS for being such a wonderful friend—for the skates at Parc Lafontaine and the poutines. Thank you also to FS for not knowing what a pigeon looks like, and for the T-rexes. You've been a great friend too.

Je regrette ne pouvoir remercier en personne ma grand-mère Liliane. Enfin, finally, I know that none of this would have been possible without the unconditional love and support of my amazing parents and sister. I am so lucky to have you as my family.

# Contents

# List of Figures

# List of Tables

# 1

# Introduction

The goal of building a universal quantum computer—a goal currently being pursued by scientists and engineers all over the world—is a challenging one. In contrast to classical computers, which store information in classical bits, the basic building blocks of quantum computers are quantum bits, a.k.a. *qubits*. These are quantum systems which (similar to classical bits) have two distinct states, but which (unlike classical bits) must be capable of existing in a superposition of both states at once. In the circuit model of quantum computing, a computation is carried out by manipulating the state of several qubits through a sequence of logical operations called gates, requiring that the qubits be able to interact and become entangled, at the end of which the "answer" is obtained by measuring the state of one or more qubits. This final measurement will collapse the state of the qubits into an eigenstate of the operator (quantum observable) being measured. In addition to the challenge of achieving precise control over these tiny quantum systems, there is also the challenge of isolating these systems from their massive classical environments, which collapse superposition states in much the same way as an intentional measurement. This loss of information to the environment is known as *decoherence*. Decoherence can be mitigated passively through qubit design: Qubits can be encoded in degrees of freedom that have a built-in resilience to certain noise processes in the environment. It can also be mitigated actively by encoding logical qubits in the collective state of many physical qubits. A quantum error correction code can then be used to detect and correct errors affecting physical qubits, thereby preventing these errors from corrupting the logically encoded information. This approach to fault tolerance typically requires many, many physical qubits per logical qubit. The questions of how to best scale up a given quantum-computing architecture to meet this requirement, and of which architectures provide the best path to scalability, are not questions with clear-cut answers in this era of noisy, intermediate-scale quantum processors.

## 1.1 Cavity quantum electrodynamics

Many different physical systems can and are being used to encode qubits. One of the leading approaches to quantum computing exploits the interaction of stationary qubits encoded in the energy levels of artificial atoms—systems engineered to have discrete, or atom-like, spectra—with quanta of the electromagnetic field (photons). These artificial atoms include gate-defined quantum dots in semiconducting heterostructures and superconducting circuits involving nonlinear elements called Josephson junctions. In such a setup, qubit-photon interactions can be used to initialize, manipulate, connect, and read out qubits. This approach has its roots in the field of cavity quantum electrodynamics (QED), which has historically studied the interaction of atoms with optical photons hosted in reflective cavities [1]. The interaction of superconducting qubits with microwave photons has come to be known as circuit QED, so called because the microwave photons with which these qubits naturally interact are the quantized excitations of LC circuits. One particular appeal of cavity/circuit QED for quantum computing is that quantum information can be coherently shared between stationary degrees-of-freedom (the artificial atoms) and photonic degrees-of-freedom (typically, states of a standing-wave mode in a microwave cavity). This not only enables qubits to interact with one another via the cavity mode, but also offers a clear path towards coherently transmitting information over long ($\gg$ cm) distances by coupling the cavities to transmission-line resonators, which serve as the microwave analog of fiber-optic cables. These transmission lines could then act as quantum interconnects between spatially separated qubit modules, allowing qubits on separate chips to interact as part of the same quantum processor.

Over the course of this introductory chapter, we will review the different regimes of operation of cavity QED and the physics characteristic to each. As the prototypical cavity-QED setup, we consider a single qubit interacting with the quantized electromagnetic field of a single cavity mode. The Hamiltonian describing this coupled system is given by (setting $\hbar = 1$)

$$H = \omega_{\mathrm{e}} |\mathrm{e}\rangle\langle\mathrm{e}| + \omega_{\mathrm{g}} |\mathrm{g}\rangle\langle\mathrm{g}| + \omega_{\mathrm{c}} a^\dagger a + H_{\mathrm{int}}, \tag{1.1}$$

where $|\mathrm{e}\rangle$ and $|\mathrm{g}\rangle$ are the two states (with associated frequencies $\omega_{\mathrm{e}}$ and $\omega_{\mathrm{g}}$) encoding the qubit, $a^\dagger$ and $a$ are bosonic creation and annihilation operators acting on the cavity mode (having frequency $\omega_{\mathrm{c}}$), and where $H_{\mathrm{int}}$ is a Hamiltonian modelling the cavity-qubit interaction. For concreteness, we assume that the coupling of the qubit to the cavity originates from the $|\mathrm{e}\rangle$ and $|\mathrm{g}\rangle$ states being associated with different charge distributions (capacitive coupling). When the potential $V(\boldsymbol{r})$ of the cavity field varies slowly on the scale of the qubit charge distribution $\rho(\boldsymbol{r})$, this coupling can be treated in a dipole approximation, giving

$$H_{\mathrm{int}} = -\boldsymbol{d} \cdot \boldsymbol{E}(\boldsymbol{r}_0), \tag{1.2}$$

where $\boldsymbol{E}(\boldsymbol{r}_0) = -\nabla V(\boldsymbol{r}_0)$ is the cavity electric field at the position $\boldsymbol{r}_0$ of the qubit and $\boldsymbol{d} = \int d^3\boldsymbol{r}\rho(\boldsymbol{r})(\boldsymbol{r} - \boldsymbol{r}_0)$ is the qubit dipole operator. We quantize the electric field as $\boldsymbol{E}(\boldsymbol{r}_0) = E_{\mathrm{RMS}}(\boldsymbol{r}_0)\hat{\boldsymbol{r}}(a + a^\dagger)$ for a field polarized along $\hat{\boldsymbol{r}}$, where here, $E_{\mathrm{RMS}}(\boldsymbol{r}_0)$ is the root-mean-squared electric field at $\boldsymbol{r}_0$ due to vacuum fluctuations in the cavity. Taking matrix elements of Eq. (1.2) with respect to the qubit eigenstates, we then have

$$H = \frac{1}{2}\omega_{\mathrm{q}}\sigma_z + \omega_{\mathrm{c}} a^\dagger a + g_x \sigma_x (a + a^\dagger) + (g_z \sigma_z + g_0)(a + a^\dagger), \tag{1.3}$$

up to a constant shift in energy, where here, $\omega_{\mathrm{q}} = (\omega_{\mathrm{e}} - \omega_{\mathrm{g}})$ is the qubit splitting, and where the transverse ($\propto \sigma_x = |\mathrm{e}\rangle\langle\mathrm{g}| + |\mathrm{g}\rangle\langle\mathrm{e}| = \sigma_+ + \sigma_-$) and longitudinal ($\propto \sigma_z = |\mathrm{e}\rangle\langle\mathrm{e}| - |\mathrm{g}\rangle\langle\mathrm{g}|$) coupling strengths are given by

$$g_x = -E_{\mathrm{RMS}}\langle\mathrm{e}|\boldsymbol{d}\cdot\hat{\boldsymbol{r}}|\mathrm{g}\rangle, \tag{1.4}$$

$$g_z = \frac{E_{\mathrm{RMS}}}{2}(\langle\mathrm{g}|\boldsymbol{d}\cdot\hat{\boldsymbol{r}}|\mathrm{g}\rangle - \langle\mathrm{e}|\boldsymbol{d}\cdot\hat{\boldsymbol{r}}|\mathrm{e}\rangle). \tag{1.5}$$

The transverse coupling strength $g_x$ is determined by the transition-dipole matrix element $\langle\mathrm{e}|\boldsymbol{d}\cdot\hat{\boldsymbol{r}}|\mathrm{g}\rangle$, describing the amplitude for electrically-driven transitions to occur between the qubit eigenstates. The longitudinal coupling strength $g_z$, by contrast, is determined by the difference of the intrinsic dipole moments $\langle\mathrm{g(e)}|\boldsymbol{d}\cdot\hat{\boldsymbol{r}}|\mathrm{g(e)}\rangle$ associated with the charge configurations defining the qubit eigenstates. In some systems, these eigenstates may possess symmetries that cause $g_z$ to vanish identically. For a system with inversion symmetry, for instance, the qubit eigenstates are simultaneous eigenstates of the parity operator $\hat{\Pi}$: $\hat{\Pi}|\mathrm{e(g)}\rangle = \pm|\mathrm{e(g)}\rangle$, where here, $\hat{\Pi}^\dagger\boldsymbol{r}\hat{\Pi} = -\boldsymbol{r}$. In this case, $g_z = 0$. This is not an unusual scenario in

atomic physics, where the relevant potential $V(\boldsymbol{r})$ is typically the inversion-symmetric Coulomb potential. Likely for this reason, the field of cavity QED has historically focused predominantly on transverse coupling. Many of the artificial atoms designed to encode qubits, however, are defined in lower-dimensional systems and may exhibit a non-zero longitudinal coupling strength, opening up a range of possibilities that are largely inaccessible in more "traditional" atomic systems. Some of these possibilities will be discussed later in this introductory chapter, as well as in Chapter 3 of this thesis.

Interestingly, the presence of the qubit also introduces a qubit-state-independent displacement of the cavity mode, proportional to the average intrinsic dipole moment

$$g_0 = -\frac{E_{\mathrm{RMS}}}{2}(\langle \mathrm{g}|\boldsymbol{d}\cdot\hat{\boldsymbol{r}}|\mathrm{g}\rangle + \langle \mathrm{e}|\boldsymbol{d}\cdot\hat{\boldsymbol{r}}|\mathrm{e}\rangle). \tag{1.6}$$

This qubit-state-independent shift can be absorbed into the definition of the cavity mode through a displacement transformation $a \to a - g_0/\omega_\mathrm{c}$ on the original Hamiltonian [Eq. (1.3)], together with a redefinition of the qubit states as eigenstates of $(\omega_\mathrm{q} - g_z g_0/\omega_\mathrm{c})\sigma_z - (g_x g_0/\omega_\mathrm{c})\sigma_x$. Typically, the Hamiltonians written in the literature omit the $\propto g_0$ term by implicitly transforming to this displaced frame, which is often justified since the $\propto g_0$ term does not lead to any qubit-state-dependent dynamics of the cavity mode. However, it is worth noting that the states $|\mathrm{g}\rangle$ and $|\mathrm{e}\rangle$ of the qubit generically couple to the cavity with unequal strengths $g_z + g_0$ and $g_z - g_0$, respectively. We return to this point in Chapter 3.

In the remainder of this section, we discuss the physics characteristic to both transverse ($\propto \sigma_x$) and longitudinal ($\propto \sigma_z$) light-matter coupling, paying attention also to the ways that such physics can be harnessed for the purpose of quantum information processing. At various points in the ensuing discussion, we invoke input-output theory [2] to relate the dynamics of the cavity mode to observables—typically the cavity reflection or transmission—that are commonly reported in experiments. The so-called input-output relation used for this purpose is simply stated as fact and applied; a derivation can be found in Section 1.2 of this introductory chapter.

### 1.1.1 Rabi and Jaynes-Cummings coupling

We begin by focusing on the transverse coupling term familiar from atomic physics, which ultimately still underpins most quantum-information processing with cavity and circuit QED. For $g_z = g_0 = 0$, Eq. (1.3) gives the Rabi Hamiltonian,

$$H_\mathrm{R} = \frac{1}{2}\omega_\mathrm{q}\sigma_z + \omega_\mathrm{c} a^\dagger a + g_x\sigma_x(a + a^\dagger). \tag{1.7}$$

The Rabi model [3, 4] was first introduced in the 1930s, but despite its apparent simplicity and success in describing nuclear magnetic resonance experiments, a full analytic solution to the Rabi model was not found until 2011 [5]. Prior to 2011, it had been conjectured that the existence of invariant subspaces associated with conserved quantities other than energy might be a requirement for a model to be analytically solvable.

We will first describe a simple, widely applicable limit of the Rabi model (the Jaynes-Cummings model) that describes most present-day cavity QED devices, before describing a few interesting effects that only emerge when this simple limit breaks down and the full Rabi Hamiltonian must be retained in order to accurately describe the system's behaviour. For $g_x \ll |\omega_\mathrm{c} + \omega_\mathrm{q}|$, the counter-rotating (number-non-conserving) terms $a^\dagger\sigma_+$ and $a\sigma_-$ in the Rabi Hamiltonian can be neglected in a rotating-wave approximation, leading to the Jaynes-Cummings Hamiltonian [6],

$$H_\mathrm{JC} = \frac{1}{2}\omega_\mathrm{q}\sigma_z + \omega_\mathrm{c} a^\dagger a + g_x(\sigma_+ a + \sigma_- a^\dagger). \tag{1.8}$$

The Jaynes-Cummings Hamiltonian admits invariant subspaces related to the continuous $U(1)$ symmetry associated with conservation of the total number $\hat{n}_\mathrm{tot} = a^\dagger a + |\mathrm{e}\rangle\langle\mathrm{e}|$ of excitations in the cavity-qubit sys-tem. Formally, this $U(1)$ symmetry exists because $U^\dagger(\phi)H_{\mathrm{JC}}U(\phi) = H_{\mathrm{JC}}$ for any $\phi \in [0, 2\pi)$, where here, $U(\phi) = e^{-i\phi\hat{n}_{\mathrm{tot}}}$. Since there are two degrees of freedom and two conserved quantities—the Hamiltonian itself together with $\hat{n}_{\mathrm{tot}}$—the Jaynes-Cummings model is integrable in the sense of Liouville [7], and the Hilbert space $\mathscr{H} = \mathbb{C}^2 \otimes L^2(\mathbb{R})$ of the qubit and cavity can be decomposed into a direct sum of dynamically invariant subspaces labelled by the total number of excitations:

$$\mathscr{H} = |\mathrm{g}, 0\rangle \oplus \sum_{n=1}^{\infty} \mathscr{H}_n. \tag{1.9}$$

Here, the non-degenerate ground state $|\mathrm{g}, 0\rangle = |\mathrm{g}\rangle \otimes |0\rangle$ is a product state of the qubit ground state $|\mathrm{g}\rangle$ and the cavity vacuum $|0\rangle$, while each $\mathscr{H}_n$ is a two-dimensional subspace spanned by the two states $|\mathrm{e}, n\rangle$ and $|\mathrm{g}, n+1\rangle$ having the same eigenvalue of $\hat{n}_{\mathrm{tot}}$. Solving the Jaynes-Cummings model then amounts to diagonalizing simple $2 \times 2$ matrices $H_n$ in the space of operators acting on each $\mathscr{H}_n$, where here

$$H_n = \begin{pmatrix} n\omega_{\mathrm{c}} + \frac{\omega_{\mathrm{q}}}{2} & g_x\sqrt{n+1} \\ g_x\sqrt{n+1} & (n+1)\omega_{\mathrm{c}} - \frac{\omega_{\mathrm{q}}}{2} \end{pmatrix}. \tag{1.10}$$

Written in terms of the detuning $\delta = \omega_{\mathrm{q}} - \omega_{\mathrm{c}}$, $H_n$ has eigenvalues $\lambda_{\pm,n} = \omega_{\mathrm{c}}(n + \frac{1}{2}) \pm \frac{1}{2}\sqrt{\delta^2 + 4g_x^2(n+1)}$, together with eigenstates

$$\begin{aligned} |+_n\rangle &= \cos\Theta_n\,|\mathrm{e}, n\rangle + \sin\Theta_n\,|\mathrm{g}, n+1\rangle\,, \\ |-_n\rangle &= -\sin\Theta_n\,|\mathrm{e}, n\rangle + \cos\Theta_n\,|\mathrm{g}, n+1\rangle\,. \end{aligned} \tag{1.11}$$

Here, the mixing angle $\Theta_n$ is defined by $\tan 2\Theta_n = 2g_x\sqrt{n+1}/\delta$. The spectrum of the Jaynes-Cummings Hamiltonian therefore consists of a non-degenerate ground state together with an infinite set of disconnected doublets $|\pm_n\rangle$ separated in energy by $\lambda_{+,n} - \lambda_{-,n}$ [Fig. 1.1]. A state initialized in a given subspace $\mathscr{H}_n$ will therefore remain in that subspace under evolution generated by $H_{\mathrm{JC}}$. This evolution can be used to coherently transfer a single excitation back and forth between the qubit and cavity mode, with clear applications for quantum computing. When this transfer occurs in the single-excitation subspace spanned by $|\pm_1\rangle$, the coherent transfer leads to so-called vacuum Rabi oscillations occurring with a characteristic frequency $2g_x$. This frequency is commonly extracted from spectroscopic measurements of the cavity as a form of system characterization, as described in greater detail below.

The $\hat{n}_{\mathrm{tot}}$-non-conserving terms $a^\dagger\sigma_+$ and $a\sigma_-$ appearing in the Rabi Hamiltonian [Eq. (1.7)] transform the continuous $U(1)$ symmetry of the Jaynes-Cummings Hamiltonian into a discrete $\mathbb{Z}_2$ symmetry associated with the parity operator $P = \sigma_z \otimes e^{i\pi a^\dagger a} = -U(\pi)$. Concretely, this is because $U^\dagger(\phi)H_{\mathrm{R}}U(\phi) = H_{\mathrm{R}}$ only for the particular choice $\phi = \pi$. In this case, the Hilbert space $\mathscr{H}$ of the qubit and cavity mode instead breaks down into a direct sum of the form

$$\mathscr{H} = \mathscr{H}_+ \oplus \mathscr{H}_-, \tag{1.12}$$

where each $\mathscr{H}_\pm$ is infinite-dimensional. It was believed for some time that the discrete symmetry and resulting infinite dimensionality of its irreducible representations $\mathscr{H}_\pm$ was insufficient for yielding a solvable model. Interestingly, however, the model was solved as recently as 2011 [5], based on the insight that the $\mathbb{Z}_2$ symmetry can be used to eliminate the discrete degree-of-freedom (qubit). This is only possible due to the number of irreducible representations of $\mathbb{Z}_2$ equalling the dimension of the Hilbert space $\mathbb{C}^2$ of the qubit. With only the continuous degree-of-freedom remaining, conservation of energy is then enough to ensure integrability [5].

The Jaynes-Cummings Hamiltonian has done and continues to do extremely well in describing a wide range of experiments. However, its validity breaks down in the ultrastrong coupling regime where $g_x$ repre-sents a sizeable fraction of $\omega_q + \omega_c$, and where counter-rotating terms can no longer be neglected. Observable consequences of these counter-rotating terms include the emergence of anti-crossings between states differing by two excitations (i.e. linked by $a^\dagger\sigma_+$ or $a\sigma_-$) [8], together with a photon-number-dependent Bloch-Siegert shift $\omega_{BS}(a^\dagger a + 1/2)$ in the qubit's resonance frequency [9], where $\omega_{BS} = g_x^2/(\omega_q + \omega_c)$. Another interesting consequence of the counter-rotating terms in Eq. (1.7) is that, unlike the Jaynes-Cummings Hamiltonian, the Rabi Hamiltonian has an entangled ground state. For the limiting case of degenerate qubit levels [$\omega_q = 0$ in Eq. (1.7)], $\sigma_x$ is preserved and the qubit-cavity interaction produces a qubit-state-dependent displacement $g_x/\omega_c$ of the cavity mode. In this case, the ground-state manifold is spanned by the degenerate states $|+\rangle|-\alpha\rangle \pm |-\rangle|\alpha\rangle$, where here, $\sigma_x|\pm\rangle = \pm|\pm\rangle$ and $|\alpha\rangle$ is a coherent state with $\alpha = g_x/\omega_c$. In the opposite limit $g_x/\omega_q \ll 1$, the ground state is a squeezed state that is weakly entangled with the qubit [10].

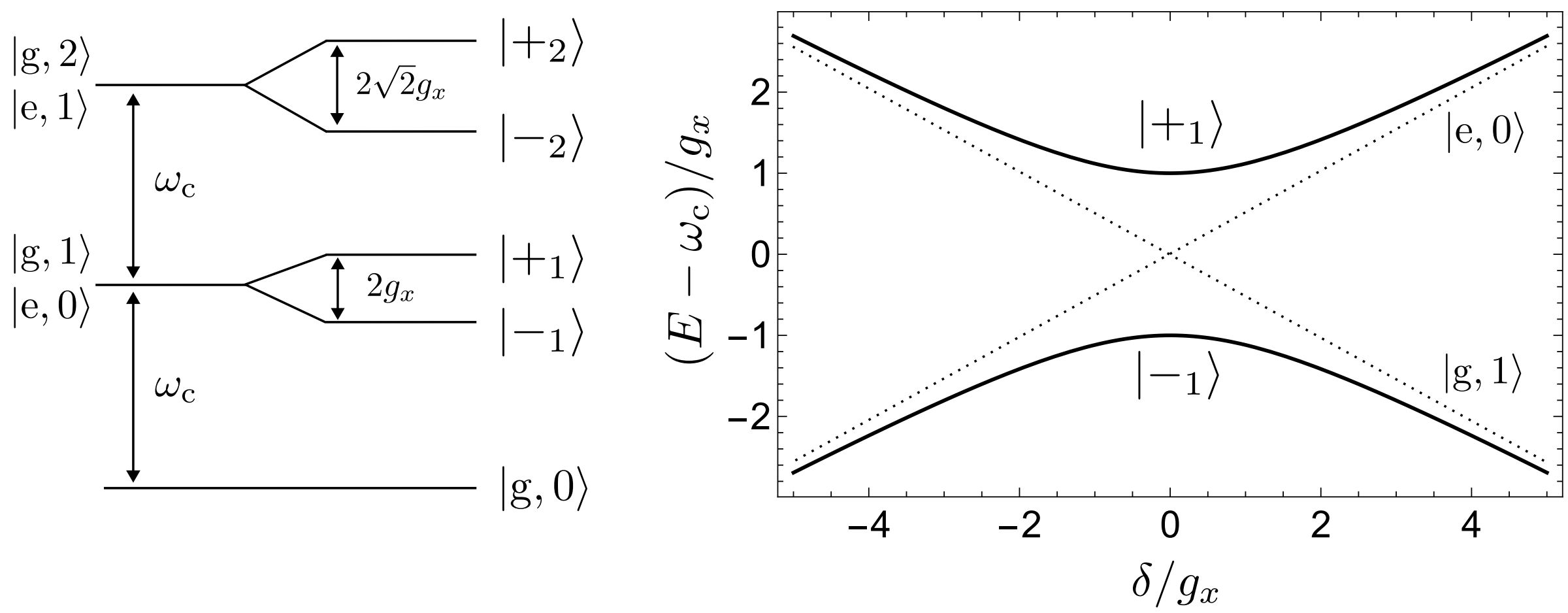

Figure 1.1: Left: At zero detuning ($\delta = \omega_q - \omega_c = 0$) and for $g_x = 0$, the product states $|e,n\rangle$ and $|g,n+1\rangle$ form degenerate doublets separated in energy from the ground state $|g,0\rangle$ by integer multiples of $\omega_c = \omega_q$. For $\delta = 0$ and $g_x \neq 0$, the eigenstates of the Jaynes-Cummings Hamiltonian are instead given by bonding ($|+_n\rangle$) and anti-bonding ($|-_n\rangle$) linear combinations of $|e,n\rangle$ and $|g,n+1\rangle$. Right: Hybridization of $|g,1\rangle$ and $|e,0\rangle$ leads to an avoided crossing at $\delta = 0$, with energy splitting $2g_x$. The dispersion of $|\pm_1\rangle$ as a function of $\delta$ is plotted in black, while the decoupled eigenstates ($g_x = 0$) are indicated by dotted lines.

The transverse coupling that can be achieved based on electric-dipole interactions has an upper limit—the fine-structure limit [11, 12]—that can be expressed in terms of parameters characterizing the size of the qubit and cavity. Denoting $d = \langle e|\boldsymbol{d}\cdot\hat{\boldsymbol{r}}|g\rangle$, the transverse coupling strength is $g_x = -dE_{RMS}(\boldsymbol{r}_0)$ [cf Eq. (1.4)], where here, $d$ has units of charge times distance. To get an estimate for the size of the root-mean-square electric field $E_{RMS}$ associated with vacuum fluctuations in the cavity, we treat the vacuum state of the cavity as having half a photon's worth of energy [11]. Half of this energy is stored in vacuum fluctuations of the electric field, allowing us to estimate that [12]

$$\hbar\frac{\omega_c}{4} = \frac{\varepsilon_0}{2}\int dV\, E^2 = \frac{\varepsilon_0}{2}E_{RMS}^2 V, \tag{1.13}$$

where $\varepsilon_0$ is the permittivity of free space and $V$ is the volume of the cavity. Although we set $\hbar = 1$ in all other parts of this thesis, we here include it explicitly in order to ensure quantitative accuracy below.

Equation (1.13) shows that for a fixed $\omega_c$, the strength $E_{RMS}$ of the vacuum fluctuations will increase as $V$ is decreased. For a three dimensional cavity, $V$ scales like the cube of the wavelength $\lambda = 2\pi c/\omega_c$: $V = O(\lambda^3)$. However, tighter confinement of the photon can be engineered by considering instead an effectively one-dimensional cavity with a transverse radius $r \ll \lambda$, in which case $V = \pi r^2\lambda/2 = \pi^2 c r^2/\omega_c$ for a cavity that is half a wavelength long. Inserting this expression for $V$ into Eq. (1.13) and re-arranging for $E_{RMS}$

gives [12]

$$E_{\text{RMS}} = \frac{1}{r}\sqrt{\frac{\hbar\omega_{\mathrm{c}}^2}{2\pi^2\varepsilon_0 c}}. \tag{1.14}$$

Finally, multiplying by $d$ allows us to express the coupling strength $g_x$ in units of the cavity frequency:

$$\frac{g_x}{\omega_{\mathrm{c}}} = -\frac{(d/e)}{r}\sqrt{\frac{2\alpha}{\pi}} \tag{1.15}$$

where here, $\alpha = e^2/4\pi\varepsilon_0\hbar c \approx 1/137$ is the fine-structure constant. The coupling strength $g_x$ is therefore set by a ratio of the length scale $(d/e)$ characterizing the size of the qubit relative to the transverse dimension $r$ of the cavity. For superconducting Cooper-pair-box (charge) qubits, $(d/e) = 2\ell_{\mathrm{JJ}}$, where $\ell_{\mathrm{JJ}}$ is a length scale characterizing the distance travelled by a Cooper pair while tunnelling through the Josephson junction. This distance $\ell_{\mathrm{JJ}}$ is typically on the order of micrometers, comparable to the transverse dimension of a superconducting transmission-line resonator, and as early as 2008, ratios $g_x/\omega_{\mathrm{c}} \approx 2.5\%$ approaching the fine-structure limit were reported [12]. Much stronger coupling strengths can be achieved for superconducting transmon qubits, for which $(d/e) \gg 2\ell_{\mathrm{JJ}}$ [13]. For a quantum-dot-based charge qubit, we instead have $(d/e) = \ell_{\mathrm{dot}}$ where $\ell_{\mathrm{dot}}$ here denotes the distance between the two dots. (Since spin-cavity coupling is typically engineered through spin-charge hybridization, as described in more detail in Appendix 3.5, $\ell_{\mathrm{dot}}$ is the relevant scale for many spin qubits as well, although constants of proportionality will vary depending on the nature of the spin-charge coupling.) With $\ell_{\mathrm{dot}}$ typically a few tens of nanometers, the coupling strengths achieved for quantum-dot-based qubits are typically weaker than those achieved for, e.g., transmons. Nevertheless, strong coupling has been reported for charge qubits [14], electron-spin qubits [15], resonant-exchange qubits [16], and hole-spin qubits [17]. Since $\alpha \ll 1$, it is often the case that $g_x \ll \omega_{\mathrm{c}}$ even for large values of $(d/e)$. However, the fine-structure limit is not a fundamental restriction on light-matter coupling strengths in general, and it should be noted that while this introductory chapter focuses on electric-dipole coupling [leading to the estimate given in Eq. (1.15)], there exist other ways of engineering stronger light-matter interactions approaching the ultrastrong regime.

Outside the ultrastrong coupling regime, the dynamics of a transversally coupled cavity-qubit system are well described by the Jaynes-Cummings Hamiltonian $H_{\mathrm{JC}}$ [Eq. (1.8)]. Due to $H_{\mathrm{JC}}$ conserving the total number of excitations, $[H_{\mathrm{JC}}, \hat{n}_{\mathrm{tot}}] = 0$, the Hilbert space $\mathscr{H} = \mathbb{C}^2 \otimes L^2(\mathbb{R})$ of the joint system can be decomposed into a direct sum of dynamically invariant subspaces $\mathscr{H}_n$ as discussed above [cf. Eq. (1.9)]. Hence, for any initial state $|\psi\rangle_0 \in \mathscr{H}_n$, evolution under $H_{\mathrm{JC}}$ will satisfy $|\psi\rangle_t = U_{\mathrm{JC}}(t)\,|\psi\rangle_0 \in \mathscr{H}_n$ for all times $t$, where here, $U_{\mathrm{JC}}(t) = e^{-iH_{\mathrm{JC}}t}$. In particular, for $\omega_{\mathrm{c}} = \omega_{\mathrm{q}}$ and for an initial product state $|\psi\rangle_0 = |\mathrm{e}, n\rangle$, the probability of measuring the qubit in its excited state at time $t$ is given by

$$P(|\mathrm{e}\rangle) = |\langle \mathrm{e}, n|\psi\rangle_t|^2 = \cos^2\left(g_x\sqrt{n+1}t\right). \tag{1.16}$$

This result [Eq. (1.16)] describes a single excitation being coherently traded back and forth between the qubit and cavity with frequency

$$\Omega_n = 2g_x\sqrt{n+1}. \tag{1.17}$$

For $n = 0$, Eq. (1.16) describes vacuum Rabi oscillations and $\Omega_0 = 2g_x$. In any realistic implementation, there will invariably be incoherent processes that degrade the coherence between $|\mathrm{e}, n\rangle$ and $|\mathrm{g}, n+1\rangle$, and which consequently reduce the visibility of these oscillations. The system is said to be in the strong coupling regime when the timescale $(2g_x)^{-1}$ for vacuum Rabi oscillations is short compared the timescales for decoherence, sources of which include cavity photon loss (with associated rate $\kappa$), qubit pure dephasing (rate $\gamma_\phi$), and qubit relaxation (rate $\gamma_1$). In this regime, a single excitation can be traded back and forth several times before being irretrievably lost to the environment.

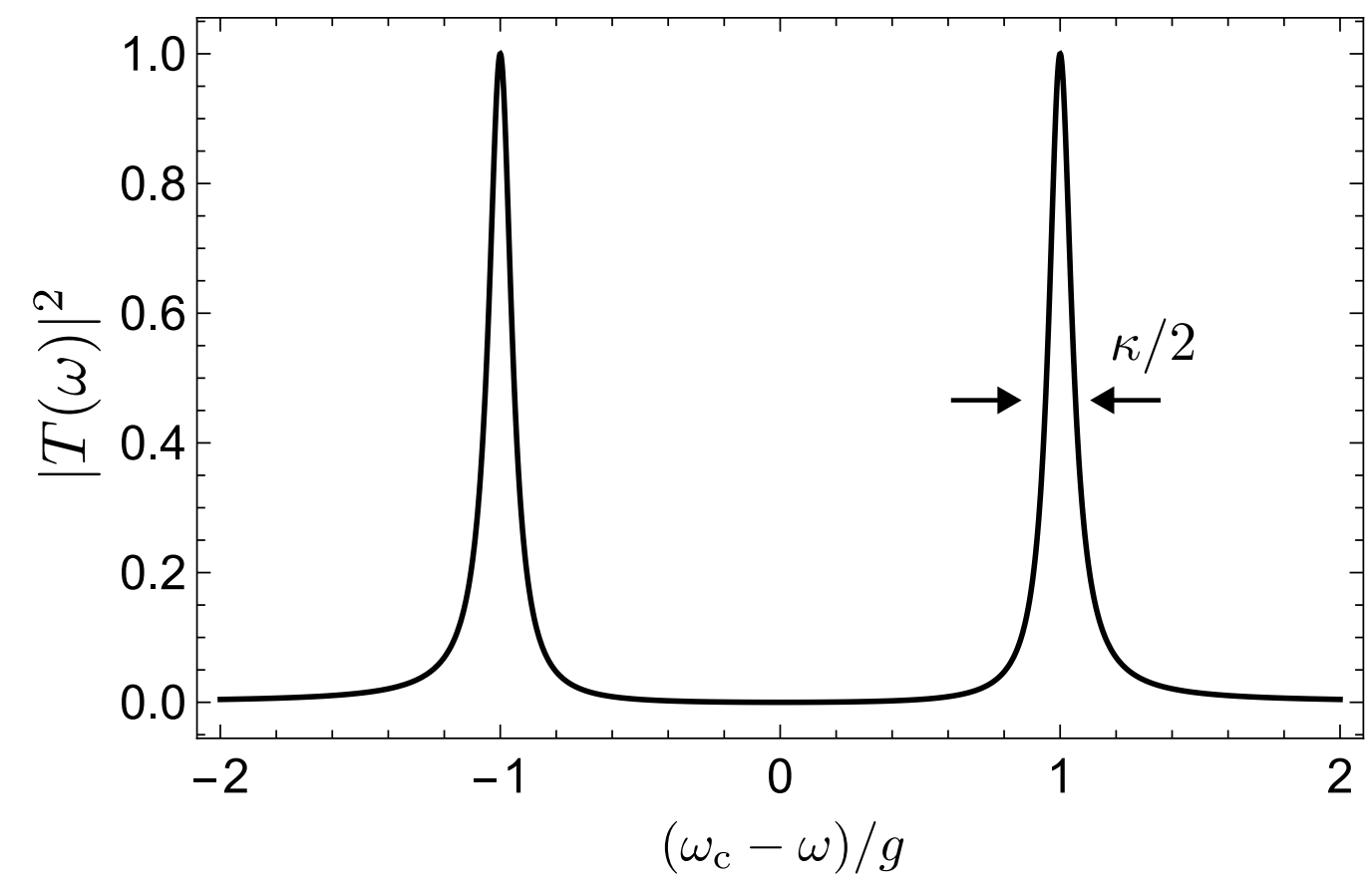

Figure 1.2: In the strong coupling regime and for weak resonant driving, the probability $|T(\omega)|^2$ of transmission through the cavity exhibits a doublet of peaks that are widely separated compared to their widths. In the absence of qubit dephasing ($\gamma_2 = 0$), these peaks are centered at $\omega_c \pm g_x$ and have widths $\kappa/2$. We have here taken $\kappa_1 = \kappa_2 = \kappa/2$.

The spectroscopic signature of vacuum Rabi dynamics consists of a doublet of Lorentzian peaks centered at (approximately) $\omega_c \pm g_x$, with peak widths $(\kappa + 2\gamma_2)/2$ that are narrow compared to their separation of $2g_x$. Here, $\gamma_2 = \gamma_\phi + 2\gamma_1$ is the total rate of qubit dephasing due to both pure dephasing and relaxation. The cavity spectrum can be measured by applying a weak, resonant input tone $r_{\rm in}(t)$ to the cavity's input port and measuring the transmission through its output port. Accounting for incoherent processes, the master equation of the joint state $\rho$ of the cavity and qubit under cavity driving is given by

$$\dot{\rho} = -i[H_{\rm JC} + H_{\rm in}(t), \rho] + \frac{\gamma_2}{2}\mathscr{D}[\sigma_z]\rho + \kappa\mathscr{D}[a]\rho, \tag{1.18}$$

where $H_{\rm in}(t) = i\sqrt{\kappa_1}(r_{\rm in}^*(t)a - \text{h.c.})$, and where the damping superoperator $\mathscr{D}[\mathscr{O}]$ acts like $\mathscr{D}[\mathscr{O}]\rho = \mathscr{O}\rho\mathscr{O}^\dagger - \{\mathscr{O}^\dagger\mathscr{O}, \rho\}/2$ for operator $\mathscr{O}$. In Eq. (1.18), the term $\propto \gamma_2\mathscr{D}[\sigma_z]$ describes qubit dephasing with rate $\gamma_2$, while the term $\kappa\mathscr{D}[a]$ describes damping of the cavity field with rate $\kappa = \kappa_1 + \kappa_2$ through the cavity input ($\kappa_1$) and output ($\kappa_2$) ports.

From the master equation given above [Eq. (1.18)], the equations of motion for the cavity annihilation operator $a$ and spin-lowering operator $\sigma_-$ can be found as $\langle\dot{\mathscr{O}}\rangle_t = \text{Tr}\{\mathscr{O}\dot{\rho}(t)\}$, giving

$$\langle\dot{a}\rangle_t = -\left(i\omega_c + \frac{\kappa}{2}\right)\langle a\rangle_t - ig_x\langle\sigma_-\rangle_t - \sqrt{\kappa_1}r_{\rm in}(t), \tag{1.19}$$

$$\langle\dot{\sigma}_-\rangle_t = -(i\omega_q + \gamma_2)\langle\sigma_-\rangle_t + ig_x\langle a\sigma_z\rangle_t. \tag{1.20}$$

For sufficiently weak input driving, the dynamics of the cavity-qubit system can reasonably be restricted to the lowest three rungs $|g,0\rangle, |\pm_1\rangle$ of the Jaynes-Cummings ladder (Fig. 1.1) under the assumption that the cavity contains at most one photon at any given time. For any state in the span of $|g,0\rangle$ and $|\pm_1\rangle$, the expectation value $\langle a\sigma_z\rangle$ in Eq. (1.20) is equal to $\langle a\sigma_z\rangle = -\langle a\rangle$. By restricting dynamics to the subspace $|g,0\rangle \oplus \mathscr{H}_1$ [cf. Eq. (1.9)], the equations of motion for $\langle a\rangle$ and $\langle\sigma_-\rangle$ can therefore be decoupled from the equation of motion for the joint observable $\langle a\sigma_z\rangle$.

The field $r_{\rm out}(\omega)$ transmitted through the output port is equal to $r_{\rm out}(\omega) = \sqrt{\kappa_2}\langle a\rangle_\omega$ [2]. By invoking the restricted-subspace approximation $\langle a\sigma_z\rangle = -\langle a\rangle$ in Eq. (1.20), we can easily solve for the cavity field $\langle a\rangle_\omega = \int dt\, e^{i\omega t}\langle a\rangle_t$ by eliminating $\langle\sigma_-\rangle$ from Eqs. (1.19) and (1.20), giving

$$r_{\rm out}(\omega) = \sqrt{\kappa_2}\langle a\rangle_\omega = T(\omega)r_{\rm in}(\omega), \tag{1.21}$$

where here, we have introduced the transmission amplitude

$$T(\omega) = \frac{i\sqrt{\kappa_1\kappa_2}(\omega_\mathrm{q} - \omega - i\gamma_2)}{(\omega_\mathrm{c} - \omega - i\frac{\kappa}{2})(\omega_\mathrm{q} - \omega - i\gamma_2) + g_x^2}. \tag{1.22}$$

For $\gamma_2 = 0$ and in the strong coupling regime, $2g_x > \kappa$, the transmission probability $|T(\omega)|^2$ exhibits a doublet of Lorentzian peaks at $\omega_\mathrm{c} \pm g_x$, having full-widths-at-half-maximum equal to $\kappa/2$ [Fig. 1.2]. The measurement of such a spectrum is frequently given in experiment as proof that the cavity-qubit system is operating in the strong-coupling regime. While the peak separation is almost always quoted as $2g_x$ in the literature (see, e.g., the review Ref. [13]), it is worth noting that qubit dephasing $\gamma_2 \neq 0$ will reduce the peak separation. This is ultimately a consequence of level attraction of the Jaynes-Cummings eigenstates $|\pm_1\rangle$ due to dressing by the qubit's environment.

Whereas spontaneous decay of a qubit into the continuum is typically irreversible, the dominant coupling of the qubit to a single cavity mode converts the decay of the qubit into a coherent and reversible process that can be used as a means of transferring quantum information from a stationary degree-of-freedom into a photonic degree-of-freedom. In particular, vacuum-Rabi dynamics can be used for quantum state transfer and entanglement generation among nodes in a quantum network [18], based on the emission and subsequent re-absorption of single photons. This is generally done by engineering a time-dependent effective coupling of the form $\Omega(t)(\sigma_- a^\dagger + \mathrm{h.c.})$. A single photon can then be released from the cavity and sent to another location by preparing the qubit in $|\mathrm{e}\rangle$ and turning on $\Omega(t)$ just long enough for the system to undergo half a vacuum Rabi oscillation. The photon can later be re-absorbed with unit probability by another cavity-qubit system located "downstream" by time-reversing the pulse $\Omega(t)$ used to emit the photon from the "upstream" cavity. This pitch-and-catch strategy was recently used to generate a Bell state of qubits separated by 30 meters [19], and in Chapter 6, we show how it can be used to mediate long-range gates using a Fock-state or time-bin encoding for the single photon. Chapters 3 and 5 of this thesis will explore ways of generating and distributing entanglement in quantum networks without the use of single photons.

### 1.1.2 The dispersive Hamiltonian

On resonance ($\omega_\mathrm{c} = \omega_\mathrm{q}$), the eigenstates of the Jaynes-Cummings Hamiltonian exhibit maximal qubit-cavity hybridization, and addressing the qubit independently becomes challenging since its decoupled eigenstates are no longer eigenstates of the joint system. A conceptual simplification is afforded by operating in the dispersive regime, where the qubit-cavity detuning $\delta = \omega_\mathrm{q} - \omega_\mathrm{c}$ far exceeds the strength of the transverse coupling, $|\delta| \gg g_x$. In this regime, the decoupled qubit and cavity eigenstates remain, to a good approximation, eigenstates of the coupled system due to off-resonant suppression of the term $(a^\dagger\sigma_- + \mathrm{h.c.})$ describing the coherent exchange of excitations between the qubit and cavity. For $g_x \ll |\delta|$, a Schrieffer-Wolff transformation of the Jaynes-Cummings Hamiltonian can be used to derive an effective Hamiltonian $H_\mathrm{disp}$ that approximately describes the cavity-qubit system in this regime.

Formally, a Schrieffer-Wolff transformation is a unitary transformation chosen to diagonalize the original Hamiltonian to first order in some perturbation. In this case, the original Hamiltonian is the Jaynes-Cummings Hamiltonian $H_\mathrm{JC}$, while the perturbation is the cavity-qubit coupling $\propto g_x$. We therefore decompose $H_\mathrm{JC}$ as the sum of an unperturbed Hamiltonian $H_0$ and a perturbation $V$ according to

$$H_\mathrm{JC} = H_0 + V, \tag{1.23}$$

$$H_0 = \frac{\omega_\mathrm{q}}{2}\sigma_z + \omega_\mathrm{c} a^\dagger a, \tag{1.24}$$

$$V = g_x(\sigma_+ a + \sigma_- a^\dagger). \tag{1.25}$$

Schrieffer-Wolff transformations are conventionally expressed in terms of an anti-Hermitian operator $S$. In this case,

$$H_{\text{disp}} = e^{S} H_{\text{JC}} e^{-S}, \tag{1.26}$$

where $S$ is chosen to eliminate $V \propto g_x$ to first order. The required $S$ can be found by expanding $H_{\text{disp}}$ [Eq. (1.26)] using the Baker-Campbell-Haussdorff formula:

$$H_{\text{disp}} = H_0 + V + [S, H_0] + [S, V] + \frac{1}{2}[S, [S, H_0]] + \frac{1}{2}[S, [S, V]] + O(S^3). \tag{1.27}$$

From Eq. (1.27), we recognize that by choosing $S$ such that

$$[S, H_0] = -V, \tag{1.28}$$

we can ensure that $H_{\text{eff}}$ is diagonal in the eigenbasis of $H_0$ to first order in $V$. Taking $S = (\sigma_+ a - \sigma_- a^\dagger) \propto [H_0, V]$ as an ansatz for the generator of the transformation, we then find by enforcing Eq. (1.28) that

$$S = \frac{g_x}{\delta}(\sigma_+ a - \sigma_- a^\dagger), \tag{1.29}$$

giving

$$H_{\text{disp}} = H_0 + \frac{1}{2}[S, V] + O(V^3) \tag{1.30}$$

$$= \frac{\omega_{\text{q}}}{2}\sigma_z + \chi \left|\text{e}\right\rangle\!\left\langle\text{e}\right| + \omega_{\text{c}} a^\dagger a + \chi \sigma_z a^\dagger a + O(V^3), \tag{1.31}$$

where the dispersive shift $\chi$ is given by $\chi = g_x^2/\delta$. The shift $\chi$ on the energy of the excited state $|\text{e}\rangle$ is often incorporated into the qubit splitting, $\omega_{\text{q}} \to \omega_{\text{q}} + \chi$, by simply redefining the zero of energy, $H_{\text{disp}} \to H_{\text{disp}} - \frac{\chi}{2}$. The operators in Eq. (1.31) are in a dressed basis, so although the eigenstates of $H_{\text{disp}}$ [neglecting $O(V^3)$] can be labelled using the same quantum numbers $|\text{e}(\text{g}), n\rangle$ as the decoupled states, the joint qubit-cavity eigenstates still exhibit weak hybridization. This can be seen by transforming the eigenvalue equation $H_{\text{disp}}|\psi\rangle = \varepsilon|\psi\rangle$ back to the original frame, since from Eq. (1.26), we know that for the same fixed $\varepsilon$, we have $H_{\text{JC}}|\tilde{\psi}\rangle = \varepsilon|\tilde{\psi}\rangle$ with $|\tilde{\psi}\rangle = e^{-S}|\psi\rangle$. However, since the amount of entanglement is small in the perturbative parameter $g_x/|\delta| \ll 1$, the cavity and qubit can be treated as decoupled to a good approximation. The qubit can therefore be used to encode quantum information without the information "delocalizing" into the cavity mode too significantly.

In the dispersive regime, transverse cavity-qubit coupling manifests as a qubit-state-dependent shift of the cavity resonance frequency: For a qubit in state $|\text{e}\rangle$, the cavity frequency is $\omega_{\text{c}} + \chi$, while for a qubit in state $|\text{g}\rangle$, it is instead $\omega_{\text{c}} - \chi$. This shift can be used for qubit readout, as will be described shortly. It can also be used to measure the photon-number parity of the cavity mode. This is accomplished via a Ramsey sequence where the qubit is first prepared in the state $|+\rangle \propto |\text{e}\rangle + |\text{g}\rangle$, then allowed to evolve in the presence of the cavity for a fixed time $t = \pi/|\chi|$. At the end of the interaction time, the qubit is measured in the $X$-basis. For a cavity containing an even (odd) number of photons, a measurement of the qubit will ideally return $|+\rangle$ ($|-\rangle$). This photon-number-parity measurement is widely used as an error-syndrome readout for quantum error correcting codes where qubits are encoded not in the state of a two-level system (i.e. artificial atom), but in the state of the cavity itself. These include binomial codes [20], where information is encoded in Fock-state superpositions of the cavity mode; cat codes [21, 22], where changes in the photon-number-parity amount to Pauli errors on the encoded information; and cavity-based dual-rail qubits [23], where the basis states of the qubit are given by the single-photon states of two coupled cavities. The driving force behind this cavity-centric approach is the observation that cavity modes suffer predominantly from a single, well-understood error: photon loss. Quantum error-correcting codes can then be designed to protect against this dominant error source, potentially reducing the physical-qubit overhead required for fault tolerance relative to codes that aim to detect and correct multiple types of error.

Both single- and two-qubit gates can be realized in the dispersive regime, enabling universal control. Some notable two-qubit gates that can be realized through dispersive coupling to a common cavity mode include the resonator-induced phase gate [24, 25], where off-resonant driving of the cavity mode mediates a *ZZ* interaction between two qubits, as well as the cross-resonance gate [26], where driving of one qubit at the frequency of the other results in a *ZX* interaction. Working in the dispersive regime also enables a form of approximately quantum non-demolition readout—dispersive readout—based on the qubit-state-dependent shift of the cavity resonance frequency: By driving the cavity with a classical probe field $r_{\text{in}}(t)$, the state of the qubit can be inferred from a phase-sensitive measurement of the field $r_{\text{out}}(t)$ reflected from the cavity [27]. Taking the cavity to be single-sided ($\kappa_1 = \kappa$, $\kappa_2 = 0$), the equation of motion for the cavity mode during this readout, conditioned on a particular state $|\text{e}\rangle, |\text{g}\rangle$ of the qubit, is given by

$$\langle \dot{a} \rangle = -i\left(\omega_{\text{c}} + \chi s - i\frac{\kappa}{2}\right)\langle a \rangle - \sqrt{\kappa} r_{\text{in}}, \tag{1.32}$$

where here, $s = \pm 1$ is the $\sigma_z$ eigenvalue of the qubit. Combined with the input output relation $r_{\text{out}}(t) = r_{\text{in}}(t) + \sqrt{\kappa}\langle a \rangle_t$ describing the outgoing field $r_{\text{out}}(t)$ in terms of the input field $r_{\text{in}}(t)$ and intracavity field $a(t)$ [2], Eq. (1.32) can be used to calculate the reflection coefficient $R(\omega) = r_{\text{out}}(\omega)/r_{\text{in}}(\omega)$ describing the phase change acquired by $r_{\text{in}}(\omega)$ upon reflection from the cavity:

$$R_s(\omega) = \frac{i(\omega - \omega_{\text{c}} + s\chi) - \frac{\kappa}{2}}{i(\omega - \omega_{\text{c}} + s\chi) + \frac{\kappa}{2}}. \tag{1.33}$$

A detailed derivation of the input-output relation and quantum Langevin equation [Eq. (1.32)] will be presented in Sec. 1.2 of this introductory chapter.

For a probe field resonant with the bare cavity frequency $\omega_{\text{c}}$, Eq. (1.33) predicts a qubit-state-conditioned phase shift of

$$\phi_s = \arg R_s(\omega_{\text{c}}) = s \arctan \frac{4\kappa\chi}{4\chi^2 - \kappa^2}, \tag{1.34}$$

which can be measured via homodyne detection to infer the qubit state. The measurement signal can be maximized by tuning the dispersive shift so that $|\chi| = \kappa/2$, in which case $|\phi_+ - \phi_-|$ assumes its maximal value of $\pi$. A two-qubit parity check can similarly be realized by dispersively coupling two qubits to a common cavity mode, ideally with dispersive shifts of the same magnitude but opposite signs [28]. Chapter 5 of this thesis will present a way of performing multiqubit parity checks of qubits coupled to *different* cavity modes using qubit-state-conditioned phase shifts on propagating pulses of light—an approach designed to enable error correction across modular architectures. These light pulses are analogous to $r_{\text{in}}(t)$ in the preceding discussion, and the required qubit-state-conditioned phase shifts can be realized in several different ways, one of which involves dispersive coupling with $\chi$ chosen so that $|\chi| = \kappa/2$.

### 1.1.3 Longitudinal light-matter coupling

The application of cavity and circuit QED to quantum computing has focused predominantly on realizing logical operations (gates, measurement, etc.) using transverse cavity-qubit coupling. In the past decade, however, there has been a growing interest in considering the possibilities enabled by longitudinal coupling, which preserves the bare qubit eigenstates $|\text{e}\rangle$ and $|\text{g}\rangle$. Chapter 3 of this thesis will show how longitudinal coupling can be used to generate entanglement between the state of a cavity-coupled qubit and the path taken by an incoming multiphoton wavepacket. Such entanglement could be used for entanglement distribution

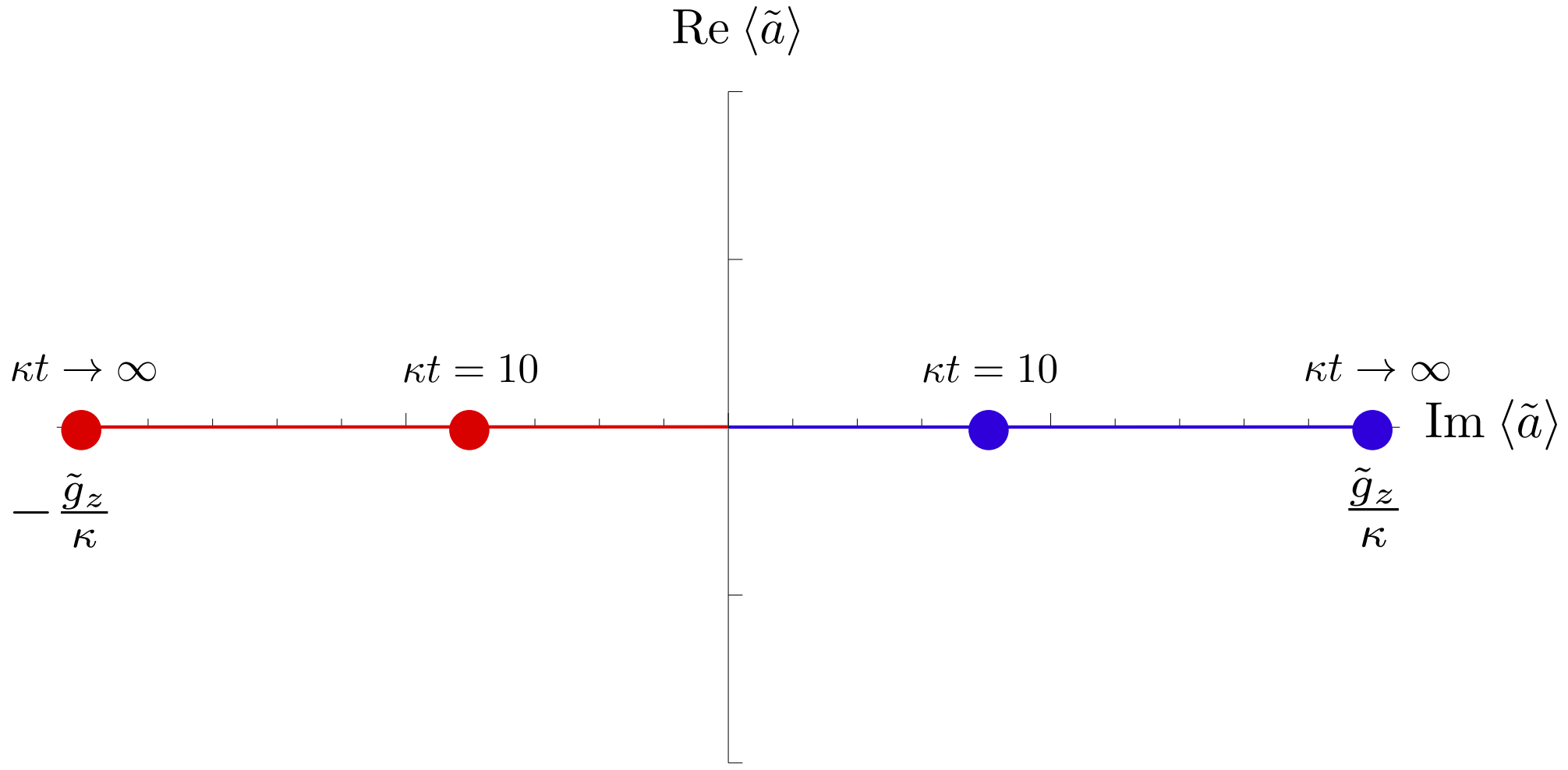

Figure 1.3: Phase-space trajectories generated by a parametrically modulated longitudinal coupling: Within a rotating-wave approximation valid for $\bar{g}_z, \tilde{g}_z \ll \omega_c$, the qubit-state-conditioned phase-space trajectories of the cavity mode diverge linearly and eventually reach a steady-state displacement of $\pm\tilde{g}_z/\kappa$ [cf. Eq. (1.37)]. The trajectory conditioned on the qubit being in state $|e\rangle$ ($|g\rangle$) is plotted in red (blue).

or quantum-enhanced precision measurements (discussed in more detail in Chapter 4).

Although dispersive $\sim \sigma_z a^\dagger a$ and longitudinal $\sim \sigma_z(a^\dagger + a)$ interactions are both proportional to $\sigma_z$, the resulting dynamics of the cavity mode are very different. As discussed above in Sec. 1.1.2, a dispersive interaction produces a qubit-state-dependent shift $\pm\chi$ of the cavity resonance frequency. In a frame rotating at the bare cavity frequency $\omega_c$, a cavity mode prepared in a finite-amplitude coherent state will therefore precess counterclockwise (clockwise) in phase-space for a qubit in state $|e\rangle$ ($|g\rangle$). A longitudinal interaction by contrast, produces no such shift of the cavity resonance frequency, and no such qubit-state-dependent precession. Rather, evolution under a Hamiltonian of the form

$$H_{\text{long}} = \frac{\omega_q}{2}\sigma_z + \omega_c a^\dagger a + g_z \sigma_z (a + a^\dagger) \tag{1.35}$$

will instead lead to a qubit-state-dependent *displacement* of the cavity mode in phase space, in this case along the out-of-phase quadrature $p = i(a^\dagger - a)$. For a static value of $g_z = \bar{g}_z$, this displacement is on the order of $\bar{g}_z/\omega_c$, which is typically $\ll 1$ and often neglected in a rotating-wave approximation. However, for a modulated longitudinal coupling $g_z(t) = \bar{g}_z + \tilde{g}_z \cos(\omega_c t)$, the size of the qubit-state-dependent displacement instead saturates at $\mp\tilde{g}_z/\kappa$ for $\sigma_z$ eigenvalue $s = \pm 1$. This can be seen from the equation of motion for the cavity annihilation operator $\langle\tilde{a}\rangle_t = e^{i\omega_c t}\langle a\rangle_t$, which we write in a frame rotating at the cavity frequency:

$$\langle\dot{\tilde{a}}\rangle_t = -ise^{i\omega_c t} g_z(t) - \frac{\kappa}{2}\langle\tilde{a}\rangle_t. \tag{1.36}$$

Integrating Eq. (1.36) under the assumption that $\langle a\rangle_0 = 0$ gives

$$\langle\tilde{a}\rangle_t = -is\int_0^t dt'\, e^{-\frac{\kappa}{2}(t-t')} e^{i\omega_c t'} g_z(t'), \tag{1.37}$$

from which it may be seen that (i) for $g_z(t) = \bar{g}_z$, the amount of displacement is $|\langle\tilde{a}\rangle_t| \sim \bar{g}_z/\omega_c$, limited by $\omega_c$, and that (ii) for $g_z(t) = \bar{g}_z + \tilde{g}_z\cos(\omega_c t)$, the resonance introduced by the modulation of $g_z(t)$ at the cavity frequency gives $e^{i\omega_c t} g_z(t) \simeq \tilde{g}_z/2$ in Eq. (1.37), leading to a displacement limited instead by $\kappa$ [Fig. 1.3].

Such parametric modulation of the longitudinal coupling strength has been considered theoretically as a means of engineering a fast, quantum non-demolition qubit readout—longitudinal parametric readout [29]. Longitudinal parametric readout has been realized experimentally using transmon qubits by engineering a synthetic longitudinal coupling starting from a transverse interaction [30, 31]. An alternative approach to generating a synthetic longitudinal interaction is described in Ref. [32], which considers the simultaneous application of a Rabi drive with strength $\Omega_\mathrm{R}$ on the qubit, together with cavity sideband driving at frequencies $\omega_\mathrm{c} \pm \Omega_\mathrm{R}$, resulting in an effective longitudinal interaction in a rotating frame. Ref. [32] also demonstrates a key advantage of longitudinal interactions for qubit readout: Since a cavity mode evolving under a longitudinal interaction [Eq. (1.35)] does not undergo any qubit-state-dependent phase-space precession, as it would for an interaction of the form $\chi \sigma_z a^\dagger a$, the acquisition time required to reach a target signal-to-noise ratio (SNR) using longitudinal parametric readout can be reduced by making use of injected squeezing [29]. No such advantage can be gained by using squeezed light for dispersive readout, since the quadrature along which the squeezing is applied will rotate in different directions conditioned on the qubit state. The fluctuations of the reflected field encoding the qubit state will then reflect the amplified fluctuations along the antisqueezed quadrature of the input field, thereby reducing the SNR.

Although superconducting circuits realizing natively longitudinal interactions have been studied theoretically [29, 33, 34], experimental demonstrations involving longitudinal coupling and superconducting (transmon) qubits tend to engineer the longitudinal interaction using natively transverse coupling, together with driving of the qubit [30, 31], or as discussed in the preceding paragraph, qubit and cavity [32]. However, it is also possible for transverse interactions to vanish identically, leaving a dominant longitudinal coupling. Consider, for instance, a double-quantum-dot (DQD) singlet-triplet qubit for which the qubit eigenstates are $|\mathrm{g}\rangle \propto (|\uparrow\downarrow\rangle - |\downarrow\uparrow\rangle)\,|\Phi_\mathrm{s}\rangle$ and $|\mathrm{e}\rangle \propto (|\uparrow\downarrow\rangle + |\downarrow\uparrow\rangle)\,|\Phi_\mathrm{a}\rangle$. Here, $|\Phi_\mathrm{s,a}\rangle$ are orbital wavefunctions that are either symmetric ($|\Phi_\mathrm{s}\rangle$) or anti-symmetric ($|\Phi_\mathrm{a}\rangle$) under exchange of the two electrons. For these qubit states, the transition-dipole matrix element $\langle \mathrm{e}|\boldsymbol{d}\cdot\hat{\boldsymbol{r}}|\mathrm{g}\rangle \propto g_x$ vanishes due to orthogonality of the spin states. The intrinsic dipole moments $\langle \mathrm{e(g)}|\boldsymbol{d}\cdot\hat{\boldsymbol{r}}|\mathrm{e(g)}\rangle$ associated with the charge distributions of the qubit eigenstates can be different, however, leading to a nonzero $g_z$ [cf. Eq. (1.5)]. This is a consequence of the orbital state $|\Phi_\mathrm{a}\rangle$ being associated with a $(1,1)$ double-dot charge occupation: $|\Phi_\mathrm{a}\rangle \propto |\mathrm{L,R}\rangle - |\mathrm{R,L}\rangle$, where here, $(n,m)$ gives the numbers $n$ and $m$ of electrons in the left and right dots having ground-state orbital wavefunctions $|\mathrm{L}\rangle$ and $|\mathrm{R}\rangle$, respectively. For a nonzero tunnel coupling between the two dots, $|\Phi_\mathrm{s}\rangle$ is given instead by a superposition of $(1,1)$, $(2,0)$, and $(0,2)$ charge occupations corresponding respectively to the orbital states $|\mathrm{L,R}\rangle + |\mathrm{R,L}\rangle$, $|\mathrm{L,L}\rangle$, and $|\mathrm{R,R}\rangle$. This difference relative to the fixed $(1,1)$ occupation of the triplet state ultimately follows from the Pauli exclusion principle, which dictates an overall antisymmetry of the total electronic wavefunction accounting for both spin and orbital components.

A parametrically modulated longitudinal coupling $g_z(t) \propto \tilde{g}_z \cos(\omega_\mathrm{c} t)$ is readily engineered for qubits whose splitting is sensitive to the zero-point voltage fluctuations of the cavity, as is typically the case for spin qubits in gate-defined quantum dots. For the singlet-triplet qubit described above, for instance, the energy splitting between the two qubit states depends on the strength of the exchange interaction $J(\varepsilon)$ between electrons in the two dots, which can be controlled by tuning the difference in chemical potential $\varepsilon$ between the dots via applied gate voltages. This double-dot detuning $\varepsilon$ can also be made sensitive to the voltage $V_\mathrm{cav} = V_\mathrm{RMS}(a + a^\dagger)$ of the cavity, which alters the qubit's electrostatic environment in much the same way as would an applied gate voltage. With this setup, a parametrically modulated longitudinal coupling can be generated by modulating nearby gate voltages at the cavity frequency so that the double-dot detuning acquires a time dependence according to $\varepsilon(t) = \varepsilon_0 + \tilde{\varepsilon}\cos(\omega_\mathrm{c} t) + e\alpha_\mathrm{L} V_\mathrm{cav}$ [35]. Here, $\alpha_\mathrm{L}$ is the lever arm of the cavity-dot coupling. A Hamiltonian describing the cavity-qubit interaction can then be derived by expanding $J(\varepsilon(t))\sigma_z$ about $\varepsilon = \varepsilon_0$ to second order in $\varepsilon(t) - \varepsilon_0$, with the required modulation of the longitudinal coupling arising from a cross term $\propto V_\mathrm{cav}\cos(\omega_\mathrm{c} t)$ appearing at second order. This cross-term is proportional to the second derivative (curvature) of the qubit energy with respect to the double-dot detuning, evaluated at $\varepsilon_0$, giving $\tilde{g}_z \propto \partial^2 J/\partial\varepsilon^2|_{\varepsilon_0}$ [35].

This concludes our brief introduction to the various regimes of operation of cavity QED mentioned throughout this thesis. In the following section, we review a formalism—input-output theory—that is commonly used to relate the damping of the cavity mode to the state of the environment causing the damping. In scenarios where the cavity mode is being deliberately measured, this environment could be a wide-bandwidth transmission line whose state is being measured (e.g., via homodyne detection) in order to learn about the dynamics of the cavity.

## 1.2 Input-output theory

Input-output theory, introduced by Collett and Gardiner in 1985 [2], provides a theoretical framework for describing the interaction of a quantum system with its surrounding environment. The formalism is based on quantizing the modes comprising the environment, enabling the identification of input and output fields interacting with the quantum system of interest. These fields will satisfy a boundary condition (given by the well-known input-output relation) at the interface of the quantum system and environment. Knowledge of the input and output fields can then be used to infer the dynamics of the quantum system. The derivation of the input-output relation presented in this section closely follows the pedagogical approach of Ref. [13].

We consider the paradigmatic setup where the environment interacting with the system consists of a coplanar waveguide resonator having a length far exceeding that of the system, here taken to be an LC oscillator ("cavity"). In this case, the resonator can be modelled as a semi-infinite, one-dimensional transmission line with Hamiltonian

$$H_{\mathrm{TL}} = \int_0^\infty d\omega\, \omega r_\omega^\dagger r_\omega, \tag{1.38}$$

where $r_\omega$ is a bosonic annihilation operator satisfying $[r_\omega, r_{\omega'}^\dagger] = \delta(\omega - \omega')$. Taking $x \geq 0$ to be the position coordinate along the transmission line, Eq. (1.38) can equivalently be expressed in terms of position-dependent conjugate charge ($Q_{\mathrm{TL}}$) and flux ($\Phi_{\mathrm{TL}}$) operators:

$$H_{\mathrm{TL}} = \int_{-\infty}^\infty dx\, \theta(x) \left[\frac{1}{2c} Q_{\mathrm{TL}}^2(x) + \frac{1}{2l}[\partial_x \Phi_{\mathrm{TL}}(x)]^2\right]. \tag{1.39}$$

Here, $\theta(x)$ is the Heaviside step function, while $c$ and $l$ are the capacitance and inductance per unit length of the transmission line, respectively. In terms of the bosonic operators $r_\omega$ introduced above, the conjugate fields appearing in Eq. (1.39) are given by

$$\begin{aligned} Q_{\mathrm{TL}}(x) &= i\int_0^\infty d\omega \sqrt{\frac{\omega c}{\pi v}} \cos\left(\frac{\omega x}{v}\right)(r_\omega^\dagger - r_\omega), \\ \Phi_{\mathrm{TL}}(x) &= \int_0^\infty d\omega \sqrt{\frac{1}{\pi \omega c v}} \cos\left(\frac{\omega x}{v}\right)(r_\omega^\dagger + r_\omega), \end{aligned} \tag{1.40}$$

where $v = 1/\sqrt{lc}$ is the speed of light in the transmission line. In the Heisenberg picture, where time evolution of operators is generated by Eq. (1.38), the charge and flux operators satisfy $Q_{\mathrm{TL}}(x,t) = c\dot{\Phi}_{\mathrm{TL}}(x,t) = cV_{\mathrm{TL}}(x,t)$, where $\dot{\Phi}_{\mathrm{TL}} = V_{\mathrm{TL}}$ is the voltage across the transmission line. That $Q_{\mathrm{TL}}$ and $\Phi_{\mathrm{TL}}$ [Eq. (1.40)] are written in terms of cosine transforms (with wavevector $k = \omega/v$) comes from considering a semi-infinite ($0 \leq x < \infty$) rather than infinite transmission line [36].

In order to identify the fields moving towards and away from the boundary at $x = 0$, we decompose the voltage $V_{\mathrm{TL}}(x,t) = \dot{\Phi}_{\mathrm{TL}}(x,t)$ into left- (L) and right- (R) moving components:

$$V_{\mathrm{TL}}(x,t) = V_{\mathrm{L}}(x,t) + V_{\mathrm{R}}(x,t), \tag{1.41}$$

where

$$V_{\mathrm{L/R}}(x,t) = i\int_0^\infty d\omega \sqrt{\frac{\omega}{4\pi c v}} e^{i\omega\left(t \pm \frac{x}{v}\right)} r_{\mathrm{L/R},\omega}^\dagger + \mathrm{h.c.} \tag{1.42}$$

Here, we have introduced bosonic operators $r^{\dagger}_{\mathrm{L/R},\omega}$ that act on the vacuum to create left- and right-moving modes, and which satisfy the commutation relations $[r_{\mu,\omega}, r^{\dagger}_{\mu',\omega'}] = \delta_{\mu,\mu'}\delta(\omega-\omega')$ for $\mu,\mu' = \mathrm{L},\mathrm{R}$. Due to the boundary condition at $x=0$, the left- and right-moving components of $V_{\mathrm{TL}} = \dot{\Phi}_{\mathrm{TL}}$ are not independent. Indeed, it may be verified directly from the form of $\Phi_{\mathrm{TL}}(x,t)$ that

$$v\partial_x \Phi_{\mathrm{TL}}(x,t) = V_{\mathrm{L}}(x,t) - V_{\mathrm{R}}(x,t). \tag{1.43}$$

To further simplify the boundary condition (1.43), we now consider the total Hamiltonian $H_{\mathrm{tot}}$ of the LC oscillator and transmission line:

$$H_{\mathrm{tot}} = H_{\mathrm{S}} + H_{\mathrm{TL}} + H_{\mathrm{int}}, \tag{1.44}$$

where here, $H_{\mathrm{S}} = Q^2/2C + \Phi^2/2L$ is the Hamiltonian of the LC oscillator. Under the assumption that the interaction between the transmission line and the LC oscillator at $x=0$ arises from capacitive coupling with strength $C_\kappa$, the interaction Hamiltonian $H_{\mathrm{int}}$ is given by $H_{\mathrm{int}} = C_\kappa (cC)^{-1}\int dx\,\delta(x) Q Q_{\mathrm{TL}}(x)$. Hamilton's equations for the field in the transmission line, given by

$$\begin{aligned} \dot{\Phi}_{\mathrm{TL}}(x) &= \theta(x)\frac{Q_{\mathrm{TL}}(x)}{c} + \delta(x)\frac{C_\kappa}{cC}Q, \\ \dot{Q}_{\mathrm{TL}}(x) &= \partial_x\left[\theta(x)\frac{\partial_x\Phi_{\mathrm{TL}}(x)}{l}\right], \end{aligned} \tag{1.45}$$

can then be combined to yield a wave equation for $\Phi_{\mathrm{TL}}$:

$$\ddot{\Phi}_{\mathrm{TL}}(x) = v^2\theta(x)\partial_x^2\Phi_{\mathrm{TL}}(x) + \delta(x)\left[v^2\partial_x\Phi_{\mathrm{TL}}(x) + \frac{C_\kappa}{cC}\dot{Q}\right]. \tag{1.46}$$

Note that for $x>0$, Eq. (1.46) reduces to the familiar form $\ddot{\Phi}_{\mathrm{TL}} = v^2\partial_x^2\Phi_{\mathrm{TL}}$. Finally, integrating Eq. (1.46) over a small region centered about zero gives the boundary condition relating the field in the transmission line to the current injected into the transmission line by the LC oscillator:

$$v^2\partial_x\Phi_{\mathrm{TL}}(x=0) = -\frac{C_\kappa}{cC}\dot{Q}. \tag{1.47}$$

Inserting Eq. (1.47) back into Eq. (1.43) gives

$$\frac{V_{\mathrm{out}}(t) - V_{\mathrm{in}}(t)}{Z_{\mathrm{TL}}} = \frac{C_\kappa}{C}\dot{Q}, \tag{1.48}$$

where $V_{\mathrm{out(in)}}(t) = V_{\mathrm{R(L)}}(x=0,t)$ are the left- and right- moving fields evaluated at $x=0$, and where $Z_{\mathrm{TL}} = \sqrt{l/c}$ is the impedance of the transmission line. Equation (1.48) is the input-output relation and can be understood as Kirchoff's current law, applied to the interface between the LC oscillator and transmission line. To re-write the input-output relation in a more conventional form, we begin by expressing $Q = i(a^\dagger - a)/\sqrt{2Z}$ in terms of mode operators $a, a^\dagger$, where here, $Z = \sqrt{L/C}$ is the impedance of the LC oscillator. Using Eq. (1.42), Eq. (1.48) can then be written as

$$i\int_0^\infty d\omega\sqrt{\frac{\omega}{4\pi c v}}e^{-i\omega t}(r_{\mathrm{L},\omega} - r_{\mathrm{R},\omega}) + \mathrm{h.c.} = -iZ_{\mathrm{TL}}\frac{C_\kappa}{C}\frac{1}{\sqrt{2Z}}\dot{a} + \mathrm{h.c.} \tag{1.49}$$

We now apply a "slowly varying envelope" approximation to the LC-oscillator dynamics by defining the slowly varying quantity $\tilde{a}(t) = e^{i\omega_{\mathrm{c}}t}a(t)$, where here, $\omega_{\mathrm{c}} = 1/\sqrt{LC}$ is the decoupled frequency of the LC oscillator. For cavity decay occurring with rate $\kappa$, we then have $\dot{a}(t) = -i\omega_{\mathrm{c}}a(t) + e^{-i\omega_{\mathrm{c}}t}\dot{\tilde{a}}(t) = -i\omega_{\mathrm{c}}a(t) +$ $O(\kappa/\omega_{\rm c})$. In the high-Q regime $\kappa \ll \omega_{\rm c}$, we can then use Eq. (1.49) to obtain the input-output relation in its more familiar form [2]

$$\hat{r}_{\rm out}(t) - \hat{r}_{\rm in}(t) = \sqrt{\kappa} a(t), \tag{1.50}$$

where in terms of the parameters introduced above, the cavity decay rate $\kappa$ is given by $\kappa = Z_{\rm TL} C_\kappa^2 \omega_{\rm c}^2/C$ [13], and where we have introduced the input and output field operators

$$\hat{r}_{\rm in(out)}(t) = \frac{i}{\sqrt{2\pi}} \int_0^\infty d\omega\, e^{-i\omega t} r_{{\rm L(R)},\omega}. \tag{1.51}$$

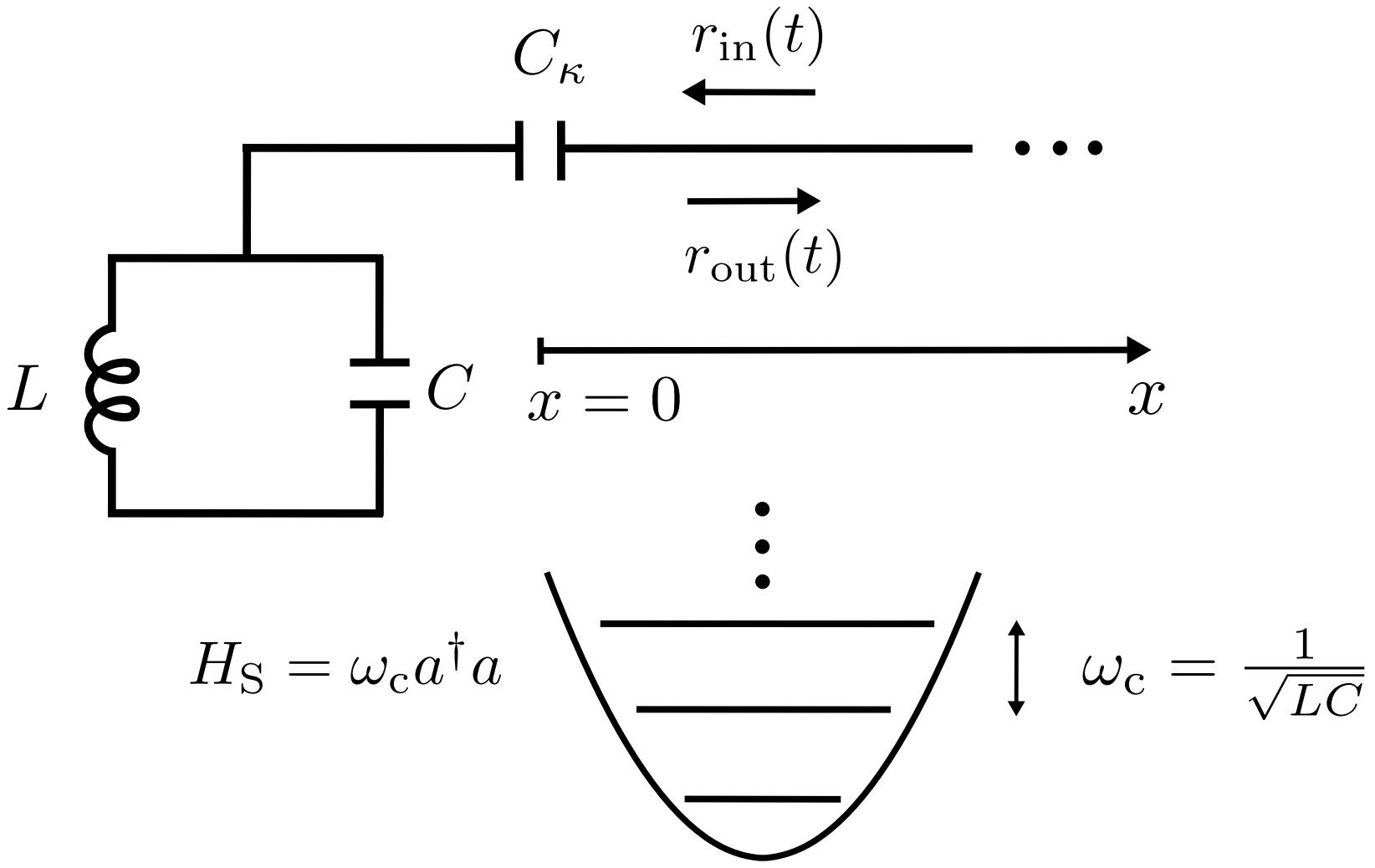

Figure 1.4: The incoming $[r_{\rm in}(t)]$ and outgoing $[r_{\rm out}(t)]$ fields in the transmission line can be related to the field of the LC oscillator through a boundary condition called the input-output relation [Eq. (1.50)]. In terms of circuit parameters, the resonance frequency $\omega_{\rm c}$ of the LC oscillator is given by $\omega_{\rm c} = 1/\sqrt{LC}$.

Since we assumed $\kappa \ll \omega_{\rm c}$ in deriving Eq. (1.50), self-consistency requires that $\kappa/\omega_{\rm c} = Z_{\rm TL} C_\kappa^2 \omega_{\rm c}/C \ll 1$. Equations (1.50) and (1.51) also assume that the oscillator response is strongly peaked about $\omega \simeq \omega_{\rm c}$, allowing us to replace $\sqrt{\omega}$ in Eq. (1.49) by $\sqrt{\omega_{\rm c}}$ for a wide-bandwidth transmission line. In cases where $\tilde{a}(t) = e^{i\omega_{\rm c} t} a(t)$ evolves on another timescale in addition to $\kappa^{-1}$ (due to coupling to a qubit, for instance), validity of Eq. (1.50) requires that this timescale also be long relative to $\omega_{\rm c}^{-1}$.

### 1.2.1 Derivation of the quantum Langevin equation

The input-output relation [Eq. (1.50)] relates the dynamics of the output field $\hat{r}_{\rm out}(t)$ to the dynamics of the LC oscillator, which typically encode the quantity of interest. To derive an equation of motion for $a(t)$, we start from the Heisenberg equation

$$\dot{a} = i[H_{\rm tot}, a] = i[H_{\rm S}, a] + \frac{C_\kappa}{c}\sqrt{\frac{\omega_{\rm c}}{2C}} Q_{\rm TL}(x=0). \tag{1.52}$$

Since $Q_{\rm TL}(x=0,t) = Q_{\rm in}(t) + Q_{\rm out}(t) = c(V_{\rm in}(t) + V_{\rm out}(t))$, we know from Eq. (1.48) that

$$Q_{\rm TL}(x=0,t) = 2Q_{\rm in}(t) + \frac{1}{v}\frac{C_\kappa}{C}\dot{Q}(t). \tag{1.53}$$

Under the same wide-bandwidth approximation as made above in deriving Eq. (1.50), we then find after some algebra that

$$\dot{a} = i[H_{\mathrm{S}}, a] + \sqrt{\kappa}(\hat{r}_{\mathrm{in}}^{\dagger} - \mathrm{h.c.}) + i\frac{\kappa}{2\omega_{\mathrm{c}}}(\dot{a}^{\dagger} - \mathrm{h.c.}). \tag{1.54}$$

The last term in Eq. (1.54) describes cavity decay. This can be seen by considering once again the slowly varying quantity $\tilde{a}(t) = e^{i\omega_{\mathrm{c}}t}a(t)$, in terms of which $\dot{a} = -i\omega_{\mathrm{c}}a + e^{-i\omega_{\mathrm{c}}t}\dot{\tilde{a}}$. By substituting this expression for $\dot{a}$ into the last term of Eq. (1.54), we recover the familiar Langevin equation for the LC-oscillator (cavity) mode,

$$\dot{a} = i[H_{\mathrm{S}}, a] - \sqrt{\kappa}\hat{r}_{\mathrm{in}} - \frac{\kappa}{2}a, \tag{1.55}$$

valid up to corrections $O(\kappa/\omega_{\mathrm{c}})$, as well as corrections due to counter-rotating terms oscillating with frequencies $\pm 2\omega_{\mathrm{c}}$. Although we previously took $H_{\mathrm{S}} = \omega_{\mathrm{c}}a^{\dagger}a$ in deriving the input-output relation [Eq. (1.50)], the Hamiltonian $H_{\mathrm{S}}$ can generally include other terms as well, such as coupling to a qubit. For a system with dynamics governed by $H_{\mathrm{S}}$, solving for the output field typically involves solving Eq. (1.55), to be used in conjunction with Eq. (1.50).

## 1.3 Non-Markovian qubit dephasing

In the master equation used to calculate the cavity transmission [Eq. (1.22)], qubit dephasing—described by the term $\gamma_2\mathscr{D}[\sigma_z]$ in Eq. (1.18)—is modelled as Markovian, consistent with an *exponential* reduction with rate $\gamma_2$ of the qubit coherence $\langle\sigma_-\rangle_t$. Whether or not a Markovian description is appropriate depends on both the spectral content of the qubit's environment and the strength of the coupling of the qubit to this environment. The distinction between Markovian and non-Markovian dephasing will play an important role in Chapter 2 of this thesis, so to emphasize this point, consider the following Hamiltonian model for qubit dephasing:

$$H_{\mathrm{Q}}(t) = \frac{1}{2}\left(\omega_{\mathrm{q}} + \eta(t)\right)\sigma_z. \tag{1.56}$$

Here, $\eta(t)$ is a classical time-dependent function describing drift of the qubit splitting. In general, $\eta(t)$ will be sampled on a shot-to-shot basis from some probability distribution $P_{\eta(t)}$, and it is this randomness that ultimately leads to qubit dephasing. This can be seen by considering the equation of motion

$$\begin{aligned} \dot{\sigma}_-(t) &= i[H_{\mathrm{Q}}(t), \sigma_-(t)] \\ &= -i\left(\omega_{\mathrm{q}} + \eta(t)\right)\sigma_-(t) \end{aligned} \tag{1.57}$$

for the spin-lowering operator. In any experiment, observables depending on $\sigma_-$ (e.g. $\sigma_x$) will be measured over many shots and will therefore depend on

$$\langle\!\langle \langle\sigma_-\rangle_t \rangle\!\rangle = e^{-i\omega_{\mathrm{q}}t}\langle\!\langle e^{-i\int_0^t dt'\,\eta(t')}\rangle\!\rangle\langle\sigma_-\rangle_0, \tag{1.58}$$

where here, $\langle\!\langle\,\rangle\!\rangle$ denotes a functional average $\langle\!\langle\cdots\rangle\!\rangle = \int d\eta(t)\,P_{\eta(t)}(\cdots)$ over the probability distribution $P_{\eta(t)}$ governing the distribution of realizations of $\eta(t)$. Equation (1.58) can be re-written in a more illuminating form by making a few (typically realistic) assumptions about the noise $\eta(t)$. First, we assume that the noise has zero mean: $\langle\!\langle\eta(t)\rangle\!\rangle = 0$. Second, we take $P_{\eta(t)}$ to be Gaussian—this is justifiable under the assumption that the noise can be attributed to a large number of uncorrelated constituents in the qubit's environment, in which case $P_{\eta(t)}$ approaches a Gaussian by the central-limit theorem. Third, we assume that the noise is stationary, in which case the two-time autocorrelation function $\langle\!\langle\eta(t)\eta(t')\rangle\!\rangle = \langle\!\langle\eta(t-t')\eta\rangle\!\rangle$ depends only on the difference $t - t'$. This assumption can be justified as long as the noise characteristics do not exhibit any special dependence on the time at which the experiment is begun.

For zero-mean, stationary, Gaussian noise, performing a cumulant expansion to second order in $\eta(t)$

allows us to rewrite Eq. (1.58) as [37]

$$\langle\!\langle \langle \sigma_- \rangle_t \rangle\!\rangle = e^{-i\omega_q t} e^{-K(t)} \langle \sigma_- \rangle_0, \tag{1.59}$$

where in terms of the spectral density $S(\omega) = \int dt\, e^{-i\omega t} \langle\!\langle \eta(t)\eta \rangle\!\rangle$ of the noise,

$$K(t) = \int \frac{d\omega}{2\pi} \frac{S(\omega)}{\omega^2} F(\omega,t), \tag{1.60}$$

$$F(\omega,t) = \frac{\omega^2}{2} \left| \int_0^t dt'\, e^{i\omega t'} \right|^2. \tag{1.61}$$

For a classical random variable like $\eta(t)$, $K(t)$ is purely real and variation $\propto \eta(t)\sigma_z$ in the qubit splitting leads only to a change in the magnitude $|\langle\!\langle \langle \sigma_- \rangle_t \rangle\!\rangle|$ of the qubit coherence. If, instead, we were to consider coupling $\propto \hat{\eta}(t)\sigma_z$ to a time-dependent operator, then the spectral density $S(\omega)$ would acquire an imaginary part depending on the commutator $[\hat{\eta}(t), \hat{\eta}]$, leading to an extra noise-dependent phase in Eq. (1.59). This generalization from purely classical noise to quantum noise is discussed in greater detail in Chapter 2.

As highlighted by Eq. (1.60), the dynamics of the qubit coherence $\langle\!\langle \langle \sigma_- \rangle_t \rangle\!\rangle$ depend on the noise spectral density $S(\omega)$. To illustrate the major qualitative difference between Markovian and non-Markovian dephasing, we consider two limiting cases: First, suppose the noise $\eta(t)$ has a very *short* correlation time, which we can model by setting $\langle\!\langle \eta(t)\eta \rangle\!\rangle = \frac{\gamma}{2}\delta(t)$, where here, $\gamma$ has units of frequency. In this case, the spectral density is constant, $S(\omega) = \gamma/2$, and from Eq. (1.60), we find that $\langle\!\langle \langle \sigma_- \rangle_t \rangle\!\rangle$ decreases exponentially:

$$\frac{\langle\!\langle \langle \sigma_- \rangle_t \rangle\!\rangle}{\langle \sigma_- \rangle_0} = e^{-i\omega_q t} e^{-\frac{\gamma}{2}t}. \tag{1.62}$$

This exponential suppression is a hallmark of Markovian dephasing and can be modelled in a master equation via the Lindbladian dissipator $\frac{\gamma}{2}\mathscr{D}[\sigma_z]$. In general, dephasing can be modelled as Markovian (history-independent) when the qubit's environment has a correlation time that is short compared to the timescale on which the qubit evolves—put colloquially, the environment does not "remember" past qubit dynamics.

For the opposite limiting case, suppose the noise has an extremely *long*, effectively infinite correlation time. The noise at time $t$ then remains strongly correlated with the noise at time $t = 0$, which we model by setting $\langle\!\langle \eta(t)\eta \rangle\!\rangle = \gamma^2/2$. The spectral density is then strongly peaked at zero frequency, $S(\omega) = \pi\gamma^2\delta(\omega)$, giving

$$\frac{\langle\!\langle \langle \sigma_- \rangle_t \rangle\!\rangle}{\langle \sigma_- \rangle_0} = e^{-i\omega_q t} e^{-\frac{1}{2}\gamma^2 t^2}. \tag{1.63}$$

In this case, the decay of $\langle\!\langle \langle \sigma_- \rangle_t \rangle\!\rangle$ is non-exponential and cannot be modelled via the usual Lindblad dissipator. Many realistic noise sources must be treated as non-Markovian—this includes the (in)famous $1/f$ noise currently limiting qubit coherence times in many solid-state systems [38]. Because non-Markovian environments retain some memory of past qubit dynamics, their impact can be mitigated by applying dynamical decoupling pulses designed to partially reverse the noise-induced qubit evolution, leading to the same expression for $K(t)$, but with a modified filter function $F(\omega,t)$ that can be engineered to reduce the sensitivity of the qubit coherence to noise at certain frequencies. Chapter 2 of this thesis, to which we now turn our attention, will discuss an approach for probing non-Markovian qubit dynamics using dynamical decoupling pulses in conjunction with measurements of the cavity field.

# Preface to Chapter 2

In the previous chapter, we discussed one way—dispersive readout—in which the state of a qubit can be probed via measurements of the cavity. Dispersive readout consists of driving the cavity with a probe field and measuring the field's change in phase upon reflection. This probe field will gradually dephase the qubit, projecting it onto an eigenstate of the measured operator, $\sigma_z$.

Dispersive readout is of course a measurement performed deliberately, but along similar conceptual lines, a qubit's environment can act like a measurement apparatus and lead to undesired dephasing—an undesired loss of information to the environment. Chapter 2 gives a strategy, which we call transient spectroscopy, for learning about environment-induced qubit dephasing via measurements of the cavity. In contrast to dispersive readout, which involves a continuous driving field applied to the cavity, transient spectroscopy leverages information contained in transient cavity dynamics—dynamics obtained in the absence of continuous driving, but in the presence of initial qubit coherence. These dynamics are due predominantly to the qubit's environment, allowing us to learn about the environment via the decay of the qubit coherence.

The previous chapter also discussed the distinction between Markovian and non-Markovian processes, and touched upon the notion of reversing qubit evolution due to the environment via the use of dynamical decoupling. In Chapter 2, we consider a cavity-coupled qubit undergoing a sequence of dynamical decoupling pulses. The intuitive picture is that every pulse applied to the qubit leads to a transient response in the cavity field. These transient responses, taken together and processed correctly, can then reveal information about the dephasing kernel $K(t)$ given in Eq. (1.60), allowing us to learn about the spectral content of the qubit's environment with the ultimate goal of mitigating its impact.

# 2

# Non-Markovian transient spectroscopy

*This chapter is the integral text from*

**Non-Markovian transient spectroscopy in cavity QED**
Z. McIntyre and W. A. Coish
Phys. Rev. Research **4**, L042039 (2022)

We theoretically analyze measurements of the transient field leaving a cavity as a tool for studying non-Markovian dynamics in cavity quantum electrodynamics (QED). Combined with a dynamical decoupling pulse sequence, transient spectroscopy can be used to recover spectral features that may be obscured in the stationary cavity transmission spectrum due to inhomogeneous broadening. The formalism introduced here can be leveraged to perform *in situ* noise spectroscopy, revealing a robust signature of quantum noise arising from non-commuting observables, a purely quantum effect.

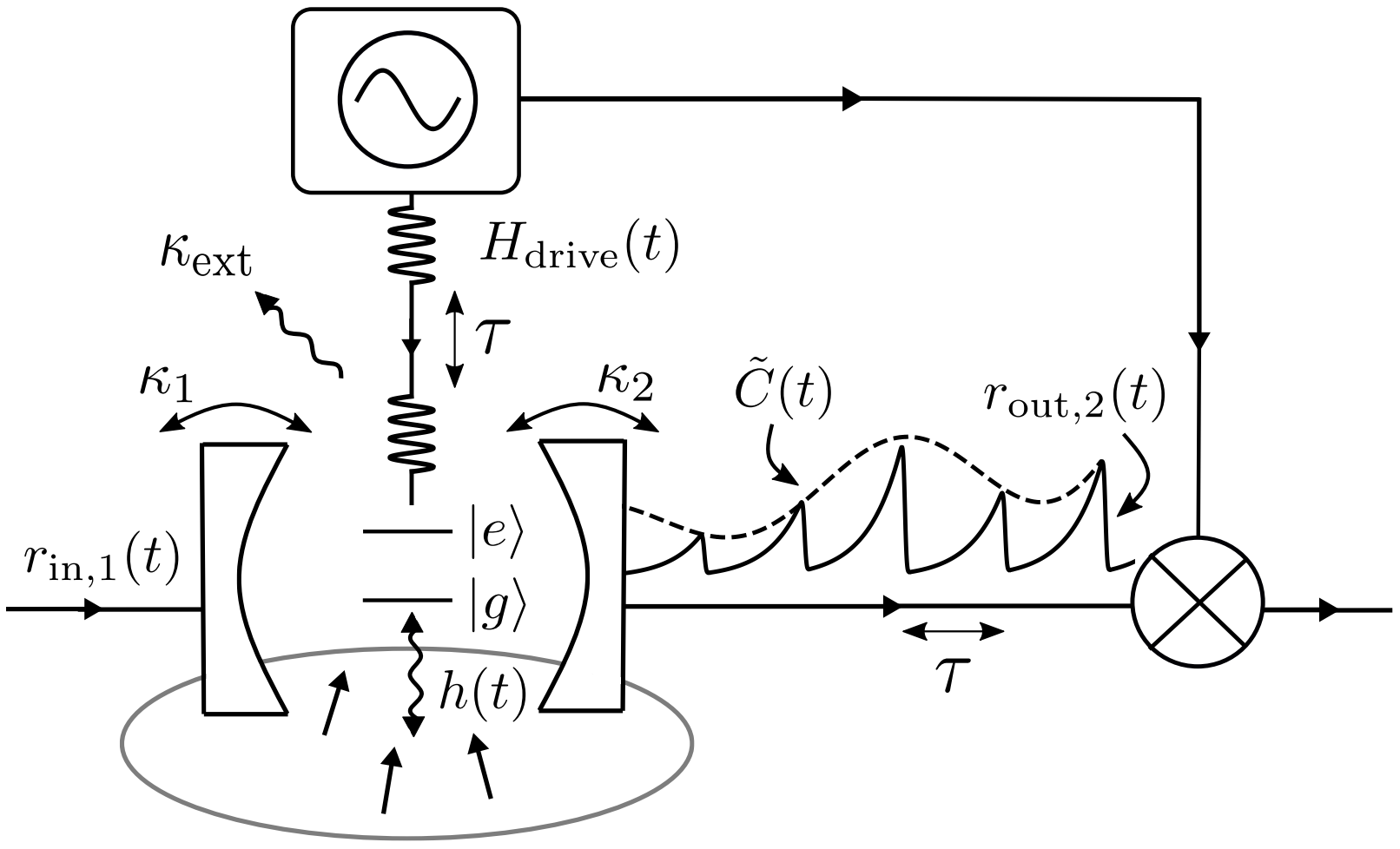

Figure 2.1: A typical cavity-QED setup. The transmission spectrum is obtained from the linear response of the output signal $r_{\text{out},2}(t)$ to a monochromatic input tone, $r_{\text{in},1}(t)$. In contrast, the transient spectrum is found, for $r_{\text{in},1}(t) = 0$, by generating a sequence of control pulses via a qubit drive $H_{\text{drive}}(t)$. These pulses induce non-Markovian coherence revivals with envelope $\tilde{C}(t)$ for a qubit coupled to its environment through some interaction $h\sigma_z/2$. A phase-sensitive measurement of $r_{\text{out},2}(t)$ can then be used to determine $\tilde{C}(t)$. The decay rates at the input and output ports of the cavity are denoted $\kappa_1$ and $\kappa_2$, respectively, while extrinsic decay is denoted $\kappa_{\text{ext}}$.

## 2.1 Introduction

Significant effort has recently gone towards reaching the strong-coupling regime of cavity quantum electrodynamics (QED) for individual long-lived spin and charge qubits, with the goals of achieving long-range coupling [1], performing fundamental studies of many-body phenomena [2], and realizing other exotic effects arising from hybrid systems [3]. Strong coupling has been observed between microwave photons and charge qubits in GaAs [4], spin qubits in silicon [5–7], resonant-exchange qubits in GaAs triple quantum dots [8], and spin qubits in carbon nanotube double quantum dots [9, 10] (DQDs). Two-qubit, photon-mediated interactions have been observed between charge qubits in GaAs DQDs [11] and between spin qubits in silicon DQDs [12, 13]. As these devices reach a progressively higher level of sophistication and quality, it is increasingly important to characterize the qubits and their local environments *in situ*, together with the components that define the cavity.

*In-situ* characterization of a two-level emitter (qubit) coupled to a cavity is often done by measuring a transmission or reflection spectrum [4, 5, 7, 14–16] in a setup similar to that shown in Fig. 2.1. In this setup, an input tone $r_{\text{in},1}(t) = (2\pi)^{-1}\int d\omega e^{-i\omega t} r_{\text{in},1}(\omega)$ is introduced, and after a time long compared to the cavity decay time $\kappa^{-1}$, the output field $r_{\text{out},2}(t)$ reaches a steady state. The stationary transmission $A_{\text{T}}(\omega) = r_{\text{out},2}(\omega)/r_{\text{in},1}(\omega)$ then carries information about the qubit accounting for its interaction with the environment and resulting decay processes. To interpret the transmission, it is common to make the simplifying assumption that the qubit dynamics are generated by a Markovian master equation, with parameters characterizing dephasing and relaxation rates. The standard tools of input-output theory [16–19] can then be applied. The Markovian assumption is often an excellent approximation for single-atom emitters [14, 15] and for the superconducting transmon qubits commonly used in circuit QED [16]. With some exceptions [20, 21], these systems typically have coherence times $T_2$ limited by the exponential energy relaxation time $T_1$: $T_2 \simeq 2T_1$. In stark contrast, spin and charge qubits defined using semiconductor nanostructures almost universally undergo non-exponential (non-Markovian) pure dephasing on a time scale $T_2^* \ll T_1$ arising from inhomogeneous broadening due to low-frequency charge noise or slow nuclear-spin environments. A different approach is required for these and many other non-Markovian systems.

A dynamical decoupling pulse sequence can help mitigate the effects of strong inhomogeneous broadening, but the result is a train of manifestly non-Markovian collapses and revivals (echoes) in qubit coherence. Although these revivals (of duration $\sim T_2^*$) can be mapped to the transient output field $r_{\text{out},2}(t)$ (see Fig. 2.1), their effect on the stationary transmission spectrum is negligible since they are, by definition, transient. Similar revivals have already been exploited for measurement in spin-echo experiments on ensembles [22, 23]. For a low-$Q$ resonator, the relationship between spin coherence and the output field is relatively simple and time-local [22]. By contrast, it is a nontrivial problem to relate the complex pattern of revivals arising from, e.g., a dynamical decoupling sequence, to real-time non-Markovian coherence dynamics for a high-$Q$ cavity. We perform this analysis here. Although our focus is on individual spin and charge qubits under a widely used dynamical decoupling sequence, the ideas presented here are generally applicable to ensembles and to a wide range of non-Markovian systems in cavity QED, a topic of significant recent interest [24–28].

## 2.2 Model

We start from a typical cavity-QED setup (see Fig. 2.1), with dynamics governed by the quantum master equation (taking $\hbar = 1$):

$$\dot{\rho} = -i[H(t),\rho] + \frac{\gamma_\phi}{2}\mathscr{D}[\sigma_z]\rho + \kappa_{\text{ext}}\mathscr{D}[a]\rho. \tag{2.1}$$

Here, $\rho = \rho(t)$ is the joint state of the qubit, cavity, quantum environment, and transmission lines. The qubit (with Pauli operator $\sigma_z = |e\rangle\langle e| - |g\rangle\langle g|$) undergoes Markovian pure dephasing at a rate $\gamma_\phi$, while photons in the cavity mode (annihilated by $a$) decay at an extrinsic rate $\kappa_{\text{ext}}$. The damping superoperator acts like $\mathscr{D}[\mathscr{O}]\rho = \mathscr{O}\rho\mathscr{O}^\dagger - \{\mathscr{O}^\dagger\mathscr{O},\rho\}/2$ for any operator $\mathscr{O}$; in addition to damping processes, the density operator $\rho$ evolves under the Hamiltonian

$$H(t) = \frac{1}{2}[\Delta + \Omega(t)]\sigma_z + \frac{1}{2}h(t) + \omega_c a^\dagger a + g\sigma_x(a + a^\dagger) + H_{\text{drive}}(t) + \sum_{i=1,2}\sum_k(\lambda_{k,i}e^{i\omega_k t}r_{k,i}^\dagger a + \text{h.c.}), \tag{2.2}$$

where $\Delta$ is the qubit resonance frequency, $\omega_c$ is the cavity frequency, and $g$ is the qubit-cavity coupling. The term $H_{\text{drive}}(t)$ describes a drive acting on the qubit, while noise is generated by $\Omega(t) = \eta(t) + h(t)$, where $\eta(t)$ is a classical noise parameter and $h(t) = e^{i(H_{\text{E}}-h/2)t}he^{-i(H_{\text{E}}-h/2)t}$ acts on the environment alone. The time dependence of $h(t)$ arises from a lab-frame Hamiltonian $H_{\text{E}} + h\sigma_z/2$ together with the assumption that the environment is prepared in a steady state while coupled to the qubit held in $|g\rangle$ (see below and Ref. [29]). We take $\eta(t)$ to be generated by a stationary Gaussian process with zero mean ($\langle\langle\eta(t)\rangle\rangle = 0$) and spectral density

$$S_\eta(\omega) = \int dt e^{-i\omega t}\langle\langle\eta(t)\eta(0)\rangle\rangle, \tag{2.3}$$

where here, the double angle brackets $\langle\langle\rangle\rangle$ represent an average over noise realizations. The terms $\propto \lambda_{k,i}$ in Eq. (2.2) describe coupling of the cavity mode to the input (output) [for $i = 1(2)$] transmission-line mode annihilated by $r_{k,i}$ and having frequency $\omega_k$. For modes propagating in one dimension, $r_{\text{in},i}(t) = [c/L]^{1/2}\sum_k e^{-i\omega_k t}\langle r_{k,i}\rangle_0$ and $r_{\text{out},i}(t) = [c/L]^{1/2}\sum_k\langle r_{k,i}\rangle_t$, where $L$ is the length of the transmission line and $c$ is the speed of light. The notation $\langle\mathscr{O}\rangle_t$ indicates an average with respect to the state $\rho(t)$, together with an average over realizations of the classical noise $\eta(t)$: $\langle\mathscr{O}\rangle_t = \langle\langle\text{Tr}\{\mathscr{O}\rho(t)\}\rangle\rangle$.

## 2.3 Transient spectroscopy

In order to accurately monitor qubit dynamics through the transient output field $r_{\text{out},2}(t)$, we consider the following protocol: (i) An undriven single-sided ($\kappa_1 = 0$) cavity is prepared in a vacuum state $|0\rangle$ far-detuned from (or decoupled from) the qubit. (ii) The qubit is prepared in its ground state $|g\rangle$, and the environment is allowed to reach a steady-state $\bar{\rho}_{\text{E}}$ in contact with the qubit: $[H_{\text{E}} - h/2, \bar{\rho}_{\text{E}}] = 0$. (iii) At $t = 0$, the qubit and cavity are tuned close to resonance (or the coupling $g$ is turned on), and a finite drive $H_{\text{drive}}(t \geq 0)$ generates qubit coherence $\langle\sigma_x\rangle_t$. This coherence is related to the cavity field $\langle\tilde{a}\rangle_t = e^{i\Delta t}\langle a\rangle_t$ via direct integration of Eq. (2.1):

$$\langle\tilde{a}\rangle_t = -ig\int_{-\infty}^{\infty} dt'\,\chi_c(t-t')e^{i\Delta t'}\langle\sigma_x\rangle_{t'}, \tag{2.4}$$

where $\chi_c(t) = e^{-i\delta t-\kappa t/2}\Theta(t)$ for a cavity-qubit detuning $\delta = \omega_c - \Delta$ and total cavity decay rate $\kappa = \kappa_{\mathrm{ext}} + \kappa_2$. Neglecting retardation effects, the measured output field is then given by the input-output relation $r_{\mathrm{out},2}(t) = \sqrt{\kappa_2}\langle a\rangle_t$ [17]. For a single cavity-coupled qubit, the protocol [(i)-(iii)] is limited to gathering $\lesssim 1$ bit of information per cycle, similar to many early cavity-QED schemes [14]. These steps must therefore be repeated many times to estimate the expectation value $\langle\sigma_x\rangle_t$.

## 2.4 Dynamical decoupling

For concreteness, we consider an $N$-pulse Carr-Purcell-Meiboom-Gill (CPMG) sequence, where coherence preparation at $t = 0$ [such that $\langle\sigma_-\rangle_0 = \frac{1}{2}\langle\sigma_x\rangle_0 \neq 0$] is followed by $\pi_x$-pulses at times $t = \tau/2, 3\tau/2, ..., (N-1/2)\tau$, leading to coherence revivals at times $t = n\tau$, $n = 1, 2, \ldots, N$ (see the supplement [29] for the general formalism, valid for other pulse sequences). We further specialize to the regime $g < \kappa \ll \tau^{-1}$, where cavity backaction effects can be treated as a small correction. In this regime, the coherence factor $C(t) = \langle\sigma_-\rangle_t/\langle\sigma_-\rangle_0$ can be written in terms of a comb of revivals (echoes) with peaks $\sim G_n(t-n\tau)$ centered at $t = n\tau$ and an echo envelope $\tilde{C}(t)$:

$$C(t) = \sum_n e^{-i\Delta(t-n\tau)}G_n(t-n\tau)\mathscr{K}^n\tilde{C}(n\tau). \tag{2.5}$$

Here, $\mathscr{K}z = z^*$ for all $z \in \mathbb{C}$. If $T_2^* \ll \tau$ [where $2/(T_2^*)^2 = (2\pi)^{-1}\int d\omega\, S_\eta(\omega)$], and if $\tilde{C}(n\tau)$ is slowly varying on the timescale $T_2^*$, then we find

$$G_n(t) = e^{\sqrt{\gamma_{\mathrm{P}} n\tau}}\left\langle\!\!\left\langle e^{-\Gamma_{\mathrm{P}}(\eta)n\tau/2}e^{-i\eta t}\right\rangle\!\!\right\rangle, \tag{2.6}$$

where $\eta = \eta(0)$ is the low-frequency contribution to $\eta(t)$, $\Gamma_{\mathrm{P}}(\eta) = g^2\kappa/[(\eta-\delta)^2 + (\kappa/2)^2]$ is the Purcell decay rate at fixed $\eta$, and $\gamma_{\mathrm{P}} = (gT_2^*)^2\kappa/2$. The echo envelope is

$$\tilde{C}(n\tau) = e^{-\sqrt{\gamma_{\mathrm{P}} n\tau}-\gamma_\phi n\tau}\langle\!\langle \mathrm{Tr}\{U_-^\dagger(n\tau)U_+(n\tau)\bar{\rho}_E\}\rangle\!\rangle, \tag{2.7}$$

where $U_\pm(n\tau) = \mathscr{T}\exp\{-\frac{i}{2}\int_0^{n\tau}[h(t') \pm s(t')\Omega(t')]\}$ are evolution operators acting on the environment, conditioned on the $\sigma_z$-eigenvalue ($\pm$) of the qubit. Here, $\mathscr{T}$ is the time-ordering operator and $s(t) = (-1)^{n(t)}$ for $n(t)$ $\pi$-pulses having taken place up to time $t$. For $n < 1/(\gamma_{\mathrm{P}}\tau)$, backaction due to Purcell decay is negligible and the revivals are well approximated by $G_n(t) \simeq G_0(t) = e^{-(t/T_2^*)^2}$. The effects of backaction will be further discussed below.

Taking the Fourier transform of Eq. (2.4) gives $\langle\tilde{a}\rangle_\omega = -ig\chi_c(\omega)\langle\sigma_x\rangle_{\omega+\Delta}$; the cavity susceptibility acts as a filter, $\chi_c(\omega) = [i(\delta-\omega)+\kappa/2]^{-1}$. In the high-$Q$ limit ($Q = \omega_c/\kappa \gg 1$), $\chi_c(\omega)$ suppresses the counter-rotating component $\langle\sigma_+\rangle_t$, allowing us to replace $\langle\sigma_x\rangle_t \simeq \langle\sigma_-\rangle_t = C(t)\langle\sigma_-\rangle_0$ in Eq. (2.4) for $|\delta| \ll |\Delta|$. Under the assumptions laid out above, we find a general expression relating $\langle\tilde{a}\rangle_\omega$ to the echo envelope $\tilde{C}(n\tau)$ [29]. For a narrow cavity resonance, $\kappa T_2^* \ll 1$, $\langle\tilde{a}\rangle_\omega$ will be sharply peaked around $\omega = \delta$, leading to:

$$\langle\tilde{a}\rangle_{\omega=\delta} \simeq -i\langle\sigma_x\rangle_0\frac{\sqrt{\pi}gT_2^*}{\kappa}\left[\tilde{C}_{N,\tau}(\delta) - \frac{1}{2}C(0)\right], \tag{2.8}$$

where we neglect corrections smaller by $O(g/\kappa)$, $O(\kappa/\Delta)$. Here,

$$\tilde{C}_{N,\tau}(\omega) = \sum_{n=0}^{N} e^{in(\omega+\Delta)\tau} \bar{G}_n \mathscr{K}^n \tilde{C}(n\tau), \tag{2.9}$$

where $\bar{G}_n = (\sqrt{\pi}T_2^*)^{-1} \int_{-\infty}^{\infty} dt G_n(t)$. The echo envelope $\tilde{C}(t)$ can thus be reconstructed by measuring $r_{\mathrm{out},2}(\omega)$ to infer $\langle \tilde{a} \rangle_{\omega=\delta}$ over some detuning interval $O(2\pi/\tau)$ and then inverting the discrete Fourier transform $\tilde{C}_{N,\tau}(\omega)$. The revival amplitudes $\mathscr{K}^n \tilde{C}(n\tau)$ depend alternately on $\tilde{C}(n\tau)$ (for $n$ even) and $\tilde{C}^*(n\tau)$ (for $n$ odd); this even/odd alternation is a direct result of the high-$Q$ limit, which (as we now show) can be exploited to identify a purely quantum effect.

## 2.5 Quantum noise

For the secular coupling $h\sigma_z/2$ considered here, we find a generic expression for the echo envelope $\tilde{C}(n\tau)$: Without loss of generality, we take $\langle h \rangle_t = 0$, in which case a Magnus expansion to second order in $h(t)$ followed by a Gaussian approximation (valid for a large uncorrelated environment [30]—see Ref. [31] for the non-Gaussian generalization) gives

$$\tilde{C}(n\tau) \simeq e^{-\sqrt{\gamma_p n\tau} - \gamma_\phi n\tau} \exp\{-i\Phi_{\mathrm{q}}(n\tau) - \chi(n\tau)\}, \tag{2.10}$$

where

$$\Phi_{\mathrm{q}}(n\tau) = \int \frac{d\omega}{2\pi} \frac{F_{\mathrm{q}}(\omega, n\tau)}{\omega^2} S_{\mathrm{q}}(\omega), \tag{2.11}$$

$$\chi(n\tau) = \int \frac{d\omega}{2\pi} \frac{F_{\mathrm{c}}(\omega, n\tau)}{\omega^2} S_{\mathrm{c}}(\omega). \tag{2.12}$$

Here, $F_{\mathrm{c}}(\omega, n\tau) = (\omega^2/2) \left| \int_0^{n\tau} dt\, e^{i\omega t} s(t) \right|^2$ is the usual filter function for classical noise [32], $F_{\mathrm{q}}(\omega, n\tau) = \omega \int_0^{n\tau} dt \sin(\omega t) s(t)$ is a new quantum-noise filter function, and

$$\begin{aligned} S(\omega) &= S_{\mathrm{c}}(\omega) + iS_{\mathrm{q}}(\omega) \\ &= \lim_{\varepsilon \to 0^+} \int_{-\infty}^{\infty} dt\, e^{-i\omega t - \varepsilon|t|} \langle \Omega(|t|)\Omega \rangle \end{aligned} \tag{2.13}$$

is the spectral density [29]. The magnitude of $\tilde{C}(n\tau)$ is then determined by the classical part of the noise spectrum $S_{\mathrm{c}}(\omega) = \mathrm{Re}[S(\omega)] = S_\eta(\omega) + S_h(\omega)$, where $S_h(\omega)$ depends only on the symmetrized correlation function $\langle \{h(|t|), h\} \rangle$. For a quantum environment, the envelope $\tilde{C}(n\tau)$ generally has a phase $\Phi_{\mathrm{q}}(n\tau)$ [Eq. (2.11)] determined by the quantum noise $S_{\mathrm{q}}(\omega) = \mathrm{Im}[S(\omega)]$, due to the antisymmetrized correlation function $\langle [h(|t|), h] \rangle$ [29]. This phase will manifest itself in the alternation between $\tilde{C}$ and $\tilde{C}^*$ in the discrete Fourier transform in Eq. (2.8) for $n$ even/odd. The importance of quantum noise due to non-commuting observables has long been recognized in the mesoscopic-physics community [33]. In addition, it has been measured in CPMG experiments performed on nitrogen-vacancy center spin qubits in diamond, leading to a phase shift $\Phi_{\mathrm{q}}(n\tau) \sim \pi$ [34].

Despite this recognition of quantum noise in other communities, a common simplification in noise spectroscopy is to assume a frequency-symmetric, real-valued spectrum, $S(\omega) = S(-\omega) = S^*(\omega)$, as would arise from a classical fluctuating field [35–40] (although quantum noise has been incorporated into some recent theory works [41, 42]). By leveraging the sensitivity of the cavity field to the phase of qubit coherence revivals in the high-$Q$ regime [Eq. (2.8)], we identify a robust even-odd modulation of revivals arising from $S_q(\omega) = \mathrm{Im}\, S(\omega)$, unique to quantum environments. Notably, this even-odd effect would not appear for coupling of the form $h\sigma_z/2$ when $\bar{\rho}_{\mathrm{E}}$ is stationary with respect to $H_{\mathrm{E}}$ alone ($[H_{\mathrm{E}}, \bar{\rho}_{\mathrm{E}}] = 0$) [41, 42], as may occur for an environment prepared in the absence of the qubit. The quantum-noise phase $\Phi_{\mathrm{q}}(n\tau)$ [Eq. (2.11)] thus appears as a direct consequence of the initial condition, $[H_{\mathrm{E}}-h/2,\bar{\rho}_{\mathrm{E}}]=0$ [29].

## 2.6 Characterizing a single nuclear spin

As a concrete application, we consider an electron spin qubit in a silicon double quantum dot (DQD) [5–7], exposed to a spatially varying magnetic field. The magnetic field is assumed to have an $x$-component $B_x(\boldsymbol{r})$ that averages to zero over the DQD and a uniform $z$-component $B_z$ [30], a setup that is commonly used to generate spin-photon coupling [44]. This leads to a secular coupling and "environment" Hamiltonian given by

$$\frac{1}{2}h\sigma_z = \frac{1}{2}AI_z\sigma_z, \quad H_{\mathrm{E}} = \gamma(B_xI_x + B_zI_z). \tag{2.14}$$

Here, $B_x = B_x(\boldsymbol{r}_0)$ for a $^{29}$Si nuclear spin located at position $\boldsymbol{r}_0$, $\boldsymbol{I}$ is a spin-I operator ($I = 1/2$ for $^{29}$Si), $A$ is the hyperfine coupling, and $\gamma = -5.319\times10^7\,\mathrm{rad\,T^{-1}\,s^{-1}}$ is the gyromagnetic ratio. The same model also applies to a spin qubit in a uniform **B**-field, provided the spin has an anisotropic $g$-tensor, leading to non-collinear quantization axes for the qubit and nuclear spin. Alternatively, this model can describe a qubit having a finite charge dipole interacting with the electric field produced by a two-level charge fluctuator, where $\gamma B_z$ and $\gamma B_x$ are replaced by the fluctuator bias and tunnel splitting, respectively [20, 45, 46].

For an electron-spin qubit in isotopically enriched silicon, coherence times may be limited by a small number of $^{29}$Si nuclear spins [47]. Extracting parameters for individual nuclear spins could facilitate decoherence suppression through a notch-filter dynamical decoupling sequence [48], or allow for a transfer of information from the electron spin to the nuclear spin for a long-lived quantum memory. The problem of characterizing the spin state of a single $^{31}$P donor nuclear spin (with hyperfine coupling $A/2\pi \approx 25$ MHz) was recently considered theoretically in Ref. [49] in the context of transmission spectroscopy (described by input-output theory). For a $^{29}$Si nuclear spin coupled to a quantum-dot-bound electron spin, however, the hyperfine coupling is orders of magnitude weaker ($A/2\pi \approx -0.25$ MHz has been measured, for instance [43]). Spectral information in the transmission $A_{\mathrm{T}}(\omega) = \langle\!\langle A_{\mathrm{T}}(\omega,\eta)\rangle\!\rangle$ might be entirely obscured due to inhomogeneous broadening as a result [Fig. 2.2(a)]. Here, we have averaged the transmission $A_{\mathrm{T}}(\omega,\eta)$ for a time-independent but random value of $\eta$ [29]. Even when spectral information about the nuclear spin is completely obscured in a more conventional measurement of the transmission spectrum, it can still be recovered in the transient spectrum resulting from a spin-echo sequence.

A finite value $B_x \neq 0$ leads to echo envelope modulations following a Hahn-echo sequence (CPMG with $N=1$). For $\kappa T_2^* > 1$ but $Q = \omega_c/\kappa \gg 1$, the cavity-field revival is modulated by the echo envelope according to $\langle a\rangle_\tau \simeq -i\frac{2g}{\kappa}\tilde{C}(\tau)\langle\sigma_-\rangle_0$ [see Fig. 2.2(b)]. [In the opposite regime, $\kappa T_2^* < 1$, it is instead modulated according to $\langle a\rangle_\tau \simeq -i\sqrt{\pi}gT_2^*\tilde{C}(\tau)\langle\sigma_-\rangle_0$.] In the case of a single environmental spin, the Gaussian approximation cannot be justified, but this model can be solved exactly: For a fully randomized (infinite-temperature) initial condition for the nuclear spin, the Hahn-echo amplitude at $t=\tau$ is given by $\tilde{C}(\tau) = e^{-\gamma_\phi\tau}[1-\delta\tilde{C}(\tau)]$, where

$$\delta\tilde{C}(\tau) = 2\sin^2(\Delta\phi)\sin^2\left(\frac{\omega_+\tau}{4}\right)\sin^2\left(\frac{\omega_-\tau}{4}\right). \tag{2.15}$$

Here, $\omega_\pm = [(\gamma B_x)^2 + (\gamma B_z \pm A/2)^2]^{1/2}/2$ and $\Delta\phi = \phi_+ - \phi_-$, where $\phi_\pm = \arctan[2\gamma B_x/(2\gamma B_z \pm A)]$. For the infinite-temperature environmental initial condition considered here, $\tilde{C}(\tau)$ is real and there is no quantum-noise contribution. A polarized initial condition would, however, lead to a complex-valued $\tilde{C}(\tau)$, a signature of a quantum environment and of quantum noise [1]. In this illustrative case of coupling to a single spin, the frequencies $\omega_\pm$ and angular difference $\Delta\phi$ can be extracted independently from the peak positions and peak heights in a Fourier transform of Eq. (2.15) [Fig. 2.2(b), inset], allowing for recovery of both components of the local magnetic field $B_x, B_z$, and of the local hyperfine coupling $A$. For the value of $A$ used in Fig. 2.2, the visibility $\sin^2(\Delta\phi)$ of the echo-envelope oscillations [c.f. Eq. (2.15)] is maximized for $B_z = B_x = 15$ mT. While this combination of values is possible, it would be fortuitous and would require a relatively low cavity frequency, $\omega_c/2\pi \simeq 0.4\,\mathrm{GHz}$. Away from these values, $\sin^2(\Delta\phi) \simeq [A\gamma B_x/(\gamma B_z)^2]^2$ for $A, \gamma B_x < \gamma B_z$. When $[A\gamma B_x/(\gamma B_z)^2]^2 \ll 1$, the amplitude of the modulations can be enhanced through a large-$N$ CPMG sequence: As is well known, a multi-pulse CPMG sequence can be used to amplify specific Fourier components of the noise [34, 38, 50]. A noise contribution $S(\omega) \sim |\beta_0|^2\delta(\omega-\omega_0)$, for instance, leads for $n = N$ to an amplitude $\tilde{C}(N\pi/\omega_0) \simeq e^{-(N/N_0)^2}$, where $N_0 = \sqrt{\pi}\omega_0/(2|\beta_0|)$, giving a visibility $\propto (N/N_0)^2$ that increases with $N$ for $N < N_0$. As $N$ increases, however, the ability to extract information about the qubit coherence may become limited by cavity-induced backaction.

[1] For $h(t) = e^{i(H_{\mathrm{E}}-h/2)t}he^{-i(H_{\mathrm{E}}-h/2)t} = \sum_\alpha c_\alpha(t)I_\alpha$, the quantum-noise term produces a phase arising from $\langle[h(t),h]\rangle = i\sum_{\alpha\beta}\varepsilon_{\alpha\beta\gamma}c_\alpha(t)c_\beta(0)\langle I_\gamma\rangle$, which is nonzero for a polarized initial state, $\langle I_\gamma\rangle = \mathrm{Tr}\left(\bar{\rho}_{\mathrm{E}}I_\gamma\right) \neq 0$. This analysis applies within the Gaussian approximation, which can be justified even for a single spin at short times.

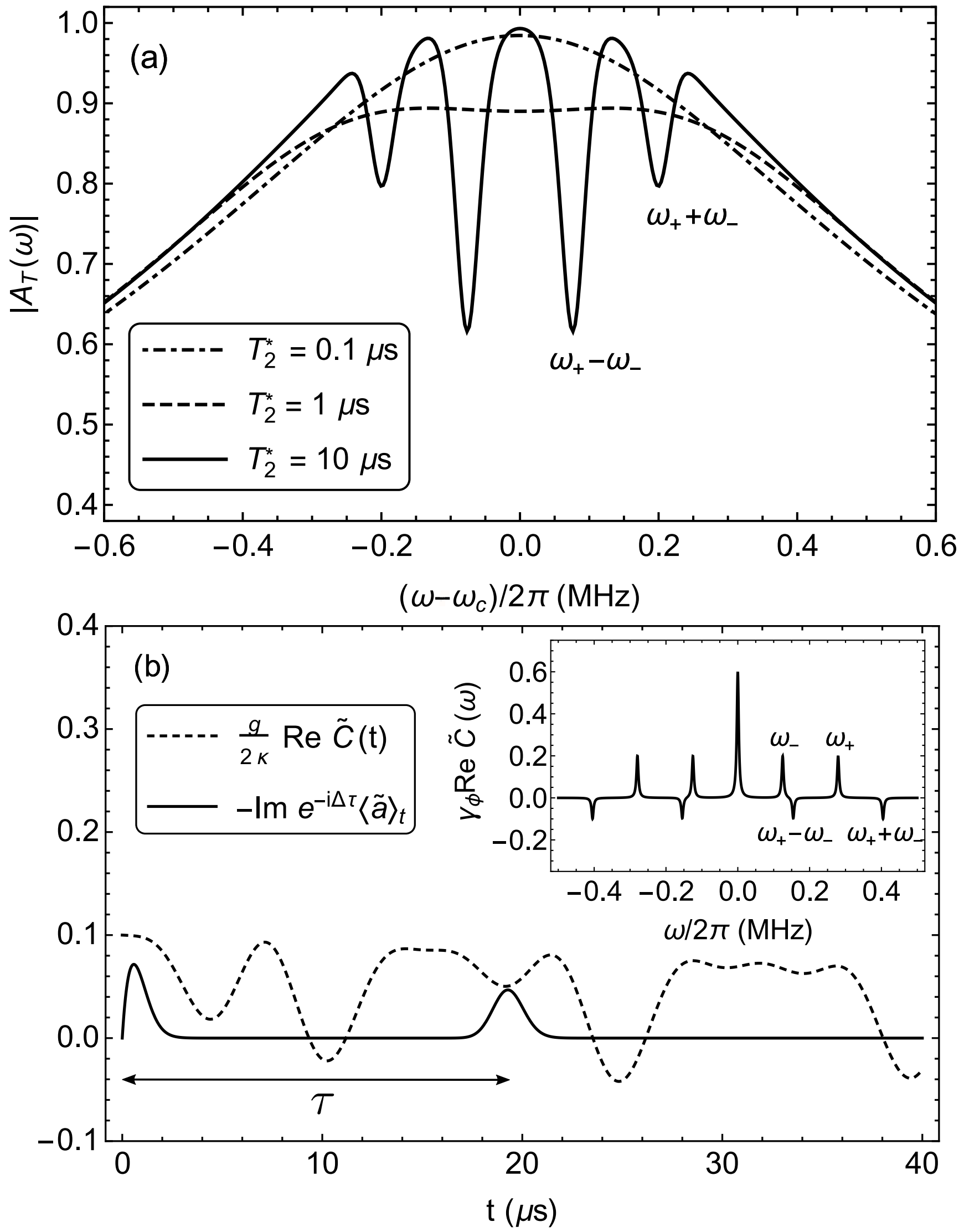

Figure 2.2: (a) Inhomogeneously broadened cavity transmission at $\delta = 0$ for three values of $T_2^*$ ($T_2^* = 0.1\ \mu$s, 1 $\mu$s, and 10 $\mu$s). We take $A/2\pi = -0.250\,$MHz [43], $\gamma B_z = \gamma B_x = A/2$, $\gamma_\phi^{-1} = 100\ \mu$s, $\kappa/2\pi = 1\,$MHz, and $g/\kappa = 0.2$. For $B_z = 15$ mT, $\omega_c/2\pi = g^*\mu_B B_z/2\pi = 0.4$ GHz, where $g^* = 2$. We assume an infinite-temperature initial state for the nuclear spin. (b) Revivals in the cavity field, modulated by the echo envelope, assuming the same parameters as in (a) with $T_2^* = 1\ \mu$s. Once the echo envelope has been mapped out (e.g. by varying $\tau$), it can be Fourier transformed [inset] to recover the frequencies $\omega_\pm$ obscured in (a).

## 2.7 Backaction

A protocol that requires continuous monitoring of a qubit in a driven cavity may suffer from backaction induced by qubit dephasing due to cavity-photon shot noise [51], as well as measurement-induced backaction that necessitates a continuous update of the quantum state [52]. In contrast, the protocol presented here involves no direct cavity driving, and the qubit is re-prepared in each measurement cycle. However, coupling to the cavity will still induce unwanted backaction on the qubit through Purcell decay, beyond the minimum backaction required to extract information about the qubit coherence dynamics: For a CPMG sequence, and for $\kappa\tau \gg 1$, we find that inhomogeneously broadened Purcell decay leads to a stretched-exponential decay, $\tilde{C}(n\tau) \propto e^{-\sqrt{\gamma_P n\tau}}$. For $n > 1/(\gamma_P\tau)$, it also gives rise to a simultaneous broadening (in time) and modulation of the echo revivals,

$$G_n(t) \simeq e^{-\left(\frac{t}{2T_2^*}\right)^2} \cos\left[\sqrt{2}(\gamma_P n\tau)^{1/4}\frac{t}{T_2^*}\right], \tag{2.16}$$

leading to an additional suppression of large-$n$ cavity revivals by a factor $\bar{G}_n \simeq 2e^{-2\sqrt{\gamma_P n\tau}}$ [29]. This suppression limits the number of revivals (echoes) that can be measured before coherence decays to zero, and hence, the signal that can be extracted in each cycle from measurements on the transmission line.

Coherence is transferred from the qubit to the cavity via Eq. (2.4), and from the cavity to the output transmission line via $\dot{r}_{k,2}(t) = -i\omega_k r_{k,2}(t) - i\eta_{k,2}a(t)$. After (i) integrating these equations of motion, (ii) tracing out the cavity, qubit, and environment, and (iii) averaging over realizations of the noise $\eta(t)$, we obtain the reduced density matrix $\rho_{\mathrm{TL}}$ of the output transmission line. Provided there is at most one photon in the transmission line (this limit can always be reached by reducing $\kappa_2/\kappa$),

$$\rho_{\mathrm{TL}} = (1-S)\rho_{\mathrm{inc}} + S|\psi\rangle\langle\psi|, \tag{2.17}$$

where $\rho_{\mathrm{inc}}$ is the incoherent part of the density matrix [$\mathrm{Tr}(r_{k,2}\rho_{\mathrm{inc}}) = 0\ \forall k$] and $|\psi\rangle = \frac{1}{\sqrt{2}}(|0\rangle + |1\rangle)$. Here, $|1\rangle = 2S^{-1}\sum_k \langle r_{k,2}\rangle r_{k,2}^\dagger|0\rangle$, where $S = 2[\sum_k |\langle r_{k,2}\rangle|^2]^{1/2}$. For $S \to 1$, information about the qubit coherence dynamics is fully transferred into the pure state $|\psi\rangle$ of a two-level system, allowing, in principle, up to one bit of information to be extracted per measurement cycle. Typically, however, $S \ll 1$ will be realized, yielding $\ll 1$ bit of information per cycle. For example, we find that a Hahn echo sequence (CPMG with $N = 1$) leads to [29]

$$S \leq S_{\mathrm{Hahn}} = \frac{\sqrt{5\pi}}{2} g T_2^* \left(\frac{\kappa_2}{\kappa}\right)^{1/2}, \tag{2.18}$$

limited by $gT_2^* \ll 1$. For a large-$N$ CPMG sequence, by contrast, we find a significantly larger bound,

$$S \lesssim S_{\mathrm{CPMG}} = \frac{2\sqrt{\pi}}{3}\left[\left(\frac{\kappa_2}{\kappa}\right)\left(\frac{1}{\kappa\tau}\right)\right]^{1/2}, \tag{2.19}$$

still limited by the small parameter $1/\kappa\tau \ll 1$. The CPMG signal is limited because Purcell decay is always active, while coherence is only transferred from the qubit to the transmission line for a small fraction of the time $\sim T_2^*/\tau \ll 1$. Since the times $t = n\tau$ of the revivals are known, we can improve on this limit if the coupling $g = g(t)$ or the detuning $\delta = \delta(t)$ is pulsed to eliminate Purcell decay for $|t - n\tau| \gtrsim T_2^*$. In this case, we find a maximum achievable signal

$$S \lesssim S_{\max} = \left(\frac{\kappa_2}{\kappa}\right)^{1/2} \tag{2.20}$$

that approaches $S_{\max} \simeq 1$ for $\kappa_2 \simeq \kappa$ [29]. Transient spectroscopy can therefore achieve the same efficiency as a single-shot readout (one bit per cycle).

A central finding of this Letter is an even/odd modulation of echo revivals under dynamical decoupling. This modulation (a unique signature of quantum noise) results from the non-stationary analog of a Lamb shift arising from quantum fluctuations of an environment, an important indicator of non-classical and non-Markovian dynamics [53, 54]. When the correlation time of the environment is short compared to the typical observation time (Markovian limit), the quantum-noise phase $\Phi_q(t)$ will advance approximately linearly, $\Phi_q(t) \simeq \Delta\omega_{\text{Lamb}}t$, reflecting a simple frequency shift. In contrast, for a non-Markovian system, $\Phi_q(t)$ may have a highly nontrivial time dependence, reflecting a complex quantum dynamics. This phase may be amplified under repeated fast qubit rotations, which have the effect of stroboscopically driving the environment away from stationarity. Ignoring this effect during a quantum computation may then lead to an accumulation of phase errors that could otherwise be fully corrected.

*Note added*—Following submission, we became aware of Ref. [55], which considers the influence of low-frequency qubit dephasing noise on the transient cavity transmission. In addition to the free-induction decay considered in Ref. [55], we also consider (i) dynamical decoupling sequences applied to the qubit, (ii) quantum noise, (iii) strategies for maximizing the signal, and (iv) cavity-induced backaction (an effect that is higher-order in $g$). As shown here, cavity-induced backaction ultimately sets the limiting timescale (determined by the inhomogeneously broadened Purcell decay time rather than by the cavity decay time $1/\kappa$) for monitoring qubit coherence through the cavity.

**Acknowledgments**—We thank A. Blais for useful discussions and Ł. Cywiński for both pointing out an error in our expression for the quantum filter function in an earlier manuscript and for bringing Refs. [41, 42] to our attention. We also acknowledge funding from the Natural Sciences and Engineering Research Council (NSERC) and from the Fonds de Recherche–Nature et Technologies (FRQNT).

# Appendices to Chapter 2

*These appendices contain the original text of the supplementary material to*
Z. McIntyre and W. A. Coish, Phys. Rev. Research **4**, L042039 (2022)

This supplement provides derivations of several results from the main text. In Sec. A1, we introduce a toggling-frame transformation that accounts for the effects of qubit driving. This toggling frame will be used to perform subsequent calculations. In Sec. A2, we relate the cavity field to qubit coherence, leading to Eq. (2.4). In Sec. A3, we analyze the effects of cavity-induced backaction on the qubit and find that inhomogeneously broadened Purcell decay leads to a stretched-exponential decay of the qubit echo envelope as well as a modification of the shape of coherence revivals. We additionally consider qubit coherence under an $N$-pulse Carr-Purcell-Meiboom-Gill (CPMG) dynamical-decoupling sequence, leading to the expression for the cavity field in terms of the CPMG echo envelope presented in Eq. (2.8). Section A4 gives explicit expressions for the quantum and classical noise contributions to Eq. (2.13), while Sec. A5 presents the expression for the cavity transmission $A_{\mathrm{T}}(\omega,\eta)$ used to generate Fig. 2.2(a). Finally, in Sec. A6, we quantify limits on the maximum signal and consider a modified protocol for which the signal is not limited by either inhomogeneous broadening or cavity-induced backaction.

## A1 The toggling frame

We consider a qubit ("system") interacting with a quantum environment through the lab-frame Hamiltonian

$$H_{\mathrm{lab}}(t)=\frac{1}{2}[\Delta+\eta(t)+h]\sigma_z+H_{\mathrm{E}}+H_{\mathrm{drive}}(t), \tag{A1}$$

where $\Delta$ is the bare qubit splitting, $h$ is an operator acting exclusively on the environment (with free Hamiltonian $H_{\mathrm{E}}$), and $H_{\mathrm{drive}}(t)$ is an arbitrary drive acting only on the qubit [having Pauli-z operator $\sigma_z=|e\rangle\langle e|-|g\rangle\langle g|$ for excited (ground) state $|e(g)\rangle$]. Classical dephasing noise is described by $\eta(t)$, which we assume is generated by stationary, zero-mean Gaussian noise fully described by the noise spectrum given in Eq. (2.3),

$$S_\eta(\omega)=\int dt\, e^{-i\omega t}\langle\!\langle\eta(t)\eta(0)\rangle\!\rangle, \tag{A2}$$

where $\langle\!\langle\cdots\rangle\!\rangle$ indicates an average over realizations of $\eta(t)$. Inhomogeneous broadening [low-frequency fluctuations in $\eta(t)$] will then lead to Gaussian free-induction decay of the qubit on a time scale $T_2^*$ given by

$$\frac{2}{(T_2^*)^2}=\int_{-\infty}^{\infty}\frac{d\omega}{2\pi}S_\eta(\omega). \tag{A3}$$

We assume that prior to the start of the experiment (at $t=0$), the environment is allowed to reach a steady-state in contact with the qubit in its ground state. The initial state of the qubit and environment is therefore taken to be of the form $|g\rangle\langle g|\otimes\bar{\rho}_{\mathrm{E}}$, where $\bar{\rho}_{\mathrm{E}}$ is stationary with respect to the Hamiltonian conditioned on the qubit being in $|g\rangle$:

$$\left[H_{\rm E}-\frac{h}{2},\bar{\rho}_{\rm E}\right]=0. \tag{A4}$$

This initial condition, which should be realized generically in experiment, is crucial for our analysis of quantum noise in Sec. A4, below. The more common "weak coupling" initial condition, $[H_{\rm E},\bar{\rho}_{\rm E}]=0$, yields no quantum noise term for coupling $h\sigma_z/2$ [31, 42].

We treat the deviation from this conditioned Hamiltonian $H_{\rm E}-h/2$ as a perturbation to the environment dynamics. In an interaction picture defined with respect to $H_{\rm E}-h/2$, the system-environment Hamiltonian then reads

$$\begin{aligned}H_{\rm SE}(t)&=U_{\rm E}^{\dagger}(t)H_{\rm lab}(t)U_{\rm E}(t)-iU_{\rm E}^{\dagger}(t)\dot{U}_{\rm E}(t),\quad U_{\rm E}(t)=e^{-i(H_{\rm E}-h/2)t}\\&=\frac{1}{2}[\Delta+\Omega(t)]\sigma_z+\frac{1}{2}h(t)+H_{\rm drive}(t),\end{aligned} \tag{A5}$$

where we have introduced $\Omega(t)=\eta(t)+h(t)$.

We now consider a master equation for the joint state $\rho(t)$ of the driven qubit and quantum environment, coupled to a cavity mode (with annihilation operator $a$), together with a quasi-continuum of transmission-line modes coupled to the cavity input and output ports, all of which evolve via a time-dependent Hamiltonian $H(t)$. In addition, we assume the qubit has a dephasing rate $\gamma_\phi$ independent of the quantum environment, and that the occupation of the cavity mode decays at an extrinsic rate $\kappa_{\rm ext}$ independent of coupling to the input and output transmission lines [Eq. (2.1) and Fig. 2.1]:

$$\dot{\rho}(t)=-i[H(t),\rho(t)]+\frac{\gamma_\phi}{2}\mathscr{D}[\sigma_z]\rho(t)+\kappa_{\rm ext}\mathscr{D}[a]\rho(t). \tag{A6}$$

Here, the damping superoperator acts according to $\mathscr{D}[\mathscr{O}]\rho=\mathscr{O}\rho\mathscr{O}^{\dagger}-\{\mathscr{O}^{\dagger}\mathscr{O},\rho\}/2$ for an arbitrary operator $\mathscr{O}$, and the Hamiltonian $H(t)$ can be written in terms of the system-environment Hamiltonian [Eq. (A5)] derived above:

$$H(t)=H_{\rm SE}(t)+\omega_c a^{\dagger}a+g\sigma_x(a^{\dagger}+a)+\sum_{i=1,2}\sum_k(\lambda_{k,i}e^{i\omega_k t}r_{k,i}^{\dagger}a+\text{h.c.}), \tag{A7}$$

where the qubit couples to the cavity mode with a Rabi coupling of strength $g$, and the cavity mode couples to the transmission-line modes with strengths $\{\lambda_{k,i}\}$. The mode of the input ($i=1$) or output ($i=2$) transmission line having freqency $\omega_k$ is associated with an annihilation operator $r_{k,i}$.

We now transform to a toggling frame to account for the effect of the qubit drive $H_{\rm drive}(t)$. Unitary evolution $U(t)$ under the full Hamiltonian $H(t)$ is related to the toggling-frame unitary $\tilde{U}(t)$ through

$$U(t)=\mathscr{T}e^{-i\int_0^t dt' H(t')}=U_{\rm TF}(t)\tilde{U}(t), \tag{A8}$$

where $\mathscr{T}$ is the time-ordering operator, and where

$$U_{\rm TF}(t)=U_{\rm drive}(t)R(t). \tag{A9}$$

Here, $U_{\rm drive}(t)=\mathscr{T}e^{-i\int_0^t dt' H_{\rm drive}(t')}$ eliminates evolution under $H_{\rm drive}(t)$, and $R(t)$ defines the rotating frame subject to $U_{\rm drive}(t)$: $R(t)=\mathscr{T}e^{-i\Delta\int_0^t dt'[a^{\dagger}a+U_{\rm drive}^{\dagger}(t')\sigma_z U_{\rm drive}(t')/2]}$. This transformation allows for a simpler analysis of observables $\tilde{\mathscr{O}}(t)$ evolving under the action of the toggling-frame Hamiltonian $\tilde{H}(t)$:

$$\tilde{H}(t)=U_{\rm TF}^{\dagger}(t)H(t)U_{\rm TF}(t)-iU_{\rm TF}^{\dagger}(t)\dot{U}_{\rm TF}(t), \tag{A10}$$

$$\tilde{\mathscr{O}}(t)=\tilde{U}^{\dagger}(t)\mathscr{O}\tilde{U}(t),\quad \tilde{U}(t)=\mathscr{T}e^{-i\int_0^t dt'\tilde{H}(t')}. \tag{A11}$$

The expectation value $\langle \mathscr{O} \rangle_t$ can then be related to $\langle \tilde{\mathscr{O}} \rangle_t$ through

$$\langle \mathscr{O} \rangle_t = \langle\!\langle \mathrm{Tr}\{U^\dagger(t)\mathscr{O}U(t)\rho(0)\} \rangle\!\rangle = \langle\!\langle \mathrm{Tr}\{\hat{\mathscr{O}}(t)\tilde{\rho}(t)\} \rangle\!\rangle; \quad \tilde{\rho}(t) = \tilde{U}(t)\rho(0)\tilde{U}^\dagger(t), \tag{A12}$$

where we have included both the quantum average $\mathrm{Tr}\{\cdots\rho(0)\}$ and the classical average over noise realizations $\langle\!\langle \cdots \rangle\!\rangle$ in the definition of the expectation value $\langle \cdots \rangle_t$. Furthermore, we denote by a "hat" the analog of an interaction-picture operator:

$$\hat{\mathscr{O}}(t) = U^\dagger_{\mathrm{TF}}(t)\mathscr{O}U_{\mathrm{TF}}(t). \tag{A13}$$

These definitions give, for example, $\langle a \rangle_t = \langle\!\langle \mathrm{Tr}\{\hat{a}(t)\tilde{\rho}(t)\} \rangle\!\rangle = e^{-i\Delta t}\langle\!\langle \mathrm{Tr}\{a\tilde{\rho}(t)\} \rangle\!\rangle = e^{-i\Delta t}\langle \tilde{a} \rangle_t$. The toggling-frame density operator evolves under

$$\dot{\tilde{\rho}}(t) = -i[\tilde{H}(t), \tilde{\rho}(t)] + \frac{\gamma_\phi}{2}\mathscr{D}[\hat{\sigma}_z(t)]\tilde{\rho}(t) + \kappa_{\mathrm{ext}}\mathscr{D}[a]\tilde{\rho}(t), \tag{A14}$$

where, in writing the transformed damping superoperators, we have used $\hat{\sigma}_z^2(t) = \sigma_z^2 = 1$ and $\hat{a}(t) = e^{i\Delta t}a$. In terms of the cavity-qubit detuning $\delta = \omega_c - \Delta$, the toggling-frame Hamiltonian is now given by

$$\tilde{H}(t) = \tilde{H}_{\mathrm{SE}}(t) + \delta a^\dagger a + g\hat{\sigma}_x(t)[e^{i\Delta t}a^\dagger + \mathrm{h.c.}] + \sum_{i=1,2}\sum_k (\lambda_{k,i}e^{i(\omega_k-\Delta)t}r^\dagger_{k,i}a + \mathrm{h.c.}), \tag{A15}$$

where

$$\tilde{H}_{\mathrm{SE}}(t) = \frac{1}{2}\Omega(t)\hat{\sigma}_z(t) + \frac{1}{2}h(t). \tag{A16}$$

## A2 Relating the output field to qubit coherence

We can recover the well-known input-output relation [17] by integrating the Heisenberg equation of motion, $\dot{r}_{k,i}(t) = i[H(t), r_{k,i}(t)]$, resulting in

$$\langle r_{k,i} \rangle_t = e^{-i\omega_k t}\langle r_{k,i} \rangle_0 - i\lambda_{k,i}\int_0^t dt'\, e^{-i\omega_k(t-t')}e^{-i\Delta t'}\langle \tilde{a} \rangle_{t'}, \quad i = 1,2. \tag{A17}$$

Summing Eq. (A17) over a quasi-continuous set of modes $k$ and performing a Markov approximation for wide-bandwidth transmission lines gives the input-output relation

$$r_{\mathrm{out},i}(t) = r_{\mathrm{in},i}(t) - i\sqrt{\kappa_i}e^{-i\Delta t}\langle \tilde{a} \rangle_t, \tag{A18}$$

where

$$r_{\mathrm{out},i}(t) = \sqrt{\frac{c}{L}}\sum_k \langle r_{k,i} \rangle_t, \quad r_{\mathrm{in},i}(t) = \sqrt{\frac{c}{L}}\sum_k e^{-i\omega_k t}\langle r_{k,i} \rangle_0, \quad \kappa_i = \frac{L}{c}|\lambda_i(\omega_c)|^2 \tag{A19}$$

for $\lambda_i(\omega = \omega_k) = \lambda_{k,i}$. Here, we have assumed one-dimensional transmission lines of length $L$ supporting linearly-dispersing modes ($\omega_k = c|k|$) with speed of light $c$. In order to relate the transmission-line dynamics more transparently to qubit coherence dynamics, we consider the quantum Langevin equation for the cavity field $\tilde{a}(t)$. Within the same Markov approximation used to obtain Eq. (A18), this equation reads

$$\dot{\tilde{a}}(t) = -\left(i\delta + \frac{\kappa}{2}\right)\tilde{a}(t) - ige^{i\Delta t}\sigma_x(t), \tag{A20}$$

where $\kappa = \kappa_1 + \kappa_2 + \kappa_{\mathrm{ext}}$. Equation (A20) provides a useful relation between $\sigma_x(t)$ (the lab-frame qubit coherence) and the cavity field $\tilde{a}(t) = e^{i\Delta t}a(t)$. This relationship is valid for an arbitrary qubit drive and for arbitrarily large qubit-cavity coupling $g$.

The goal is now to understand dynamics of the cavity field evolving under Eq. (A20) due to some specific driven qubit dynamics $\sigma_x(t)$. As explained in Sec. A1, above, we assume that the joint state of the qubit, cavity, environment, and transmission line is given, for $t \leq 0$, by $\rho(t \leq 0) = |g,0,0\rangle\langle g,0,0| \otimes \bar{\rho}_{\mathrm{E}}$, where $|\sigma, n_c, \nu\rangle$ denotes the state of the qubit ($\sigma = g, e$), the cavity mode containing $n_c$ photons, and the transmission line ($\nu = 0$ is the vacuum for all $i,k$). The initial state of the environment, $\rho_{\mathrm{E}}(t \leq 0) = \bar{\rho}_{\mathrm{E}}$, is assumed to be stationary for $t < 0$. This is true provided (i) the qubit drive is not turned on until $t = 0$, $H_{\mathrm{drive}}(t < 0) = 0$, (ii) the qubit-environment interaction $h\sigma_z/2$ is secular ($[h\sigma_z, \sigma_z] = 0$, as assumed above), and (iii) the environment has reached a steady-state in contact with the qubit, $[H_{\mathrm{E}} - h/2, \bar{\rho}_{\mathrm{E}}] = 0$ [Eq. (A4)]. The state $|g,0,0\rangle$ is stationary provided either the qubit-cavity coupling vanishes [$g(t) = 0$ for $t < 0$], or the qubit and cavity are far detuned until $t = 0$, so that $|g,0,0\rangle$ is an eigenstate of $H(t)$ with small corrections. Integrating and averaging Eq. (A20) under these assumptions recovers Eq. (2.4),

$$\langle \tilde{a} \rangle_t = -ig \int_{-\infty}^{\infty} dt' \, \chi_c(t-t') e^{i\Delta t'} \langle \sigma_x \rangle_{t'}, \quad \chi_c(t) = e^{-i\delta t - \frac{\kappa}{2} t} \Theta(t), \quad \langle \sigma_x \rangle_t \propto \Theta(t), \tag{A21}$$

where $\Theta(t)$ is a Heaviside function. A finite qubit coherence, $\langle \sigma_x \rangle_t \neq 0$, can be introduced at $t = 0$ with, e.g., a rapid $\pi/2$-pulse, after which the cavity field will evolve according to Eq. (A21).

In terms of the Fourier transform, $\langle \mathscr{O} \rangle_\omega = \int dt \, e^{i\omega t} \langle \mathscr{O} \rangle_t$, Eq. (A21) reads

$$\langle \tilde{a} \rangle_\omega = -ig\chi_c(\omega) \langle \sigma_x \rangle_{\omega+\Delta}, \quad \chi_c(\omega) = \int_{-\infty}^{\infty} dt \, e^{i\omega t} \chi_c(t) = \frac{1}{i(\delta - \omega) + \kappa/2}. \tag{A22}$$

In the limit of low $Q = \omega_c/\kappa < 1$, we can take $\chi_c(\omega) \sim 2/\kappa$ to be flat on the scale of variation of $\langle \sigma_x \rangle_{\omega+\Delta}$ for $\Delta \sim \omega_c$. The cavity field consequently mirrors the dynamics of the qubit time-locally: $\langle a \rangle_t \propto \langle \sigma_x \rangle_t$ [22]. Notably, Eqs. (A21) and (A22) accurately reflect dynamics even in the regime of high $Q$. This is typically the regime of interest for the devices (see, e.g., [4–10]) designed to reach the strong-coupling regime of cavity QED. The high-$Q$ regime also admits a cavity-filter approximation, in which we replace $\langle \sigma_x \rangle_t \simeq \langle \sigma_- \rangle_t$ in the convolution:

$$\langle \tilde{a} \rangle_t \simeq -ig \int_{-\infty}^{\infty} dt' \, \chi_c(t-t') e^{i\Delta t'} \langle \sigma_- \rangle_{t'} \quad [\text{high}-Q: \ \max(|\delta|, \kappa) \ll |\Delta|]. \tag{A23}$$

A direct consequence of this cavity-filter (high-$Q$) approximation is that the cavity field $\langle \tilde{a} \rangle_t$ [and hence, the output field $r_{\mathrm{out},2}(t)$ via Eq. (A18)] will show a unique signature of quantum noise when we consider a dynamical decoupling sequence in Sec. A3, below.

Equation (A23), together with the input-output relation, Eq. (A18), provides a direct link between the dynamics of the output field $r_{\mathrm{out},2}(t)$ and qubit coherence dynamics $\langle \sigma_- \rangle_t$. No stationarity assumption, weak-coupling approximation, or Markov approximation has been made on $\langle \sigma_- \rangle_t$ up to this point. Provided an accurate model of non-Markovian dynamics can be found for $\langle \sigma_- \rangle_t$ (under, say, a dynamical decoupling sequence), this model can be directly tested from a measurement of $r_{\mathrm{out},2}(t)$. Alternatively, non-Markovian dynamics in $\langle \sigma_- \rangle_t$ can be inferred from the transient dynamics of $r_{\mathrm{out},2}(t)$. The qubit dynamics translated to the output field will, however, depend on the effects of the cavity filter and cavity-induced backaction, as we now show.

## A3 Cavity-induced backaction

We assume that qubit coherence is created at $t = 0$ with a rapid $(-\pi/2)_y$-rotation: $\rho(0^+) = |+\rangle\langle +| \otimes \bar{\rho}_{\mathrm{E}} \otimes |0,0\rangle\langle 0,0|$, where $\sigma_x |+\rangle = |+\rangle$, and where $|0,0\rangle$ indicates the simultaneous vacuum state of the cavity and transmission line. (We use the notation $\theta_\alpha$ to indicate an infinitesimal-duration rotation of the qubit by angle $\theta$ about the $\alpha$-axis.) The initial $\pi/2$-pulse is followed by a sequence of dynamical-decoupling $\pi_x$-pulses due to $H_{\mathrm{drive}}(t)$. For such a pulse sequence, $\hat{\sigma}_z(t) = U^\dagger_{\mathrm{drive}}(t) \sigma_z U_{\mathrm{drive}}(t) = s(t)\sigma_z$, where $s(t) = (-1)^{n(t)}$ is a sign function depending on $n(t)$, the number of $\pi$-pulses having taken place up to time $t$. The lab-frame

expectation value $\langle\sigma_-\rangle_t$ in Eq. (A23) is then related to toggling-frame observables via:

$$\langle\sigma_-\rangle_t = \begin{cases} e^{-i\phi(t)}\langle\tilde{\sigma}_-\rangle_t, & n(t) \text{ even} \\ e^{i\phi(t)}\langle\tilde{\sigma}_+\rangle_t, & n(t) \text{ odd} \end{cases}; \quad \phi(t) = \int_0^t dt' s(t')\Delta. \tag{A24}$$

The phase $\phi(t)$ advances at a rate $\dot{\phi}(t) = +\Delta$ for $n(t)$ even and $\dot{\phi}(t) = -\Delta$ for $n(t)$ odd, so $\langle\sigma_-\rangle_t \sim e^{-i\Delta t}$ for all times $t$ up to corrections $\sim \langle\tilde{\sigma}_\pm\rangle_t$. To solve for $\langle\tilde{\sigma}_\pm\rangle_t$, we evaluate the Heisenberg equations of motion under $\tilde{H}(t)$ [from Eq. (A15)], which we rewrite as

$$\tilde{H}(t) = \tilde{H}_{\mathrm{SE}}(t) + \delta a^\dagger a + \sum_{i=1,2}\sum_k (\lambda_{k,i} e^{i(\omega_k-\Delta)t} r_{k,i}^\dagger a + \text{h.c.}) + \begin{cases} g(e^{i(\phi(t)-\Delta t)}\sigma_+ a + \text{h.c.}) + \text{c.r.}, & n(t) \text{ even} \\ g(e^{i(\phi(t)+\Delta t)}\sigma_+ a^\dagger + \text{h.c.}) + \text{c.r.}, & n(t) \text{ odd} \end{cases}, \tag{A25}$$

where

$$\tilde{H}_{\mathrm{SE}}(t) = \frac{1}{2}\Omega(t)s(t)\sigma_z + \frac{1}{2}h(t); \quad \Omega(t) = \eta(t) + h(t). \tag{A26}$$

In Eq. (A25), "c.r." indicates counter-rotating terms $\sim e^{\pm i2\Delta t}$ that lead to small corrections for $|g| \ll |\delta \pm \Delta|$. The usual excitation-preserving co-rotating terms ($\sim \sigma_+ a$ and $\sim \sigma_- a^\dagger$) for $n(t)$ even are replaced by excitation non-conserving terms ($\sim \sigma_- a$ and $\sigma_+ a^\dagger$) for $n(t)$ odd. These will generally lead to cavity heating, making the analysis of qubit-cavity dynamics under a dynamical decoupling sequence more challenging than for the undriven case [56]. These effects can nevertheless be controlled in the appropriate limits. Neglecting counter-rotating terms in $\tilde{H}(t)$, the equation of motion for $\tilde{\sigma}_-(t)$ is then given [within the rotating-wave approximation (RWA)] by

$$\dot{\tilde{\sigma}}_-(t) \simeq i[\tilde{H}_{\mathrm{SE}}(t), \tilde{\sigma}_-(t)] - \gamma_\phi \tilde{\sigma}_-(t) + \begin{cases} ige^{i[\phi(t)-\Delta t]}\tilde{\sigma}_z(t)\tilde{a}(t), & n(t) \text{ even} \\ ige^{i[\phi(t)+\Delta t]}\tilde{\sigma}_z(t)\tilde{a}^\dagger(t), & n(t) \text{ odd} \end{cases}. \tag{A27}$$

$$(\mathrm{RWA}: g \ll |\delta \pm \Delta|)$$

The bilinear terms $\sim \tilde{\sigma}_z(t)\tilde{a}(t)$ and $\sim \tilde{\sigma}_z(t)\tilde{a}^\dagger(t)$ make a general integration of these equations difficult. However, we note that for free-induction decay [$n(t) = 0$ for all $t$], the dynamics are restricted (under the rotating-wave approximation and for an undriven cavity with $\kappa_1 = 0$) to the subspace spanned by $\{|g,0,0\rangle, |e,0,0\rangle, |g,1,0\rangle, |g,0,k\rangle\}$, where $|g,0,k\rangle = r_{k,2}^\dagger|g,0,0\rangle$. In this case, the state of the qubit and cavity is restricted to the bottom three rungs of the Jaynes-Cummings ladder. Within this subspace, we have $\langle\tilde{\sigma}_z(t)\tilde{a}(t)\rangle = -\langle\tilde{a}\rangle_t$ for all time, allowing for a direct solution to the coupled equations for $\langle\tilde{\sigma}_-\rangle_t$ and $\langle\tilde{a}\rangle_t$. As described above, this exact replacement is no longer possible under a dynamical decoupling sequence. However, we can still justify a similar approximate replacement provided (i) that $g \ll \kappa$, so that the cavity contains at most one photon at any time, and (ii) that the minimum time $\tau$ between $\pi$-pulses is long compared to the timescale $\kappa^{-1}$ of cavity transients. Under these conditions, we perform the restricted-subspace approximation:

$$\langle\tilde{\sigma}_z(t)\tilde{a}(t)\rangle \simeq -\langle\tilde{a}(t)\rangle, \quad n(t) \text{ even}; \quad \langle\tilde{\sigma}_z(t)\tilde{a}^\dagger(t)\rangle \simeq \langle\tilde{a}^\dagger(t)\rangle, \quad n(t) \text{ odd}; \quad (g < \kappa,\ \kappa\tau \gg 1). \tag{A28}$$

The first approximate equality follows from the same logic given above for free-induction decay, $n(t) = 0$. The second approximation follows provided evolution is approximately restricted to the subspace spanned by $\{|g,0,0\rangle, |e,0,0\rangle, |e,1,0\rangle, |e,0,k\rangle\}$ for most of the time when $n(t)$ is odd. These approximations will be violated due to heating effects on a time scale $\sim \kappa^{-1}$ in the vicinity of $\pi$-pulses, but if the time between subsequent $\pi$-pulses is sufficiently long, the cumulative effect of these transients will amount to a small correction to the qubit coherence dynamics.

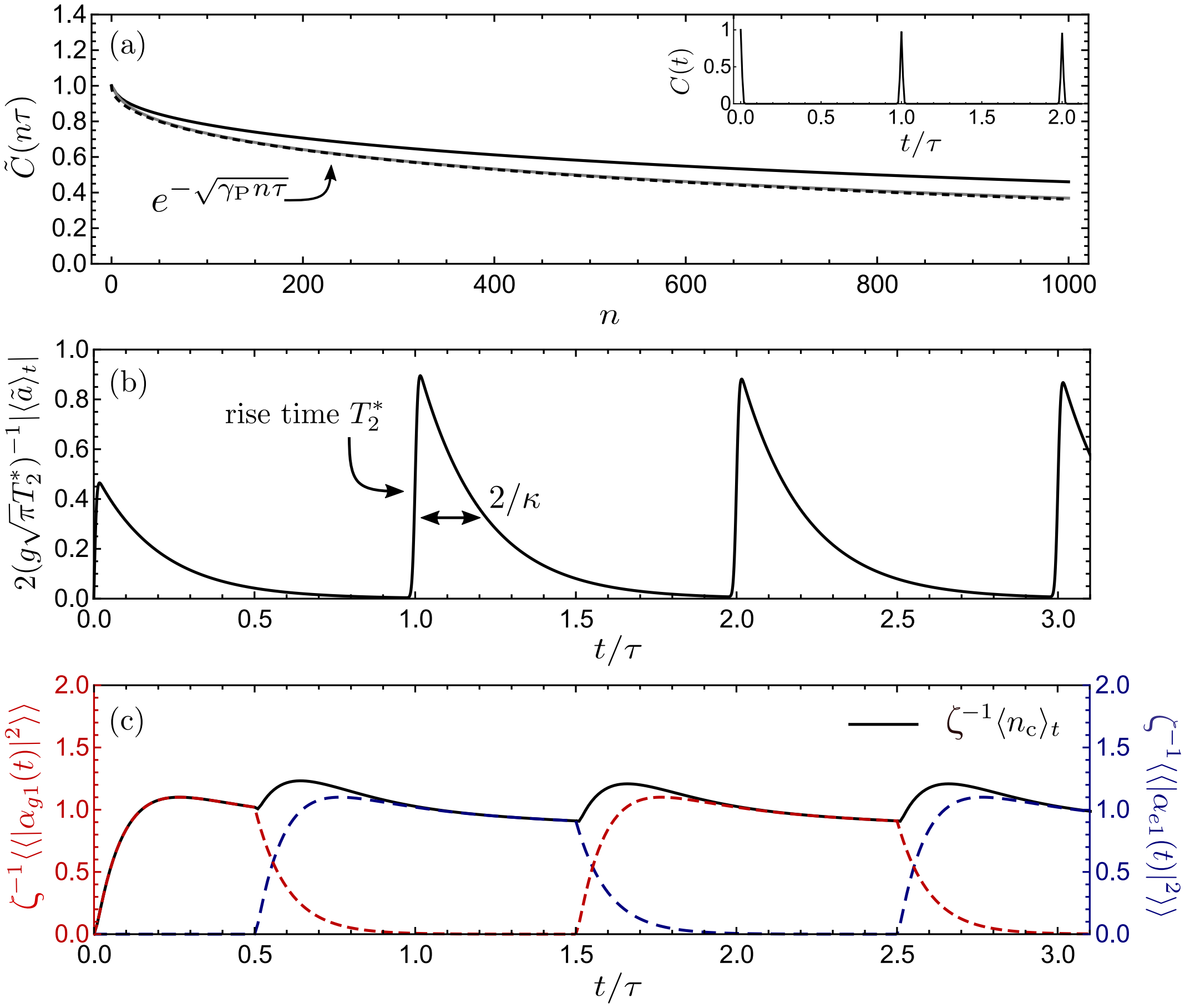

Figure A1: (a) Solid black line: The echo envelope $\tilde{C}(n\tau) = \langle\!\langle \langle\tilde{\psi}(n\tau)|\sigma_-|\tilde{\psi}(n\tau)\rangle \rangle\!\rangle/\langle\sigma_-\rangle_0$, where $|\tilde{\psi}\rangle$ evolves under $\tilde{H}(t)$ in the restricted subspace described by Eq. (A29). Gray line: $\tilde{C}(n\tau) \simeq \langle\!\langle e^{-\Gamma_{\mathrm{P}}(\eta)n\tau/2}\rangle\!\rangle$. Black dashed line: The approximate form $\tilde{C}(n\tau) \simeq e^{-\sqrt{\gamma_{\mathrm{P}} n\tau}}$. Inset: The qubit coherence $C(t)$, consisting of Gaussian revivals with width $\sim T_2^*$ centered at times $t = n\tau$, with $n$ an integer. (b) The cavity field $\langle\tilde{a}\rangle_t = \sum_{\sigma=e,g} \langle\!\langle \alpha^*_{\sigma 0}(t)\alpha_{\sigma 1}(t)\rangle\!\rangle$. (c) Red (blue) dashed line: $\langle\!\langle |\alpha_{g1}(t)|^2\rangle\!\rangle$ [$\langle\!\langle |\alpha_{e1}(t)|^2\rangle\!\rangle$], normalized by $\zeta = \sqrt{\pi} g^2 T_2^*/\kappa$. Black line: The cavity occupation $\langle n_c\rangle_t = \sum_{\sigma=e,g} \langle\!\langle |\alpha_{\sigma 1}(t)|^2\rangle\!\rangle$. We have verified that the inequality $|\langle\tilde{a}\rangle_t|^2 \leq \langle n_c\rangle_t(1-\langle n_c\rangle_t)$ is satisfied for all times $t$, as required by positivity of the cavity density matrix in the subspace of $n_c = 0, 1$. (Though unresolvable on this scale, $\langle n_c\rangle_t$ also rises initially on a timescale $T_2^*$.) For all figures, we take $g = 0.1\kappa$, $\kappa T_2^* = 0.1$ (giving $\sqrt{\pi} g T_2^* \sim 10^{-2}$, $\zeta \sim 10^{-3}$), and $\kappa\tau = 10$.

To illustrate the validity of the restricted-subspace approximation, Eq. (A28), we consider the simplified case of $\gamma_\phi = h(t) = 0$ and additionally assume that the low-frequency noise is static [$\eta(t) = \eta$]. We then directly integrate the Schrödinger equation [$\partial_t |\tilde{\psi}(t)\rangle = -i\tilde{H}(t)|\tilde{\psi}(t)\rangle$] *without* assuming Eq. (A28), but restricting to a subspace that allows for at most one photon in the cavity or transmission line (this can always be justified for $g < \kappa$ at a sufficiently short time):

$$|\tilde{\psi}(t)\rangle = \alpha_{g0}(t)|g,0,0\rangle + \alpha_{e0}(t)|e,0,0\rangle + \alpha_{g1}(t)|g,1,0\rangle + \alpha_{e1}(t)|e,1,0\rangle + \sum_k \left[\alpha_{gk}(t)|g,0,k\rangle + \alpha_{ek}(t)|e,0,k\rangle\right]. \tag{A29}$$

For a fixed value of the noise parameter $\eta$, the Schrödinger equation can be integrated analytically piecewise for each time interval between $\pi$-pulses. After averaging over a Gaussian distribution in $\eta$ values with $\langle\langle \eta^2 \rangle\rangle = 2/(T_2^*)^2$, the resulting contributions $\langle\langle |\alpha_{g1}(t)|^2 \rangle\rangle$ and $\langle\langle |\alpha_{e1}(t)|^2 \rangle\rangle$ to the number of cavity photons, $\langle n_c \rangle_t = \sum_{\sigma=e,g} \langle\langle |\alpha_{\sigma 1}(t)|^2 \rangle\rangle$, are shown in Fig. A1(c) for the case of a Carr-Purcell-Meiboom-Gill (CPMG) sequence:

$$\left(\frac{\tau}{2} - \pi_x - \frac{\tau}{2}\right)^N \quad \text{(CPMG)}. \tag{A30}$$

Here, $\tau/2$ denotes a delay of duration $\tau/2$. Up to small transient corrections on a time scale $\sim 1/\kappa$ around the times of the $\pi$-pulses, we have $\alpha_{e1}(t) \simeq 0$ for $n(t)$ even and $\alpha_{g1}(t) \simeq 0$ for $n(t)$ odd, justifying the approximation in Eq. (A28). In addition, the exact echo envelope for this case, $\tilde{C}(n\tau) = \langle\langle \langle \tilde{\psi}(n\tau)|\sigma_-|\tilde{\psi}(n\tau)\rangle \rangle\rangle / \langle \sigma_- \rangle_0$, is shown in Fig. A1(a), and the cavity field $\langle \tilde{a} \rangle_t = \langle\langle \langle \tilde{\psi}(t)|a|\tilde{\psi}(t)\rangle \rangle\rangle$ is shown in Fig. A1(b). Even in this case, where there is no external source of pure-dephasing dynamics [$\gamma_\phi = h(t) = 0$ and $\eta(t) = \eta$], qubit coherence is still lost due to inhomogeneously broadened Purcell decay, an effect that we now consider in detail.

To separate fast and slow dynamics, we write $\Omega(t) = \eta + \delta\Omega(t)$, where we assume that $\eta$ is a large static contribution (inhomogeneous broadening) and that $\delta\Omega(t) = \delta\eta(t) + h(t)$ generates pure-dephasing dynamics that are slow on the scale $\kappa^{-1}$. We now eliminate the evolution under the fast term $\sim s(t)\eta$ by defining

$$\tilde{\tilde{\sigma}}_-(t) = e^{i\int_0^t dt' s(t')\eta}\,\tilde{\sigma}_-(t). \tag{A31}$$

In terms of this quantity, and within the restricted subspace approximation [Eq. (A28)], Eq. (A27) can be written as

$$\dot{\tilde{\tilde{\sigma}}}_-(t) \simeq i[\tilde{V}(t), \tilde{\tilde{\sigma}}_-(t)] - \gamma_\phi \tilde{\tilde{\sigma}}_-(t) + ig e^{i\int_0^t dt' s(t')\eta} \begin{cases} -e^{i[\phi(t)-\Delta t]}\tilde{a}(t), & n(t)\ \text{even} \\ e^{i[\phi(t)+\Delta t]}\tilde{a}^\dagger(t), & n(t)\ \text{odd} \end{cases}, \tag{A32}$$

where

$$\tilde{V}(t) = \tilde{H}_{\mathrm{SE}}(t) - \frac{1}{2}\eta s(t)\sigma_z = \frac{1}{2}\delta\Omega(t)s(t)\sigma_z + \frac{1}{2}h(t). \tag{A33}$$

In order to rewrite Eq. (A32) as a closed equation, we insert the result for $\tilde{a}(t)$ in terms of $\sigma_-(t)$ within the cavity-filter (high-$Q$) approximation [leading to Eq. (A23) after averaging]. Neglecting contributions $\sim a(0)$, $r_{k,i}(0)$ that vanish under the average ($\langle \tilde{a} \rangle_0 = \langle r_{k,i} \rangle_0 = 0$), this gives:

$$\dot{\tilde{\tilde{\sigma}}}_-(t) \simeq i[\tilde{V}(t), \tilde{\tilde{\sigma}}_-(t)] - \gamma_\phi \tilde{\tilde{\sigma}}_-(t) - i\int_0^t dt' \Sigma(t,t')\tilde{\tilde{\sigma}}_-(t'), \tag{A34}$$

with a time-nonlocal memory kernel (self-energy) given by

$$\Sigma(t,t') = -ig^2 e^{i\int_{t'}^t dt'' s(t'')\eta} \begin{cases} \chi_c(t-t')e^{i[\phi(t)-\phi(t')-\Delta(t-t')]}, & n(t)\ \text{even} \\ \chi_c^*(t-t')e^{i[\phi(t)-\phi(t')+\Delta(t-t')]}, & n(t)\ \text{odd} \end{cases}. \tag{A35}$$

The cavity susceptibility $\chi_c(t-t')$ suppresses contributions for which the times $t,t'$ are well separated; major contributions to the integral thus occur for $t-t' \lesssim \kappa^{-1} \ll \tau$. Except for small intervals of width $\sim 1/\kappa$ around the time of the $\pi$-pulses, we thus have $n(t)=n(t')$ wherever $\chi_c(t-t')$ has significant weight, giving $\phi(t)-\phi(t') \simeq s(t)\Delta(t-t')$. This justifies the following replacements, with small corrections for $\kappa\tau \gg 1$:

$$e^{i[\phi(t)-\phi(t')-\Delta(t-t')]} \simeq 1, \quad [n(t)\,\text{even}]; \quad e^{i[\phi(t)-\phi(t')+\Delta(t-t')]} \simeq 1, \quad [n(t)\,\text{odd}], \qquad (\kappa\tau \gg 1). \tag{A36}$$

With these replacements, the self-energy becomes

$$\Sigma(t,t') \simeq \Sigma_0(t,t-t'), \quad \Sigma_0(t_1,t_2) = -ig^2 e^{-\frac{\kappa}{2}t_2} e^{is(t_1)(\eta-\delta)t_2}, \quad (\kappa\tau \gg 1). \tag{A37}$$

If $\tilde{\tilde{\sigma}}_-(t)$ evolves slowly on the timescale $\kappa^{-1}$, then we can write the equation of motion for $\tilde{\tilde{\sigma}}_-$ in terms of the dispersive shift $\Delta\omega(\eta)$ and Purcell decay rate $\Gamma_P(\eta)$ as

$$\dot{\tilde{\tilde{\sigma}}}_-(t) \simeq i[\tilde{V}(t), \tilde{\tilde{\sigma}}_-(t)] - \left\{ is(t)\Delta\omega(\eta) + \gamma_\phi + \frac{1}{2}\Gamma_{\mathrm{P}}(\eta) \right\} \tilde{\tilde{\sigma}}_-(t), \tag{A38}$$

where

$$i\int_0^\infty dt' \Sigma_0(t,t') = \frac{g^2\left[\kappa/2 + is(t)(\eta-\delta)\right]}{(\eta-\delta)^2 + (\kappa/2)^2} = \frac{1}{2}\Gamma_{\mathrm{P}}(\eta) + is(t)\Delta\omega(\eta). \tag{A39}$$

Integrating Eq. (A38) and transforming back to $\tilde{\sigma}_-(t)$ [via Eq. (A31)] then gives

$$\langle \tilde{\sigma}_- \rangle_t \simeq e^{-\gamma_\phi t} \langle e^{-\frac{\Gamma_{\mathrm{P}}(\eta)}{2}t} e^{-i\int_0^t dt' s(t')[\Delta\omega(\eta)+\eta]} \tilde{U}_-^\dagger(t)\tilde{U}_+(t) \rangle \langle \sigma_- \rangle_0, \tag{A40}$$

where

$$\tilde{U}_\pm(t) = \mathscr{T} e^{-i\int_0^t dt' \tilde{V}_\pm(t')}, \quad \tilde{V}_\pm(t) = \langle e,g|\tilde{V}(t)|e,g\rangle = \frac{1}{2}[h(t) \pm \delta\Omega(t)s(t)]. \tag{A41}$$

As explained in Sec. A1, $\langle \cdots \rangle$ includes both a quantum average and an average over realizations of the noise $\eta(t)$. If we assume that the inhomogeneous broadening $\eta$ is approximately statistically independent of the dynamical contribution $\delta\eta(t) = \eta(t) - \eta$ over the short timescale $\sim T_2^*$ of the coherence revivals, $\langle\langle \delta\eta(t)\eta \rangle\rangle \simeq 0$, then we can write

$$\langle \tilde{\sigma}_- \rangle_t \simeq \left\langle\!\!\left\langle e^{-\frac{\Gamma_{\mathrm{P}}(\eta)}{2}t} e^{-i\int_0^t dt' s(t')[\Delta\omega(\eta)+\eta]} \right\rangle\!\!\right\rangle \tilde{C}_0(t) \langle \sigma_- \rangle_0, \tag{A42}$$

$$\tilde{C}_0(t) = e^{-\gamma_\phi t} \langle \tilde{U}_-^\dagger(t)\tilde{U}_+(t) \rangle, \tag{A43}$$

where $\tilde{C}_0(t)$ describes the contribution to the slowly varying envelope of qubit coherence due to the environment and low-frequency noise, in the absence of cavity coupling.

Provided the restricted-subspace approximation [Eq. (A28)] holds, the result given in Eqs. (A42) and (A43) can be used to describe qubit coherence dynamics under an arbitrary dynamical decoupling sequence. For concreteness, we now specialize to an $N$-pulse CPMG sequence with pulse spacing $\tau$ [Eq. (A30)]. In this case, $\int_0^{n\tau} dt s(t) = 0$, leading to a perfect cancellation of the inhomogeneous broadening $\eta$ and dispersive shift $\Delta\omega(\eta)$ at revival/echo times $t = n\tau$. These echoes are suppressed by an overall decay envelope set by $\langle\langle e^{-\Gamma_{\mathrm{P}}(\eta)n\tau/2} \rangle\rangle$, which gives rise to an asymptotic stretched-exponential behavior arising from the inhomogeneously broadened Purcell decay:

$$\langle\langle e^{-\frac{\Gamma_{\mathrm{P}}(\eta)}{2}n\tau} \rangle\rangle = \int_{-\infty}^\infty d\eta \, \frac{T_2^*}{\sqrt{4\pi}} e^{-\frac{1}{4}\eta^2(T_2^*)^2} e^{-\frac{\Gamma_{\mathrm{P}}(\eta)}{2}n\tau} \sim e^{\left(\frac{\kappa T_2^*}{4}\right)^2} e^{-\sqrt{\gamma_{\mathrm{P}} n\tau}}, \quad n\tau \to \infty, \tag{A44}$$

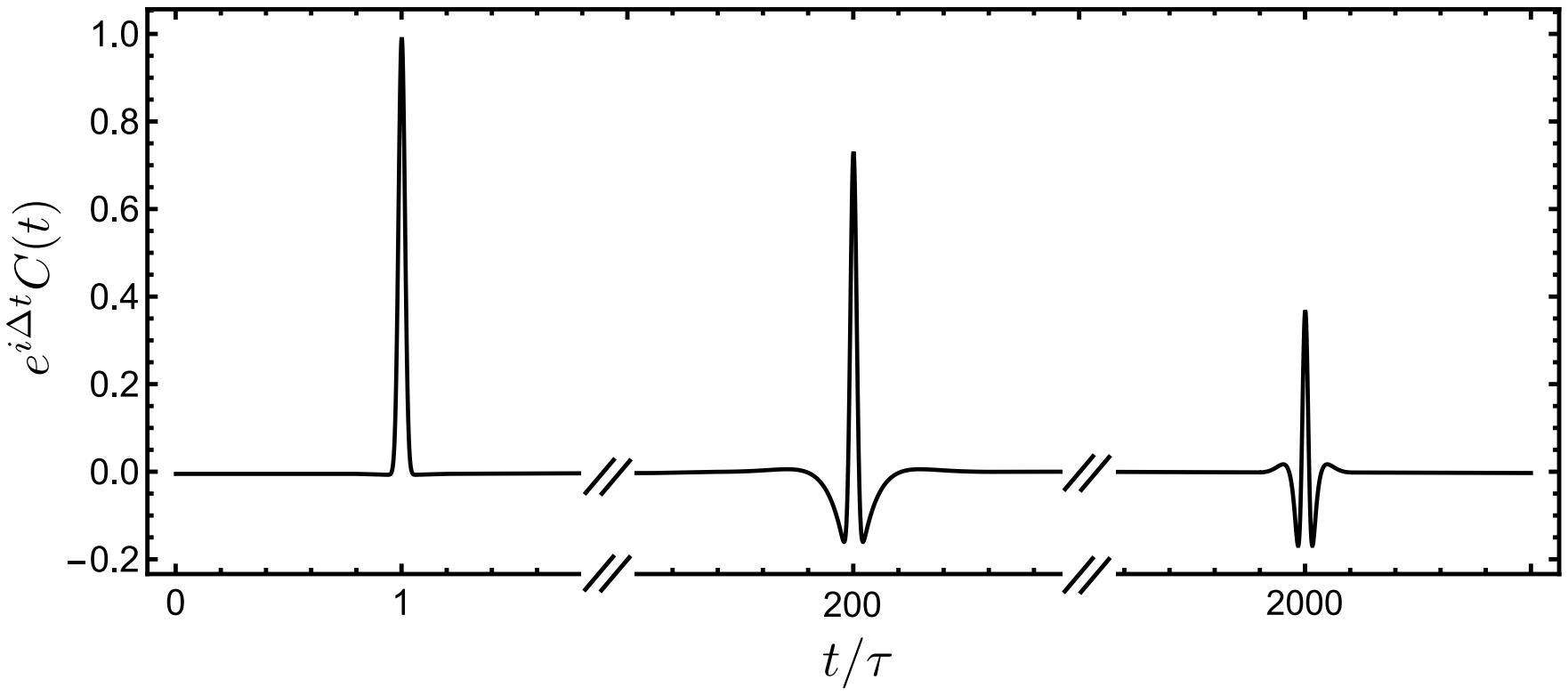

Figure A2: Solid line: Revivals described by $e^{i\Delta t}C(t) = e^{i\Delta t}\langle\sigma_-\rangle_t/\langle\sigma_-\rangle_0$ [evaluated from Eq. (A47) by numerically integrating $G_n(t) = e^{\sqrt{\gamma_P n\tau}}\langle\!\langle e^{-\Gamma_P(\eta)n\tau/2}e^{-i\eta t}\rangle\!\rangle$, assuming $\Delta\tau = 2\pi m$ for some $m \in \mathbb{Z}$ so that $e^{i\Delta t}C(t)$ is real] for three different values of $n$: $n = 1, 200, 2000$. We assume as in Fig. A1 that $g = 0.1\kappa$, $\kappa T_2^* = 0.1$, and $\kappa\tau = 10$, giving $\gamma_P^{-1} = 2000\tau$. For $\gamma_P n\tau \gtrsim 1$, the revivals are suppressed by a stretched exponential. In addition, the Gaussian envelope broadens and becomes modulated by a cosine due to an interplay of inhomogeneous broadening and Purcell decay.

where

$$\gamma_P \simeq (gT_2^*)^2\frac{\kappa}{2}, \quad T_2^*|\delta| \ll 1. \tag{A45}$$

The results displayed in Eqs. (A44) and (A45) were obtained by noting that the integrand ($\propto \exp[-F(\eta)]$) approaches the sum of two Gaussians centered at $\eta = \pm\eta_0$ in the limit $n\tau \to \infty$, where $\eta_0$ is a stationary point $[F'(\eta_0) = 0]$.

Combining Eqs. (A42) and (A44), the total echo envelope is thus given at any echo $t = n\tau$ (for $\kappa T_2^* < 1$) by

$$\tilde{C}(n\tau) = \langle\tilde{\sigma}_-\rangle_{n\tau}/\langle\tilde{\sigma}_-\rangle_0 \simeq e^{-\sqrt{\gamma_P n\tau}}\tilde{C}_0(n\tau), \quad (\kappa T_2^* < 1). \tag{A46}$$

If $\tilde{C}_0(t)$ is slowly-varying on the timescale $\sim T_2^*$ of each echo, this gives:

$$\langle\tilde{\sigma}_-\rangle_t \simeq \langle\sigma_-\rangle_0 \sum_n G_n(t - n\tau)\tilde{C}(n\tau), \tag{A47}$$

where $G_n(t) = e^{\sqrt{\gamma_P n\tau}}\langle\!\langle e^{-\Gamma_P(\eta)n\tau/2}e^{-i\eta t}\rangle\!\rangle$. The same asymptotic analysis described above can be used to find the shape of $G_n(t)$ at short and long times (for $\kappa T_2^* < 1$):

$$G_n(t) \simeq \begin{cases} e^{-\left(\frac{t}{T_2^*}\right)^2}, & \gamma_P n\tau < 1 \\ e^{-\left(\frac{t}{2T_2^*}\right)^2}\cos\left[\sqrt{2}(\gamma_P n\tau)^{1/4}\frac{t}{T_2^*}\right], & \gamma_P n\tau > 1 \end{cases}. \tag{A48}$$

Remarkably, the shape of the echoes changes as $n$ increases: For $n\tau < 1/\gamma_P$, the echoes are simply Gaussians broadened by $T_2^*$, but for $n\tau > 1/\gamma_P$, the width of the echo peaks doubles, and moreover, the peaks show a cosine modulation due to a combination of Purcell decay and inhomogeneous broadening [Fig. A2].

Recalling the relation between the lab-frame expectation value $\langle\sigma_-\rangle_t$ and the toggling-frame observables $\langle\tilde{\sigma}_\pm\rangle_t$ [Eq. (A24)], we use Eq. (A47) to determine that

$$e^{i\Delta t}\langle\sigma_-\rangle_t \simeq \frac{1}{2}\langle\sigma_x\rangle_0 \sum_{n=0}^{N} G_n(t - n\tau)e^{i\Delta n\tau}\begin{cases}\tilde{C}(n\tau), & n \text{ even} \\ \tilde{C}^*(n\tau), & n \text{ odd}\end{cases}, \tag{A49}$$

where we have used the fact that for an initial $(-\pi/2)_y$ pulse on the qubit, $\langle\sigma_-\rangle_0 = \frac{1}{2}\langle\sigma_x\rangle_0$. Substituting Eq. (A49) into Eq. (A23) gives

$$\langle\tilde{a}\rangle_t \simeq \frac{1}{2}\langle\sigma_x\rangle_0\left[\frac{1}{2}f_0(t)+\sum_{n=1}^{N} f_n(t-n\tau)e^{i\Delta n\tau}\mathscr{K}^n\tilde{C}(n\tau)\right], \quad \kappa\tau \gg 1, \tag{A50}$$

where we have introduced the complex conjugation operator $\mathscr{K}$ ($\mathscr{K}z = z^*, z \in \mathbb{C}$), and where the wavepackets $f_n(t)$ are given by

$$f_n(t) = -ig\int_{-\infty}^{t} dt'\chi_c(t-t')G_n(t'). \tag{A51}$$

The first term, $(1/2)f_0(t)$, in Eq. (A50) reflects the fact that revivals with $n \geq 1$ have twice as much "area" as the initial free-induction decay. For $\kappa T_2^* \ll 1$, we can replace $G_n(t)$ in Eq. (A51) by a delta function with appropriate weight:

$$G_n(t) \simeq \delta(t)\sqrt{\pi}T_2^*\bar{G}_n; \quad \bar{G}_n = \frac{1}{\sqrt{\pi}T_2^*}\int_{-\infty}^{\infty} dt\, G_n(t). \tag{A52}$$

In this case, we have the simplified form

$$f_n(t) \simeq -i\sqrt{\pi}gT_2^*\bar{G}_n\chi_c(t) = -i\sqrt{\pi}gT_2^*\bar{G}_n e^{-i\delta t-\frac{\kappa}{2}t}\Theta(t), \quad \kappa T_2^* \ll 1. \tag{A53}$$

The amplitude of the $n^{\text{th}}$ revival in the cavity field due to a CPMG sequence is thus suppressed by the dimensionless factor $\bar{G}_n$ *in addition* to any suppression due to coherence decay. From the asymptotic forms given in Eq. (A48), we have

$$\bar{G}_n \simeq \begin{cases} 1, & n \ll 1/(\gamma_P\tau) \\ 2e^{-2\sqrt{\gamma_P n\tau}}, & n \gg 1/(\gamma_P\tau) \end{cases}. \tag{A54}$$

Even in the absence of external sources of dephasing [$\gamma_\phi = h(t) = 0$], Purcell decay in $\tilde{C}(n\tau) \simeq e^{-\sqrt{\gamma_P n\tau}}$ together with the additional suppression of $\bar{G}_n$ due to the cosine modulation of echoes [Eq. (A48)] results in an asymptotic suppression of cavity revivals:

$$\bar{G}_n\tilde{C}(n\tau) \sim 2e^{-3\sqrt{\gamma_P n\tau}}, \quad \left[\gamma_\phi = h(t) = 0, n\tau \gg 1/\gamma_P\right]. \tag{A55}$$

Taking the Fourier transform of Eq. (A50) and applying the approximations given above for $\kappa T_2^* \ll 1$ gives

$$\langle\tilde{a}\rangle_\omega \simeq -i\frac{\langle\sigma_x\rangle_0}{2}\sqrt{\pi}gT_2^*\chi_c(\omega)\left[\frac{1}{2}+\sum_{n=1}^{N} e^{i(\omega+\Delta)n\tau}\bar{G}_n\mathscr{K}^n\tilde{C}(n\tau)\right], \quad T_2^*\kappa \ll 1, \tag{A56}$$

where $\chi_c(\omega) = [i(\delta-\omega)+\kappa/2]^{-1}$. Since $\chi_c(\omega)$ is peaked at $\omega = \delta$, we can maximize the signal by considering $\omega = \delta$. This gives Eq. (2.8),

$$\langle\tilde{a}\rangle_{\omega=\delta} \simeq -i\langle\sigma_x\rangle_0\frac{\sqrt{\pi}gT_2^*}{\kappa}\left[\tilde{C}_{N,\tau}(\delta)-\frac{1}{2}\right], \tag{A57}$$

where

$$\tilde{C}_{N,\tau}(\omega) = \sum_{n=0}^{N} e^{in\omega\tau}[e^{in\delta_\Delta\tau}\bar{G}_n\mathscr{K}^n\tilde{C}(n\tau)] \tag{A58}$$

is the discrete Fourier transform of $e^{in\delta_\Delta\tau}\bar{G}_n\mathscr{K}^n\tilde{C}(n\tau)$. We have written Eq. (A58) in terms of $\delta_\Delta$, where $\delta_\Delta \equiv \Delta \pmod{2\pi/\tau}$, in order to emphasize that, due to the periodicity of $e^{i\Theta}$ ($\Theta \in \mathbb{R}$), the shift by $\Delta$ of the frequency content of the echo envelope is equivalent to a phase shift due to a smaller quantity $\delta_\Delta$ bounded above by $2\pi/\tau$. [In the case where $\tau = 2\pi m/\Delta$ for some $m \in \mathbb{Z}$, $\delta_\Delta = 0$.] Provided both $\Delta$ and $\tau$ are known, the revival amplitudes $\bar{G}_n \mathscr{K}^n \tilde{C}(n\tau)$ can be recovered by sweeping the detuning over some interval $O(2\pi/\tau)$, inverting the discrete Fourier transform, multiplying by $\bar{G}_n^{-1}$, and, in the case where $\tau \neq 2\pi m/\Delta$, multiplying by the appropriate phase factor ($e^{-i\delta_\Delta n\tau}$).

## A4 Quantum noise

We can find a generic expression for the echo envelope $\tilde{C}_0(n\tau)$ [Eq. (A43)] within a leading-order Magnus expansion and Gaussian approximation. These approximations are generic and conditions for their validity will be satisfied for a broad class of non-Markovian environments.

At $t = 0$, the qubit is subjected to a fast $(-\pi/2)_y$-pulse followed by a dynamical decoupling sequence (starting at $t = 0^+$) consisting of fast $\pi_x$ pulses. As shown in Sec. A3 [Eq. (A43)], the echo envelope is described by

$$\tilde{C}_0(t) = e^{-\gamma_\phi t} \langle \tilde{U}_-^\dagger(t) \tilde{U}_+(t) \rangle, \tag{A59}$$

where [recalling Eq. (A41)] $\tilde{U}_\pm(t) = \mathscr{T} e^{-i\int_0^t dt' \tilde{V}_\pm(t')}$ for $\tilde{V}_\pm(t) = [h(t) \pm \delta\Omega(t)s(t)]/2$. Provided the Magnus expansion converges, the unitaries $\tilde{U}_\pm(t)$ can in turn be expressed using the Magnus expansion as

$$\tilde{U}_\pm(t) = e^{-i\Lambda_\pm(t)}, \quad \Lambda_\pm(t) = \sum_{k=1}^{\infty} \Lambda_\pm^{(k)}(t), \tag{A60}$$

where, in terms of $h(t) = e^{i(H_{\mathrm{E}}-h/2)t} h e^{-i(H_{\mathrm{E}}-h/2)t}$, the first two terms in the sum are given by

$$\Lambda_\pm^{(1)}(t) = \int_0^t dt' \, \frac{1}{2}[(1 \pm s(t'))h(t') \pm \delta\eta(t')s(t')], \tag{A61}$$

$$\Lambda_\pm^{(2)}(t) = -\frac{i}{8} \int_0^t dt' \int_0^{t'} dt'' \, (1 \pm s(t'))(1 \pm s(t''))[h(t'), h(t'')]. \tag{A62}$$

[A sufficient condition for convergence of the Magnus expansion is $\int_0^t dt' |\tilde{V}_\pm(t')| < \pi$ [57], which can be satisfied for weak coupling to an environment, e.g. spins or charge fluctuators, having a bounded spectrum.] Substituting Eq. (A60) into Eq. (A59), we then find using the Baker-Campbell-Hausdorff formula that, neglecting $O(h^3)$ and higher,

$$\tilde{C}_0(t) \simeq e^{-\gamma_\phi t} \langle e^{-i\Lambda_{\mathrm{eff}}(t)} \rangle, \tag{A63}$$

where

$$\Lambda_{\mathrm{eff}}(t) = \Lambda_+(t) - \Lambda_-(t) + \frac{1}{2i}[\Lambda_+(t), \Lambda_-(t)]. \tag{A64}$$

$\Lambda_{\mathrm{eff}}(t)$ is Hermitian, which can easily be verified using the fact that $\Lambda_\pm^\dagger(t) = \Lambda_\pm(t)$. Making a Gaussian approximation, i.e. neglecting third- and higher-order cumulants in $h(t)$ [30], and, without loss of generality, taking $\langle \Lambda_\pm^{(1)}(t) \rangle = 0$ gives

$$\tilde{C}_0(t) \simeq e^{-\gamma_\phi t} e^{-i\Phi_{\mathrm{q}}(t) - \chi(t)}, \tag{A65}$$

where, in terms of $\delta\Omega(t) = \delta\eta(t) + h(t)$, the reduction in amplitude is described by

$$
\begin{aligned}
\chi(t) &= \frac{1}{2}\langle[\Lambda_+^{(1)}(t) - \Lambda_-^{(1)}(t)]^2\rangle \\
&= \frac{1}{4}\int_0^t dt' \int_0^t dt'' s(t')s(t'')\langle\{\delta\Omega(t'), \delta\Omega(t'')\}\rangle \\
&= \frac{1}{2}\int_0^t dt' \int_0^t dt'' s(t')s(t'')\mathrm{Re}\langle\delta\Omega(t')\delta\Omega(t'')\rangle \\
&= \frac{1}{2}\int_0^t dt' \int_0^t dt'' s(t')s(t'')\mathrm{Re}\langle\delta\Omega(|t'-t''|)\delta\Omega\rangle.
\end{aligned}
\tag{A66}
$$

In the last line above, we used stationarity of the correlation function $\langle\delta\Omega(t')\delta\Omega(t'')\rangle$ for our choice of initial state, $[H_\mathrm{E} - h/2, \bar{\rho}_\mathrm{E}] = 0$, as well as symmetry of $\mathrm{Re}\langle\delta\Omega(t')\delta\Omega(t'')\rangle$ under interchange of $t' \leftrightarrow t''$, so that $\mathrm{Re}\langle\delta\Omega(t')\delta\Omega(t'')\rangle = \mathrm{Re}\langle\delta\Omega(|t'-t''|)\delta\Omega\rangle$. For CPMG and $t = n\tau$, this result for $\chi$ can equivalently be written in terms of $\Omega(t) = \eta + \delta\Omega(t)$ [by replacing $\delta\Omega$ with $\Omega$ in Eq. (A66)] since $\int_0^{n\tau} dt s(t)\eta = 0$.

In addition to the reduction in amplitude, there is also a phase given by

$$
\Phi_\mathrm{q}(t) = \langle\Lambda_+^{(2)}(t) - \Lambda_-^{(2)}(t)\rangle + \frac{1}{2i}\langle[\Lambda_+^{(1)}(t), \Lambda_-^{(1)}(t)]\rangle. \tag{A67}
$$

Evaluating the two terms separately gives

$$
\langle\Lambda_+^{(2)}(t) - \Lambda_-^{(2)}(t)\rangle = \frac{1}{2}\int_0^t dt' \int_0^{t'} dt''[s(t') + s(t'')]\mathrm{Im}\langle h(t'-t'')h\rangle \tag{A68}
$$

and

$$
\frac{1}{2i}\langle[\Lambda_+^{(1)}(t), \Lambda_-^{(1)}(t)]\rangle = \frac{1}{2}\int_0^t dt' \int_0^{t'} dt''[s(t') - s(t'')]\mathrm{Im}\langle h(t'-t'')h\rangle, \tag{A69}
$$

where we have replaced $(2i)^{-1}\langle[h(t'), h(t'')]\rangle = \mathrm{Im}\langle h(t')h(t'')\rangle = \mathrm{Im}\langle h(t'-t'')h\rangle$. Adding Eqs. (A68) and (A69) then gives the total phase,

$$
\Phi_\mathrm{q}(t) = \int_0^t dt' \int_0^{t'} dt'' s(t')\mathrm{Im}\langle h(|t'-t''|)h\rangle = \int_0^t dt' \int_0^{t'} dt'' s(t')\mathrm{Im}\langle\delta\Omega(|t'-t''|)\delta\Omega\rangle, \tag{A70}
$$

where introducing the absolute value in $\delta\Omega(|t'-t''|)$ has no effect on the integral since $t' \geq t''$ over the entire region of integration. [It does however allow us to write the classical and quantum parts in a symmetric way in Eq. (A71), below.] Given that $\Phi_\mathrm{q}(t)$ depends on $\mathrm{Im}\langle\delta\Omega(|t|)\delta\Omega\rangle = \langle[\delta\Omega(|t|), \delta\Omega]\rangle/(2i)$, it can *only* arise from a quantum environment. This phase was also derived in Refs. [41, 42], but for the case where $[H_\mathrm{E}, \bar{\rho}_\mathrm{E}] = 0$ and for coupling of the form $h|e\rangle\langle e|$. $\Phi_\mathrm{q}$ was however found to vanish identically for the case where $[H_\mathrm{E}, \bar{\rho}_\mathrm{E}] = 0$ and for coupling $h\sigma_z/2$ [41, 42], highlighting the importance of the initial state (here arising from preparation of $\bar{\rho}_\mathrm{E}$ with the qubit in $|g\rangle$) in determining whether or not a particular form of the coupling leads to a dependence on quantum noise.

We now introduce the spectral density

$$
S(\omega) = S_\mathrm{c}(\omega) + iS_\mathrm{q}(\omega) = \lim_{\varepsilon\to 0^+}\int_{-\infty}^{\infty} dt\, e^{-i\omega t - \varepsilon|t|}\langle\Omega(|t|)\Omega\rangle, \tag{A71}
$$

where the infinitesimal parameter $\varepsilon$ ensures convergence of the integral. Equation (A71) corresponds to Eq. (2.13). In general, there are two contributions to $S(\omega)$: a classical part $S_\mathrm{c}(\omega) = \mathrm{Re}\, S(\omega) = S_h(\omega) + S_\eta(\omega)$ that sets the magnitude ($\sim e^{-\chi}$) of $\tilde{C}_0(t)$, and a quantum part $S_\mathrm{q}(\omega) = \mathrm{Im}\, S(\omega)$ that contributes a phase. The classical and quantum parts $S_h(\omega)$ and $S_{\rm q}(\omega)$ depend, respectively, on symmetrized and anti-symmetrized correlation functions:

$$S_h(\omega) = \frac{1}{2}\int \frac{dt}{2\pi} e^{-i\omega t}\langle\{h(|t|),h\}\rangle, \tag{A72}$$

$$S_{\rm q}(\omega) = \lim_{\varepsilon\to 0^+}\frac{1}{2i}\int \frac{dt}{2\pi} e^{-i\omega t-\varepsilon|t|}\langle[h(|t|),h]\rangle. \tag{A73}$$

For $\int_0^t dt' s(t') = 0$ (as would occur for CPMG and $t = n\tau$), the quantities $\chi(t)$ and $\Phi_{\rm q}(t)$ appearing in $\tilde{C}_0(t)$ [Eq. (A65)] can be written in terms of $S_{\rm c,q}(t)$ as

$$\chi(t) = \int \frac{d\omega}{2\pi}\frac{F_{\rm c}(\omega,t)}{\omega^2}S_{\rm c}(\omega) \tag{A74}$$

and

$$\Phi_{\rm q}(t) = \int \frac{d\omega}{2\pi}\frac{F_{\rm q}(\omega,t)}{\omega^2}S_{\rm q}(\omega), \tag{A75}$$

corresponding to Eqs. (2.11) and (2.12). Here, we have introduced classical-noise (c) and quantum-noise (q) filter functions given by

$$F_{\rm c}(\omega,t) = \frac{\omega^2}{2}\left|\int_0^t dt'\, e^{i\omega t'}s(t')\right|^2 \tag{A76}$$

and

$$F_{\rm q}(\omega,t) = \omega\int_0^t dt' \sin(\omega t')s(t'). \tag{A77}$$

In noise spectroscopy, it is commonly assumed that $S_{\rm q}(\omega) = 0$ and consequently, that $\tilde{C}_0(n\tau)$ is real-valued and positive [35–39]. However, ignoring the quantum-noise correction when it is significant can lead to erroneous conclusions. In these cases, the simple generalization provided by Eq. (A75) can be used to perform accurate noise spectroscopy, even when the quantum noise is significant.

## A5 Effect of inhomogeneous broadening on the cavity transmission spectrum

In order to calculate the cavity transmission spectrum $A_{\rm T}(\omega) = r_{\rm out,2}(\omega)/r_{\rm in,1}(\omega)$ using input-output theory [17], one must consider the coupled quantum Langevin equations of the cavity and (qubit-environment) system. We assume no qubit driving, $H_{\rm drive}(t) = 0$. Within a rotating-wave approximation (RWA) requiring that $g$, $|\delta| \ll |\Delta+\omega_c|$, these equations can be decoupled in linear-response with respect to $g$ [49, 58], resulting in a qubit susceptibility [Eq. (A79), below] that depends only on the eigenstates $|\sigma,m\rangle$ of $H_\sigma = \langle\sigma|\frac{1}{2}h\sigma_z|\sigma\rangle + H_{\rm E}$ for a fixed value of $\sigma\in\{e,g\}$. The effect of inhomogeneous broadening can then be modeled by averaging $A_{\rm T}(\omega) = \langle\!\langle A_{\rm T}(\omega,\eta)\rangle\!\rangle$ over the distribution $\langle\!\langle\cdots\rangle\!\rangle$ of noise realizations. For a time-independent $\eta$, this gives

$$A_{\rm T}(\omega) \simeq \left\langle\!\!\left\langle \frac{-\sqrt{\kappa_1\kappa_2}}{i(\omega_c-\omega)+ig^2\chi_\eta(\omega)+\kappa/2}\right\rangle\!\!\right\rangle, \tag{A78}$$

where

$$\chi_\eta(\omega) = i\sum_{mn}\frac{(p_{en}-p_{gm})|\langle g,m|e,n\rangle|^2}{i(\Delta-\omega+\eta-(\varepsilon_{gm}-\varepsilon_{en}))+\gamma_\phi}. \tag{A79}$$

In Eq. (A79), we denote by $\varepsilon_{\sigma m}$ the eigenenergies of the Hamiltonian $H_\sigma$: $H_\sigma|\sigma,m\rangle = \varepsilon_{\sigma m}|\sigma,m\rangle$. This result also assumes an initial state $\rho(0)$ that is diagonal in the eigenbasis of $\sum_\sigma H_\sigma|\sigma\rangle\langle\sigma|$, so that $\rho(0) = \sum_{\sigma,m}p_{\sigma m}|\sigma\rangle\langle\sigma|\otimes|\sigma,m\rangle\langle\sigma,m|$.

## A6 Quantifying the signal

After each measurement cycle, typically involving an $N$-pulse dynamical decoupling sequence, the transmission line coupled to the output port will be in a quantum state $\rho_{\mathrm{TL}}$ that encodes information about the coherence dynamics that occurred throughout the dynamical decoupling sequence. The amount of information that can be gained per measurement cycle will be limited by both the nature of the (generally mixed) state $\rho_{\mathrm{TL}}$ and by the inference procedure used to extract information from it. Here, we characterize a measure of the signal that depends only on $\rho_{\mathrm{TL}}$.

Provided $\rho_{\mathrm{TL}}$ can be described in the subspace of zero or one photons (this limit can always be reached by taking $\kappa_2/\kappa$ sufficiently small), we can write it in the general form

$$\rho_{\mathrm{TL}} = (1-S)\rho_{\mathrm{inc}} + S|\Psi\rangle\langle\Psi|, \tag{A80}$$

where the incoherent part $\rho_{\mathrm{inc}}$ satisfies $\mathrm{Tr}\{r_{k,2}\rho_{\mathrm{inc}}\} = 0\;\forall k$. In Eq. (A80), the size of the signal is characterized by $S \in [0,1]$, and the coherence is fully described by the state $|\Psi\rangle$ of an effective two-level system:

$$|\Psi\rangle = \frac{1}{\sqrt{2}}(|0\rangle + |1\rangle); \quad |1\rangle = \frac{2}{S}\sum_k \langle r_{k,2}\rangle_t r_{k,2}^{\dagger}|0\rangle; \quad S = 2[\sum_k |\langle r_{k,2}\rangle_t|^2]^{1/2}. \tag{A81}$$

To interpret the meaning of the signal $S$, it is useful to consider an extreme example. We consider a qubit prepared in an initial state determined by some fixed (but initially undetermined) phase $\phi_0$: $\langle\sigma_-\rangle_0 = e^{-i\phi_0}/2$. The qubit is then coupled to the cavity, but otherwise has no source of dephasing: $\gamma_\phi = h(t) = \eta(t) = 0$. We assume for this example that there is no additional dynamics induced through dynamical decoupling $[H_{\mathrm{drive}}(t) = 0]$ and furthermore take $\kappa_1 = \kappa_{\mathrm{ext}} = 0$. After a time $t \gg \Gamma_{\mathrm{P}}^{-1}$, the state of the qubit will have been transferred, via a Wigner-Weisskopf decay process, to a definite pure state $|\Psi\rangle = |\Psi(\phi_0)\rangle$ of the output transmission line, giving $S = 1$. The initial phase $\phi_0$ of the qubit can then be inferred (in principle) through a phase estimation procedure by performing measurements on a well defined two-level subspace of transmission-line states, yielding up to one bit per measurement. In general, the state of the transmission line may be correlated with the state of the qubit, environment, and cavity. These correlations, together with the average $\langle\!\langle\rangle\!\rangle$ over realizations of the random noise parameter $\eta(t)$, will lead to a mixed state $\rho_{\mathrm{TL}}$ with $S < 1$. Having a reduced value $S < 1$ thus sets a fundamental limitation on the information that can be extracted from the complete state of the transmission line. It is straightforward to characterize the maximum achievable signal $S$ given the coefficients $\langle r_{k,2}\rangle_t$ arising from a CPMG sequence, fully accounting for correlations with other degrees of freedom and accounting for random noise. We now proceed with this task.

Limits on the signal $S$ can be found in the present context from the sequence of derivations given above. We substitute Eq. (A53) for the approximate form of the wavepackets $f_n(t)$ (for $\kappa T_2^* \ll 1$) into Eq. (A50) for the cavity field $\langle\tilde{a}\rangle_t$. This result is then substituted into Eq. (A17) for $\langle r_{k,2}\rangle_t$, which, for $\langle r_{k,2}\rangle_0 = 0$, gives

$$\langle r_{k,2}\rangle_t \simeq -\frac{1}{2}\langle\sigma_x\rangle_0\sqrt{\pi}gT_2^*\eta_{k,2}\left[\frac{1}{2}X_{0k}(t) + \sum_{n=1}^{N}\bar{G}_n X_{nk}(t)\mathscr{K}^n\tilde{C}(n\tau)\right], \tag{A82}$$

where (recalling the relation $\delta = \omega_c - \Delta$),

$$X_{nk}(t) = e^{i\omega_k(t-n\tau)}\int_0^t dt' e^{-i(\omega_k-\omega_c)t'-\frac{\kappa}{2}t'}\Theta(t'-n\tau). \tag{A83}$$

For times $t - n\tau \gg \kappa^{-1}$ long compared to the timescale over which cavity transients die out, this object takes

the simple form

$$X_{nk}(t) \simeq \frac{e^{-i\omega_k(t-n\tau)}}{\frac{\kappa}{2} - i(\omega_k - \omega_c)}\Theta(t-n\tau); \quad t - n\tau \gg \kappa^{-1}. \tag{A84}$$

To evaluate $S$, we substitute the expression for $\langle r_{k,2}\rangle_t$ [Eq. (A82)] into $\sum_k |\langle r_{k,2}\rangle_t|^2$. In addition to terms arising from the same echo/revival, proportional to

$$\sum_k |\eta_{k,2}|^2 |X_{nk}(t)|^2 \simeq \sum_k \frac{|\eta_{k,2}|^2}{(\omega_k - \omega_c)^2 + (\kappa/2)^2}\Theta(t-n\tau) \simeq \frac{\kappa_2}{\kappa}, \quad t - n\tau \gg \kappa^{-1}, \tag{A85}$$

there will also be cross terms associated with distinct echoes at times $n\tau$, $m\tau$, with $n \neq m$. These cross-terms will, however, be suppressed exponentially for $\kappa\tau \gg 1$:

$$\sum_k |\eta_{k,2}|^2 X_{nk}(t) X_{mk}^*(t) \simeq \frac{\kappa_2}{\kappa} e^{-|n-m|\kappa\tau/2}\Theta(t-n\tau)\Theta(t-m\tau) \simeq 0; \quad (n \neq m, \kappa\tau \gg 1). \tag{A86}$$

Neglecting these cross terms for $\kappa\tau \gg 1$, we then find that

$$S = \left[|\langle\sigma_x\rangle_0|^2 \pi (gT_2^*)^2 \frac{\kappa_2}{\kappa} N_{\text{eff}}\right]^{1/2}, \tag{A87}$$

where the parameter $N_{\text{eff}}$ scales with the number of revivals/echoes that can be achieved before coherence is lost:

$$N_{\text{eff}} = \frac{1}{4} + \sum_{n=1}^{N} \left|\bar{G}_n \tilde{C}(n\tau)\right|^2, \quad t - N\tau > \kappa^{-1}. \tag{A88}$$

For a Hahn echo sequence ($N = 1$), the maximum signal is achieved for $|\langle\sigma_x\rangle_0| = 1$ and $|\bar{G}_1 \tilde{C}(\tau)| = 1$, giving

$$S \leq S_{\text{Hahn}} = \frac{\sqrt{5\pi}}{2} gT_2^* \sqrt{\frac{\kappa_2}{\kappa}}. \tag{A89}$$

In this case, the total recoverable signal $S$ per cycle is thus limited by $gT_2^* \ll 1$. In the case of a CPMG sequence, we expect the product $|\bar{G}_n \tilde{C}(n\tau)|$ to be upper-bounded in the best case by the asymptotic form given in Eq. (A55), resulting in

$$\begin{aligned} N_{\text{eff}} &\leq \frac{1}{4} + \sum_{n=1}^{\infty} 4e^{-6\sqrt{\gamma_{\text{P}} n\tau}} & \text{(A90)} \\ &\simeq \frac{4}{\gamma_{\text{P}}\tau}\int_0^\infty dx e^{-6\sqrt{x}} = \frac{2}{9\gamma_{\text{P}}\tau}, \quad (\gamma_{\text{P}}\tau \ll 1). & \text{(A91)} \end{aligned}$$

Inserting this result into the definition for $S$ and using the relation $\gamma_{\text{P}} = (gT_2^*)^2\kappa/2$ gives an approximate upper bound on the signal that can be achieved with a CPMG sequence:

$$S \lesssim S_{\text{CPMG}} = \frac{2\sqrt{\pi}}{3}\sqrt{\frac{\kappa_2}{\kappa}\frac{1}{\kappa\tau}}. \tag{A92}$$

For the CPMG sequence, the recoverable signal is not limited by $gT_2^* \ll 1$, but it is still small in the parameter $1/\kappa\tau \ll 1$.

We can improve on the result given in Eq. (A92) by modulating the coupling $g \to g(t)$ [or the detuning $\delta \to \delta(t)$] as a function of time so that $g(t) \neq 0$ [$\delta(t) \lesssim (T_2^*)^{-1}$] only for times $|t - n\tau| \leq t_{\text{on}}$, where $t_{\text{on}} < T_2^*$ is short compared to the duration of a revivial. This has the effect of reducing cavity-induced backaction on the qubit by eliminating incoherent Purcell decay at a rate $\Gamma_P = g^2\kappa/[(\eta-\delta)^2+(\kappa/2)^2]$ for times when the qubit coherence is already suppressed by inhomogeneous broadening. In this case, the self-energy [Eq. (A35)] is given by

$$\Sigma(t,t') \simeq -ig^2\sum_n \Theta_n(t)\Theta_n(t'), \tag{A93}$$

where $\Theta_n(t) = \Theta\left(t-\left(n\tau-\frac{t_{\rm on}}{2}\right)\right) - \Theta\left(t-\left(n\tau+\frac{t_{\rm on}}{2}\right)\right)$. Substituting this result into the equation of motion for $\tilde{\tilde{\sigma}}_-(t)$ [Eq. (A34)] then gives

$$\tilde{C}(n\tau) = \left[1-(gt_{\rm on})^2\right]^n \tilde{C}_0(n\tau), \tag{A94}$$

leading to wavepackets of the form

$$f_n(t) \simeq -igt_{\rm on}\chi_c(t) = -igt_{\rm on}e^{-i\delta t-\frac{\kappa}{2}t}\Theta(t). \tag{A95}$$

Following the same reasoning that led to the limits $S_{\rm Hahn}$ and $S_{\rm CPMG}$ above, we assume the ideal case where $\tilde{C}_0(n\tau)=1$ [in Eq. (A94)] and $|\langle\sigma_x\rangle_0|=1$. This gives an upper bound

$$S \leq \left[(gt_{\rm on})^2\frac{\kappa_2}{\kappa}N_{\rm eff}\right]^{1/2}, \tag{A96}$$

where

$$N_{\rm eff} \leq \frac{1}{4}+\sum_{n=1}^{\infty}\left[1-(gt_{\rm on})^2\right]^n \simeq \frac{1}{(gt_{\rm on})^2}, \quad |gt_{\rm on}| \ll 1. \tag{A97}$$

The signal, being $T_2^*$-independent, is therefore no longer limited by inhomogeneous broadening:

$$S \lesssim S_{\rm max} = \sqrt{\frac{\kappa_2}{\kappa}}. \tag{A98}$$

In this case, for $\kappa_2/\kappa \to 1$, it is possible (at least in principle) to extract one bit of information per cycle, similar to the case of a single-shot readout.

# Preface to Chapter 3

In Chapter 2, we considered a cavity-QED setup with transversal cavity-qubit coupling $\propto \sigma_x(a+a^\dagger)$ and saw how a direct relation between the qubit coherence $\langle\sigma_x\rangle_t$ and cavity field $\langle a\rangle_t$ can be used to perform *in-situ* qubit noise spectroscopy via measurements of the field leaking out of the cavity. Whereas this application falls predominantly in the realm of system characterization or noise mitigation, cavity-QED architectures also hold great appeal for quantum information processing. In the remainder of this thesis, the focus shifts towards controlling and designing light-matter systems with the goal of accomplishing subroutines — such as entanglement generation and stabilizer measurements — that are useful within the broader context of fault-tolerant quantum computing.

As discussed in the introductory chapter, longitudinal coupling $\propto \sigma_z(a+a^\dagger)$ leads to qualitatively different physics as compared to transverse coupling due to its preservation of the bare qubit ($\sigma_z$) eigenstates. The resulting cavity dynamics have been considered in the existing literature as providing a mechanism enabling a fast, quantum-non-demolition qubit readout. In Chapter 3, we show how a parametrically modulated longitudinal coupling can be used to engineer a novel quantum-optical interference effect that allows the state of a qubit to be entangled with the path taken by a propagating wavepacket. The entanglement between the qubit and wavepacket can subsequently be converted into entanglement between two or more distant qubits, allowing for preparation of distant qubits in Bell or GHZ states.

# 3

# A photonic which-path entangler

*This chapter is the integral text from*

**A photonic which-path entangler based on longitudinal cavity-qubit coupling**

Z. M. McIntyre and W. A. Coish

We show that a modulated longitudinal cavity-qubit coupling can be used to control the path taken by a multiphoton coherent-state wavepacket conditioned on the state of a qubit, resulting in a qubit–which-path (QWP) entangled state. QWP states can generate long-range multipartite entanglement using strategies for interfacing discrete- and continuous-variable degrees-of-freedom. Using the approach presented here, entanglement can be distributed in a quantum network without the need for single-photon sources or detectors.

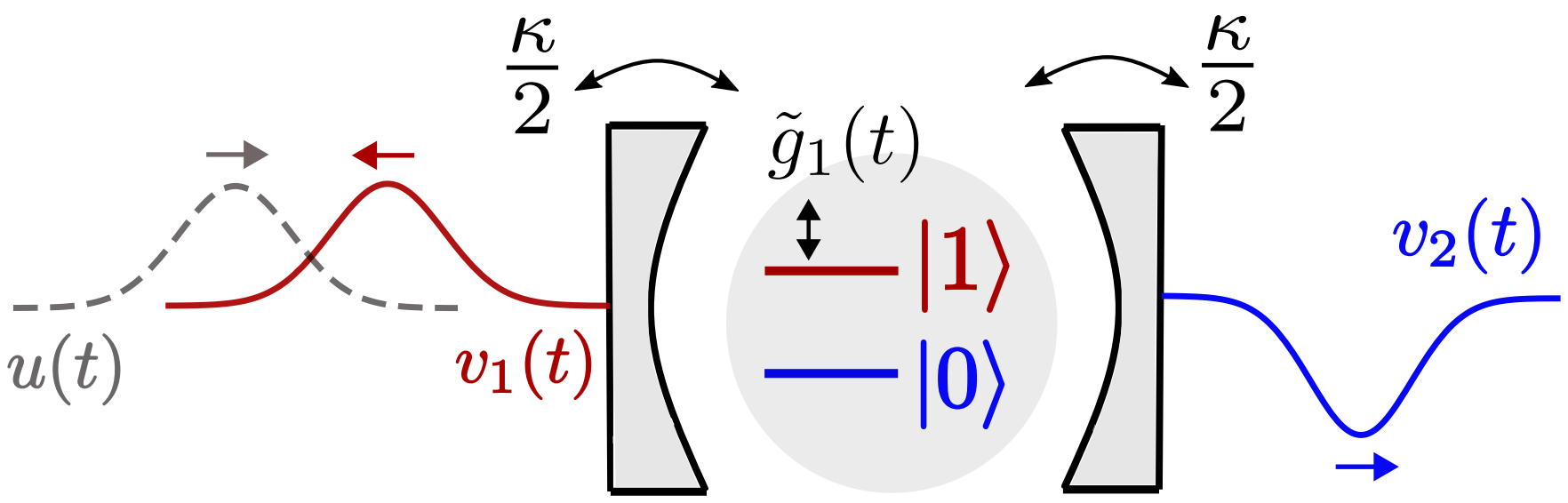

Figure 3.1: In the presence of a longitudinal cavity-qubit coupling modulated at the cavity frequency with envelope $\tilde{g}_1(t)$, an incoming coherent state with waveform $u(t)$ is reflected (transmitted) into a coherent state with waveform $v_1(t)$ [$v_2(t)$] for control-qubit state $|1\rangle$ ($|0\rangle$). This effect requires symmetric decay rates $\kappa_1 = \kappa_2 = \kappa/2$ for cavity ports $i = 1, 2$.

## 3.1 Introduction

Fault-tolerant quantum computing will require redundancy to identify and correct errors during a computation. In most architectures, the physical qubits will therefore vastly outnumber the logical qubits. The need to scale up existing architectures has motivated a network approach where remote qubits, grouped into nodes, are connected by quantum-photonic interconnects [1–5]. These quantum networks naturally require entanglement distribution across nodes. Consequently, significant effort has gone towards generating both heralded [6–12] and deterministic [13–20] qubit-photon entanglement.

In this Letter, we present a photonic which-path entangler that correlates the path of an incoming multi-photon coherent-state wavepacket with the state of a cavity-coupled control qubit (Fig. 3.1). The resulting which-path degree of freedom, consisting of a coherent-state wavepacket traveling in one of two transmission lines, can be re-encoded in the photon-number parity of a continuous-variable degree-of-freedom, then used to generate entanglement with one or more distant qubits. The entangler presented here therefore provides a natural interface between discrete- and continuous-variable approaches to hybrid quantum computation [21–26]. The qubit–which-path (QWP) state generated by the entangler also has greater potential sensitivity for phase measurements than either the comparable entangled coherent state (ECS) [27, 28] (consisting of a superposition of coherent states, one in each interferometer arm) or NOON state [29–31] (an analogous superposition of $N$-photon Fock states). Quantum-enhanced interferometry has applications in, e.g., biological imaging [32–34] and gravitational wave detection [35–38].

A key requirement for the entangler is a modulated longitudinal (qubit-eigenstate preserving) cavity-qubit coupling. Longitudinal coupling has attracted significant theoretical and experimental attention in recent years, as it is currently being realized and leveraged in a number of promising quantum-computing architectures [39–49]. Though many current implementations of, e.g., cavity-coupled spin, charge, and superconducting qubits make use of the (transverse) Rabi coupling, longitudinal cavity-qubit couplings are no less fundamental. They can be engineered for single-electron-spin qubits in double quantum dots (DQDs), which can be coupled to cavity electric fields via magnetic-field gradients [42], as well as for hole-spin qubits in semiconductors [47, 49, 50], which interact with electric fields via their large intrinsic spin-orbit coupling. Two-electron-spin singlet-triplet qubits in DQDs can be longitudinally coupled to electric fields by modulating a gate voltage controlling the strength of the exchange interaction [45, 48]. Longitudinal coupling can also be engineered in various superconducting-qubit architectures [39–41, 43, 44]. Moreover, even in systems where the dominant source of cavity-qubit coupling is intrinsically transverse, an effectively longitudinal interaction can be engineered (in some rotating frame) by modulating the coupling strength at both the cavity and qubit frequencies [46], making the theory presented here widely applicable.

## 3.2 Model

A longitudinal cavity-qubit interaction arises, e.g., from the DC Stark shift due to electric dipole coupling ($\propto Ey$ for a cavity electric field $E$ polarized along $\hat{y}$). Quantizing the cavity field (focusing on a single cavity mode of frequency $\omega_c$ and annihilation operator $a$), and in adiabatic perturbation theory, we find an interaction proportional to the product of $E \propto i(a^\dagger - a)$ and the dipole matrix element $\langle s(t)|y|s(t)\rangle$ taken with respect to the instantaneous qubit energy eigenstate $|s(t)\rangle$ ($s = 0,1$) [51]. This gives an effective Hamiltonian $H_{\text{eff}}(t) = \sum_s ig_s(t)(a^\dagger - a)|s\rangle\langle s|$, where $|s\rangle$ is a time-independent state in the adiabatic frame, and where $g_s(t) = g_s[\{x_j(t)\}] \propto \langle s(t)|y|s(t)\rangle$ inherits time dependence from a collection of control parameters $\{x_j\}$ that determine $|s(t)\rangle$. For spin qubits, the spin-dependent electric dipole matrix elements [and consequently $g_s(t)$] can be modulated through external electric fields or gate voltages [42, 47, 49]. An analogous mechanism exists for flux-tunable superconducting transmon qubits, in which the couplings $g_s(t)$ can instead be tuned by modulating a flux [41]. In what follows, we assume a time-independent value $g_0(t) = \bar{g}_0$ and a sinusoidal modulation of $g_1(t)$ at the cavity frequency, $g_1(t) = \bar{g}_1 + 2|\tilde{g}_1(t)|\cos[\omega_c t - \vartheta(t)]$, where $\tilde{g}_1(t) = e^{i\vartheta(t)}|\tilde{g}_1(t)|$ is a slowly varying envelope with $\tilde{g}_1(0) \simeq 0$ and duration $\tau$. (See the Supplementary Material, Ref. [51], for a sufficient condition on the parameters $\{x_j\}$ in general, as well as specific conditions to achieve this coupling modulation for double-quantum-dot charge and spin qubits.) A polaron transformation can then be used to eliminate the term $\propto \bar{g}_0$ by incorporating a small shift $\sim \bar{g}_0^2/\omega_c$ in the qubit frequency $\omega_q$. Going to an interaction picture with respect to the decoupled Hamiltonian $\omega_c a^\dagger a + \omega_q \sigma_z/2$ ($\sigma_z = |0\rangle\langle 0| - |1\rangle\langle 1|$), and within a rotating-wave approximation requiring that $|\bar{g}_1|, |\tilde{g}_1(t)| \ll \omega_c$, the cavity-qubit Hamiltonian is then given by [51]

$$H_0(t) = \frac{\xi(t)}{2}\sigma_z + i|1\rangle\langle 1|\left[\tilde{g}_1(t)a^\dagger - \text{h.c.}\right], \tag{3.1}$$

where we have introduced a stochastic noise parameter $\xi(t)$ leading to qubit dephasing. In general, the dipole approximation also produces a transverse Rabi term $[ig_\perp \sigma_x(a^\dagger - a)]$, which, in the regime $|g_\perp| < |\delta|$ ($\delta = \omega_q - \omega_c$), leads to a dispersive coupling $\chi\sigma_z a^\dagger a$, where $\chi = g_\perp^2/\delta$. Any effects due to transverse coupling can be suppressed by operating in the regime $|g_\perp| \ll |\delta|$.

The longitudinal interaction $\propto \tilde{g}_1(t)$ displaces the cavity vacuum into a finite-amplitude coherent state for $s = 1$. A similar effect is studied in Ref. [41] to design a fast quantum non-demolition (QND) qubit readout. Relative to Ref. [41], we additionally consider driving of the cavity by an input field. In particular, we assume that the cavity is coupled to external transmission lines at input ($i = 1$) and output ($i = 2$) ports (Fig. 3.1). An input spatiotemporal mode (wavepacket) with normalized waveform $u(t)$ [$\int dt|u(t)|^2 = 1$] can be represented by the mode operator $b_u = \int dt\, u^*(t) r_{\text{in},1}(t)$ [52, 53], where $r_{\text{in},i}(t)$ satisfies the input-ouput relation $r_{\text{out},i}(t) = r_{\text{in},i}(t) + \sqrt{\kappa_i}a(t)$ [54]. Here, $r_{\text{out},i}(t)$ is the output field, and $\kappa_i$ is the rate of decay from cavity port $i$. We assume that the quantum state of the incoming wavepacket is a coherent state with initial amplitude $\langle b_u\rangle = \alpha_0$, giving $\langle r_{\text{in},1}\rangle_t = u(t)\alpha_0$. Where it appears, the notation $\langle \mathscr{O}\rangle_t$ indicates an average with respect to the initial state $\rho(0)$: $\langle \mathscr{O}\rangle_t = \text{Tr}\{\mathscr{O}(t)\rho(0)\}$ for operator $\mathscr{O}$. The reflected and transmitted waveforms are given in the frequency domain by $v_1(\omega) = R(\omega)u(\omega)$ and $v_2(\omega) = T(\omega)u(\omega)$, respectively, where $R(\omega) = \langle r_{\text{out},1}\rangle_\omega / \langle r_{\text{in},1}\rangle_\omega$ and $T(\omega) = \langle r_{\text{out},2}\rangle_\omega / \langle r_{\text{in},1}\rangle_\omega = \sqrt{\kappa_2}\langle a\rangle_\omega / \alpha_0 u(\omega)$ are the reflection and transmission coefficients with $\langle \mathscr{O}\rangle_\omega = \int dt\, e^{i\omega t}\langle \mathscr{O}\rangle_t$.

To derive the transmission $T(\omega)$ conditioned on the qubit state $|s\rangle$, we now find $\langle a\rangle_\omega$ from the quantum Langevin equation for $\langle a\rangle_t$,

$$\langle \dot{a}\rangle_t = -\frac{\kappa}{2}\langle a\rangle_t + \tilde{g}_1(t)s - \sqrt{\kappa_1}\alpha_0 u(t). \tag{3.2}$$

The displacement of the cavity vacuum due to the interaction $\propto \tilde{g}_1(t)$ can therefore be canceled exactly, conditioned on the qubit being in state $|1\rangle$, by ensuring that $\sqrt{\kappa_1}\alpha_0 u(t) = \tilde{g}_1(t)$. Destructive interference then precludes a transfer of photons to the output transmission line, leading to perfect reflection of the input field. Evidence of such destructive interference was recently observed experimentally in Ref. [55], where a modulated longitudinal coupling and a cavity drive were both generated with a common voltage source (acting as a common phase reference). Because the input state is a coherent state [and coherent states are eigenstates of $r_{\mathrm{in},1}(t)$], there are no quantum fluctuations about the average dynamics $\langle r_{\mathrm{in},1}\rangle_t = \alpha_0 u(t)$. For a non-ideal input, however, fluctuations about the average ($\alpha_0 \to \alpha_0 + \delta\alpha$) lead to imperfect cancellation for $s = 1$.

For a cavity that is initially empty, we have $\langle a\rangle_0 = 0$. Integrating the quantum Langevin equation [Eq. (3.2)] with this initial condition gives

$$\langle a\rangle_\omega = \chi_c(\omega)[\tilde{g}_1(\omega)s - \sqrt{\kappa_1}\alpha_0 u(\omega)], \tag{3.3}$$

where $\chi_c(\omega) = (\kappa/2 - i\omega)^{-1}$. For $\sqrt{\kappa_1}\alpha_0 u(\omega) = \tilde{g}_1(\omega)$, the transmission can then be written as

$$T(\omega) = (1-s)\frac{\sqrt{\kappa_1\kappa_2}}{i\omega - \kappa/2}. \tag{3.4}$$

The input pulse $u(\omega)$ has support for $\omega \lesssim 1/\tau$ localized about the cavity frequency (corresponding to $\omega = 0$ in the rotating frame). Near-perfect transmission can then be achieved for $s = 0$ and $\kappa_1 = \kappa_2 = \kappa/2$ by operating in the regime of large $\kappa\tau$, where $\chi_c(\omega)$ is much broader in frequency than $u(\omega)$. Finite-bandwidth effects for a Gaussian input waveform $u(t)$ may be neglected provided [51] $N = |\alpha_0|^2 \ll (\kappa\tau)^4$. Given $T(\omega)$ for a fixed value of $s$, $R(\omega)$ is related to $T(\omega)$ through the input-output relation, $\sqrt{\kappa_2}[R(\omega) - 1] = \sqrt{\kappa_1}T(\omega)$.

An alternative way to condition the cavity transmission on the state of a qubit would be to engineer a qubit-state-dependent shift of the cavity frequency using dispersive coupling $\chi\sigma_z a^\dagger a$, where $2|\chi| \gg \kappa$. A narrow-band input tone at frequency $\chi$ would then be transmitted conditioned on state $|0\rangle$ and reflected for state $|1\rangle$. However, in the dispersive regime ($\varepsilon = |g_\perp/\delta| < 1$), this necessarily requires (very) strong coupling $|g_\perp| \gg \kappa/\varepsilon$. The entangler presented here, by contrast, can be operated even if $|\tilde{g}_1| \lesssim \kappa$. Dipole-induced transparency [56] and reflection [57] also result in perfect steady-state transmission or reflection of a weak input pulse conditional on the presence of a resonant, transversally coupled dipole. These effects are not, however, QND in the state of the decoupled dipole and furthermore require that the cavity be driven with an average of $N_{\mathrm{cav}} \lesssim 1$ intracavity photon [58]. For the entangler presented here, by contrast, the transmission vs reflection of a transient pulse is QND; it is conditioned on the decoupled qubit state $|s\rangle$. Moreover, the entangler works in the regime $N_{\mathrm{cav}} \sim N/(\kappa\tau) > 1$, provided the finite-bandwidth condition is satisfied [$N \ll (\kappa\tau)^4$ for a Gaussian waveform].

The qubit-state-conditioned transmission [Eq. (3.4)] can be used to generate entangled states. To describe the states associated with the reflected and transmitted fields, we use the virtual-cavity formalism of Refs. [52, 53] to recast the input, reflected, and transmitted wavepackets as the fields emitted from—or absorbed into—fictitious (virtual) single-sided cavities coupled to the transmission lines via time-dependent couplings. This formalism allows for an efficient description of the scattering of an input pulse into pre-specified spatiotemporal modes, which can be modeled as the modes of virtual cavities. Accounting for the input pulse, cavity field, reflected pulse, and transmitted pulse, the evolution of the cavity and transmission lines is then fully described by an effective four-mode model. The quantum master equation governing this evolution is [52, 53]

$$\dot{\rho}_\xi = -i[H_0(t) + V(t), \rho_\xi] + \sum_{i=1,2}\mathscr{D}[L_i]\rho_\xi, \tag{3.5}$$

where $\rho_\xi$ is the density matrix conditioned on a realization of the noise $\xi(t)$, and where

$$V(t) = \frac{i}{2}[\sqrt{\kappa_1}\lambda_u^*(t)a_u^\dagger a + \sum_{i=1,2}\sqrt{\kappa_i}\lambda_{v_i}(t)a^\dagger a_{v_i} + \lambda_u^*(t)\lambda_{v_1}(t)a_u^\dagger a_{v_1} - \mathrm{h.c.}]. \tag{3.6}$$

In Eq. (3.5), $\mathscr{D}[L]\rho = L\rho L^\dagger - \frac{1}{2}\{L^\dagger L, \rho\}$ is a damping superoperator. The operators $L_1 = \sum_{\mu=u,v_1} \lambda_\mu(t) a_\mu + \sqrt{\kappa_1} a$ and $L_2 = \lambda_{v_2}(t) a_{v_2} + \sqrt{\kappa_2} a$ model decay from the virtual cavity modes $a_\mu$, as well as decay out of the cavity mode $a$ with rate $\kappa = \kappa_1 + \kappa_2$. The couplings to the virtual cavities are given by $\lambda_u(t) = u(t)/(\int_t^\infty dt' |u(t')|^2)^{1/2}$ and $\lambda_{v_i}(t) = (-1)^i v_i(t)/(\int_0^t dt' |v_i(t')|^2)^{1/2}$ [52, 53, 59]. The $i$-dependent sign in $\lambda_{v_i}$ reflects the fact that in our model, a transmitted pulse undergoes a $\pi$ phase shift [cf Eq. (3.4)]. The couplings $\lambda_u(t)$ and $\lambda_{v_i}(t)$ have singularities at $t \to \infty$ and $t = 0$, respectively, and, for real cavities, both $\lambda_u(t)$ and $\lambda_{v_i}(t)$ would have to be truncated to finite values to realize absorption (emission) into (out of) a chosen spatiotemporal mode [59]. In the virtual-cavity formalism, however, these unphysical couplings are abstractions used to calculate the dynamics into/out of a chosen mode. No additional (real) cavities or time-dependent couplings are required to realize the entangler.

We assume a fast $\pi/2$ pulse can be used to prepare the qubit in the $(t = 0)$ initial state $|+\rangle = (|1\rangle + |0\rangle)/\sqrt{2}$, with the cavity in the vacuum state ($\langle a \rangle_0 = 0$). A coherent-state wavepacket then evolves from the input mode $a_u$ to the two output modes $a_{v_i}$ ($i = 1, 2$). For times $t \gg \tau$ exceeding the duration of the input pulse, the quantum states associated with the reflected and transmitted waveforms $v_i(t)$, conditioned on $s$, will have been fully transferred into their respective fictitious cavities: $\alpha_{is} = \lim_{t\to\infty} \langle a_{v_i} \rangle_t = (-1)^{i-1} \alpha_0 \int \frac{d\omega}{2\pi} |v_i(\omega)|^2$. The joint state of the qubit and transmission lines, found from a direct integration of Eq. (3.5), is then $\rho(t) = \langle\!\langle \rho_\xi(t) \rangle\!\rangle$, where $\langle\!\langle\rangle\!\rangle$ denotes an average over realizations of $\xi(t)$, and where $\rho_\xi(t) = |\Psi_{\xi(t)}\rangle\langle\Psi_{\xi(t)}|$ with

$$|\Psi_{\xi(t)}\rangle = \frac{1}{\sqrt{2}} \left( e^{\frac{i}{2}\theta_\xi(t)} |1, \psi_1\rangle + e^{-\frac{i}{2}\theta_\xi(t)} |0, \psi_0\rangle \right). \tag{3.7}$$

Here, $\theta_\xi(t) = \int_0^t dt' \xi(t')$ is a random phase, $|\psi_s\rangle = \prod_{i=1,2} D_i(\alpha_{is}) |\text{vac}\rangle$ is the state of the transmission lines conditioned on $s$, $|\text{vac}\rangle$ is the vacuum, and $D_i(\alpha) = \exp\{\alpha a_{v_i}^\dagger - \text{h.c.}\}$ is a displacement operator. For $\kappa_1 = \kappa_2 = \kappa/2$ and up to corrections in $N/(\kappa\tau)^4 \ll 1$, only one of $\alpha_{is}$ is nonzero for each value of $s$: For $s = 1$, $\alpha_{11} = \alpha_0$ and $\alpha_{21} = 0$, while for $s = 0$, $\alpha_{10} = 0$ and $\alpha_{20} = -\alpha_0$. Equation (3.7) therefore describes a photonic which-path qubit entangled with the control qubit (a QWP state). Under the same finite-bandwidth conditions, imperfections in the input source ($\alpha_0 \to \alpha_0 + \delta\alpha$) will lead instead to $\alpha_{11} = \alpha_0$, $\alpha_{21} = -\delta\alpha$, $\alpha_{10} = 0$, and $\alpha_{20} = -(\alpha_0 + \delta\alpha)$. This follows from integrating the Langevin equation [Eq. (3.2)] with $\alpha_0 \to \alpha_0 + \delta\alpha$ and solving for the reflected and transmitted fields. If we take $\delta\alpha$ to be a complex-valued, zero-mean Gaussian random variable, then the fidelity of the ideal QWP state [Eq. (3.7) with $\xi = 0$] with respect to the mixed state obtained by averaging over $\delta\alpha$ is $e^{-\langle|\delta\alpha|^2\rangle_{\delta\alpha}}$, where here, $\langle\rangle_{\delta\alpha}$ describes an average over $\delta\alpha$. High-fidelity QWP states therefore require a stable coherent-state source with a low absolute noise level, below one photon per pulse ($\langle|\delta\alpha|^2\rangle_{\delta\alpha} \ll 1$).

### 3.3 Entanglement distribution

The which-path degree-of-freedom can be entangled with a second ("target") qubit for long-range entanglement distribution. Crucially, this also provides a direct avenue for quantifying entanglement in QWP states through measurements of stationary qubits only. The setup is illustrated schematically in Fig. 3.2. By interfering the reflected and transmitted fields at a 50:50 beamsplitter, the output modes can be mapped to new modes $a_\pm$ such that $\langle a_{\pm,s} \rangle = (\alpha_{1s} \pm \alpha_{2s})/\sqrt{2}$. This gives $\langle a_{-,s} \rangle = \alpha$ independent of $s$, where $\alpha = \alpha_0/\sqrt{2}$. The $a_+$ mode, by contrast, has $s$-dependence given by $\langle a_{+,s} \rangle = (2s - 1)\alpha$. The beamsplitter consequently re-encodes the which-path degree-of-freedom in the phase of the coherent-state amplitude: $\langle a_{+,1} \rangle = +\alpha$ and $\langle a_{+,0} \rangle = -\alpha$.

For clarity, we now set $\xi(t) = 0$ for the purpose of explaining the protocol. The effects of dephasing [$\xi(t) \neq 0$] are included in the relevant result, Eq. (3.8), below. Since the $a_-$ mode does not share entanglement with either the qubit or $a_+$ mode, it can be measured (traced over) without disturbing the state of the qubit and $a_+$ mode. The post-measurement state is then $(1/2)\sum_{\lambda=\pm} \sqrt{\mathscr{N}_\lambda} |\lambda, C_\lambda\rangle$, where $|C_\pm\rangle = (|+\alpha\rangle \pm |-\alpha\rangle)/\sqrt{\mathcal{N}_\pm}$ are cat states (consisting of only even or odd photon-number states) and $\mathcal{N}_\pm$ are normalization factors. With the target qubit initialized in $|+\rangle$, the control and target qubits can be entangled through a phase flip on the target, $|+\rangle \to |-\rangle$, conditioned on an odd photon-number parity [51, 60–63]. Following such a phase flip, the state of the qubits and electric field is $(1/2)\sum_{\lambda=\pm}\sqrt{\mathcal{N}_\lambda}\,|\lambda,\lambda,C_\lambda\rangle$. In the limit $\langle\alpha|-\alpha\rangle \to 0$, a final quadrature measurement of the electric field can then be used to project the qubits into the Bell state $(|{+}{+}\rangle \pm |{-}{-}\rangle)/\sqrt{2}$, conditioned on outcome $|\pm\alpha\rangle$. Multi-qubit Greenberger-Horne-Zeilinger (GHZ) states $|+,+,...,+\rangle \pm |-,-,...,-\rangle$ can also be generated by allowing the field to interact sequentially with a series of potentially distant qubits. In contrast to the well-established single-photon pitch-and-catch approach to long-range entanglement distribution, involving the emission and destructive reabsorption of a single photon [13–19], the approach presented here is QND in the photon number and therefore provides a clear path towards the generation of multipartite entangled states.

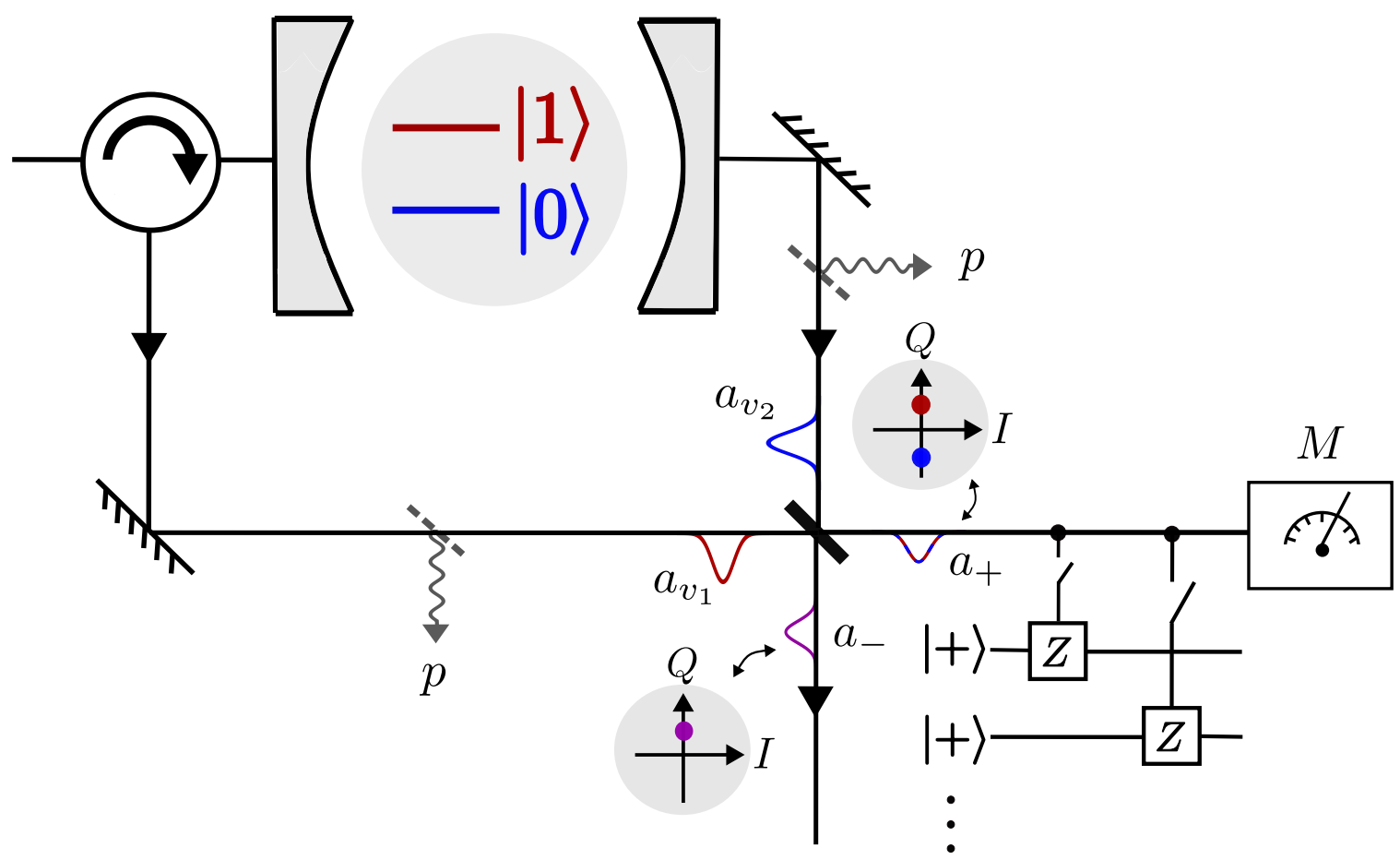

Figure 3.2: An interferometry setup can be used to entangle the which-path degree of freedom with a target qubit initialized in $|+\rangle$ via a conditional phase shift: $Z|+\rangle = |-\rangle$. This can be accomplished by re-encoding the which-path degree-of-freedom in the photon-number parity of a cat state (composed of a superposition of two distinct coherent states occupying the spatiotemporal mode annihilated by $a_+$) propagating to the right of the 50:50 beamsplitter. The symbol labeled $M$ represents a measurement of the quadrature $Q$, which can be used to prepare an $m$-qubit GHZ state (or Bell state for $m = 2$) involving the control qubit and $(m-1)$ target qubits. Photon loss occurring with probability $p$ is modeled with fictitious beamsplitters having reflectivity $p$.

We can quantify the effects of photon loss on the amount of distributed entanglement by inserting a fictitious beamsplitter, described by the unitary $B_{c,c'}(\varphi) = e^{i\varphi(c^\dagger c' + \mathrm{h.c.})}$ [28], into each arm of the interferometer (Fig. 3.2). With $c = a_{v_i}$, this describes scattering from $a_{v_i}$ into loss mode $a'_{v_i}$ (in the environment) with probability $p = \cos^2\varphi$. Tracing over the loss modes $a'_{v_{1,2}}$ then yields a reduced density matrix for the state of the qubit and modes $a_{v_{1,2}}$. In the presence of such loss, and for finite coherent-state overlap $\langle\alpha|-\alpha\rangle \neq 0$, the post-measurement state of the qubits is not an ideal Bell state, but is instead mixed: For a measurement of the electric field along the quadrature $Q$ of coherent-state displacement [64], the post-measurement state of the qubits (conditioned on an inferred displacement along $\pm Q$) is an X-state [65] whose concurrence $C$ [66–68] can easily be computed [51]:

$$C(t) = \max\{0, \mathrm{erf}(\sqrt{N_\eta})e^{-N_p-\chi_\xi(t)} - \mathrm{erfc}(\sqrt{N_\eta})\}, \tag{3.8}$$

where $N_p = pN = p|\alpha_0|^2$ and $N_\eta = \eta(1-p)N$ are controlled by the average number of photons lost and detected, respectively. Here, $\eta \in (0,1]$ is the detector efficiency, and

$$\chi_\xi(t) = \int \frac{d\omega}{2\pi}\frac{4\sin^2\left(\frac{\omega t}{2}\right)}{\omega^2}S(\omega) \tag{3.9}$$

results from an average $\langle\!\langle\rangle\!\rangle$ over realizations of the noise $\xi(t)$, here taken to be stationary, zero-mean Gaussian noise with spectral density $S(\omega) = \int dt\, e^{-i\omega t}\langle\!\langle \xi(t)\xi(0)\rangle\!\rangle$. The concurrence [Eq. (3.8)] quantifies the amount of entanglement that can be distributed to a second qubit in the presence of (symmetric) photon loss in the interferometer, qubit dephasing, and imperfect assignment fidelity at the final measurement of the electric field. In particular, the expression for $C(t)$ indicates that for fixed values of $p$ and $\eta$, there is an optimal $N$ that maximizes the entanglement [51]. In the presence of asymmetric losses with probabilities $p_1$ and $p_2$, Eq. (3.8) with $p = \max\{p_1, p_2\}$ provides a lower bound on the achievable concurrence.

The requirement $\sqrt{\kappa/2}\alpha_0 u(t) = \tilde{g}_1(t)$ limits the average number of photons in the input coherent state. For a Gaussian $u(t)$, $N = |\alpha_0|^2 = 2\sqrt{\pi}(\tilde{g}_1^{\max})^2\tau/\kappa$, where $\tilde{g}_1^{\max} = \max_t|\tilde{g}_1(t)|$. Together with the bandwidth requirement $N \ll (\kappa\tau)^4$, this implies that $N \ll N_{\max} \equiv (\tilde{g}_1^{\max}\tau)^{8/5}$. A larger $\tau$ therefore increases $N_{\max}$. However, the pulse duration is also subject to the requirement $\tau < T_2^*$, where $T_2^*$ is the dephasing time of the qubit [defined by $\chi_\xi(T_2^*) = 1$]. For example, if $g_1^{\max}/2\pi = 1\,\mathrm{MHz}$ and $\tau = 1\,\mu\mathrm{s}$, then $N_{\max} \simeq 19$. For $\tilde{g}_1/\kappa = 1/8$ (weak coupling), we then have $N \simeq 3$, close to the value that maximizes the two-qubit concurrence ($C \simeq 0.95$) for $p = 0.01$ [51]. This scenario may be realistic for, e.g., an electron-spin qubit in a silicon DQD with a magnetic field gradient. The longitudinal coupling for this case can be comparable to the transverse coupling $\sim 1-10\,\mathrm{MHz}$ [42]. Dephasing times for electron-spin qubits in $^{28}$Si quantum dots reach $T_2^* \sim 100\,\mu\mathrm{s} \gg \tau$ [69]. The same values of $N$ and $N_{\max}$ could also be realized for flux-tunable transmons, with a longitudinal coupling $\sim 10\,\mathrm{MHz}$ [41] and pulse duration $\tau \sim 100\,\mathrm{ns} \ll T_2^*$ (for transmons, coherence times reach $T_2^* \simeq 100\,\mu\mathrm{s}$ [70]).

## 3.4 Precision Metrology

The entangler described above can also be used to perform quantum-enhanced precision measurements of a phase $\phi$ acquired by the field reflected from the cavity as it propagates along arm 1 of an interferometer (Fig. 3.2). The fundamental precision bound for estimation of $\phi$ (the quantum Cramér-Rao bound [71, 72]) is better for QWP states than for either NOON states (superpositions of $N$-photon Fock states, one in each interferometer arm) or entangled coherent states [27] (similarly, superpositions of coherent states) having the same average number $N$ of photons [73].

## 3.5 Outlook

The entangler presented here could also be used to perform measurements of the phase acquired by the control qubit. Specifically, a modulated longitudinal coupling, followed by a rapid reset [74–77] $|0\rangle \to |1\rangle$, can be used to map the relative phase $|0\rangle + e^{i\theta}|1\rangle$ of the initial qubit state onto the state $|-\alpha\rangle + e^{i\theta}|\alpha\rangle$ of the $a_+$ mode. A projective measurement of $|C_\pm\rangle$ then yields a single bit of information about $\theta$ (the maximum achievable for a single-shot qubit readout). This may be useful in situations where $\theta$ encodes information about dynamics induced by a classical or quantum environment [78, 79].

**Acknowledgments**—We thank A. Blais and K. Wang for useful discussions. We also acknowledge funding from the Natural Sciences and Engineering Research Council (NSERC) and from the Fonds de recherche du Québec–Nature et technologies (FRQNT).

# Appendices to Chapter 3

*These appendices contain the original text of the supplementary material to*
Z. M. McIntyre and W. A. Coish, Phys. Rev. Lett. **132**, 093603 (2024)

## B1 Cavity-qubit Hamiltonian

In general, the wavefunctions defining the ground and excited states of a qubit can be tuned through one or more parameters. For spin qubits in gate-defined double quantum dots, these could be gate voltages; for superconducting qubits, they could be fluxes. When these parameters are modulated in time, the Hamiltonian $H_{\mathrm{q}}$ of the qubit acquires a time dependence. The full cavity-qubit Hamiltonian can then be written as

$$H(t) = H_{\mathrm{q}}(t) + \omega_{\mathrm{c}} a^{\dagger} a + H_{\mathrm{int}}, \tag{B1}$$

where $H_{\mathrm{q}}(t)\,|s(t)\rangle = \varepsilon_s(t)\,|s(t)\rangle$ is the Hamiltonian whose low-energy instantaneous eigenstates (labelled by $s = 0,1$) are used to encode the qubit, and where $a$ annihilates an excitation in the cavity mode (whose frequency is denoted $\omega_{\mathrm{c}}$). In many architectures (involving e.g. atoms, excitons, or spins together with spin-orbit coupling), the cavity-qubit interaction $H_{\mathrm{int}}$ has its origins in the electric-dipole interaction $H_{\mathrm{int}} = e\boldsymbol{E}\cdot\boldsymbol{r}$, where $e > 0$ is the magnitude of the electron charge, $\boldsymbol{E}$ is the electric field of the cavity at the location of the dipole (qubit), and $\boldsymbol{r}$ is the position operator of an electron.

We transform $H(t)$ [Eq. (B1)] via the unitary transformation $U(t) = \sum_s |s\rangle\langle s(t)|$ to the instantaneous adiabatic eigenbasis, in which

$$\tilde{H}(t) = U(t)H(t)U^{\dagger}(t) - iU(t)\dot{U}^{\dagger}(t). \tag{B2}$$

So far, this transformation is exact. Treating the evolution of the system within an adiabatic approximation,

$$\tilde{H}(t) \simeq \tilde{H}_{\mathrm{adiabatic}}(t) = U(t)H(t)U^{\dagger}(t), \tag{B3}$$

requires that the usual adiabaticity condition be satisfied,

$$\frac{|\langle \bar{s}(t)|\partial_t|s(t)\rangle|}{|\varepsilon_0(t) - \varepsilon_1(t)|} \ll 1, \tag{B4}$$

where $s \in \{0,1\}$, $\bar{1} = 0$, and $\bar{0} = 1$. A similar condition must also be obeyed by proximal excited states. In particular, since we consider modulation at the cavity frequency $\omega_c$, there should be no coupled excited states at (or near) the cavity resonance. The cavity electric field is quantized $\boldsymbol{E} = i\boldsymbol{E_0}(a^{\dagger} - a)$, giving

$$\tilde{H}_{\mathrm{adiabatic}}(t) = \varepsilon_0(t)\,|0\rangle\langle 0| + \varepsilon_1(t)\,|1\rangle\langle 1| + \omega_{\mathrm{c}} a^{\dagger} a + i\sum_{s=0,1} g_s(t)\,|s\rangle\langle s|\,(a^{\dagger} - a) + i\sum_{s} g_{\perp}(t)\,|s\rangle\langle \bar{s}|\,(a^{\dagger} - a), \tag{B5}$$

where the longitudinal coupling $g_s(t) = e\boldsymbol{E_0}\cdot\langle s(t)|\boldsymbol{r}|s(t)\rangle$ depends on the $s$-dependent electric dipole moment $-e\langle s(t)|\boldsymbol{r}|s(t)\rangle$, and where the transverse coupling $g_{\perp}(t) = e\boldsymbol{E_0}\cdot\langle s(t)|\boldsymbol{r}|\bar{s}(t)\rangle$ depends on the transition dipole moment $-e\langle s(t)|\boldsymbol{r}|\bar{s}(t)\rangle$. When the transition dipole is weak, or for a qubit strongly detuned from the cavity, the transverse term may be strongly suppressed, yielding an interaction that is predominantly longitudinal.

We assume that $g_0(t)$ and $g_1(t)$ can both be tuned by varying parameters $x_j(t)$. (These could be local potentials or electric fields for atomic, exciton, or spin qubits, or they could be fluxes for flux-biased superconducting qubits.) To linear order,

$$g_s(t) = \bar{g}_s + \delta g_s(t) \simeq \bar{g}_s + \sum_j \frac{\partial g_s}{\partial x_j}\delta x_j(t), \tag{B6}$$

where $\delta g_s(t)$ describes a time-dependent modulation of $g_s(t) = \bar{g}_s + \delta g_s(t)$ [and similarly for $\delta x_i(t)$].

When one of $g_s(t)$ (for $s = 0,1$) can be modulated independently of the other with a single $x_j(t)$, the required time-dependent modulation can be realized directly. More generally, it may be necessary to consider control through a minimum of two parameters $x_j(t)$ ($j = 0,1$). Realizing the time-dependence of $g_s(t)$ required for the entangler then calls for a solution to the system of equations

$$\begin{pmatrix}\delta x_0\\ \delta x_1\end{pmatrix} = \boldsymbol{J}^{-1}\begin{pmatrix}\delta g_0(t)\\ \delta g_1(t)\end{pmatrix} = \boldsymbol{J}^{-1}\begin{pmatrix}0\\ 2|\tilde{g}_1(t)|\cos[\omega_c t - \vartheta(t)]\end{pmatrix}, \tag{B7}$$

where here, $\boldsymbol{J}$ is a $2\times 2$ matrix with elements $J_{sj} = \partial g_s/\partial x_j$, and where $\tilde{g}_1(t) = e^{i\vartheta(t)}|\tilde{g}_1(t)|$ varies slowly relative to the timescale $\omega_c^{-1}$. A sufficient condition for Eq. (B7) to have a nontrivial solution is that $\boldsymbol{J}^{-1}$ be defined, i.e. that the determinant of $\boldsymbol{J}$ be nonvanishing.

We assume that the modulations of $\delta x_j(t)$ prescribed by Eq. (B7) have a negligible impact on $\varepsilon_s(t) \simeq \varepsilon_s$, so that, neglecting the transverse coupling term ($\propto g_\perp$) and up to a constant shift in energy,

$$\tilde{H}_{\mathrm{adiabatic}}(t) \simeq \frac{1}{2}\omega_q'\sigma_z + \omega_c a^\dagger a + i g_0\,|0\rangle\langle 0|\,(a^\dagger - a) + i g_1(t)\,|1\rangle\langle 1|\,(a^\dagger - a), \tag{B8}$$

where here, $\sigma_z = |0\rangle\langle 0| - |1\rangle\langle 1|$ and $\omega_q' = \varepsilon_0 - \varepsilon_1$.

We now perform a polaron transformation $\tilde{\tilde{H}}(t) = e^S \tilde{H}_{\mathrm{adiabatic}}(t) e^{-S}$ on Eq. (B8), where

$$S = i\frac{g_0}{\omega_c}\,|0\rangle\langle 0|\,(a^\dagger - a). \tag{B9}$$

Up to a constant shift in energy, the result of this transformation is

$$\tilde{\tilde{H}}(t) = \frac{1}{2}\omega_q\sigma_z + \omega_c a^\dagger a + i g_1(t)\,|1\rangle\langle 1|\,(a^\dagger - a), \tag{B10}$$

where here, $\omega_q = \omega_q' - g_0^2/2\omega_c$ is the qubit frequency accounting for the polaron shift. To model the effects of dephasing, we assume that the qubit splitting $\omega_q$ in Eq. (B10) is subject to classical fluctuations described by a stochastic noise process $\xi(t)$: $\omega_q \to \omega_q + \xi(t)$. Transforming Eq. (B10) to an interaction picture with respect to $\omega_q\sigma_z/2 + \omega_c a^\dagger a$ and performing a rotating-wave approximation then recovers Eq. (3.1) of the main text.

### B1.1 Physical examples

In this section, we derive the form of the longitudinal coupling between a microwave cavity constructed from a superconducting stripline resonator and a qubit encoded in either the charge or spin states of an electron in a double quantum dot (Fig. B1). In each case (charge or spin qubits), we find the conditions under which the coupling can be modulated for only one of the qubit states [giving a form $\propto g_1(t)\,|1\rangle\langle 1|$ with $\delta g_0(t) = 0$], as required for the which-path entangler.

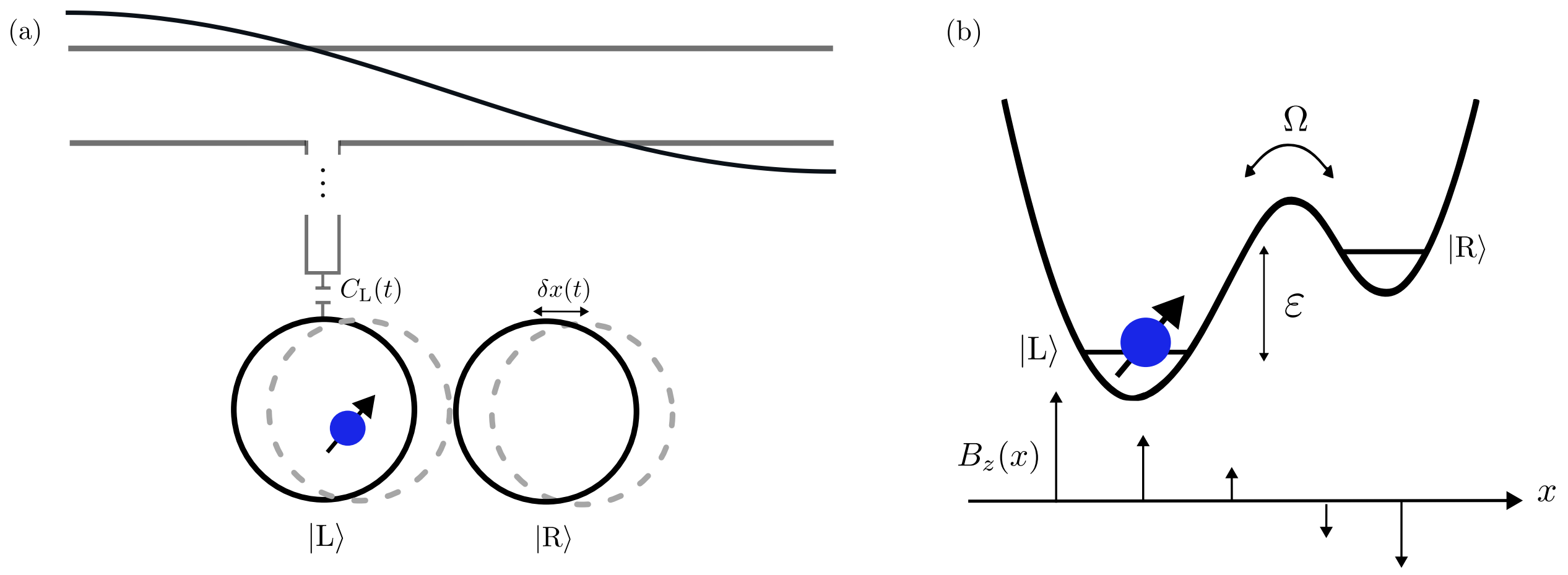

Figure B1: (a) An electron in a double quantum dot can be coupled capacitively to a microwave resonator via a metallic finger (vertical gray rectangle) located near one of the dots. We assume that the dot-to-finger lever arm $\alpha_\mathrm{L} \propto C_\mathrm{L}$ of the left dot can be modulated by shifting the position $x(t)$ of the orbitals $\psi_l(\boldsymbol{r},t)$ ($l = \mathrm{L},\mathrm{R}$) of the individual dots through a gate voltage modulation. (b) A spin qubit in a double quantum dot: A magnetic field gradient $B_z(x)$ across the double-dot axis can be used to couple the electron spin and charge (orbital) degrees of freedom. The electron spin can then couple to the resonator via an interaction of the form given in Eq. (B16).

The interaction between the charge density operator $\rho(\boldsymbol{r})$ for a qubit and the electric potential $V(\boldsymbol{r},t)$ due to a resonator can be written as

$$H_\rho = \int d^3r\rho(\mathbf{r})V(\mathbf{r},t). \tag{B11}$$

When the potential $V(\mathbf{r},t)$ is a slowly-varying function of $\boldsymbol{r}$ on the scale of the qubit charge distribution, it can be expanded about the location $\mathbf{r}_0$ of the qubit, giving the electric dipole approximation (in the length gauge):

$$H_\rho \simeq QV(\mathbf{r}_0,t) - \boldsymbol{d}\cdot\boldsymbol{E}(\mathbf{r}_0,t). \tag{B12}$$

Here, $\boldsymbol{E}(\mathbf{r}_0,t) = -\boldsymbol{\nabla} V(\mathbf{r}_0,t)$ is the cavity electric field at the position of the qubit. The quantities $Q$ and $\boldsymbol{d}$ are the total charge and dipole operators, respectively:

$$Q = \int d^3r\rho(\boldsymbol{r}); \quad \boldsymbol{d} = \int d^3r\rho(\boldsymbol{r})(\boldsymbol{r}-\boldsymbol{r}_0). \tag{B13}$$

For qubit states constructed from systems with overall charge neutrality (e.g., neutral atoms and excitons), $Q \simeq 0$. An effective qubit-cavity interaction can then be derived by quantizing the cavity field and forming matrix elements of the dipole operator, as described in the main text and in Sec. B1, above. Since the dipole operator is odd under inversion about $\boldsymbol{r}_0$ ($\boldsymbol{d} \to -\boldsymbol{d}$), the longitudinal coupling will vanish for these systems (within the dipole approximation) if the qubit eigenstates $|s\rangle$ have definite parity under inversion. This will be the case for high-symmetry free-atom eigenstates, but it will not generally be true for lower-symmetry designed systems like the quantum dots considered here.

#### B1.1.1 Charge qubit in a double quantum dot

In contrast to the far-field approach described above, many experiments that couple quantum dots to superconducting resonators instead exploit the strong near-field coupling of a quantum-dot charge to a metallic gate that extends from the resonator to the vicinity of the quantum dot [80–82] (Fig. B1). For these systems, it is important to work directly from Eq. (B11) to find an accurate microscopic description of the coupling.

To simplify the effective coupling, we now consider only a single quantized resonator mode. We further project the charge density operator onto the lowest two (left/right) orbital states of a double quantum dot. Thus, we approximate

$$V(\boldsymbol{r},t) \simeq i\phi_0(\boldsymbol{r})V_0(a^\dagger e^{i\omega_c t} - ae^{-i\omega_c t}), \tag{B14}$$

$$\rho(\boldsymbol{r},t) \simeq -e|\psi_{\mathrm{L}}(\boldsymbol{r},t)|^2\,|\mathrm{L}\rangle\langle\mathrm{L}| - e|\psi_{\mathrm{R}}(\boldsymbol{r},t)|^2\,|\mathrm{R}\rangle\langle\mathrm{R}|, \tag{B15}$$

where the dimensionless mode function $\phi_0(\boldsymbol{r})$ solves Poisson's equation subject to the device geometry. Here, $V_0$ is the amplitude of zero-point voltage fluctuations in the resonator itself, and $\psi_l(\boldsymbol{r},t)$ $(l = \mathrm{L},\mathrm{R})$ is the envelope function for quantum dot $l$. We assume that the quantum-dot orbitals can be manipulated via external gate voltages, leading to adiabatically varying time-dependent envelope functions. This has been achieved in experiments on shuttling electrons between quantum dots [83].

The coupling to the right dot $|\mathrm{R}\rangle$ is negligible if $\phi_0(\boldsymbol{r})$ is vanishingly small wherever the envelope function $\psi_{\mathrm{R}}(\boldsymbol{r},t)$ has significant weight. In this limit, we insert Eqs. (B14) and (B15) into Eq. (B11), giving the coupling between a double-dot charge qubit and a microwave resonator [84],

$$e^{-i\omega_c a^\dagger a t}H_\rho e^{i\omega_c a^\dagger a t} \simeq H_{\mathrm{int}} = ig_c(t)\,|\mathrm{L}\rangle\langle\mathrm{L}|\,(a^\dagger - a), \quad g_c(t) = -e\alpha_{\mathrm{L}}(t)V_0. \tag{B16}$$

The coupling $g_c(t)$ between the charge and the resonator can thus be controlled through the lever arm $\alpha_{\mathrm{L}}(t)$, which can itself be controlled via the left dot orbital:

$$\alpha_{\mathrm{L}}(t) = \int d^3r\phi_0(\boldsymbol{r})|\psi_{\mathrm{L}}(\boldsymbol{r},t)|^2, \tag{B17}$$

$$\alpha_{\mathrm{R}}(t) = \int d^3r\phi_0(\boldsymbol{r})|\psi_{\mathrm{R}}(\boldsymbol{r},t)|^2 \simeq 0. \tag{B18}$$

By manipulating the shape, size, and position of the left quantum dot via gate voltages, $\alpha_{\mathrm{L}}$ can therefore be modulated. The lever arm $\alpha_{\mathrm{L}}$ can equivalently be written in terms of the capacitance $C_{\mathrm{L}}$ between the left dot and the resonator [Fig. B1(a)] and the total capacitance $C_\Sigma$ of the left dot [84]:

$$\alpha_{\mathrm{L}} = \frac{C_{\mathrm{L}}}{C_\Sigma}. \tag{B19}$$

For a charge qubit encoded in the left/right basis ($|\mathrm{L}\rangle \to |1\rangle$, $|\mathrm{R}\rangle \to |0\rangle$), Eq. (B16) directly achieves the form of longitudinal coupling required for the which-path entangler, where $g_c(t) \to g_1(t)$, and where $g_0(t) \simeq 0$ provided the cross-capacitance between the resonator and the right dot is negligible [equivalently, provided Eq. (B18) is satisfied]. Charge qubits of this type are sensitive to electric field fluctuations, leading to coherence times typically in the range of $T_2^* \sim 100\,\mathrm{ps}$-$10\,\mathrm{ns}$ [85]. These short coherence times would likely limit the applicability of pure charge qubits to the entangler scheme presented in the main text. For this reason, in the next section we consider longitudinal coupling for a spin qubit, since spin qubits typically have longer coherence times.

#### B1.1.2 Spin qubit in a double quantum dot

An interaction of the form given in Eq. (B16) can also be used to generate coupling between a microwave resonator and an electron spin. This generally requires coupling of the spin and charge degrees-of-freedom, since direct magnetic coupling of the electron spin magnetic moment to the resonator magnetic field is typically weak, on the order of a few tens of Hz [86, 87]. In the absence of intrinsic spin-orbit coupling, a synthetic spin-orbit interaction can be generated from a magnetic field gradient across the two dots [88] [Fig. B1(b)]. For a magnetic field oriented along the $z$ axis, the Hamiltonian describing spin-charge hy-bridization is given by [42]

$$H_{\rm q}(t) = \frac{1}{2}[\varepsilon\tau_z + \Omega(t)\tau_x] + \frac{1}{2}(b_{\rm L}\,|{\rm L}\rangle\langle{\rm L}| + b_{\rm R}\,|{\rm R}\rangle\langle{\rm R}|)\sigma_z, \tag{B20}$$

where $\sigma_z$ is the Pauli-Z operator of the spin, $\varepsilon$ is the double-dot detuning controlling the relative dot potentials [Fig. B1(b)], and where $b_{\rm L,R} = g^*\mu_{\rm B}\int d^3r|\psi_{\rm L,R}(\boldsymbol{r})|^2 B_z(\boldsymbol{r})$. Here, $g^*$ is the (material-dependent) electron $g$-factor and $\mu_{\rm B}$ is the Bohr magneton. The tunnel splitting $\Omega(t)$ can be controlled through a gate voltage and will be used to modulate the coupling. In Eq. (B20), we have also introduced the orbital pseudospin Pauli operators

$$\tau_x = |{\rm L}\rangle\langle{\rm R}| + |{\rm R}\rangle\langle{\rm L}|, \tag{B21}$$

$$\tau_z = |{\rm R}\rangle\langle{\rm R}| - |{\rm L}\rangle\langle{\rm L}|. \tag{B22}$$

Equation (B20) can be diagonalized conditioned on the spin state $|\sigma\rangle$ (where $\sigma = \uparrow,\downarrow$ with $\sigma_z|\sigma\rangle = \pm|\sigma\rangle$), giving

$$\begin{aligned} |+,\sigma\rangle_t &= \cos\frac{\theta_\sigma(t)}{2}|{\rm R}\rangle|\sigma\rangle + \sin\frac{\theta_\sigma(t)}{2}|{\rm L}\rangle|\sigma\rangle, \\ |-,\sigma\rangle_t &= -\sin\frac{\theta_\sigma(t)}{2}|{\rm R}\rangle|\sigma\rangle + \cos\frac{\theta_\sigma(t)}{2}|{\rm L}\rangle|\sigma\rangle, \end{aligned} \tag{B23}$$

where $\tan\theta_{\uparrow,\downarrow}(t) = \Omega(t)/[\varepsilon \pm \Delta b_z]$ for $\Delta b_z = (b_{\rm R} - b_{\rm L})/2$. When the Zeeman energy is small compared to the double-dot orbital energy, the lowest two energy eigenstates are predominantly distinguished by spin $\sigma = \uparrow,\downarrow$, but they will also have slightly different charge distributions. The two states $|-,\sigma\rangle_t$ then encode the qubit:

$$|0(t)\rangle = |-,\uparrow\rangle_t, \tag{B24}$$

$$|1(t)\rangle = |-,\downarrow\rangle_t. \tag{B25}$$

In order to obtain an effective qubit-resonator interaction, we now project Eq. (B16) (with a time-independent $g_c$) into the qubit subspace and transform to the instantaneous adiabatic eigenbasis via the unitary $U = \sum|s\rangle\langle s(t)|$ (as described in Sec. B1), giving

$$\begin{aligned} H_{\rm int}^{\rm eff}(t) &= UP(t)H_{\rm int}P(t)^\dagger U^\dagger, \quad P(t) = \sum_\sigma |-,\sigma\rangle_t\,\langle -,\sigma| \\ &= ig_0(t)\,|0\rangle\langle 0|\,(a^\dagger - a) + ig_1(t)\,|1\rangle\langle 1|\,(a^\dagger - a), \end{aligned} \tag{B26}$$

where

$$\begin{aligned} g_0(t) &= g_c\cos^2\frac{\theta_\uparrow(t)}{2} = \frac{g_c}{2}\left(1 + \frac{\varepsilon + \Delta b_z}{\sqrt{(\varepsilon+\Delta b_z)^2 + \Omega^2(t)}}\right), \\ g_1(t) &= g_c\cos^2\frac{\theta_\downarrow(t)}{2} = \frac{g_c}{2}\left(1 + \frac{\varepsilon - \Delta b_z}{\sqrt{(\varepsilon-\Delta b_z)^2 + \Omega^2(t)}}\right). \end{aligned} \tag{B27}$$

For $\varepsilon = -\Delta b_z$, we then have $g_0 = g_c/2$ (time-independent) for all finite values of the tunnel splitting $\Omega(t)$. Hence for this choice of parameters ($\varepsilon = -\Delta b_z$), modulating $\Omega(t)$ only affects the coupling of the resonator

to state $|1\rangle$:

$$g_0 = \frac{g_c}{2}, \tag{B28}$$

$$g_1(t) = \frac{g_c}{2}\left(1 - \frac{2\Delta b_z}{\sqrt{4\Delta b_z^2 + \Omega^2(t)}}\right). \tag{B29}$$

For $\Omega(t) = \bar{\Omega} + \delta\Omega(t)$ with $\bar{\Omega} \gg |\Delta b_z|, |\delta\Omega(t)|$, we then have $g_1(t) = \bar{g}_1 + \delta g_1(t)$, where

$$\delta g_1(t) \simeq \frac{g_c \Delta b_z}{\bar{\Omega}^2}\delta\Omega(t). \tag{B30}$$

## B2 Finite-bandwidth corrections to ideal QWP states

In this section, we quantify finite-bandwidth effects due to a finite duration of the input wavepacket $u(t)$. These corrections lead to imperfect transmission and reflection of the incident wavepacket for a qubit prepared in the state $|0\rangle$, leading to deviations from the idealized QWP states considered in the main text.

For a qubit initialized in $|+\rangle = (|1\rangle + |0\rangle)/\sqrt{2}$, the QWP state generated for an incident coherent state with amplitude $\alpha_0$ is given by Eq. (3.7) of the main text (here we consider the case $\xi = 0$ since dephasing does not affect the present discussion):

$$|\Psi\rangle = \frac{1}{\sqrt{2}}\left(|1, \psi_1\rangle + |0, \psi_0\rangle\right), \tag{B31}$$

where for $s = 0, 1$,

$$|\psi_s\rangle = \prod_{i=1,2} D_i(\alpha_{is})\,|\text{vac}\rangle\,. \tag{B32}$$

Here, $|\text{vac}\rangle$ is the vacuum, $D_i(\alpha) = e^{\alpha a_{v_i}^\dagger - \text{h.c.}}$ is a displacement operator that generates a coherent state with amplitude $\alpha$ in mode $a_{v_i}$, and

$$\alpha_{is} = (-1)^{i-1}\alpha_0 \int_0^\infty dt\,|v_i(t)|^2 = (-1)^{i-1}\alpha_0 \int \frac{d\omega}{2\pi}\,|v_i(\omega)|^2 \tag{B33}$$

is the qubit-state-dependent stationary amplitude of the coherent state in mode $a_{v_i}$. The reflected ($v_1$) and transmitted ($v_2$) pulses are given by $v_1(\omega) = R(\omega)u(\omega)$ and $v_2(\omega) = T(\omega)u(\omega)$, respectively, where $R(\omega)$ and $T(\omega)$ are the reflection and transmission coefficients, and where $u(\omega)$ is the waveform of the input pulse. The input pulse is assumed to have a duration $\sim \tau$.

Up to corrections that vanish for $\kappa\tau \to \infty$, only one of $\alpha_{is}$ is nonzero for each value of $s$, provided

$$\kappa_1 = \kappa_2 = \frac{\kappa}{2}. \tag{B34}$$

With this choice of $\kappa_{1,2}$, Eq. (B31) describes a coherent state whose which-path degree-of-freedom is entangled with the qubit: For a qubit in state $|1\rangle$, the incident coherent state is fully reflected, whereas for a qubit in state $|0\rangle$, it is fully transmitted, up to corrections that vanish for $\kappa\tau \to \infty$. In order to quantify these corrections, we consider the fidelity $F = |\langle\Psi_\infty|\Psi_{\kappa\tau}\rangle|^2$ of the idealized QWP state $|\Psi_\infty\rangle$ having ideal which-path character, obtained by taking $\kappa\tau \to \infty$, relative to the QWP state $|\Psi_{\kappa\tau}\rangle$ obtained for finite $\kappa\tau$. This is easily accomplished using the relation $R(\omega) - 1 = T(\omega)$ that follows from input-output theory after setting $\kappa_1 = \kappa_2 = \kappa/2$, together with Eq. (3.4) of the main text:

$$T(\omega) = (1-s)\frac{\kappa/2}{i\omega - \kappa/2}. \tag{B35}$$

From Eq. (B33), we then have $\alpha_{11} = \alpha_0$, $\alpha_{21} = 0$, as well as the non-trivial coherent-state amplitudes

$$\alpha_{10} = \alpha_0 \int \frac{d\omega}{2\pi} |u(\omega)|^2 \times \frac{\omega^2}{\omega^2 + \kappa^2/4}, \tag{B36}$$

$$\alpha_{20} = -\alpha_0 \int \frac{d\omega}{2\pi} |u(\omega)|^2 \times \frac{\kappa^2/4}{\omega^2 + \kappa^2/4}, \tag{B37}$$

which depend on $u(\omega)$. Since the state $|1, \psi_1\rangle$ appearing in Eq. (B31) describes a perfectly reflected wavepacket for any $\kappa\tau$, the fidelity $F$ is controlled entirely by $|0, \psi_0\rangle$, where ideally (i.e. for $\kappa\tau \to \infty$), $|\psi_0\rangle = D_2(\alpha_0)|\text{vac}\rangle$. The fidelity is therefore given by

$$F = \frac{1}{4}\left|1 + e^{-\frac{1}{2}|\alpha_0|^2} e^{\alpha_0^* \alpha_{20}} \prod_{i=1,2} e^{-\frac{1}{2}|\alpha_{i0}|^2}\right|^2, \tag{B38}$$

where $\alpha_{10}$ and $\alpha_{20}$ are given by Eqs. (B36) and (B37). As an example, we consider an input waveform $u(\omega)$ having a Gaussian envelope with width $\sim 1/\tau$:

$$|u(\omega)|^2 = 2\sqrt{\pi}\tau e^{-\omega^2\tau^2}. \tag{B39}$$

Here, the normalization of $u(\omega)$ is set by the requirement that $\int dt |u(t)|^2 = 1$. For this choice of $u(\omega)$, Eqs. (B36) and (B37) give $\alpha_{10} = \alpha_0(1 - \sqrt{\pi} y e^{y^2} \operatorname{erfc} y)$ and $\alpha_{20} = -\alpha_0 \sqrt{\pi} y e^{y^2} \operatorname{erfc} y$, where $y = \kappa\tau/2$. Taylor expanding in $(\kappa\tau)^{-1} \ll 1$, we find that $\alpha_{10} \simeq \alpha_0[2(\kappa\tau)^{-2} - 12(\kappa\tau)^{-4}]$ and $\alpha_{20} \simeq -\alpha_0[1 - 2(\kappa\tau)^{-2} + 12(\kappa\tau)^{-4}]$, giving

$$F = \frac{1}{4}\left(1 + e^{-4\frac{|\alpha_0|^2}{(\kappa\tau)^4} + \text{h.o.t.}'\text{s}}\right)^2 \simeq 1 - 4\frac{|\alpha_0|^2}{(\kappa\tau)^4}, \tag{B40}$$

where h.o.t.′s designates terms that are higher order in $1/\kappa\tau$. For a Gaussian wavepacket, neglecting the subleading corrections therefore requires that $N = |\alpha_0|^2 \ll (\kappa\tau)^4$. Throughout the rest of this supplement, we neglect corrections due to finite $\kappa\tau$ and consider only the limit $\kappa\tau \to \infty$.

## B3 Two-qubit concurrence

In this section, we derive a formula for the concurrence accounting for pure-dephasing noise on the control qubit, photon loss in the transmission lines, and detection errors arising from both imperfect coherent-state distinguishability and imperfect detector efficiency. The result derived here corresponds to Eq. (3.8) of the main text.

Starting from the QWP state [cf. Eq. (3.7) of the main text]

$$|\Psi_{\xi(t)}\rangle = \frac{1}{\sqrt{2}}(e^{\frac{i}{2}\theta_\xi(t)}|1, \psi_1\rangle + e^{-\frac{i}{2}\theta_\xi(t)}|0, \psi_0\rangle), \tag{B41}$$

where $\theta_\xi(t) = \int_0^t dt' \xi(t')$ is a noise-dependent random phase, we model the effect of photon loss by inserting a fictitious beamsplitter into each interferometer arm. These beamsplitters are modeled by the unitary operator [89]

$$B_{c,c'}(\varphi) = e^{i\frac{\varphi}{2}(c^\dagger c' + \text{h.c.})}, \tag{B42}$$

which describes scattering from mode $c$ into loss mode $c'$ with probability $p = \cos^2(\varphi/2)$. Although the loss from the two arms may be asymmetric in reality (with probability $p_i$ for loss from arm $i$), the expression for the concurrence derived here provides a lower bound for the concurrence achieved with asymmetric losses, if we take $p = \max\{p_1, p_2\}$. (Increasing the incoherent loss rate in either arm can only decrease entanglement.)

Acting on Eq. (B41) with $\prod_{i=1,2} B_{a_{v_i},a'_{v_i}}(\varphi_i)$, then tracing out modes $a'_{v_1}$ and $a'_{v_2}$ gives the reduced density matrix [still conditioned on a single realization of $\xi(t)$]

$$\rho_\xi(t) = \frac{1}{2}\left[|\Phi_1\rangle\langle\Phi_1| + |\Phi_0\rangle\langle\Phi_0| + e^{-pN}(e^{i\theta_\xi(t)}|\Phi_1\rangle\langle\Phi_0| + \text{h.c.})\right], \tag{B43}$$

where $N = |\alpha_0|^2$, and where we have introduced

$$|\Phi_1\rangle = D_1(\alpha_0\sqrt{1-p})\,|1,\text{vac}\rangle, \quad |\Phi_0\rangle = D_2(-\alpha_0\sqrt{1-p})\,|0,\text{vac}\rangle. \tag{B44}$$

Photon loss therefore leads to a suppression of the off-diagonal terms in Eq. (B43) (dephasing). This effect is accompanied by a reduction of the coherent-state amplitude: $\alpha_0 \to \alpha_0\sqrt{1-p}$. Note that our ability to write $|\Phi_0\rangle$ in terms of a single displacement operator is a consequence of neglecting finite-$\kappa\tau$ corrections.

In order to re-encode the which-path degree-of-freedom in a parity degree-of-freedom, the coherent states comprising the which-path qubit are interfered at a beamsplitter (described by unitary $U_{\text{BS}}$) that transforms the output modes $a_{v_i}$ into new modes $a_\pm = (a_{v_1} \pm a_{v_2})/\sqrt{2}$. Following this operation, the resulting field in the $a_-$ mode is unentangled with either the qubit or $a_+$ mode. Tracing out the $a_-$ mode ($\text{Tr}_{a_-}\{\}$) and averaging ($\langle\!\langle\rangle\!\rangle$) over realizations of the noise $\xi(t)$ then gives the following reduced density matrix for the state of the qubit and $a_+$ mode:

$$\rho(t) = \langle\!\langle \text{Tr}_{a_-}\{U_{\text{BS}}\rho_\xi(t)U_{\text{BS}}^\dagger\}\rangle\!\rangle = \frac{1}{2}\sum_{s=0,1}|\Psi_\sigma\rangle\langle\Psi_\sigma| + \frac{1}{2}e^{-pN-\chi_\xi(t)}(|\Psi_1\rangle\langle\Psi_0| + \text{h.c.}), \tag{B45}$$

where $|\Psi_0\rangle = |0, -\alpha_p\rangle$ and $|\Psi_1\rangle = |1, +\alpha_p\rangle$ for $\alpha_p = \alpha_0[(1-p)/2]^{1/2}$, and where for zero-mean, Gaussian, stationary noise having spectral density $S(\omega) = \int dt\, e^{-i\omega t}\langle\!\langle\xi(t)\xi\rangle\!\rangle$,

$$\chi_\xi(t) = \int \frac{d\omega}{2\pi}\frac{4\sin^2\left(\frac{\omega t}{2}\right)}{\omega^2}S(\omega). \tag{B46}$$

The photonic degree-of-freedom can then be entangled with a second qubit, initialized in $|+\rangle$, through an interaction that imparts a phase shift sending $|+\rangle \to |-\rangle$, conditioned on the electric field having an odd photon-number parity [60–63]. The photon-number parity can be identified by re-expressing $|\pm\alpha_p\rangle$ in terms of the even- and odd-parity cat states $|C_\pm\rangle = N_\pm(|+\alpha_p\rangle \pm |-\alpha_p\rangle)$, where $N_\pm = (2 \pm 2e^{-2|\alpha_p|^2})^{-1/2}$. In addition to the approaches presented in Refs. [60–63], the required parity-conditioned phase flip could also be implemented by reflecting the incoming field off a single-sided cavity dispersively coupled to a qubit, i.e., coupled through an interaction of the form $\chi\sigma_z a^\dagger a$. In that case, the reflection coefficient (where $\omega$ is relative to the bare cavity frequency) is given by

$$R(\omega) = \frac{i(z\chi - \omega) - \kappa/2}{i(z\chi - \omega) + \kappa/2}, \tag{B47}$$

where $z = +1$ ($z = -1$) for a qubit in state $|1\rangle$ ($|0\rangle$). In this setup, an input coherent state occupying a spatiotemporal mode of duration $\tau$, resonant with $\chi$ (corresponding to the cavity frequency conditioned on the qubit being in state $|1\rangle$), will acquire a $\pi$ phase shift (sending $|\pm\alpha\rangle \to |\mp\alpha\rangle$) provided $\kappa\tau$ is large. With the qubit in state $|0\rangle$, however, and provided $2|\chi| \gg \kappa$, where $\kappa$ is the cavity decay rate (requiring that $\chi \gg \kappa \gg 1/\tau$), the same field will be far off resonance and will therefore be reflected without a phase shift.

Homodyne detection of the electric field quadrature along the axis of coherent-state displacement can be described by the two-element positive operator-valued measure (POVM) $P_\pm = (1 \pm C_\theta)/2$, where $\theta$ (not to be confused with $\theta_\xi$, defined above) is the phase of $\alpha_p = e^{i\theta}|\alpha_p|$, and where [64]

$$C_\theta = \int_0^\infty dx H_\theta(x) - \int_{-\infty}^0 dx H_\theta(x), \quad H_\theta(x) = \frac{1}{\sqrt{\pi(1-\eta)}}\exp\left\{ -\frac{(x/\sqrt{\eta} - \hat{x}_\theta)^2}{1/\eta - 1} \right\}. \tag{B48}$$

Here, $\eta$ is the detection efficiency and $\hat{x}_\theta = (e^{i\theta}a_+^\dagger + e^{-i\theta}a_+)/\sqrt{2}$. The operator $C_\theta$ [Eq. (B48)] has the following symmetries with respect to $|\pm\alpha_p\rangle$ [64]:

$$\langle \alpha_p | C_\theta | \alpha_p \rangle = -\langle -\alpha_p | C_\theta | -\alpha_p \rangle = \mathrm{erf}(\sqrt{2\eta}|\alpha_p|) \tag{B49}$$

$$\langle \alpha_p | C_\theta | -\alpha_p \rangle = \langle -\alpha_p | C_\theta | \alpha_p \rangle = 0. \tag{B50}$$

We define Krauss operators $K_\pm$ given by $K_\pm = U\sqrt{1 \pm C_\theta}$, where $U$ is a unitary. This can be done since $||C_\theta|| \leq 1$. The post-measurement state of the qubits can then be written as

$$\tilde{\rho}_\pm(t) = \frac{\mathrm{Tr}_{a_+}\{K_\pm \rho(t) K_\pm^\dagger\}}{\mathrm{Tr}\{K_\pm^\dagger K_\pm \rho(t)\}}, \tag{B51}$$

where $\mathrm{Tr}_{a_+}\{\cdots\}$ denotes a partial trace over the state of the $a_+$ mode. Using the symmetries of $C_\theta$ given in Eqs. (B49) and (B50), we then find that

$$\tilde{\rho}_\pm(t) = p_\pm \rho_+(t) + p_\mp \rho_-(t), \tag{B52}$$

where $p_\pm = \frac{1}{2}[1 \pm \mathrm{erf}(\sqrt{2\eta}|\alpha_p|)]$, and where the states $\rho_\pm(t)$ are given by

$$2\rho_\pm(t) = (1 \pm \sigma_z\sigma_z) + e^{-pN-\chi_\xi(t)}(\sigma_x\sigma_x \mp \sigma_y\sigma_y). \tag{B53}$$

In the computational basis $\{|0,0\rangle, |0,1\rangle, |1,0\rangle, |1,1\rangle\}$, these can be written in matrix form as

$$\rho_+(t) = \frac{1}{2}\begin{pmatrix} 1 & 0 & 0 & e^{-pN-\chi_\xi(t)} \\ 0 & 0 & 0 & 0 \\ 0 & 0 & 0 & 0 \\ e^{-pN-\chi_\xi(t)} & 0 & 0 & 1 \end{pmatrix}, \quad \rho_-(t) = \frac{1}{2}\begin{pmatrix} 0 & 0 & 0 & 0 \\ 0 & 1 & e^{-pN-\chi_\xi(t)} & 0 \\ 0 & e^{-pN-\chi_\xi(t)} & 1 & 0 \\ 0 & 0 & 0 & 0 \end{pmatrix}. \tag{B54}$$

Note that in the absence of both dephasing and photon loss ($p = \xi = 0$), the above states are Bell states: $\rho_\pm = (|+,+\rangle \pm |-,-\rangle)/\sqrt{2}$ for $|\pm\rangle = (|1\rangle \pm |0\rangle)/\sqrt{2}$.

Given Eq. (B54), it is easily seen that $\tilde{\rho}_\pm(t)$ is an X state [65], so called because its nonzero entries form an 'X' shape. For an X state of the form

$$\begin{pmatrix} a & 0 & 0 & w \\ 0 & b & z & 0 \\ 0 & z^* & c & 0 \\ w^* & 0 & 0 & d \end{pmatrix}, \tag{B55}$$

where $a+b+c+d = 1$, the concurrence $C$ [66–68] is easy to calculate [65]: $C = 2\max\{0, |z| - \sqrt{ad}, |w| - \sqrt{bc}\}$. For $\tilde{\rho}_\pm(t)$, this gives $C(t) = \max\{0, (1-\delta)e^{-pN-\chi_\xi(t)} - \delta\}$, where $\delta = p_- = \mathrm{erfc}(\sqrt{2\eta}|\alpha_p|)/2$. In

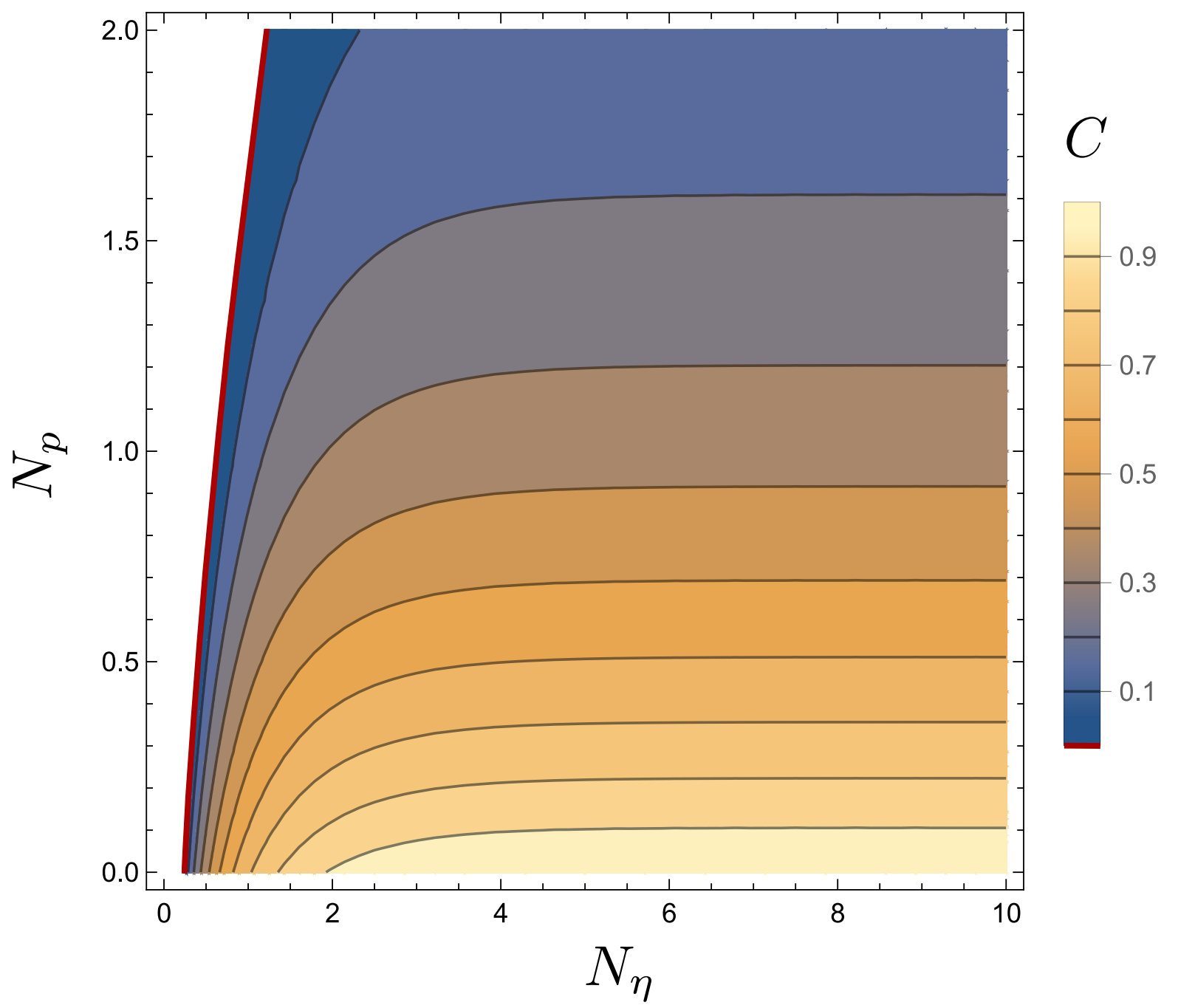

Figure B2: Sudden death of entanglement: For each value of $N_\eta = \eta(1-p)N$, there is a critical threshold in $N_p = pN$ beyond which the two-qubit concurrence vanishes. The contour $C = 0$ is indicated in red.

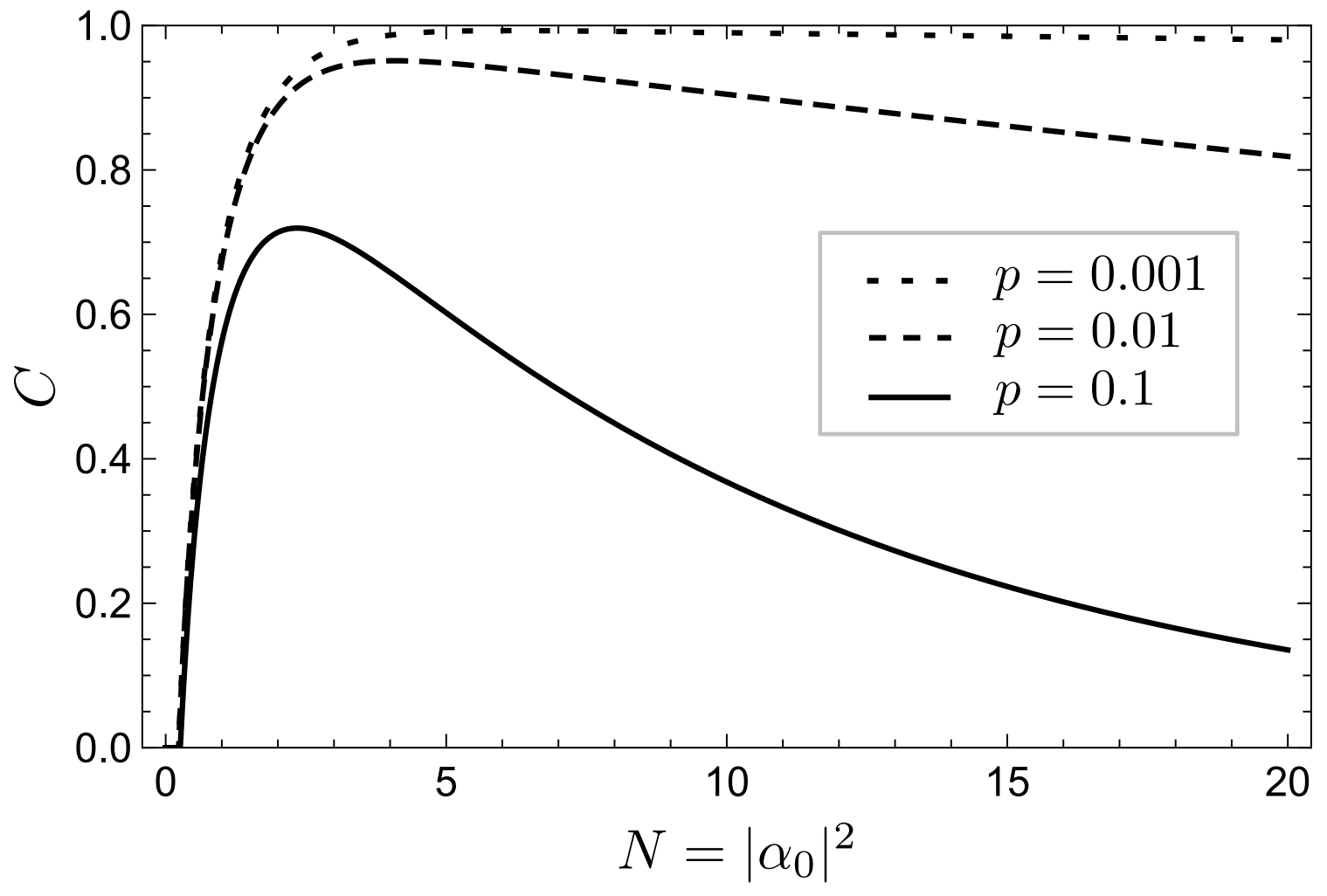

Figure B3: The concurrence [Eq. (B56)] with $\xi(t) = 0$ and $\eta = 1$ for $p = 0.001$ (dotted line), $p = 0.01$ (dashed line), and $p = 0.1$ (solid line).

terms of $N_\eta = \eta(1-p)N$ and $N_p = pN$, this result can be written as

$$C(t) = \max\{0, \operatorname{erf}(\sqrt{N_\eta})e^{-N_p-\chi_\xi(t)} - \operatorname{erfc}(\sqrt{N_\eta})\}. \tag{B56}$$

This recovers Eq. (3.8) of the main text. We plot the concurrence as a function of $N_\eta$ and $N_p$ with $\xi(t) = 0$ (Fig. B2) in order to illustrate the sudden death of entanglement. In Fig. B3, we plot $C$ as a function of $N$ for $\eta = 1$ in order to illustrate that for any fixed value of $p$, there exists a value of $N$ such that the concurrence is maximized.

# Preface to Chapter 4

Chapter 3 presented an approach for generating entanglement between the state of a qubit and the path taken by a multi-photon wavepacket. Given the which-path nature of this entangled state, a natural application to consider is precision interferometry, where the goal is to estimate some unknown parameter $\phi$ by sending quantum states of light through different paths in an interferometer. The best precision that can be achieved in estimating $\phi$ based on measurements of some output state $\rho(\phi)$ is set by the quantum Cramér-Rao bound (CRB), which gives a lower bound on the variance $\Delta\phi^2$ of estimates of $\phi$. The quantum CRB itself depends on a quantity called the quantum Fisher information, given by the supremum of the classical Fisher information associated with all possible measurements of $\rho(\phi)$. As such, the quantum CRB can be understood as giving the best possible sensitivity to $\phi$ allowed by the axioms of quantum mechanics.

The quantum Fisher information is commonly used as a means of quantifying the utility of a quantum state for the purpose of quantum-enhanced parameter estimation. Coherent states of light, which behave classically, give a quantum CRB that scales like $\Delta\phi^2 \propto N^{-1}$, where $N$ is the average number of photons in the coherent state. Nonclassical states can be used to achieve an improved (Heisenberg) scaling $\Delta\phi^2 \propto N^{-2}$, which provides a precision advantage over classical states as $N$ gets large. As alluded to in Chapter 3 and reported in Chapter 4, we find that qubit–which-path (QWP) states give rise to a Heisenberg-limited quantum CRB, signalling the potential usefulness of such states for quantum-enhanced parameter estimation. That being said, Chapter 3 does contain an incorrect statement: In Sec. 3.4, it was asserted that QWP states have a better fundamental precision bound (a.k.a. quantum Fisher information) than entangled coherent states for *the same average number N of photons*. This is not true and was stated erroneously because the statement *is* true when comparing at a fixed amplitude $\alpha$—a quantity which, for entangled coherent states, is not related to the average photon number via the simple relationship $N = |\alpha|^2$ that one might naively expect (this point resulted in an inconsistent comparison). The incorrect statement in Sec. 3.4 was supported by a paper in preparation, whose content we now give in Chapter 4. We would like to reassure the reader that the result is stated correctly in Chapter 4.

A Heisenberg-limited quantum CRB does not, however, come with a prescription for achieving this advantage over classical resources *in a practical experimental setting*. For a wide class of so-called path-symmetric states, it is known that the quantum CRB could be saturated with number-resolving measurements, where one counts the number of photons in the two output arms of the interferometer. One potential downside to this strategy is that photon counting requires specialized, cryogenically cooled detectors. In the following chapter, we present optimal measurement schemes for quantum interferometry with qubit–which-path states and entangled coherent states. These schemes are based on homodyne detection, which requires significantly less specialized detectors. We hope that the ideas contained in this chapter provide one possible avenue towards alleviating the technical challenges associated with quantum-enhanced interferometry.

# 4

# Optimal measurements for quantum interferometry

*This chapter is the integral text from*

**Homodyne detection is optimal for quantum interferometry with path-entangled coherent states**
Z. M. McIntyre and W. A. Coish
Phys. Rev. A **110**, L010602 (2024)

We present measurement schemes that do not rely on photon-number resolving detectors, but that are nevertheless optimal for estimating a differential phase shift in interferometry with either an entangled coherent state or a qubit–which-path state (where the path taken by a coherent-state wavepacket is entangled with the state of a qubit). The homodyning schemes analyzed here achieve optimality (saturate the quantum Cramér-Rao bound) by maximizing the sensitivity of measurement outcomes to phase-dependent interference fringes in a reduced Wigner distribution. In the presence of photon loss, the schemes become suboptimal, but we find that their performance is independent of the phase to be measured. They can therefore be implemented without any prior information about the phase and without adapting the strategy during measurement, unlike strategies based on photon-number parity measurements or direct photon counting.

## 4.1 Introduction

Interferometry for phase estimation is one of the fundamental tasks of quantum metrology [1], with applications in fields ranging from biophysics [2–4] to gravitational wave detection [5–7]. The ultimate goal of quantum-enhanced interferometry is to determine an unknown phase $\phi$ with a precision better than the standard quantum limit (shot-noise limit) for uncorrelated photons, given by $\Delta\phi \geq \delta\phi_{\rm SQL} = N^{-1/2}$, where $\Delta\phi$ is the standard deviation of $\phi$ and $N$ is the number of photons that pass through the interferometer in a single measurement. Correlations arising from non-classical states can, however, lead to better phase sensitivity, with improved scaling at the Heisenberg limit, $\Delta\phi \propto N^{-1}$.

The first study of quantum-enhanced interferometry considered phase estimation with a Mach-Zehnder interferometer (Fig. 4.1) fed by a coherent state mixed on a beamsplitter with a squeezed vacuum state [8]. In this configuration, Heisenberg-limited precision can be achieved by counting the precise number of photons arriving at each of two output ports of the interferometer [9, 10]. Photon counting is in fact optimal for this state, in the sense that it enables the best precision allowed by quantum mechanics, $\delta\phi_{\rm min}$, given by the quantum Cramér-Rao bound (CRB) [11–13],

$$\Delta\phi \geq \delta\phi_{\rm min} = \frac{1}{\sqrt{MI_{\rm Q}(\rho_\phi)}}, \tag{4.1}$$

where here, $M$ is the number of independent measurements and $I_{\rm Q}(\rho_\phi)$ is the quantum Fisher information of $\rho_\phi = e^{-i\phi A}\rho(0)e^{i\phi A}$ with respect to $A$, the generator of $\phi$. Formally, the quantum Fisher information is given by $I_{\rm Q}(\rho_\phi) = {\rm Tr}\{\rho_\phi \mathscr{L}^2\}$, with the symmetric logarithmic derivative operator $\mathscr{L}$ defined implicitly through the relation $\partial_\phi \rho_\phi = (\mathscr{L}\rho_\phi + \rho_\phi \mathscr{L})/2$ [14].

Photon counting provides an optimal strategy not just for a coherent state mixed with squeezed vacuum [9, 10], but for any path-symmetric pure state [18]. The class of path-symmetric states includes many of the states most commonly considered for quantum metrology, such as N00N states [19–21], twin Fock states [22], two-mode squeezed vacuum states [23], and entangled coherent states (ECS's) [16]. This optimal measurement strategy may, however, be associated with additional technological complexity: Photon-number resolving detectors (typically, superconducting transition-edge sensors) must be kept at cryogenic temperatures, and state-of-the-art number-resolving detectors have only now demonstrated the ability to resolve up to $\sim 100$ photons [24], while phase-sensitive (quadrature) measurements like homodyne and heterodyne detection require less-specialized equipment.

## 4.2 Optimal homodyning schemes

In this Letter, we present homodyne-detection-based schemes that are optimal (in the absence of photon loss, $p = 0$ in Fig. 4.1) for quantum interferometry with either of two path-entangled coherent states—an ECS [25] or a qubit–which-path (QWP) state [17]:

$$|{\rm ECS}\rangle = \mathscr{N}_\alpha(|\alpha,0\rangle + |0,\alpha\rangle), \tag{4.2}$$

$$|{\rm QWP}\rangle = \frac{1}{\sqrt{2}}\left(|\uparrow\rangle|\alpha,0\rangle + |\downarrow\rangle|0,\alpha\rangle\right), \tag{4.3}$$

where $\mathscr{N}_\alpha = [2(1+e^{-|\alpha|^2})]^{-1/2}$. Here, $|\alpha_1,\alpha_2\rangle = \prod_{i=1,2} {\rm D}_i(\alpha_i)|0\rangle$, with vacuum state $|0\rangle$ and displacement operator ${\rm D}_i(\alpha) = e^{\alpha a_i^\dagger - {\rm h.c.}}$. This is a two-mode coherent state with amplitude $\alpha_i$ in the traveling-wave mode that is annihilated by $a_i$, located in arm $i = 1,2$ of the interferometer. In Eq. (4.3), the states $|\uparrow\rangle, |\downarrow\rangle$ are energy eigenstates of a two-level system (qubit).

As light passes through the interferometer, the initially prepared state acquires a dependence on the differential phase $\phi = \phi_1 - \phi_2$ through unitary evolution generated by $U_\phi = \prod_{i=1,2} e^{-i\phi_i n_i}$, where $n_i = a_i^\dagger a_i$ is the number operator for mode $i$. For a pure state, the quantum Fisher information $I_{\rm Q}$ of $|S_\phi\rangle = U_\phi|S\rangle$ is given in terms of the variance of $J_3 = (n_1 - n_2)/2$ with respect to $|S_\phi\rangle$ as $I_Q(|S_\phi\rangle) = 4\mathrm{Var}_{|S_\phi\rangle}(J_3)$ [14]. Evaluating the variance gives

$$I_Q(|\mathrm{ECS}_\phi\rangle) = \bar{n}^2 + [1 + w(\bar{n}e^{-\bar{n}})]\bar{n}, \tag{4.4}$$

$$I_Q(|\mathrm{QWP}_\phi\rangle) = \bar{n}^2 + \bar{n}, \tag{4.5}$$

where $\bar{n} = \langle n_1 + n_2\rangle$ is the total average number of photons, and where $w(z)$ is the Lambert $W$ function. For a QWP state, $\bar{n} = |\alpha|^2$. For an ECS, however, $\bar{n} = |\alpha|^2/(1+e^{-|\alpha|^2})$. Inverting this relation is what produces a dependence on $w(\bar{n}e^{-\bar{n}})$. The term $\propto w$ in $I_Q(|\mathrm{ECS}_\phi\rangle)$ provides a small advantage over the QWP state at small $\bar{n}$. For large $\bar{n}$, however, the advantage is exponentially suppressed since $w(\bar{n}e^{-\bar{n}}) \simeq \bar{n}e^{-\bar{n}}$ for $\bar{n} \gg 1$. At large $\bar{n}$, both ECS's and QWP states provide Heisenberg-limited scaling $\propto \bar{n}^2$. Both states (ECS and QWP) also have a small precision advantage over N00N states consisting of superpositions $|N00N\rangle \propto |N,0\rangle + |0,N\rangle$ of $N$-photon Fock states, for which $I_Q(|N00N_\phi\rangle) = N^2$ [19–21]. An analogous expression for the quantum Fisher information of an ECS [Eq. (4.4)] was derived in Ref. [26] for estimation of the total phase shift $\phi_1$ in mode 1 (generated by $a_1^\dagger a_1$), rather than estimation of the differential phase shift $\phi = \phi_1 - \phi_2$ (generated by $J_3$).

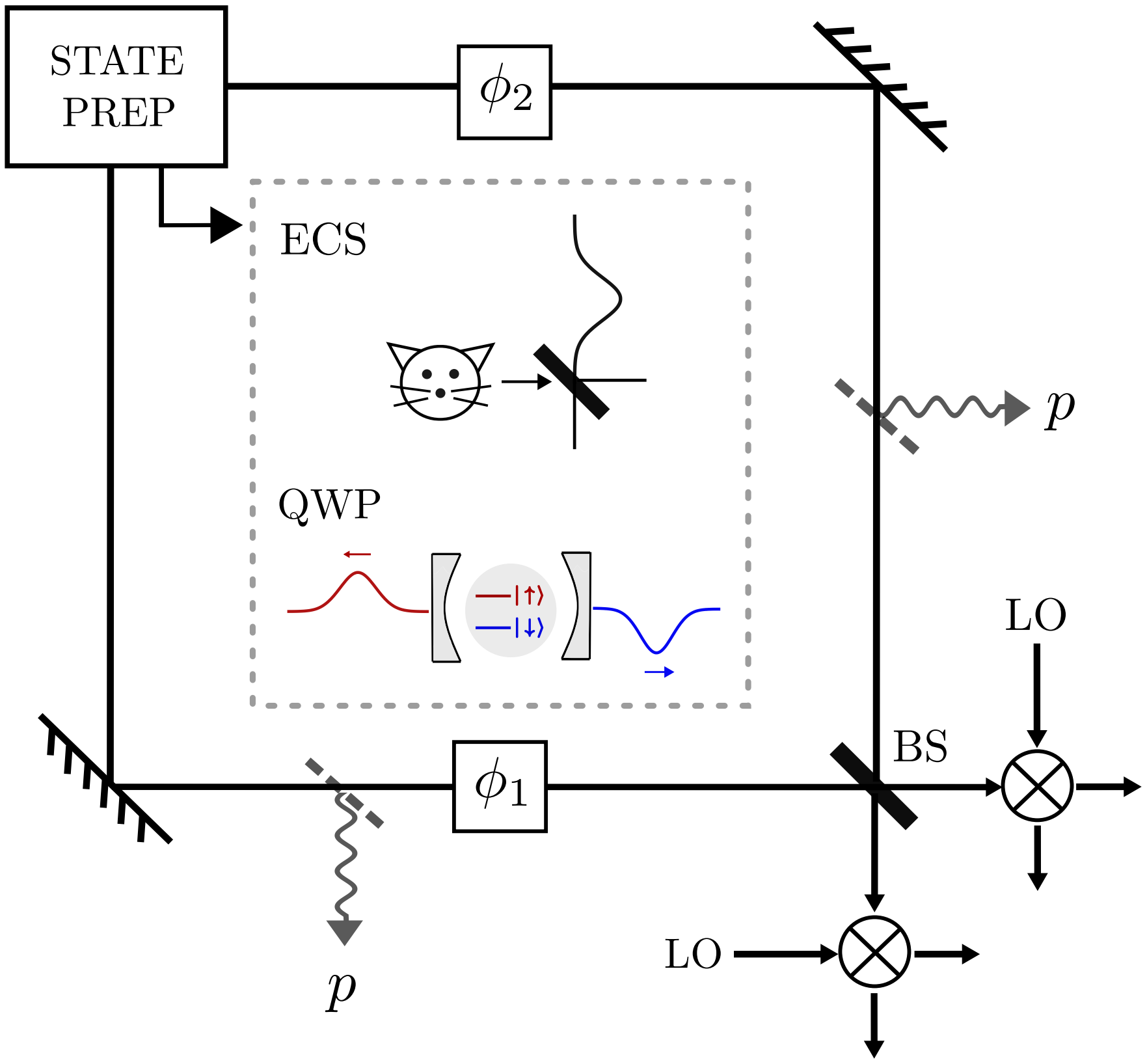

Figure 4.1: A Mach-Zehnder interferometer can be used to estimate the differential phase shift $\phi = \phi_1 - \phi_2$. Photon loss from the interferometer occurs with a probability-per-photon $p$. Homodyne detection is implemented by mixing the output of the interferometer with a local oscillator (LO) field. State preparation: An ECS can be produced by mixing an even cat state $\propto |\alpha/\sqrt{2}\rangle + |-\alpha/\sqrt{2}\rangle$ with a coherent state $|\alpha/\sqrt{2}\rangle$ on a 50:50 beamsplitter [15, 16]. In a cavity-QED setup, a QWP state can be generated by feeding a coherent-state wavepacket into the input port of a cavity containing a qubit prepared in $|+\rangle \propto |\uparrow\rangle + |\downarrow\rangle$, while also modulating the strength of an asymmetric longitudinal ($\propto |\uparrow\rangle\langle\uparrow|$) cavity-qubit coupling [17].

Not every measurement scheme can be used to saturate the quantum CRB. For a scheme where $\phi$ is estimated by measuring some quantity $\mathscr{O}$ having outcomes $x$, described by the positive-operator valued measure (POVM) $\{\hat{\Pi}_x\}$, the standard deviation $\Delta\phi$ of any unbiased estimator $\hat{\phi}(x)$ has a lower bound given by the classical CRB [27],

$$\Delta\phi \geq \delta\phi = \frac{1}{\sqrt{MI_C(\phi)}}. \tag{4.6}$$

Here, the classical Fisher information $I_C(\phi)$ is given by

$$I_C(\phi) \equiv I_C[p(x|\phi)] = \int dx \left(\partial_\phi \ln p(x|\phi)\right)^2 p(x|\phi), \tag{4.7}$$

where $p(x|\phi) = \mathrm{Tr}\{\rho_\phi \hat{\Pi}_x\}$. Under some regularity conditions [requiring, for instance, that $p(x|\phi)$ have a unique global maximum], the maximum-likelihood estimator $\hat{\phi}_{\mathrm{MLE}}(\boldsymbol{x}) = \mathrm{argmax}_\phi\, p(\boldsymbol{x}|\phi)$ saturates the CRB in the asymptotic limit $M \to \infty$, where here, $\boldsymbol{x} = \{x_i\}_{i=1}^M$ is a set of observations sampled from $p(x|\phi)$ [27]. Since the classical CRB can be saturated in principle, a measurement scheme is optimal when its classical Fisher information $I_C(\phi)$ is equal to $I_Q(\rho_\phi)$, in which case $\delta\phi = \delta\phi_{\min}$.

We now explain how homodyne detection can be used to achieve an optimal measurement for ECS's and QWP states in the absence of photon loss. After light passes through the interferometer, the initially prepared state $|S\rangle$ is mapped to $|S_\phi\rangle$. The light is then passed through a 50:50 beamsplitter BS (Fig. 4.1) that maps the interferometer modes $a_i$ $(i = 1,2)$ to output modes $a_\pm = (a_1 \pm a_2)/\sqrt{2}$ via a unitary operation $U_{\mathrm{BS}}$. The resulting state $|\widetilde{S}_\phi\rangle = U_{\mathrm{BS}}|S_\phi\rangle$ is then given by

$$|\widetilde{\mathrm{ECS}}_\phi\rangle = \mathscr{N}_\alpha(|\alpha_{\phi_1}, \alpha_{\phi_1}\rangle + |-\alpha_{\phi_2}, \alpha_{\phi_2}\rangle), \tag{4.8}$$

$$|\widetilde{\mathrm{QWP}}_\phi\rangle = \frac{1}{\sqrt{2}}(|\uparrow\rangle|\alpha_{\phi_1}, \alpha_{\phi_1}\rangle + |\downarrow\rangle|-\alpha_{\phi_2}, \alpha_{\phi_2}\rangle), \tag{4.9}$$

where $\alpha_{\phi_j} = e^{i\phi_j}\alpha/\sqrt{2}$. The measurement schemes consist of (I) measuring modes $a_\pm$ with homodyne

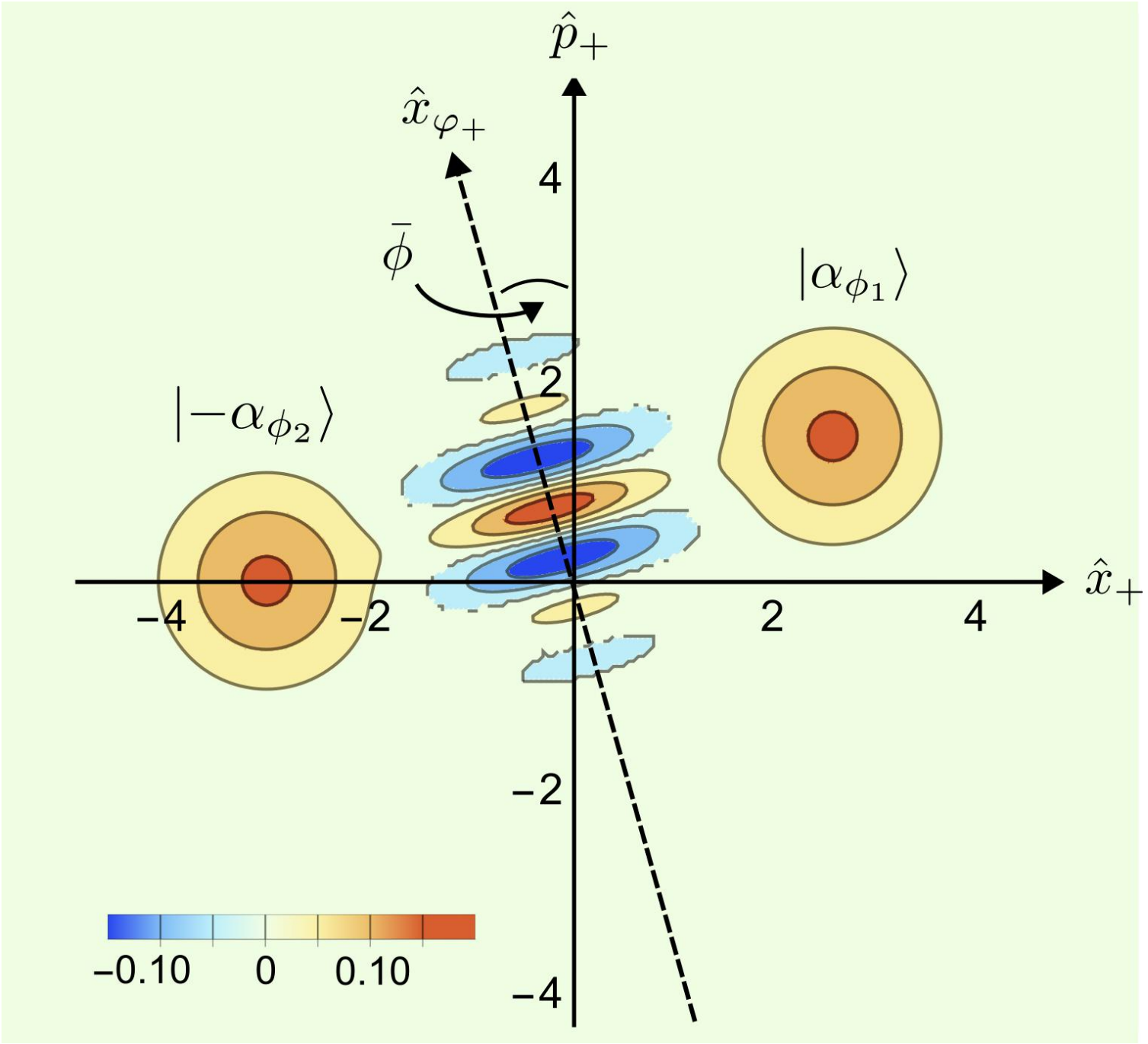

Figure 4.2: Reduced Wigner distribution $W(x_+, p_+) = \int dx_- dp_- W(x_+, p_+, x_-, p_-)$ of mode $a_+$, where here, $W(x_+, p_+, x_-, p_-)$ is the Wigner distribution of the state resulting from an initial ECS [Eq. (4.8)] for $\alpha = 3$, $\phi_1 = 0.5$, and $\phi_2 = 0$. Homodyne detection of mode $a_+$ with a local-oscillator phase of $\varphi_+ = \pi/2 + \bar{\phi}$ implements a projection onto the rotated quadrature indicated by the dashed black line. Here, $\alpha_{\phi_j} = e^{i\phi_j}\alpha/\sqrt{2}$ [cf. Eqs. (4.8) and (4.9)].

detection using local-oscillator phases $\varphi_\pm$, respectively, where

$$\begin{aligned} \varphi_+ &= \frac{\pi}{2} + \bar{\phi}, \\ \varphi_- &= \bar{\phi}, \\ \bar{\phi} &= \frac{1}{2}(\phi_1 + \phi_2). \end{aligned} \tag{4.10}$$

Prior information about the average phase $\bar{\phi}$ is therefore required. For the ECS, that completes the measurement. In the case of the QWP state, the homodyne measurements are followed by (II) a measurement of the qubit in the Pauli-$X$ basis, with outcomes $X = \pm$ for states $|\pm\rangle = (|\uparrow\rangle \pm |\downarrow\rangle)/\sqrt{2}$.

To evaluate the classical Fisher information [Eq. (4.7)] associated with the measurement schemes presented here, we derive conditional probability distributions $p_S(x|\phi)$ governing the measurement outcomes, where for both states $S = \text{ECS}, \text{QWP}$, the variable $x$ includes the two outcomes for homodyne detection of modes $a_\pm$ [Step (I)], and where for the QWP state, $x$ also includes the outcome of the $X$-basis qubit measurement [Step (II)]. We find that in the absence of photon loss ($p = 0$), the measurement schemes described by (I)-(II) are optimal,

$$I_\text{C}[p_S(x|\phi)] = I_\text{Q}(|S_\phi\rangle), \quad S = \text{ECS}, \text{QWP}. \tag{4.11}$$

The optimality is a consequence of choosing local-oscillator phases $\varphi_\pm$ [Eq. (4.10)] that make the measurement outcomes maximally sensitive to the $\phi$-dependent fringes in the Wigner distribution of $|\tilde{S}_\phi\rangle$ [Eqs. (4.8) and (4.9)] (see Fig. 4.2 for the case of $S = \text{ECS}$). The measurement schemes presented above only make use of homodyne detection and (in the case of the QWP state) single-qubit control/readout. Notably, we have found that an optimal measurement for the QWP state can be devised without the use of entangling operations, such as the controlled-phase gate considered in Ref. [28] as a way of mapping phase information from a bosonic system into the state of a qubit. Such entangling operations may be difficult to implement in an interferometer. Additionally, while the authors of Ref. [29] have argued that achieving Heisenberg-limited metrology with an ECS cannot be accomplished with homodyning, we show here that this is untrue provided we have prior information about $\bar{\phi}$.

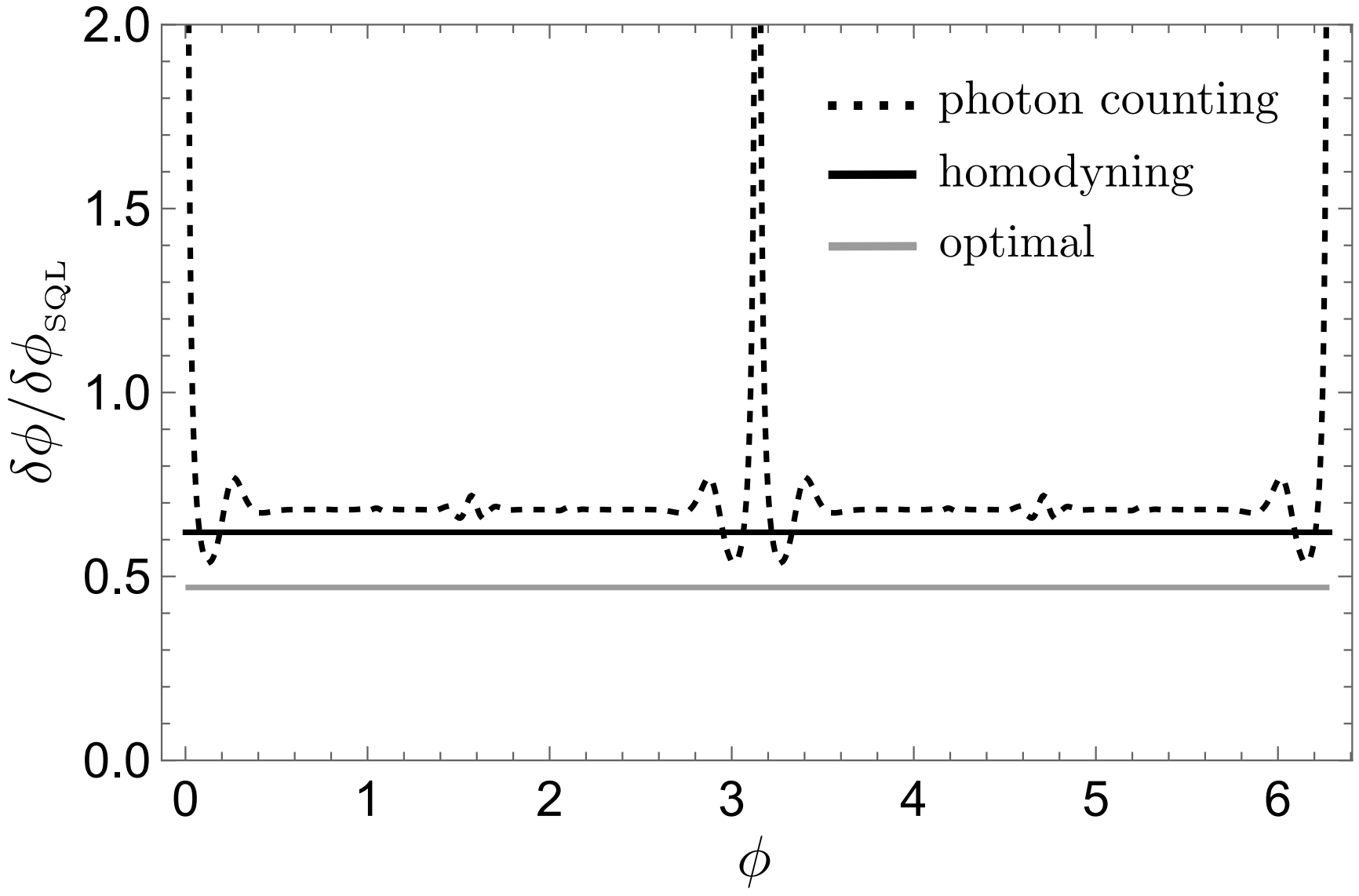

Figure 4.3: Precision $\delta\phi$ [Eq. (4.6)] as a function of $\phi$, relative to the standard quantum limit $\delta\phi_{\rm SQL} \equiv [(1-p)M\bar{n}]^{-1/2}$, for an ECS with $\bar{n} = 10$ and $p = 0.05$ (5% photon loss). The values that can be achieved with homodyne detection (solid black line) and photon counting (dashed black line) were calculated using the probability distributions given in Eqs. (4.14) and (4.15), respectively. The gray line corresponds to the optimal precision [$\delta\phi_{\rm min}$ with the quantum Fisher information given in Eq. (4.13)].

## 4.3 Phase independence is robust to photon loss

A consequence of Eq. (4.11) is that the precision $\delta\phi \propto [I_{\rm C}(\phi)]^{-1/2}$ that can be achieved using these measurements is independent of the true value of $\phi$ (since $I_{\rm Q}$ is $\phi$-independent), allowing for an optimal non-adaptive measurement without prior knowledge of $\phi$. In the case of an ECS, this can be contrasted to the scheme based on photon-number parity measurements [16], where information about $\phi$ is extracted by determining whether the number of photons in one of the output modes of the interferometer is even or odd. Although parity measurements are sub-optimal, they can nevertheless be used to achieve Heisenberg-limited scaling [16]. For $|\widetilde{\rm ECS}_\phi\rangle$, the probability $p(\text{even}|\phi)$ of measuring an even number of photons in one of the output modes exhibits $\phi$-dependent oscillations that can be used to extract information about $\phi$ [16]. However, since the visibility of these oscillations is suppressed by a factor $e^{-|\alpha|^2\sin^2(\phi/2)}$, the scheme is effective in the limit $\bar{n} \simeq |\alpha|^2 \gg 1$ only if $|\alpha|^2\phi^2 \ll 1$, requiring prior knowledge of $\phi$ with a precision $\sim 1/|\alpha|$. The need for prior characterization of $\phi$ could be eliminated by retaining the full counting statistics, as photon counting is also optimal for an ECS [18, 30]. (We find that photon counting is optimal for a QWP state as well, when supplemented by a final $X$-basis measurement of the qubit [31].) However, as soon as photon loss is introduced, the classical Fisher information associated with photon counting acquires a dependence on $\phi$, and some amount of *a priori* knowledge is required in order to avoid values of $\phi$ where the Fisher information vanishes (in which case $\delta\phi \to \infty$). As we now show, the homodyne-based measurement schemes presented here do not suffer from this drawback (Fig. 4.3).

From this point onward, we focus on the ECS and therefore dispense with the use of explicit subscripts indicating the state being considered. The results for the QWP state are qualitatively similar and are given in the Supplementary Material [31].

To investigate performance accounting for photon loss, we model losses in the interferometer by inserting a fictitious beamsplitter into each interferometer arm [26]. These beamsplitters are modeled by the operator $R_{c,c_\ell}(p) = e^{\arcsin\sqrt{p}(c_\ell^\dagger c - \text{h.c.})}$, describing scattering of photons from mode $c$ into loss mode $c_\ell$ with probability $p$. Under the action of the lossy interferometer, the initial state $|\text{ECS}\rangle$ [Eq. (4.2)] evolves to

$$\rho_\phi = \text{Tr}_\ell\{RU_\phi\rho_0U_\phi^\dagger R^\dagger\}, \quad R = \prod_{i=1,2} R_{a_i,a_{\ell_i}}(p), \tag{4.12}$$

where $\rho_0 = |\text{ECS}\rangle\langle\text{ECS}|\otimes|0\rangle\langle 0|_\ell$ is the initial state of the interferometer and loss modes (annihilated by $a_{\ell_i}$, $i = 1,2$), and where $\text{Tr}_\ell$ describes a trace over the state of both loss modes. Note that the same state $\rho_\phi$ is obtained regardless of the order in which $U_\phi$ and $R$ are applied. For a mixed state $\rho_\phi$, the quantum Fisher information of $\rho_\phi$ with respect to $J_3$ can be calculated by evaluating matrix elements of $J_3$ in the eigenbasis of $\rho_\phi$ [14]. This procedure gives

$$I_\text{Q}(\rho_\phi) = (1-p)^2\bar{n}^2e^{-2p[\bar{n}+w(\bar{n}e^{-\bar{n}})]} + (1-p)\bar{n}[1+(1-p)w(\bar{n}e^{-\bar{n}})], \tag{4.13}$$

where $w(z)$ is again the Lambert $W$ function. Photon loss therefore controls a transition from Heisenberg-limited ($\propto \bar{n}^2$) scaling to scaling at the standard quantum limit ($\propto \bar{n}$). An analogous result for estimation of the total phase shift $\phi_1$ in arm 1 (rather than $\phi = \phi_1 - \phi_2$), accounting for photon loss, was presented in Ref. [26]. A detailed derivation of the quantum Fisher information of the QWP state, accounting for photon loss and qubit dephasing, is given in the Supplementary Material [31].

Homodyne detection is performed by mixing the signal field with a local oscillator prepared in a coherent state $|\beta\rangle$, where here, we assume that $\beta \in \mathbb{R}^+$. In the strong-oscillator limit $|\beta| \gg |\alpha|$, homodyne detection of mode $a$ with a local oscillator in state $|\beta e^{i(\varphi-\pi)}\rangle$ implements a projection onto the eigenbasis $|x_\varphi\rangle = e^{-i\varphi a^\dagger a}|x\rangle$ of the rotated quadrature operator $\hat{x}_\varphi = \hat{x}\cos\varphi + \hat{p}\sin\varphi$ [32], where here, $\hat{x} = (a^\dagger + a)/\sqrt{2}$ and $\hat{p} = i(a^\dagger - a)/\sqrt{2}$ are canonically conjugate, and where $|x\rangle$ is an eigenstate of $\hat{x}$ with eigenvalue $x$. For measurement of mode $a_+$ with local-oscillator phase $\varphi_+ = \pi/2 + \bar{\phi}$ [Eq. (4.10)], this corresponds to projecting the coherent state in mode $a_+$ onto a quadrature rotated by an amount $\bar{\phi}$ relative to the out-of-phase quadrature: $\hat{x}_{\varphi_+} = -\hat{x}_+\sin\bar{\phi} + \hat{p}_+\cos\bar{\phi}$ (Fig. 4.2). For the measurement scheme presented here, the POVM element describing the measurement of the ECS [Step (I)] is therefore given by $\hat{\Pi}_x = \bigotimes_{\sigma=\pm} e^{-i\varphi_\sigma a_\sigma^\dagger a_\sigma}|x_\sigma\rangle\langle x_\sigma|e^{i\varphi_\sigma a_\sigma^\dagger a_\sigma}$. Without loss of generality, we assume that $\alpha \in \mathbb{R}$, in which case the probability distribution $p(x|\phi) = \text{Tr}\{\rho_\phi\hat{\Pi}_x\}$ governing the homodyne-measurement outcomes is given by

$$p(x|\phi) = 2\mathcal{N}_\alpha^2\left[1 + e^{-p\alpha^2}\cos\Theta_x(\phi)\right]\prod_{s=\pm} g_s(x_s,\phi), \tag{4.14}$$

where $g_s(x_s,\phi) = \pi^{-1/2}\exp\{-[x_s - \mu_s(\phi)]^2\}$, $\mu_+(\phi) = \sqrt{1-p}\alpha\sin\frac{\phi}{2}$, $\mu_-(\phi) = \sqrt{1-p}\alpha\cos\frac{\phi}{2}$, and $\Theta_x(\phi) = 2x_+\mu_-(\phi) - 2x_-\mu_+(\phi)$. Setting $p = 0$, this result [Eq. (4.14)] recovers Eq. (4.11) for $S = \text{ECS}$.

The term $\sim \cos\Theta_x(\phi)$ in Eq. (4.14) is a consequence of phase-space interference in the Wigner distribution of $\rho_\phi$. To build intuition for this, consider the single-mode cat state $|\text{C}_+\rangle \propto (|\alpha\rangle + |-\alpha\rangle)$. For $\alpha \in \mathbb{R}$, the states $|\pm\alpha\rangle$ are displaced along the $\hat{x}$ quadrature. A homodyne measurement with a local-oscillator phase $\pi/2$ (corresponding to a projection onto the $\hat{p}$ axis) then returns a displacement $x_{\pi/2}$ with probability $p(x_{\pi/2}) \propto (1 + \cos\sqrt{8}\alpha x_{\pi/2})$ [33], where here, the oscillating term is a reflection of interference fringes parallel to the $\hat{x}$-axis in the Wigner distribution of $|\text{C}_+\rangle$. For the measurement of modes $a_\pm$ proposed here, the local oscillator phases $\varphi_\pm$ [Eq. (4.10)] are both chosen so that the phase-space axis associated with the measured quadrature $\hat{x}_{\varphi_\pm}$ bisects the angle subtended by the coherent-state displacement of modes $a_\pm$ in the two branches of $|\widetilde{\text{ECS}_\phi}\rangle$ (Fig. 4.2). Measurements of displacements along these axes are therefore maximally sensitive to the interference fringes between the two branches, resulting in an optimal detection scheme in the ideal scenario of zero photon loss ($p = 0$) [cf. Eq. (4.11)]. The dependence of these interference fringes on $\phi$ is what produces Heisenberg-limited scaling $\propto \bar{n}^2$ in the classical Fisher information for this measurement scheme.

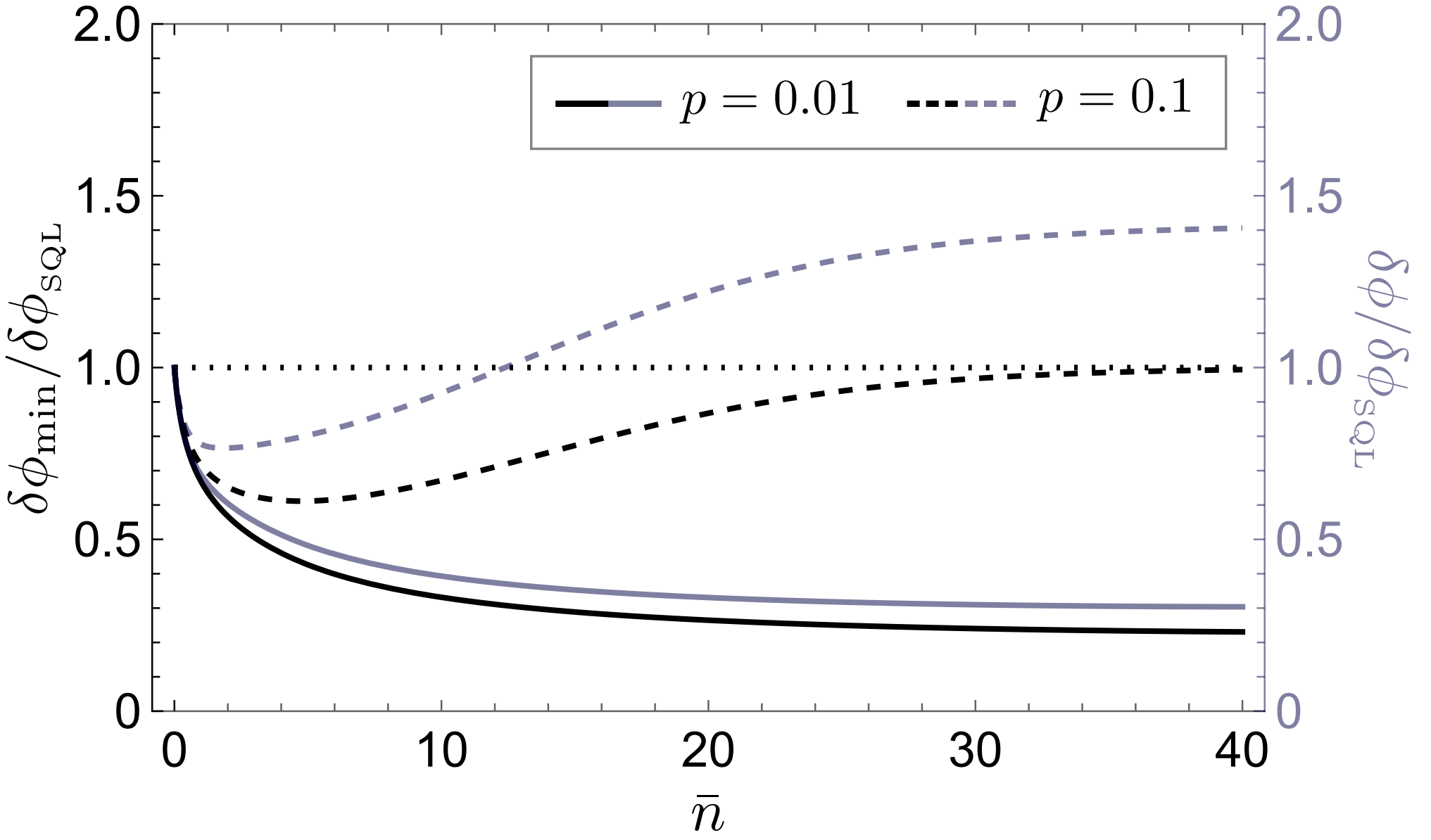

Figure 4.4: Left axis: Precision given by the quantum CRB, $\delta\phi_{\min}$, relative to the standard quantum limit $\delta\phi_{\rm SQL} \equiv [(1-p)M\bar{n}]^{-1/2}$, for an ECS with photon loss $p = 0.01$ (solid black line) and $p = 0.1$ (dashed black line). Right axis: Precision that can be attained with homodyning, $\delta\phi$, for $p = 0.01$ (solid gray line) and $p = 0.1$ (dashed gray line).

In Fig. 4.4, we compare the precision $\delta\phi$ that can be achieved using this homodyning scheme to $\delta\phi_{\min}$ [Eq. (4.1)] for two values of $p$. For $\bar{n} \gg p^{-1}$, the performance of the homodyning scheme saturates at $\delta\phi = \sqrt{2}\delta\phi_{\rm SQL}$ (Fig. 4.4). This is because for $\bar{n} \simeq \alpha^2 \gg p^{-1}$, the interference term in Eq. (4.14) is exponentially suppressed, and $p(x|\phi)$ is given approximately by the product of two Gaussians: $p(x|\phi) \approx \prod_{s=\pm} g_s(x_s,\phi)$. In this case, $I_{\rm C}[p(x|\phi)] \approx I_{\rm C}[g_+(x_+,\phi)] + I_{\rm C}[g_-(x_-,\phi)]$. Noting that $\mu_\pm(\phi)$ both oscillate with a period $4\pi$ (rather than $2\pi$), the factor of $\sqrt{2}$ relating $\delta\phi$ to $\delta\phi_{\rm SQL}$ in the limit $p\bar{n} \gg 1$ can therefore be understood as a consequence of "sub-resolution" in the Gaussian distributions, to be contrasted with super-resolution [34], where the distributions would instead depend on an amplified phase $m\phi$ with $m > 1$.

As discussed above, photon counting can also be used to saturate the quantum CRB for an ECS in the absence of photon loss [18, 30]. Accounting for photon loss, the probability $p(m,n|\phi)$ of detecting $m$ and $n$ photons in modes $a_+$ and $a_-$, respectively, is given by

$$p(m,n|\phi) = 2\mathcal{N}_\alpha^2 \left[1 + e^{-p\alpha^2} \cos\Theta_{m,n}(\phi)\right] \prod_{j=m,n} P(j;\lambda_\alpha), \tag{4.15}$$

where $\Theta_{m,n}(\phi) = (m+n)\phi + m\pi$, $P(j;\lambda) = e^{-\lambda}\lambda^j/j!$, and $\lambda_\alpha = (1-p)\alpha^2/2$. We have verified numerically that the Fisher information $I_{\rm C}[p(x|\phi)]$ is independent of $\phi$, while $I_{\rm C}[p(m,n|\phi)] = 0$ for $\phi = 0, \pi$, leading to singularities in Fig. 4.3. In the asymptotic limit $M \to \infty$, the distribution of outcomes $(\hat{\phi}_{\rm MLE} - \phi)$ associated with the maximum-likelihood estimate $\hat{\phi}_{\rm MLE}$ of $\phi$ converges to a zero-mean normal distribution with variance $\delta\phi^2 = 1/MI_{\rm C}(\phi)$ [27]. For maximum-likelihood estimation in the vicinity of $\phi = 0, \pi$, however, the maximum-likelihood estimator will not converge to the true value of $\phi$ when an estimation strategy based on photon counting is used. This caveat is not present when the homodyning scheme is used instead, due to the phase independence of $I_{\rm C}[p(x|\phi)]$ (Fig. 4.3).

Here, we have presented measurement schemes based on homodyne detection that are optimal, in the absence of photon loss, for interferometry using either an ECS or a QWP state. The schemes achieve optimality by using prior knowledge of the average phase $\bar{\phi}$ to choose local-oscillator phases that maximize the sensitivity of measurement outcomes to the $\phi$-dependent interference fringes in the states' Wigner distributions. We have also shown that the achievable precision, as given by the CRB, is independent of the true value of $\phi$, even in the presence of photon loss.

A natural extension of the strategies used here would be to investigate whether an optimal homodyning scheme can be found for the Caves state (produced by mixing a coherent state with squeezed vacuum [8]). For a coherent state combined on a beamsplitter with any other quantum state of light, it was found that the squeezed vacuum produces the largest quantum Fisher information at a fixed average photon number [10]. A Caves state therefore has greater potential sensitivity than either the ECS or the QWP state investigated here. For a Caves state, it is known that photon-number parity measurements saturate the quantum CRB in the absence of photon loss, but only in the vicinity of $\phi = 0$ [35]. In the presence of photon loss, it was found in Ref. [36] that a non-optimal homodyning scheme for the Caves state exhibited better sensitivity than parity measurements. Optimizing the homodyning scheme for this state using the ideas presented here could therefore lead to a better and more practical inference method.

**Acknowledgments**—We thank J. Sankey for useful discussions. We also acknowledge funding from the Natural Sciences and Engineering Research Council of Canada (NSERC) and from the Fonds de Recherche du Québec–Nature et technologies (FRQNT).

# Appendices to Chapter 4

*These appendices contain the original text of the supplementary material to*
Z. M. McIntyre and W. A. Coish, Phys. Rev. A **110**, L010602 (2024)

In this supplement, we provide additional results for quantum metrology with qubit–which-path (QWP) states. In Sec. C1, we give a derivation of the quantum Fisher information of the QWP state accounting for photon loss and qubit dephasing. Next, in Sec. C2, we calculate the classical Fisher information of the homodyning scheme presented in the main text, for a QWP state, and show that it is optimal in the ideal case of no photon loss or qubit dephasing. Finally, in Sec. C3, we show that in the absence of photon loss and qubit dephasing, photon counting followed by an $X$-basis qubit readout also constitutes an optimal strategy for a QWP state. We then derive the classical Fisher information for this counting-based strategy and show that, unlike the classical Fisher information of the homodyning scheme, it depends on the true value of the phase being estimated, similar to the case discussed in the main text for an ECS.

## C1 Quantum Fisher information for qubit–which-path states

Here, we provide a detailed derivation of the quantum Fisher information $I_{\mathrm{Q}}(\rho_\phi, J_3)$ of QWP states with respect to $J_3 = (a_1^\dagger a_1 - a_2^\dagger a_2)/2$, accounting for photon loss and qubit dephasing.

The qubit and modes 1,2 of the interferometer are initially prepared in the state

$$|\mathrm{QWP}\rangle = \frac{1}{\sqrt{2}}\left(|\uparrow\rangle\,|\alpha,0\rangle + |\downarrow\rangle\,|0,\alpha\rangle\right). \tag{C1}$$

The Hamiltonian describing a qubit undergoing dephasing due to a combination of classical noise and a quantum environment is given by

$$H_{\mathrm{QE}}(t) = \frac{1}{2}[\omega_{\mathrm{q}}(t) + h]Z + H_{\mathrm{E}}, \tag{C2}$$

where $Z = |\uparrow\rangle\langle\uparrow| - |\downarrow\rangle\langle\downarrow|$ is a Pauli-Z operator, $\omega_{\mathrm{q}}(t)$ is the randomly fluctuating qubit splitting, and $h$ is an operator acting on the qubit's environment (having decoupled Hamiltonian $H_{\mathrm{E}}$). As explained in the main text, we model photon loss in the interferometer via a beamsplitter-type unitary [26]

$$R_{c,c_\ell}(p) = e^{\arcsin\sqrt{p}(c_\ell^\dagger c - \mathrm{h.c.})}, \tag{C3}$$

describing scattering of photons from mode $c$ to loss mode $c_\ell$ with a per-photon probability $p$. Under the action of the lossy interferometer, the initial state $\rho(0) = |\mathrm{QWP}\rangle\langle\mathrm{QWP}| \otimes |0\rangle\langle 0|_\ell \otimes \rho_{\mathrm{E}}$ (where $|0\rangle_\ell$ denotes the common vacuum state of the loss modes and $\rho_{\mathrm{E}}$ is the initial state of the environment) evolves into the mixed state

$$\rho_\phi(t) = \langle\!\langle \langle \mathrm{Tr}_\ell\{R U_\phi(t) \rho(0) U_\phi^\dagger(t) R^\dagger\}\rangle_{\mathrm{E}} \rangle\!\rangle, \quad R = \prod_{i=1,2} R_{a_i, a_{\ell_i}}(p), \tag{C4}$$

where here, $\mathrm{Tr}_\ell$ denotes a trace over the state of the loss modes, $\langle\rangle_{\mathrm{E}}$ denotes an average over the initial state of the environment, $\langle\!\langle\,\rangle\!\rangle$ is an average over realizations of $\omega_{\mathrm{q}}(t)$, and where evolution of the qubit and modes 1,2 is described by

$$U_\phi(t) = \mathscr{T} e^{-i\int_0^t d\tau H_{\mathrm{QE}}(\tau)} \prod_{i=1,2} e^{-i\phi_i \hat{n}_i}. \tag{C5}$$

Here, $\mathscr{T}$ is the time-ordering operator and (as in the main text) $\phi_i$ is the phase acquired by photons passing through interferometer arm $i = 1,2$, with number operator $n_i$. Note that the same state $\rho_\phi(t)$ is obtained regardless of the order in which $U_\phi(t)$ and $R$ are applied.

In terms of the eigenstates $|\lambda_k\rangle$ and eigenvalues $\lambda_k$ of $\rho_\phi$ [Eq. (C4)], the quantum Fisher information of $\rho_\phi$ with respect to $J_3$ is given by [14]

$$I_{\mathrm{Q}}(\rho_\phi, J_3) = 2\sum_{k,j} \frac{(\lambda_k - \lambda_j)^2}{\lambda_k + \lambda_j} |\langle\lambda_k|J_3|\lambda_j\rangle|^2. \tag{C6}$$

To find the spectral decomposition of $\rho_\phi$, note that the action of the beamsplitter $R$ [cf. Eq. (C4)] on $|\alpha_1,\alpha_2\rangle\,|0\rangle_\ell$ ($\ell = \ell_1,\ell_2$) is given by

$$R\,|\alpha_1,\alpha_2\rangle\,|0\rangle_\ell = |\sqrt{1-p}\,\alpha_1, \sqrt{1-p}\,\alpha_2\rangle\,|\sqrt{p}\,\alpha_1, \sqrt{p}\,\alpha_2\rangle_{\ell_1,\ell_2}. \tag{C7}$$

Using Eq. (C7) and taking the trace over the loss modes, we then find that

$$\rho_\phi(t) = \frac{1}{2}\sum_{\sigma=\uparrow,\downarrow} |\Psi_\sigma\rangle\langle\Psi_\sigma| + \frac{1}{2}e^{-p|\alpha|^2-\chi(t)}\left[e^{-i\vartheta(t)}\,|\Psi_\uparrow\rangle\langle\Psi_\downarrow| + \mathrm{h.c.}\right], \tag{C8}$$

where $\left|\Psi_\uparrow\right\rangle = |\uparrow\rangle\left|e^{i\phi_1}\sqrt{1-p}\,\alpha, 0\right\rangle$ and $\left|\Psi_\downarrow\right\rangle = |\downarrow\rangle\left|0, e^{i\phi_2}\sqrt{1-p}\,\alpha\right\rangle$, and where $\chi(t)$ and $\vartheta(t)$ are defined via

$$\langle\!\langle\langle U_\uparrow(t)U_\downarrow^\dagger(t)\rangle_{\mathrm{E}}\rangle\!\rangle \equiv e^{-\chi(t)-i\vartheta(t)}, \quad U_\sigma(t) = \mathscr{T} e^{-i\int_0^t d\tau \langle\sigma|H_{\mathrm{QE}}(\tau)|\sigma\rangle}. \tag{C9}$$

Closed-form expressions for $\chi(t)$ and $\vartheta(t)$ in terms of the spectral density of the environment can be found in a weak-coupling approximation and often (but not always) assuming approximate Gaussian fluctuations. Various forms can be found depending on details of the environmental initial conditions and the specific form of the qubit-environment coupling term $h$. Notably, $\vartheta(t) = \vartheta_{\mathrm{dyn}}(t) + \vartheta_{\mathrm{q}}(t)$ generally includes a contribution $\vartheta_{\mathrm{q}}(t)$ that is unique to a quantum environment [37–40], in addition to the usual (classical) dynamical phase, $\vartheta_{\mathrm{dyn}}(t) = \int_0^t d\tau\,(\langle\!\langle\omega_q(\tau)\rangle\!\rangle + \langle h(\tau)\rangle_{\mathrm{E}})$ (with $h(t) = e^{iH_{\mathrm{E}}t}he^{-iH_{\mathrm{E}}t}$). Expressions for $\vartheta_{\mathrm{q}}(t)$ given in, e.g., Refs. [37–40] in the context of dynamical decoupling can used in Eq. (C9) by restricting to the case of free-induction decay (where no decoupling pulses are applied).

Equation (C8) is easy to diagonalize, giving eigenvalues and eigenvectors

$$\lambda_\pm(t) = \frac{1}{2}(1 \pm e^{-p|\alpha|^2-\chi(t)}), \tag{C10}$$

$$|\lambda_\pm\rangle_t = \frac{1}{\sqrt{2}}(\left|\Psi_\downarrow\right\rangle \pm e^{i\vartheta(t)}\left|\Psi_\uparrow\right\rangle). \tag{C11}$$

Calculating the quantum Fisher information using Eq. (C6) then becomes a question of keeping track of indices, which run over $k = \pm$ as well as $k \neq \pm$ (for which $\lambda_k = 0$). We break up the sum as follows:

$$I_{\mathrm{Q}}(\rho_\phi, J_3) = 4\sum_{k=\pm}\sum_{j\neq\pm} \lambda_k |\langle\lambda_j|J_3|\lambda_k\rangle|^2 + 4(\lambda_+ - \lambda_-)^2 |\langle\lambda_+|J_3|\lambda_-\rangle|^2. \tag{C12}$$

The double sum in Eq. (C12) can be rewritten as

$$\sum_{\substack{k=\pm\\ j\neq\pm}} \lambda_k |\langle\lambda_j|J_3|\lambda_k\rangle|^2 = \sum_{k=\pm} \lambda_k \langle\lambda_k|J_3 \sum_{j\neq\pm} |\lambda_j\rangle\langle\lambda_j| J_3 |\lambda_k\rangle, \tag{C13}$$

where $\sum_{j\neq\pm} |\lambda_j\rangle\langle\lambda_j| = \mathbb{1} - \sum_{k=\pm} |\lambda_k\rangle\langle\lambda_k|$. From here, Eq. (C12) can be evaluated using $|\langle\lambda_\pm|J_3|\lambda_\mp\rangle|^2 = \frac{1}{4}(1-p)^2\bar{n}^2$ and $\langle\lambda_\pm|J_3^2|\lambda_\pm\rangle = \frac{1}{4}[(1-p)^2\bar{n}^2 + (1-p)\bar{n}]$, giving

$$I_{\mathrm{Q}}(\rho_\phi, J_3) = e^{-2p\bar{n}-2\chi(t)}(1-p)^2\bar{n}^2 + (1-p)\bar{n}, \tag{C14}$$

where $\bar{n} = |\alpha|^2$ is the average number of photons in the initial QWP state.

## C2 Fisher information for homodyne detection followed by qubit readout

Next, we consider the measurement scheme proposed in the main text, consisting of (I) homodyne detection of the output modes $a_\pm$ of the interferometer with local-oscillator phases $\varphi_\pm$, followed by (II) measurement of the qubit in the $X$ basis $|\pm\rangle$, where $|\pm\rangle = (|\uparrow\rangle + |\downarrow\rangle)/\sqrt{2}$. The local-oscillator phases are given in Eq. (10) of the main text,

$$\begin{aligned} \varphi_+ &= \frac{\pi}{2} + \bar{\phi}, \\ \varphi_- &= \bar{\phi}, \\ \bar{\phi} &= \frac{1}{2}(\phi_1 + \phi_2). \end{aligned} \tag{C15}$$

In order to calculate the classical Fisher information $I_{\mathrm{C}}(\phi)$ associated with Steps (I)-(II), we first need to evaluate the conditional probability distribution

$$p(x|\phi) = \mathrm{Tr}\{U_{\mathrm{BS}}\rho_\phi U_{\mathrm{BS}}^\dagger \hat{\Pi}_x\}, \tag{C16}$$

where $\rho_\phi$ is given by Eq. (C8) and $U_{\mathrm{BS}}$ is a beamsplitter unitary that maps the annihilation operators $a_i$ ($i = 1,2$) to new modes $a_\pm = (a_1 \pm a_2)/\sqrt{2}$. In Eq. (C16), we have also introduced the POVM element $\hat{\Pi}_x$ describing homodyne detection of modes $a_\pm$ with outcomes $x_\pm$ [Step (I)], followed by qubit readout with outcome $X = \pm$ [Step (II)]:

$$\hat{\Pi}_x = |X\rangle\langle X| \bigotimes_{\sigma=\pm} |x_{\varphi_\sigma}\rangle\langle x_{\varphi_\sigma}|, \quad |x_{\varphi_\sigma}\rangle = e^{-i\varphi_\sigma a_\sigma^\dagger a_\sigma} |x_\sigma\rangle, \tag{C17}$$

where here, $|x_\sigma\rangle$ is an eigenstate of $\hat{x}_\sigma = (a_\sigma^\dagger + a_\sigma)/\sqrt{2}$ having eigenvalue $x_\sigma$.

Using Bayes' Rule, we can re-write the conditional distribution $p(x|\phi)$ as

$$p(x|\phi) = p(X|x_+, x_-, \phi)p(x_+, x_-|\phi), \tag{C18}$$

where, under the assumption that $\alpha \in \mathbb{R}$, the conditional probability distribution $p(x_+, x_-|\phi)$ governing the distribution of measurement outcomes for Step (I) is given by

$$p(x_+, x_-|\phi) = \prod_{s=\pm} g_s(x_s, \phi), \tag{C19}$$

where $g_s(x_s, \phi) = \pi^{-1/2}\exp\{-[x_s - \mu_s(\phi)]^2\}$, $\mu_+(\phi) = \sqrt{1-p}\alpha\sin\frac{\phi}{2}$, and $\mu_-(\phi) = \sqrt{1-p}\alpha\cos\frac{\phi}{2}$. In contrast to the case of the ECS [cf. Eq. (14) of the main text], the probability distribution governing the homodyne outcomes does not exhibit interference fringes. The relative phase giving rise to such interference is instead kicked back onto the state of the qubit, as we now show.

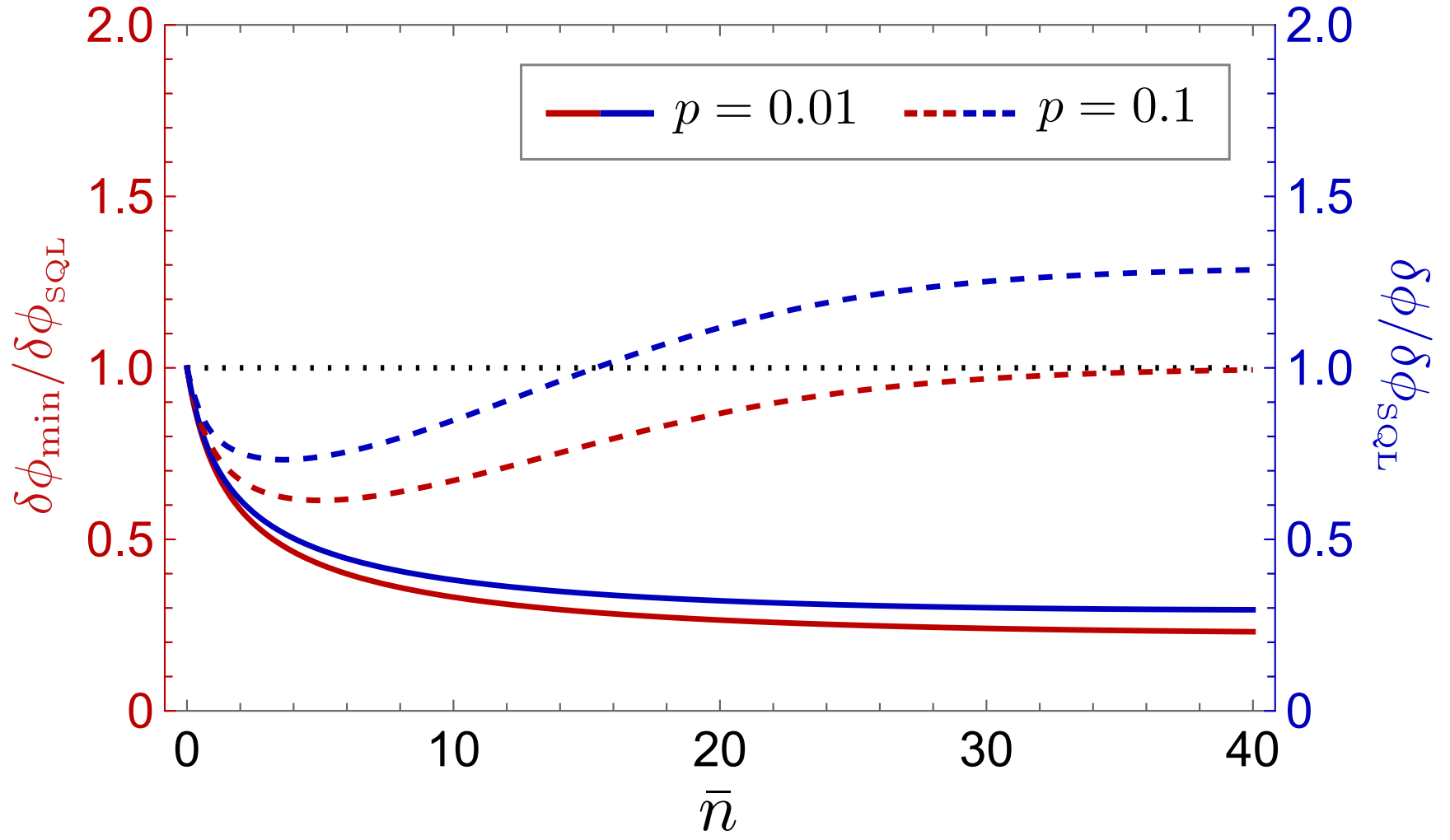

Figure C1: Left axis: Precision given by the quantum CRB $\delta\phi_{\min}$, relative to the standard quantum limit $\delta\phi_{\mathrm{SQL}} \equiv [(1-p)M\bar{n}]^{-1/2}$, for QWP states with $p = 0.01$ (solid red line) and $p = 0.1$ (dashed red line). Right axis: Precision $\delta\phi$ that can be attained with homodyning and $X$-basis qubit readout for $p = 0.01$ (solid blue line) and $p = 0.1$ (dashed blue line). Here, we neglect qubit dephasing, $\chi = \vartheta = 0$.

Using Eq. (C19), we can evaluate the post-measurement state $\rho_{\mathrm{Q}}(x_+, x_-, \phi)$ of the qubit as

$$\rho_{\mathrm{Q}}(x_+, x_-, \phi) = \frac{\mathrm{Tr}_{\mathrm{photons}}\{U_{\mathrm{BS}}\rho_\phi U_{\mathrm{BS}}^\dagger \bigotimes_{\sigma=\pm} |x_{\varphi_\sigma}\rangle\langle x_{\varphi_\sigma}|\}}{p(x_+, x_-|\phi)} \tag{C20}$$

$$= \frac{1}{2}\left[|\uparrow\rangle\langle\uparrow| + |\downarrow\rangle\langle\downarrow| + e^{-p\alpha^2-\chi(t)}\left(e^{-2i[x_+\mu_-(\phi)-x_-\mu_+(\phi)]-i\vartheta(t)}|\uparrow\rangle\langle\downarrow| + \mathrm{h.c.}\right)\right], \tag{C21}$$

where, in the first equality, $\mathrm{Tr}_{\mathrm{photons}}$ denotes a partial trace over the state of the $a_+, a_-$ modes. The probability of obtaining a measurement outcome $\pm$ for the $X$-basis qubit measurement [Step (II)], conditioned on outcomes $x_+, x_-$ for the homodyne measurement, can then be evaluated as

$$p(\pm|x_+, x_-, \phi) = \langle\pm|\rho_{\mathrm{Q}}(x_+, x_-, \phi)|\pm\rangle \tag{C22}$$

$$= \frac{1}{2} \pm \frac{1}{2}e^{-p\alpha^2-\chi(t)}\cos\left[2x_+\mu_-(\phi) - 2x_-\mu_+(\phi) + \vartheta(t)\right]. \tag{C23}$$

In the absence of photon loss ($p = 0$) and qubit dephasing [$\chi(t) = \vartheta(t) = 0$], the classical Fisher information for this measurement sequence is given by

$$I_{\mathrm{C}}[p(x|\phi)] = \bar{n}^2 + \bar{n}, \quad p = 0, \quad \chi = \vartheta = 0, \tag{C24}$$

which is equal to the quantum Fisher information of the QWP state given in Eq. (5) of the main text. When any of $p, \chi, \vartheta \neq 0$, the Fisher information can be evaluated numerically (Fig. C1).

## C3 Photon counting with QWP states

### C3.1 Photon counting saturates the quantum Cramér-Rao bound in the absence of photon loss

In this section, we show that photon counting, supplemented by a final $X$-basis measurement of the qubit, can be used to saturate the quantum CRB in the absence of photon loss and qubit dephasing.

For $p=0$, the state of the qubit and modes 1,2 just prior to the final beamsplitter $U_{\mathrm{BS}}$ is given by

$$\left|\mathrm{QWP}_{\phi}\right\rangle = \frac{1}{\sqrt{2}}\left(\left|\uparrow\right\rangle\left|e^{i\phi_1}\alpha,0\right\rangle + \left|\downarrow\right\rangle\left|0,e^{i\phi_2}\alpha\right\rangle\right). \tag{C25}$$

Under the action of $U_{\mathrm{BS}}$, the modes $a_{1,2}$ transform according to $a_{1/2}=(a_+\pm a_-)/\sqrt{2}$, giving Eq. (9) of the main text,

$$\begin{aligned}|\widetilde{\mathrm{QWP}}_{\phi}\rangle &= U_{\mathrm{BS}}\left|\mathrm{QWP}_{\phi}\right\rangle \\ &= \frac{1}{\sqrt{2}}\left(\left|\uparrow\right\rangle\left|e^{i\phi_1}\tfrac{\alpha}{\sqrt{2}},e^{i\phi_1}\tfrac{\alpha}{\sqrt{2}}\right\rangle + \left|\downarrow\right\rangle\left|-e^{i\phi_2}\tfrac{\alpha}{\sqrt{2}},e^{i\phi_2}\tfrac{\alpha}{\sqrt{2}}\right\rangle\right).\end{aligned} \tag{C26}$$

Conditioned on detecting $m$ photons in mode $a_+$ and $n$ photons in mode $a_-$, the post-measurement state of the qubit is then given by

$$\begin{aligned}|\mathrm{Q}_{\phi}^{m,n}\rangle &= \frac{\langle m,n|\widetilde{\mathrm{QWP}}_{\phi}\rangle}{\sqrt{p(m,n)}} \\ &= \frac{1}{\sqrt{2}}\left(e^{i\phi(m+n)}\left|\uparrow\right\rangle + e^{i\pi m}\left|\downarrow\right\rangle\right),\end{aligned} \tag{C27}$$

where here,

$$p(m,n) = \frac{e^{-|\alpha|^2}}{m!n!}\left(\frac{|\alpha|^2}{2}\right)^{m+n} \tag{C28}$$

is the probability of detecting $m$ and $n$ photons in modes $a_+$ and $a_-$, respectively. Due to the entanglement of the qubit with the electromagnetic field, the number-resolving measurements have the effect of kicking back a relative phase $\phi(m+n)$ onto the qubit superposition state [Eq. (C27)]. The relative phase can then be estimated by measuring the qubit.

As in the homodyning scheme, we take the final measurement of the qubit to be a measurement in the Pauli-$X$ eigenbasis. Accounting also for the number-resolving measurements of modes $a_\pm$, the classical Fisher information $I_{\mathrm{C}}(\phi)$ for this strategy is given by

$$I_{\mathrm{C}}(\phi) = \sum_{X=\pm}\sum_{m,n=0}^{\infty} p(X,m,n|\phi)\left(\partial_\phi \log p(X,m,n|\phi)\right)^2. \tag{C29}$$

From Bayes' Rule,

$$p(X,m,n|\phi) = p_{m,n}(\pm|\phi)p(m,n), \tag{C30}$$

where $p(m,n)$ is the Poisson-distributed probability of detecting $m$ photons at one output and $n$ photons at the other [Eq. (C28)], and where

$$p_{m,n}(\pm|\phi) = |\langle\pm|\mathrm{Q}_{\phi}^{m,n}\rangle|^2 \tag{C31}$$

$$= \frac{1}{2}\left(1\pm\cos[\phi(m+n)-\pi m]\right) \tag{C32}$$

is the probability of measuring an eigenvalue $X=\pm 1$, given fixed values of $m$ and $n$. Since $p(m,n)$ is independent of $\phi$, it therefore follows that

$$I_{\mathrm{C}}(\phi) = \sum_{m,n=0}^{\infty} p(m,n)I_{m,n}(\phi), \tag{C33}$$

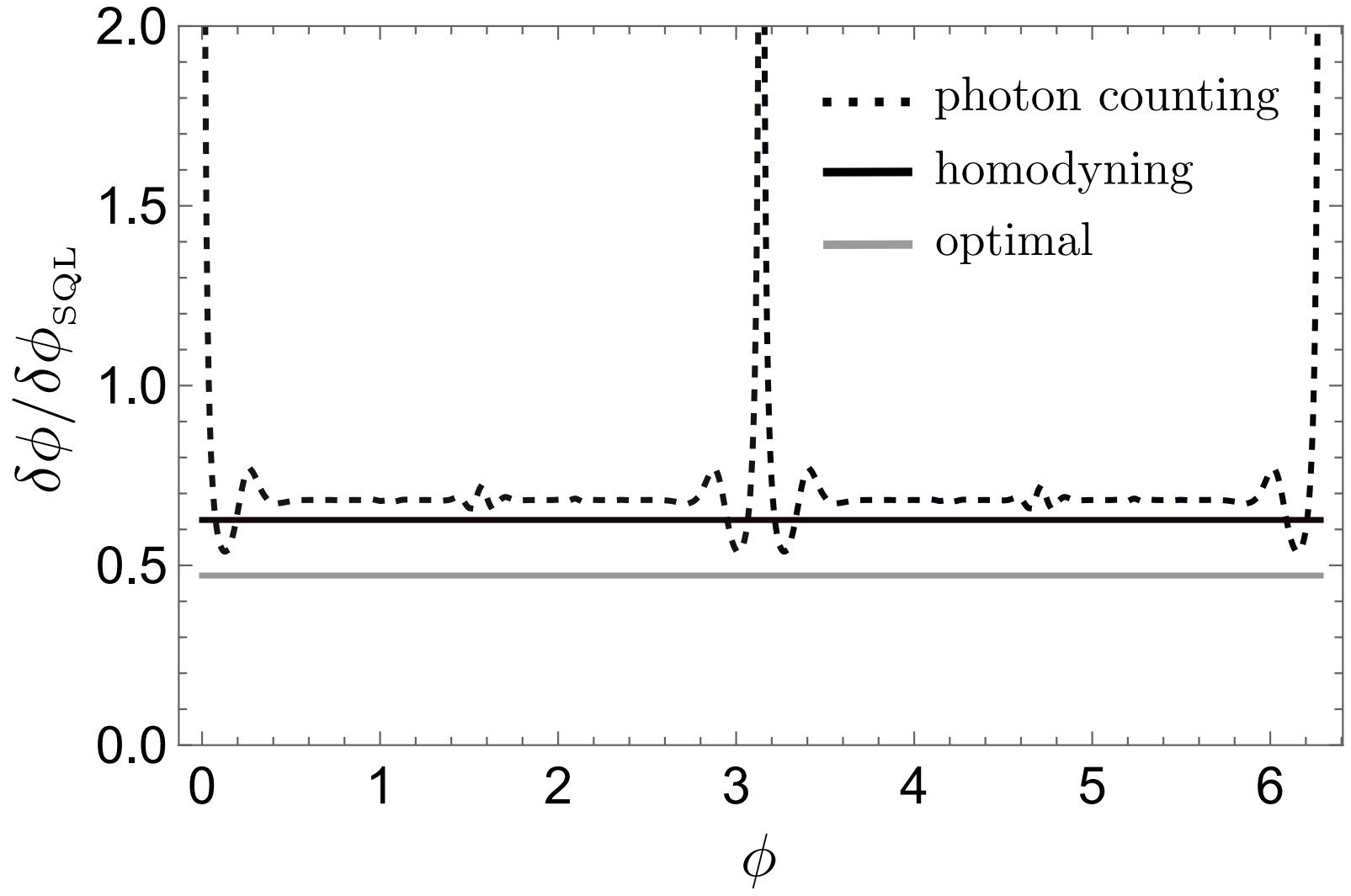

Figure C2: Precision $\delta\phi$ as a function of $\phi$, relative to the standard quantum limit $\delta\phi_{\rm SQL} \equiv [(1-p)M\bar{n}]^{-1/2}$, for a QWP state with $\bar{n} = 10$ and $p = 0.05$ (5% photon loss). The gray line indicating the performance of an optimal scheme is given by $\delta\phi_{\rm min} = 1/\sqrt{MI_{\rm Q}}$, where $I_{\rm Q}$ is the quantum Fisher information given in Eq. (C14).

where the $(m,n)$-conditioned classical Fisher information of the $X$-basis qubit measurement is given by

$$I_{m,n}(\phi) = \sum_{X=\pm} \frac{[\partial_\phi p_{m,n}(X|\phi)]^2}{p_{m,n}(X|\phi)} = (m+n)^2. \tag{C34}$$

Evaluating the average in Eq. (C33) then gives

$$I_{\rm C}(\phi) = \bar{n}^2 + \bar{n}, \tag{C35}$$

which is equal to the quantum Fisher information $I_{\rm Q}(|{\rm QWP}_\phi\rangle)$ given in Eq. (5) of the main text. Photon counting is therefore an optimal strategy in the absence of photon loss and qubit dephasing.

### C3.2 Classical Fisher information in the presence of photon loss

Here, we neglect the effects of qubit dephasing as they do not affect the central message of this subsection concerning the $\phi$-dependence of the classical Fisher information associated with photon counting in the presence of photon loss. Dephasing can be incorporated with the replacements $e^{-p|\alpha|^2} \to e^{-p|\alpha|^2-\chi(t)}$ and $(m+n)\phi \to (m+n)\phi - \vartheta(t)$.

In the presence of photon loss ($p \neq 0$), the post-measurement state of the qubit can be written as

$$\rho_{\rm Q}(m,n,\phi) = \frac{\langle m,n|U_{\rm BS}\rho_\phi U_{\rm BS}^\dagger|m,n\rangle}{p(m,n)} \tag{C36}$$

$$= \frac{1}{2}\left[|\uparrow\rangle\langle\uparrow| + |\downarrow\rangle\langle\downarrow| + e^{-p|\alpha|^2}\left(e^{i\phi(m+n)-im\pi}|\uparrow\rangle\langle\downarrow| + {\rm h.c.}\right)\right], \tag{C37}$$

where $\rho_\phi$ is given by Eq. (C8) [with $\chi = \vartheta = 0$], and where, in the first equality, the probability of detecting $m$ and $n$ photons in modes $a_+$ and $a_-$, respectively, is now given by

$$p(m,n) = \frac{e^{-(1-p)|\alpha|^2}}{m!n!}\left(\frac{(1-p)|\alpha|^2}{2}\right)^{m+n}. \tag{C38}$$

The probability $p_{m,n}(\pm|\phi)$ of measuring the qubit in state $|\pm\rangle$, conditioned on photon-counting outcomes $m$ and $n$, is then given by

$$p_{m,n}(\pm|\phi) = \frac{1}{2}\left(1 \pm e^{-p|\alpha|^2}\cos[\phi(m+n) - \pi m]\right). \tag{C39}$$

As in the $p = 0$ case, Bayes' Rule [Eq. (C30)] can be used to write the classical Fisher information in the form of Eq. (C33), where for $p \neq 0$, the Fisher information of the qubit measurement is now given by [cf. Eq. (C34)]

$$I_{m,n}(\phi) = \frac{(m+n)^2 \sin^2(m+n)\phi}{e^{2p|\alpha|^2} - \cos^2(m+n)\phi}. \tag{C40}$$

Averaging Eq. (C40) over the distribution $p(m,n)$ given in Eq. (C38) gives the classical Fisher information $I(\phi)$. As was the case with the ECS, the Fisher information associated with the homodyning scheme is independent of the true value of the phase, while the Fisher information associated with photon counting vanishes for $\phi = 0, \pi$, leading to divergences in $\delta\phi$ (Fig. C2).

# Preface to Chapter 5

As companies and experimental groups begin to demonstrate logical qubits with lifetimes exceeding those of their constituent physical qubits, the benefits of scaling to larger numbers of physical qubits become evident. The last two chapters have considered the generation and manipulation of entanglement between stationary degrees-of-freedom and classical (a.k.a. coherent-state) photonic wavepackets. These wavepackets provide a natural means of connecting stationary qubits in a modular architecutre, where small numbers of qubits are grouped into nodes ("modules") and connected via quantum-photonic interconnects. Modular architectures are widely regarded as providing a promising path towards scalability.

As we now show, entanglement between qubits and coherent-state wavepackets could also be used to perform multi-qubit measurements called parity checks, which form the bedrock of quantum error correction. These parity checks are typically performed using sequences of two-qubit gates applied between an ancilla qubit and several physical qubits, with the final ancilla measurement then providing information about the joint parity of the physical qubits. In a modular architecture, direct two-qubit gates between distant qubits may not be available. These long-range gates could alternatively be realized by first sharing entangled resource states (Bell pairs) across nodes, then consuming the Bell pairs as part of a gate-teleportation protocol. In this chapter, we show that the need for Bell-pair distribution or extra ancilla qubits for parity checks could be eliminated by instead mapping parity information onto a propagating degree-of-freedom—a coherent-state wavepacket—that interacts with each qubit in turn. Relative to parity checks implemented via gate teleportation, this approach eliminates the need for extra ancilla Bell pairs, which ultimately increase the number of qubits required to implement a code. It also replaces qubit state preparation and measurement (SPAM)—a leading error source in current devices—with coherent-state preparation and homodyne detection, both of which are considered relatively straightforward to implement.

# 5

# Long-range parity checks in a modular architecture

*This chapter is the integral text from*

Long range, multi-qubit parity checks have applications in both quantum error correction and measurement-based entanglement generation. Such parity checks could be performed using qubit-state-dependent phase shifts on propagating pulses of light described by coherent states $|\alpha\rangle$ of the electromagnetic field. We consider "flying-cat" parity checks based on an entangling operation that is quantum non-demolition (QND) for Schrödinger's cat states $|\alpha\rangle \pm |-\alpha\rangle$. This operation encodes parity information in the phase of maximally distinguishable coherent states $|\pm\alpha\rangle$, which can be read out using a phase-sensitive measurement of the electromagnetic field. In contrast to many implementations, where single-qubit errors and measurement errors can be treated as independent, photon loss during flying-cat parity checks introduces errors on physical qubits at a rate that is anti-correlated with the probability for measurement errors. We analyze this trade-off for three-qubit parity checks, which are a requirement for universal fault-tolerant quantum computing with the subsystem surface code. We further show how a six-qubit entangled "tetrahedron" state can be prepared using these three-qubit parity checks. The tetrahedron state can be used as a resource for controlled quantum teleportation of a two-qubit state, or as a source of shared randomness with potential applications in three-party quantum key distribution. Finally, we provide conditions for performing high-quality flying-cat parity checks in a state-of-the-art circuit QED architecture, accounting for qubit decoherence, internal cavity losses, and finite-duration pulses, in addition to transmission losses.

## 5.1 Introduction

The ability to scale up quantum computers is crucial for the long-term goal of performing computations fault-tolerantly. Many current proposals for scaling up take a modular approach where groups of qubits at the nodes of a network are linked by quantum photonic interconnects [1–3]. Recent experiments have shown significant progress, demonstrating entanglement distribution across nodes that are separated by tens of metres [4–6]. In a distributed architecture, the ability to perform long-range (inter-node) parity checks would also provide a clear path towards implementing quantum error correction [7, 8].

Many quantum error correcting codes, including the surface code [9–11], fall within a class of stabilizer codes [12] known as Calderbank-Shor-Steane (CSS) codes [13, 14]. In CSS codes, errors can be detected by measuring the parities of groups of qubits, given by the eigenvalues of stabilizer operators of the form $X^{\otimes n}$ and $Z^{\otimes m}$, where $X$ and $Z$ are Pauli-$X$ and Pauli-$Z$ operators, respectively. Here, $n$ and $m$ are integers giving the weight of the stabilizer, equal to the number of qubits on which the operator acts nontrivially. A parity check can be performed using a sequence of entangling gates applied between an ancilla qubit and each of several code qubits. The ancilla qubit can then be measured to infer the parity of the code qubits.

There are many strategies that can be used to implement two-qubit gates at long range. Given a source of distributed entanglement, these gates can be performed by teleportation [15–18]. Alternatively, long-range two-qubit gates can be realized in solid state systems by coupling qubits via the real or virtual excitations of superconducting [19–23], ferromagnetic [24, 25], or normal-metal [26] elements.

Rather than mapping parity onto an ancilla qubit, parity information can more broadly be extracted by measuring any observable—not necessarily corresponding to a qubit observable—whose outcome depends on parity. In a circuit quantum electrodynamics (QED) architecture, for instance, this could be a shift in the resonance frequency of a microwave resonator dispersively coupled to multiple code qubits [27–32].

In an architecture involving photonic interconnects, a natural way to connect qubits is using pulses of light. An inter-node parity check could then be performed via qubit-state-selective phase shifts $\alpha \to e^{i\phi}\alpha$ on a light pulse prepared in a coherent state $|\alpha\rangle$. Such state-selective phase shifts have been analyzed theoretically as a means of implementing quantum gates [33, 34]. In addition, small conditional phase shifts ($\phi < \pi$) generated from weak light-matter interactions have been theoretically considered as a resource for parity measurements [35, 36]. More recently, large phase shifts ($\phi \simeq \pi$) have been realized experimentally in both the optical [37, 38] and microwave [39–42] domains. These experiments demonstrate the enabling technology for parity checks based on "flying-cat" states given by superpositions of macroscopically distinct coherent states, $|\mathrm{C}_{\pm}\rangle \propto |\alpha\rangle \pm |-\alpha\rangle$, associated with propagating pulses of optical or microwave light. In ancilla-based parity checks (involving gates on a common ancilla qubit), measurement errors and gate errors typically have independent physical origins and thus occur with independent probabilities. This is not the case for flying-cat parity checks, where these two error sources are correlated with the average number of photons in the coherent state, $|\alpha|^2$. In particular, there is a trade-off between the rate of measurement errors (errors made in discriminating $|\pm\alpha\rangle$) and the rate of errors on code qubits due to backaction arising from photon loss. While the coherent-state distinguishability is controlled by a finite overlap $\langle\alpha|-\alpha\rangle$ that decreases with increasing $|\alpha|^2$, the rate of photon loss increases with increasing $|\alpha|^2$.

In this paper, we quantify the performance of flying-cat parity checks in the presence of photon loss. To do this, we derive a superoperator describing the backaction of photon loss on code qubits during a weight-3 flying-cat parity check. From this superoperator, error patterns on code qubits together with their associated error probabilities can be identified. This knowledge can then be used to optimize $\alpha$ to minimize the total error probability. High quality weight-3 parity checks have applications in both universal quantum computing with surface codes [43] and entanglement distribution in multi-node quantum networks. As a short-term demonstration of these ideas, we further show that weight-3 Pauli-$X$ and Pauli-$Z$ parity checks can be used to prepare a particular six-qubit entangled state. This state can be used for three-party controlled quantum teleportation of an arbitrary two-qubit state. It also provides a natural testbed for flying-cat parity checks in early implementations, en route to the more ambitious goal of performing fault-tolerant quantum

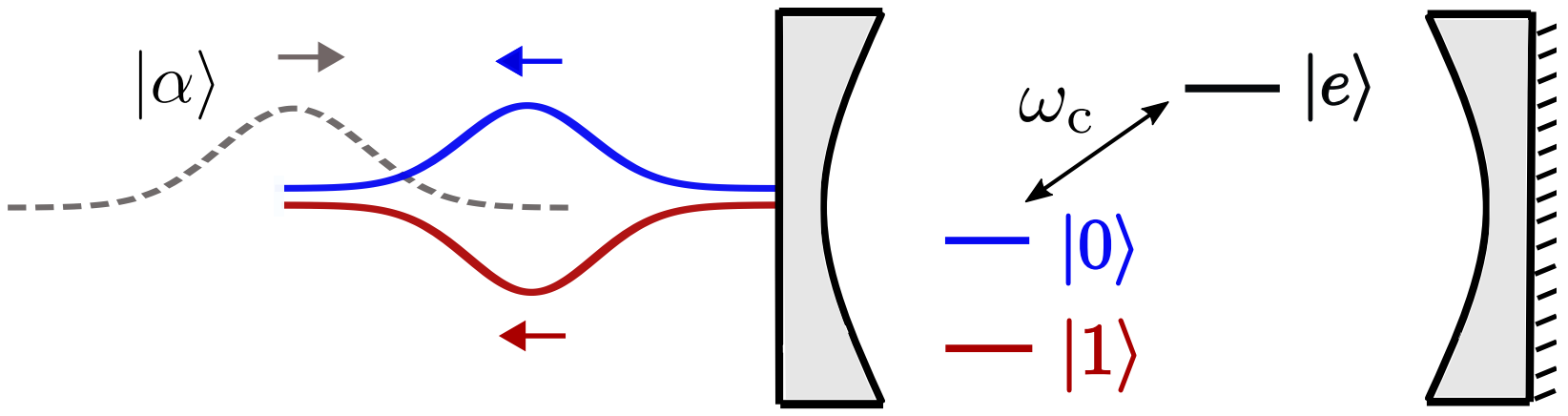

Figure 5.1: A qubit-state-selective phase shift on an incoming coherent state $|\alpha\rangle$ occupying some propagating quasimode of the electromagnetic field can be obtained by resonantly coupling a cavity mode having frequency $\omega_c$ to the transition $|0\rangle \leftrightarrow |e\rangle$ involving qubit state $|0\rangle$ and some auxiliary level $|e\rangle$ [33, 34, 38, 40, 41].

computation.

The layout of this paper is as follows. In Sec. 5.2, we describe how qubit-state-selective phase shifts imprint parity information onto propagating light pulses. In Sec. 5.3, we discuss the aforementioned trade-off between measurement errors and errors due to photon loss. Next, in Sec. 5.4, we illustrate how flying-cat parity checks can be used to generate entanglement in a multi-node network. In particular, we show how they can be used to prepare a six-qubit entangled state that provides a resource for three-party controlled quantum teleportation. Finally, in Sec. 5.5, we establish conditions for implementing flying-cat parity checks in a state-of-the-art circuit QED architecture, accounting for imperfections due to finite-bandwidth pulses, internal cavity loss, qubit decoherence, and transmission losses.

## 5.2 Coherent states as parity meters

A parity measurement of two or more physical qubits can be performed by allowing a propagating light pulse in coherent state $|\alpha\rangle$ to interact with each qubit according to the entangling operation described by

$$\begin{aligned} |0,\pm\alpha\rangle &\rightarrow |0,\pm\alpha\rangle\,, \\ |1,\pm\alpha\rangle &\rightarrow |1,\mp\alpha\rangle\,, \end{aligned} \tag{5.1}$$

where $|0\rangle$ and $|1\rangle$ are the qubit Pauli-$Z$ eigenstates [with $Z|s\rangle = (-1)^s|s\rangle$ for $s = 0,1$]. This operation can be realized by coupling a three-level system with states $|1\rangle, |0\rangle$, and $|e\rangle$ to a single quantized cavity mode, such that the $|0\rangle \leftrightarrow |e\rangle$ transition is resonantly coupled to the cavity mode [33, 34] (Fig. 5.1). An input coherent state $|\alpha\rangle$ occupying a propagating spatiotemporal mode (light pulse) that is resonant with the decoupled cavity frequency $\omega_c$ will then acquire a different phase upon reflection from the cavity, depending on the qubit state. When the qubit is in state $|1\rangle$, the cavity frequency is unaffected by the presence of the qubit, and the input pulse is reflected with a $\pi$ phase shift [44]. However, when the qubit is in state $|0\rangle$, hybridization of the qubit and cavity suppresses the cavity density of states at $\omega_c$, and the input pulse is reflected with no phase shift. This is of course not the only way of realizing the interaction given in Eq. (5.1). Another way to realize this scenario can be achieved in the dispersive regime (see Sec. 5.5 below for a detailed discussion of this second strategy).

After interacting with $n$ qubits in state $|s_1 s_2 \dots s_n\rangle$ ($s_i \in \{0,1\}$), $|\alpha\rangle$ will have acquired a total phase shift $(-1)^s = e^{is\pi}$, where $s = \sum_i s_i \,(\mathrm{mod}\, 2)$ is the $Z$-basis parity of the $n$ qubits:

$$|s_1 s_2 \dots s_n\rangle |\alpha\rangle \rightarrow |s_1 s_2 \dots s_n\rangle \left|e^{is\pi}\alpha\right\rangle. \tag{5.2}$$

The use of such an entangling interaction for continuous stabilization of a Bell state was analyzed theoretically in Ref. [45] and demonstrated experimentally (with phase shifts $< \pi$ due to a small dispersive coupling) with superconducting qubits in Ref. [46].

Switching from $Z$-basis to $X$-basis parity checks simply requires that Hadamards be applied to the qubits

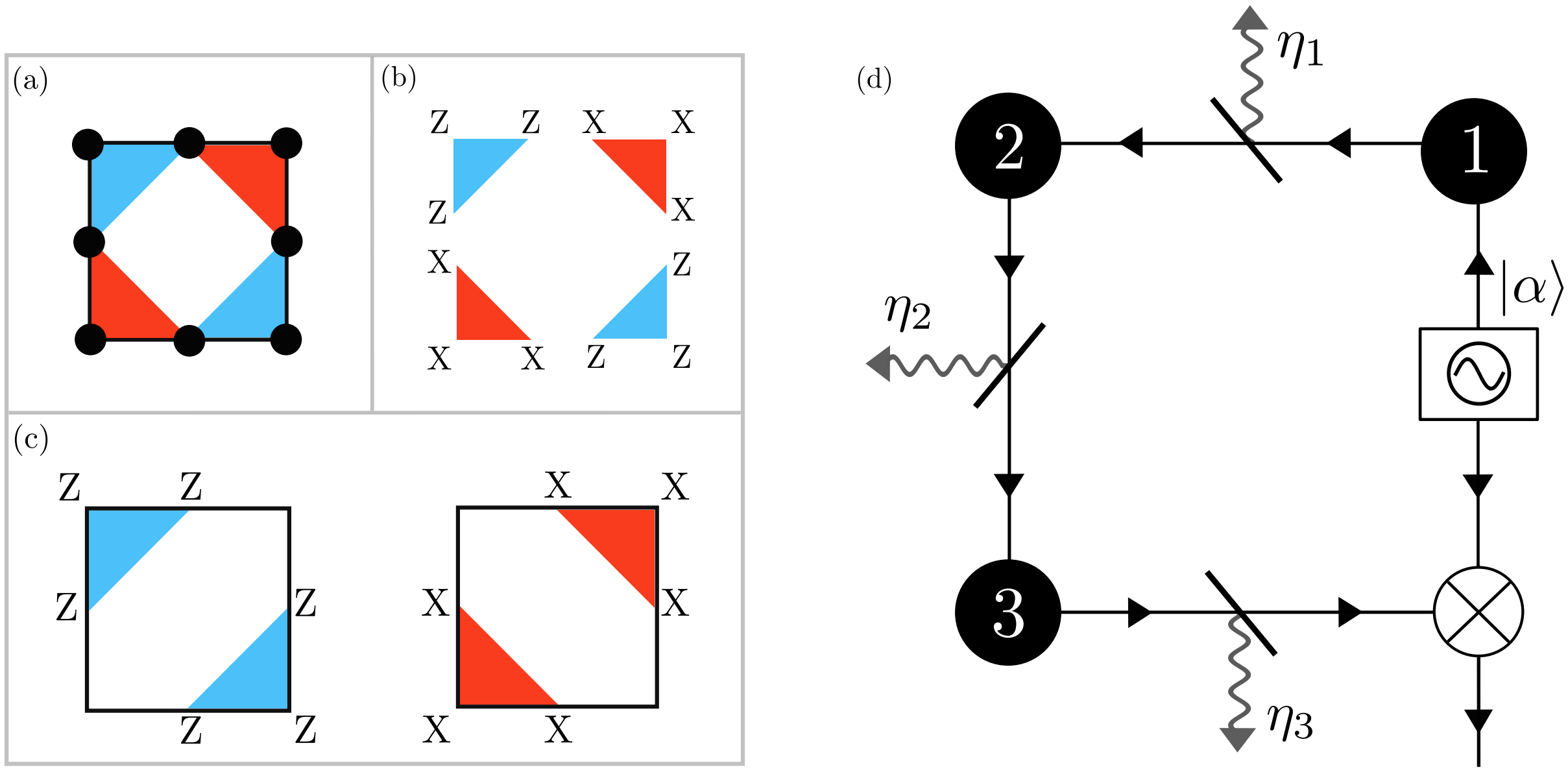

Figure 5.2: (a) The unit cell of the subsystem surface code [43]. Physical qubits represented by solid black dots are located on the edges and vertices of a square lattice. (b) Gauge operators: Each stabilizer can be represented as the product of two weight-3 gauge operators. Gauge operators do not all mutually commute, but they do commute with all stabilizers and therefore preserve the code space. (c) Stabilizers of the subsystem surface code: Each unit cell hosts two stabilizers, one of the form $Z^{\otimes 6}$ (left) and one of the form $X^{\otimes 6}$ (right). (d) A light pulse is prepared in a coherent state $|\alpha\rangle$ and interacts sequentially with three qubits. The interaction of the light pulse with each qubit results in the entangling operation of Eq. (5.1). Photon loss in transit is modeled by a series of beam splitters having reflectivities $\eta_i$ ($i = 1, 2, 3$). The parity of the code qubits is inferred by performing a phase-sensitive measurement of the coherent state amplitude. This phase-sensitive measurement could be achieved through homodyne detection by interfering the light pulse with a reference signal and measuring the resulting intensity.

both before and after the entangling operation described in Eq. (5.1):

$$\begin{aligned} |+,\pm\alpha\rangle &\xrightarrow{H} |0,\pm\alpha\rangle \to |0,\pm\alpha\rangle \xrightarrow{H} |+,\pm\alpha\rangle\,, \\ |-,\pm\alpha\rangle &\xrightarrow{H} |1,\pm\alpha\rangle \to |1,\mp\alpha\rangle \xrightarrow{H} |-,\mp\alpha\rangle\,, \end{aligned} \tag{5.3}$$

where here, $X|\pm\rangle = \pm|\pm\rangle$. Hence, for both $X$- and $Z$-basis parity checks, the parity of the code qubits can be inferred by measuring the phase of the coherent-state amplitude.

### 5.2.1 Weight-3 checks are enough

Many quantum error correcting codes are formulated as stabilizer codes [12] defined by a stabilizer group $\mathscr{S}$. The associated code space $C$ is then given by the span of the vectors $|\psi\rangle$ stabilized by $\mathscr{S}$, i.e., satisfying $S_j|\psi\rangle = |\psi\rangle$ for all stabilizers $S_j \in \mathscr{S}$. In the usual surface code [11], for instance, correctable errors can be detected by measuring weight-4 stabilizers of the form $X^{\otimes 4}$ and $Z^{\otimes 4}$.

A surface code can alternatively be operated as a subsystem code [43] [Fig. 5.2(a-c)]. A subsystem code [47–49] is defined by a non-Abelian gauge group $\mathscr{G}$ together with an Abelian stabilizer subgroup $\mathscr{S} \subseteq \mathscr{G}$. This construction allows the code space $C$ stabilized by $\mathscr{S}$ to be partitioned into subsystems, $C = A \otimes B$. Here, the subsystems $A$ and $B$ are defined by the action of gauge operators $g \in \mathscr{G}$, which may act nontrivially only on subsystem $B$: $g|\psi_A\rangle \otimes |\psi_B\rangle = |\psi_A\rangle \otimes |\psi_B'\rangle$ for $|\psi_A\rangle \in A$, $|\psi_B\rangle, |\psi_B'\rangle \in B$. Errors acting on the codespace are then defined only up to an equivalence relation involving $\mathscr{G}$. In particular, error operators $E$ and $E'$ are equivalent in their action on encoded logical information (in subsystem $A$) if $E' = Eg$ for some $g \in \mathscr{G}$. Much of the appeal of subsystem codes comes from the fact that stabilizer eigenvalues can be inferred from measurements of lower-weight gauge operators. For the subsystem surface code [43] [Fig. 5.2(a-c)], for instance, the eigenvalues of the weight-6 stabilizers $X^{\otimes 6}$ can be inferred by multiplying together the measurement outcomes of two weight-3 gauge operators of the form $X^{\otimes 3}$. Correctable errors in a surface code can therefore be detected using weight-3 parity checks only (as opposed to the usual weight-4 parity checks), with a fault-tolerance threshold of order 1% [43, 50]. Gates on logical qubits can be performed as in the usual surface code [11, 43]. With this long-term application in mind, we consider error sources—photon loss and finite coherent-state distinguishability—that are specific to weight-3 flying-cat parity checks [Fig. 5.2(d)] and show how to minimize their total impact. Further error sources (finite-duration pulses, internal losses, and qubit decoherence) are assessed in Sec. 5.5, below.

An important advantage of the subsystem surface code is that errors on ancilla qubits used for syndrome readout propagate onto at most one other qubit, modulo gauge operators [43]. The extra degree of freedom provided by the gauge group therefore eliminates the horizontal hook errors considered in Ref. [51] for the surface code, wherein ancilla-qubit errors propagate onto several qubits.

## 5.3 Trade-off between measurement errors and errors due to photon loss

Parity checks performed using coherent states exhibit an unavoidable trade-off between error sources depending on the average number of photons per pulse, $|\alpha|^2$. Due to the non-orthogonality of $|\pm\alpha\rangle$, larger field amplitudes $\alpha$ yield better measurement fidelities. However, they also produce a stronger backaction on the code qubits in the presence of photon loss. In this section, we consider these two competing error sources: measurement errors and errors due to photon loss.

In a real-world implementation, there may be a number of other error sources affecting the quality of the operation required for parity checks [Eq. (5.1)]. For example, the qubit-state-dependent conditional phase introduced may not be precisely $\pi$, phase noise or thermal light may corrupt the prepared coherent states, and photodetectors may suffer from amplifier noise, dark counts, etc. In this section, we assume that these implementation-dependent imperfections lead to a background error rate that can be controlled, leaving the effects of photon loss and imperfect coherent-state distinguishability as the dominant error sources. These effects are likely to be common to any implementation. We consider several additional sources of error in the specific context of a circuit-QED setup in Sec. 5.5.

### 5.3.1 Photon-loss-induced backaction

In this subsection, we evaluate the backaction of photon loss on the state of the three code qubits involved in the parity check. Photon loss during a flying-cat parity check can lead to two-qubit correlated errors on the code qubits (horizontal hook errors, Fig. 5.3). As is the case for parity measurements via noisy ancilla qubits [43], these horizontal hook errors are gauge-equivalent to single-qubit Pauli errors, and they can therefore be corrected as if they were single-qubit Pauli errors. This key advantage of the subsystem surface code is therefore retained for flying-cat parity checks.

To perform a weight-3 parity check, a coherent state $|\alpha\rangle$ will interact with three qubits. In what follows, it will be helpful to define the even and odd ($Z$-basis) parity subspaces given by

$$\begin{aligned} \mathscr{H}_+ &= \mathrm{span}\{|000\rangle, |110\rangle, |011\rangle, |101\rangle\}, \\ \mathscr{H}_- &= \mathrm{span}\{|001\rangle, |010\rangle, |100\rangle, |111\rangle\}. \end{aligned} \tag{5.4}$$

We write the state $|\Xi\rangle$ of the three qubits prior to the parity check as

$$|\Xi\rangle = c_+ |\Xi_+\rangle + c_- |\Xi_-\rangle, \tag{5.5}$$

where $|\Xi_\pm\rangle \in \mathscr{H}_\pm$, and where $\sum_{\sigma=\pm} |c_\sigma|^2 = 1$. In the absence of photon loss, the states $|\Xi_\pm\rangle$ are preserved by Eq. (5.1): $|\Xi\rangle |\alpha\rangle \to c_+ |\Xi_+\rangle |\alpha\rangle + c_- |\Xi_-\rangle |-\alpha\rangle$. To perform an $X$-basis parity check, Hadamards are

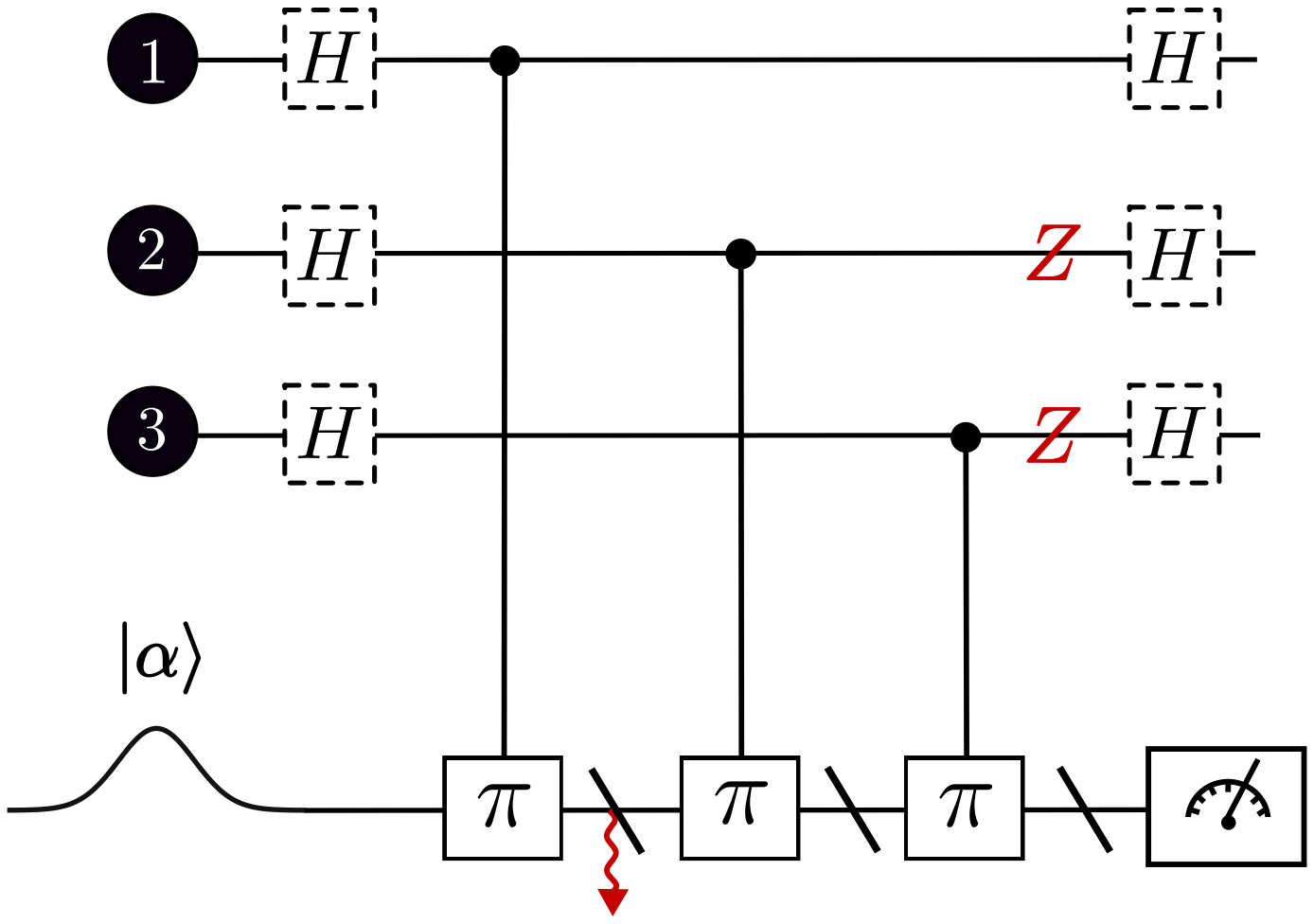

Figure 5.3: Horizontal hook error $Z_2Z_3$ due to photon loss during a parity check.

simply applied before/after the $Z$-basis parity check, as described in Eq. (5.3). In that case, we take $|\Xi\rangle$ to describe the state of the three qubits after the first set of Hadamards.

In the presence of photon loss, the parity check has a backaction on the code qubits that leads to the introduction of errors beyond the background error rate. We model the effect of photon loss to the environment via a beam splitter-type interaction [52]

$$R(\eta) = e^{\arcsin\sqrt{\eta}(a_\ell^\dagger a - \text{h.c.})}, \tag{5.6}$$

where here, $a$ is an annihilation operator defined such that $a|\alpha\rangle = \alpha|\alpha\rangle$, $a_\ell$ is associated with an environmental loss mode into which photons may be scattered, and $\eta$ is the beam splitter reflectivity. Under this operation, $R^\dagger(\eta)aR(\eta) = \sqrt{1-\eta}a + \sqrt{\eta}a_\ell$. The amplitude $\alpha$ of $|\alpha\rangle$ is therefore reduced according to

$$R(\eta)|\alpha\rangle|0\rangle_\ell = \left|\sqrt{1-\eta}\alpha\right\rangle|\sqrt{\eta}\alpha\rangle_\ell, \tag{5.7}$$

where here, $|\,\rangle_\ell$ is the state of the loss mode. Although coherent-state amplitude may be lost to many different environmental modes in principle, these modes can be treated as a single effective mode since the state of the environment will eventually be traced over. To account for the effect of loss at different points in the parity-check operation, we imagine that the light pulse undergoes such a beam splitter interaction $R(\eta_i)$, $i = 1,2,3$, after every entangling operation [Fig. 5.2(d)]. Accounting for loss in this manner, the initial state $|s_1s_2s_3\rangle|\alpha\rangle|0,0,0\rangle_\ell$ of the qubits, coherent state, and environment (here, $|0,0,0\rangle_\ell$ denotes the vacuum state of the three loss modes) evolves into

$$|s_1s_2s_3\rangle|(-1)^s\bar{\alpha}\rangle|(-1)^{s_1}\alpha_1,(-1)^\mu\alpha_2,(-1)^s\alpha_3\rangle_\ell, \tag{5.8}$$

where $s = s_1+s_2+s_3 \pmod 2$ is the $Z$-basis parity of $|s_1s_2s_3\rangle$, $\mu = s_1+s_2 \pmod 2$, and where

$$\begin{aligned} \bar{\alpha} &= \alpha\prod_{i=1}^{3}(1-\eta_i)^{1/2}, \\ \alpha_i &= \alpha\eta_i^{1/2}\prod_{j<i}(1-\eta_i)^{1/2}. \end{aligned} \tag{5.9}$$

The joint state $\rho$ of the qubits and electromagnetic field following the entangling operations (but prior to measurement) can then be obtained by tracing over the loss modes:

$$\rho = \sum_{\sigma,\sigma'=\pm} \rho_{\sigma,\sigma'} \left|\sigma\bar{\alpha}\right\rangle\!\left\langle\sigma'\bar{\alpha}\right|, \tag{5.10}$$

where $\rho_{\pm,\pm} \in \mathscr{H}_\pm$. For $|\Xi_+\rangle = \sum_{i=0,3,5,6} c_i |i\rangle$, where here, $i$ is the base-10 value of $s_1s_2s_3$ in binary, $\rho_{+,+}$ takes the following form in the basis $\{|0\rangle = |000\rangle, |3\rangle = |011\rangle, |5\rangle = |101\rangle, |6\rangle = |110\rangle\}$:

$$\frac{\rho_{+,+}}{|c_+|^2} = \begin{pmatrix} |c_0|^2 & c_0c_3^*e^{-2|\alpha_2|^2} & c_0c_5^*e^{-2(|\alpha_1|^2+|\alpha_2|^2)} & c_0c_6^*e^{-2|\alpha_1|^2} \\ c_3c_0^*e^{-2|\alpha_2|^2} & |c_3|^2 & c_3c_5^*e^{-2|\alpha_1|^2} & c_3c_6^*e^{-2(|\alpha_1|^2+|\alpha_2|^2)} \\ c_5c_0^*e^{-2(|\alpha_1|^2+|\alpha_2|^2)} & c_5c_3^*e^{-2|\alpha_1|^2} & |c_5|^2 & c_5c_6^*e^{-2|\alpha_2|^2} \\ c_6c_0^*e^{-2|\alpha_1|^2} & c_6c_3^*e^{-2(|\alpha_1|^2+|\alpha_2|^2)} & c_6c_5^*e^{-2|\alpha_2|^2} & |c_6|^2 \end{pmatrix}. \tag{5.11}$$

The block $\rho_{-,-}/|c_-|^2$ takes the same form as Eq. (5.11) under the mapping $(0,3,5,6) \mapsto (1,2,4,7)$. The off-diagonal blocks $\rho_{\pm,\mp} \propto c_\pm c_\mp^*$ in Eq. (5.10) appear as a result of considering an initial state $|\Xi\rangle$ [Eq. (5.5)] of mixed parity and will be suppressed exponentially $\sim e^{-2|\bar{\alpha}|^2}$ in the final post-measurement state.

In the absence of photon loss ($\eta_j = 0$ for $j = 1,2,3$), $\rho_{\sigma,\sigma'} = c_\sigma c_{\sigma'}^* |\Xi_\sigma\rangle\langle\Xi_{\sigma'}|$ and $\bar{\alpha} = \alpha$. With $\eta_j \neq 0$, however, $\rho_{\sigma,\sigma'} \neq c_\sigma c_{\sigma'}^* |\Xi_\sigma\rangle\langle\Xi_{\sigma'}|$ due to the photon-loss-induced backaction of the parity check. Hence, while the interaction with the electromagnetic field remains quantum non-demolition (QND) in the parity of $|s_1s_2s_3\rangle$ in the presence of photon loss, it is no longer QND in $|\Xi_\pm\rangle$ as desired. It may be verified that the backaction leading to this non-QND character can be described by the composition of two dephasing channels:

$$\frac{\rho_{\pm,\pm}}{|c_\pm|^2} = (\mathscr{E}_{p_1,E_1} \circ \mathscr{E}_{p_2,E_2})(|\Xi_\pm\rangle\langle\Xi_\pm|), \tag{5.12}$$

where $\mathscr{E}_{p,E}(\rho) = (1-p)\rho + pE\rho E^\dagger$, $E_1 = Z_2Z_3$, $E_2 = Z_3$, and where

$$\begin{aligned} p_1 &= \frac{1}{2}(1 - e^{-2|\alpha_1|^2}) \simeq \eta_1|\alpha|^2, \\ p_2 &= \frac{1}{2}(1 - e^{-2|\alpha_2|^2}) \simeq \eta_2|\alpha|^2. \end{aligned} \tag{5.13}$$

The approximations in Eq. (5.13) are valid to leading order in $\eta_j \ll 1$.

The operators $E_{1,2}$ can be understood as follows: In the cat-state basis $|\mathrm{C}_\pm\rangle \propto |\alpha\rangle \pm |-\alpha\rangle$, Eq. (5.1) describes a CZ gate for any nonzero $\alpha$. Since $|\mathrm{C}_\pm\rangle$ are eigenstates of the photon-number parity operator $e^{i\pi a^\dagger a}$, the loss of a photon constitutes an $X$ error, $|\mathrm{C}_\pm\rangle \to |\mathrm{C}'_\mp\rangle$, up to a change in coherent-state amplitude, $|\alpha\rangle \to |\alpha'\rangle$ [53]. Hence, as a result of the CZ gates involved in the parity check, photon loss leads to $Z$ errors on the code qubits. (Generally speaking, an $X$ error on qubit $a$, propagated through a CZ acting on qubits $a$ and $b$, leads to a $Z$ error on qubit $b$.) These $Z$ errors appear in the combination $E_1 = Z_2Z_3$ for photon loss occurring between qubits 1 and 2 (Fig. 5.3), and as $E_2 = Z_3$ for photon loss occurring between qubits 2 and 3.

In a surface code where stabilizer parity checks are performed using propagating pulses of light, photon loss would therefore lead to correlated (horizontal hook) errors on code qubits. However, due to the gauge degree of freedom introduced by the gauge group, the horizontal hook error $Z_2Z_3$ introduced by the parity check [cf. Eq. (5.12)] is equivalent to the single-qubit error $Z_1$ [9]. This is because $Z_2Z_3(Z_1Z_2Z_3) = Z_1$, where $Z_1Z_2Z_3 \in \mathscr{G}$. Since $X$-basis parity checks differ only by some additional Hadamards applied to $\rho$ [Eq. (5.10)], the effects of photon loss during $X$-basis parity checks are instead equivalent, up to a gauge operator, to single-qubit $X_1$ and $X_3$ errors occurring with the same probabilities (Table 5.1). These errors are not detected by the parity check in which they are introduced (since $Z$-basis parity checks only detect $X$ errors and vice versa), but may be detected in the next parity check of the other basis.

| Operator being measured | Error $E$ introduced by photon loss | Prob($E$) |
|---|---|---|
| $Z^{\otimes 3}$ | $Z_1$ | $p_1$ |
| | $Z_3$ | $p_2$ |
| $X^{\otimes 3}$ | $X_1$ | $p_1$ |
| | $X_3$ | $p_2$ |

Table 5.1: Errors arising due to photon loss during a $Z^{\otimes 3}$ gauge-operator parity check are equivalent (up to a gauge transformation) to $Z_1$ or $Z_3$ errors arising with probabilities $p_1$ and $p_2$, respectively. During an $X^{\otimes 3}$ parity check, these errors are instead equivalent to $X_1$ or $X_3$ errors.

### 5.3.2 Measurement errors

Determining the parity of two or more qubits using this approach requires a phase-sensitive measurement of the electromagnetic field (to infer the sign of $\alpha$). Such a measurement can be implemented via homodyne detection [54]. In homodyne detection, a signal field is first mixed on a 50:50 beam splitter with a local oscillator in coherent state $|\beta\rangle$, and the output intensity is then measured, giving information about the phase of the signal relative to the local oscillator. In this case, the signal field is the propagating light pulse entangled with the state of the qubits [Eq. (5.10)]. Without loss of generality, we assume that $\alpha$ is real and positive, $\alpha \in \mathbb{R}^+$. With a strong local oscillator ($\beta \in \mathbb{R} \gg \alpha$), homodyne detection implements a projective measurement $|x\rangle\langle x|$ onto the $x$ eigenbasis [54], where here, $|x\rangle$ is an eigenstate of $\hat{x} = (a^\dagger + a)/\sqrt{2}$. The sign $\pm = \mathrm{sign}(x)$ of the measured displacement can then be taken as an inference of even ($+$) or odd ($-$) parity for the state of the qubits. In the event that $\alpha$ is complex, the measured quadrature can be adjusted by introducing a phase on the local oscillator: $\beta \to \beta e^{i(\varphi-\pi)}$. In that case, the projection is instead onto an eigenstate of $\hat{x}_\varphi = \hat{x}\cos\varphi + \hat{p}\sin\varphi$, where $[\hat{x}, \hat{p}] = i$ [54].

Given $\rho$ [Eq. (5.10)], the state $\rho_x$ of the code qubits conditioned on the measurement outcome $x$ is then given by

$$\rho_x = \frac{\mathrm{Tr}_{\mathrm{EM}}\{|x\rangle\langle x|\,\rho\,|x\rangle\langle x|\}}{p(x)}, \tag{5.14}$$

where $p(x) = \mathrm{Tr}\{|x\rangle\langle x|\,\rho\}$, and where $\mathrm{Tr}_{\mathrm{EM}}\{\cdots\}$ denotes a partial trace over the state of the electromagnetic field. This quantity can be evaluated straightforwardly using the representation $\langle x|\bar{\alpha}\rangle = \pi^{-1/4}e^{-\frac{1}{2}(x-\sqrt{2}\bar{\alpha})^2}$ of $|\bar{\alpha}\rangle$. For a thresholded decision, where $x > 0$ ($x < 0$) is taken to indicate even parity (odd parity), the post-measurement state for an inference of even parity is given by

$$\rho_+ = \int_0^\infty dx\, p(x)\rho_x, \tag{5.15}$$

with a similar expression for $\rho_-$ involving an integral from $-\infty$ to 0. The joint probability $P(\text{“}\pm\text{”}, \mp)$ of the state being odd/even ($\mp$) for an inference of even/odd parity (“$\pm$”) is

$$P(\text{“}\pm\text{”}, \mp) = \mathrm{Tr}\{\hat{\Pi}_\mp \rho_\pm\} = \frac{|c_\mp|^2}{2}\mathrm{erfc}(\sqrt{2}\bar{\alpha}), \tag{5.16}$$

where $\hat{\Pi}_\pm$ is a projector onto $\mathscr{H}_\pm$ and $\mathrm{erfc}(x)$ is the complementary error function. The error $p_{\mathrm{M}}$ associated with the measurement is then given by

$$p_{\mathrm{M}} = P(\text{“}+\text{”}, -) + P(\text{“}-\text{”}, +) = \frac{1}{2}\mathrm{erfc}(\sqrt{2}\bar{\alpha}). \tag{5.17}$$

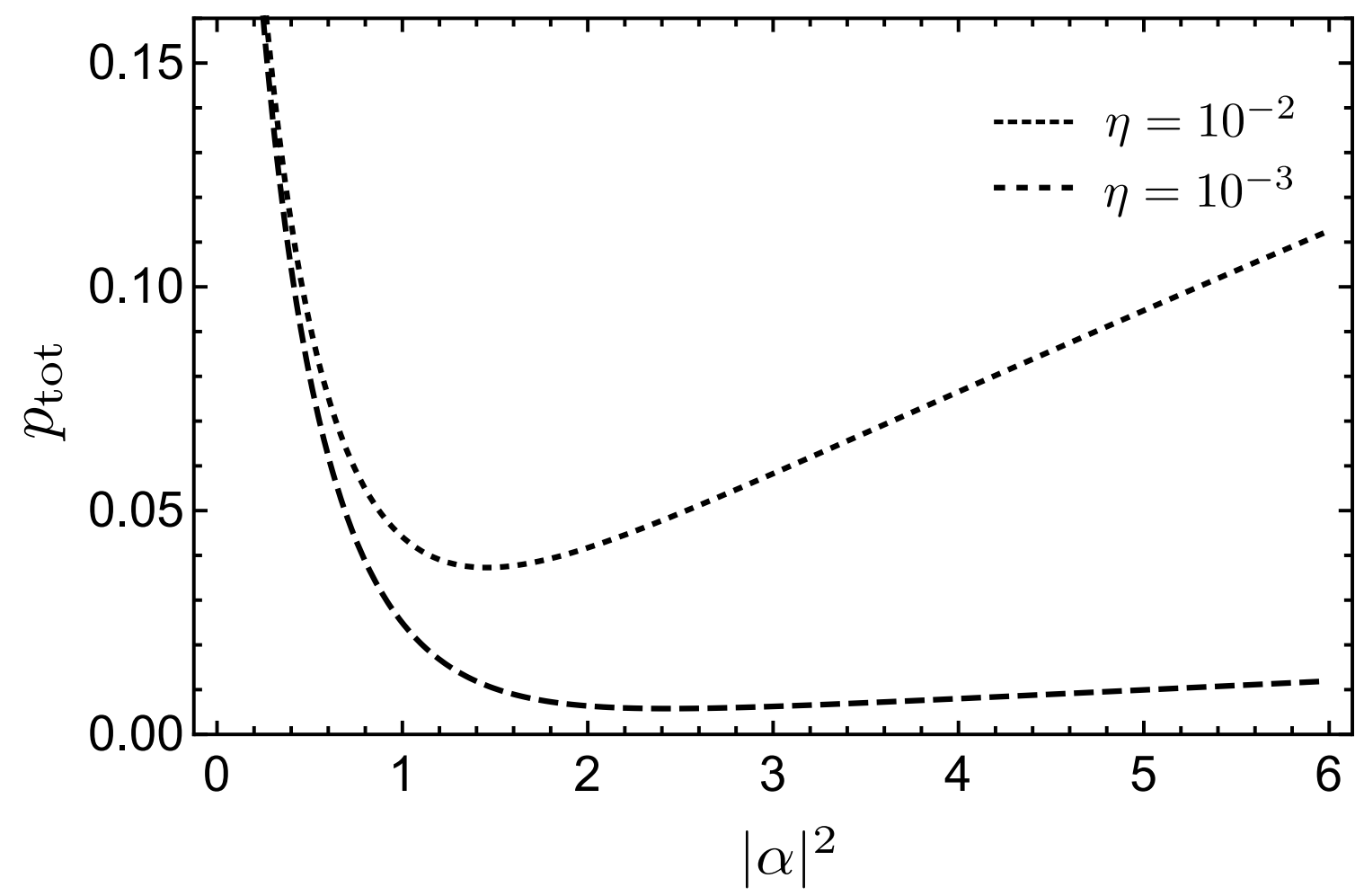

Figure 5.4: The total error $p_{\rm tot}$ [Eq. (5.18)] is given by the sum of competing terms: measurement errors, which decrease with $\alpha$, and errors due to photon loss, which increase with $\alpha$. For simplicity, we assume that $\eta_i = \eta$ for all $i$.

The total probability of an error occurring as part of a parity check—either an incorrect parity inference or a code-qubit error—is then (neglecting corrections of order $p_i p_j$ for $p_i \ll 1$):

$$p_{\rm tot} = p_{\rm M} + p_1 + p_2 \simeq \frac{1}{2}\frac{e^{-2\alpha^2}}{\alpha\sqrt{2\pi}} + \frac{1}{2}\sum_{j=1,2} \eta_j |\alpha|^2. \tag{5.18}$$

In the second, approximate equality, we have neglected subleading corrections in $\eta_j|\alpha|^2 < 1$ and approximated the error function by its asymptotic form for $|\alpha| > 1$. Since errors introduced by photon loss occur at a rate that increases with $\alpha$ ($\sim \eta_j|\alpha|^2$), while measurement errors are suppressed for large $\alpha$ ($\sim \alpha^{-1}e^{-2\alpha^2}$), there is an inherent trade-off between the two types of error as a function of coherent-state amplitude. Their total impact can therefore be minimized by optimizing $\alpha$ (Fig. 5.4).

In light of the nonlinear (exponential) suppression of $p_{\rm M}$ with $|\alpha|^2$, it is natural to consider whether repeating the measurement can lead to a smaller error. In this case, a single-shot measurement with a coherent state of amplitude $\alpha = \sqrt{N}\alpha_0$ leads to the same measurement error $p_{\rm M}$ as a sequence of $N$ measurements performed with smaller coherent state amplitudes $\alpha_0$. However, this equivalent performance can only be reached if a soft decision is made, where each measurement outcome $x_k$ $(k = 1, 2, \ldots, N)$ is assigned a confidence $p(\pm|x_k) = \mathrm{Tr}\{\hat{\Pi}_\pm \rho_{x_k}\}$, and where the individual outcomes are correlated in determining the most likely parity. If instead, each of the $N$ measurements is thresholded independently leading to a majority-vote decision, the error is less favorable, $p_{\rm M,th} \sim \sqrt{p_{\rm M}}$ (up to logarithmic corrections for large $N$) [55]. The large discrepancy between the error for a thresholded (hard) decision and the soft decision is a feature of the Gaussian distribution of measurement outcomes $x_k$ [55, 56], a model that is accurately realized for measurement of a coherent state quadrature. Exploiting this soft-decision advantage in repeated measurements with weak pulses may prove useful if the amplitude of each coherent state is limited for technical reasons (nonlinearity, unwanted qubit excitation, amplifier noise, etc.).

## 5.4 Entangled resource states

In addition to enabling quantum error correction, flying-cat parity checks could also be used to distribute entanglement in a quantum network, with applications in both quantum computing and quantum communication. In this section, we discuss the concrete example of generating three-qubit Greenberger-Horne-Zeilinger (GHZ) states. GHZ states have applications in precision sensing and multi-party quantum communication.

They could also be used as the central resource for weight-3, ancilla-based inter-node parity checks using the approach of Refs. [7, 8] (an alternative to the strategy for direct parity checks described above). We further show how to create a particular six-qubit entangled "tetrahedron" state using only $X$- and $Z$-basis weight-3 parity checks and single-qubit gates. This state can be used as a resource for three-party controlled quantum teleportation of an arbitrary two-qubit state. Its preparation also provides a convenient setting in which to benchmark the performance of the parity checks themselves.

### 5.4.1 A three-qubit GHZ state

An alternative, established approach to long-range parity checks involves the generation and consumption of four-qubit GHZ states [7, 8]. In this setup, each node houses a code qubit and (at least) one ancilla. To perform a parity check, four ancilla qubits located at four different nodes are prepared in a GHZ state. The parity of the code qubits located at these nodes can then be inferred by applying controlled-NOT gates between the code qubit and the ancilla located at each node, then measuring all ancillae. The distributed GHZ state in this proposal could be generated, in principle, via flying-cat parity checks. Although the original proposal of Refs. [7, 8] focused on weight-4 parity checks, weight-3 checks are sufficient to operate a subsystem surface code, as described above, so in this section we consider the preparation of three-qubit GHZ states. The ancilla-based approach of Refs. [7, 8] may be preferable over a direct flying-cat parity check in the case where the photon loss rate is high, leading to a high probability of errors being introduced on the code qubits. The low-fidelity GHZ states produced in the presence of high photon loss could then be purified "offline" prior to being consumed as part of a parity check, as considered in Refs. [7, 8].

A simple measurement-based strategy for three-qubit GHZ-state preparation begins with preparation of the three qubits in $|{+}{+}{+}\rangle$. Measuring $Z_1Z_2$ and $Z_2Z_3$ projects the qubits into an entangled state, which can then be transformed into $|\mathrm{GHZ}\rangle \propto |000\rangle + |111\rangle$ with an $X$-gate correction on the appropriate qubit. For a measurement outcome $Z_1Z_2 = -1$, this is qubit 1; for $Z_2Z_3 = -1$, it is qubit 3; if both outcomes are $-1$, it is qubit 2. However, as described above in Sec. 5.3, a $Z$-basis flying-cat parity check may introduce $Z$ errors on the qubits. For a measurement of $Z_iZ_j$, these are single-qubit $Z_j$ errors occurring with probability $p_{ij} = (1 - e^{-2\eta_{ij}|\alpha|^2})/2$, where here, $\eta_{ij}$ is the beam splitter reflectivity quantifying the strength of the losses incurred while traveling between qubits $i$ and $j$. Measurement errors occurring with probability $q_{ij} = \mathrm{erfc}[\sqrt{2}(1-\eta_{ij})\alpha]/2$ will also result in the wrong correction operator being applied. In the presence of photon loss and accounting for measurement errors, the two parity checks and correction step described above will lead to preparation of the following mixed state:

$$\sigma = (1-p)\,|\mathrm{GHZ}\rangle\langle\mathrm{GHZ}| + \sum_i p_i\,|i_\perp\rangle\langle i_\perp|\,, \tag{5.19}$$

where $p = \sum_i p_i = p_{12} + p_{23} - p_{12}p_{23} + q_{12} + q_{23} + q_{12}q_{23}$, and where $\langle \mathrm{GHZ}|i_\perp\rangle = 0$ for all $i$.

Since $|\mathrm{GHZ}\rangle$ can be described as the simultaneous $+1$ eigenstate of a set of CSS stabilizer generators—namely, $Z_1Z_2$, $Z_2Z_3$, and $X_1X_2X_3$—several noisy copies of $\sigma$ distributed across three nodes ("Alice", "Bob", and "Charlie") could be purified using any CSS error-correcting code, e.g., $C_4$ [57]. In the case of $C_4$, this protocol would require that Alice, Bob, and Charlie share four copies of $\sigma$. They would each measure the stabilizers of $C_4$ on their four qubits, and then use local operations and classical communication to produce a less noisy, encoded version of $\sigma$ [57]. Since the six-qubit "tetrahedron" state discussed in the following subsection also has a description in terms of a set of CSS stabilizer generators, the same protocol [57] could be used to purify this state as well.

### 5.4.2 A six-qubit entangled state

In this section, we show how weight-3 parity checks can be used to prepare a "tetrahedron" state—a six-qubit entangled state where each qubit can be associated with one edge of a tetrahedron [Fig. 5.5(a)]. This state provides a source of shared randomness with potential applications for three-party quantum key distribution. A tetrahedron state can also be used as a resource for controlled quantum teleportation of an arbitrary two-qubit state. Independent of these potential applications, the tetrahedron state provides a natural testbed for benchmarking the performance of weight-3 flying-cat parity checks (since its preparation requires both $X$- and $Z$-basis parity checks).

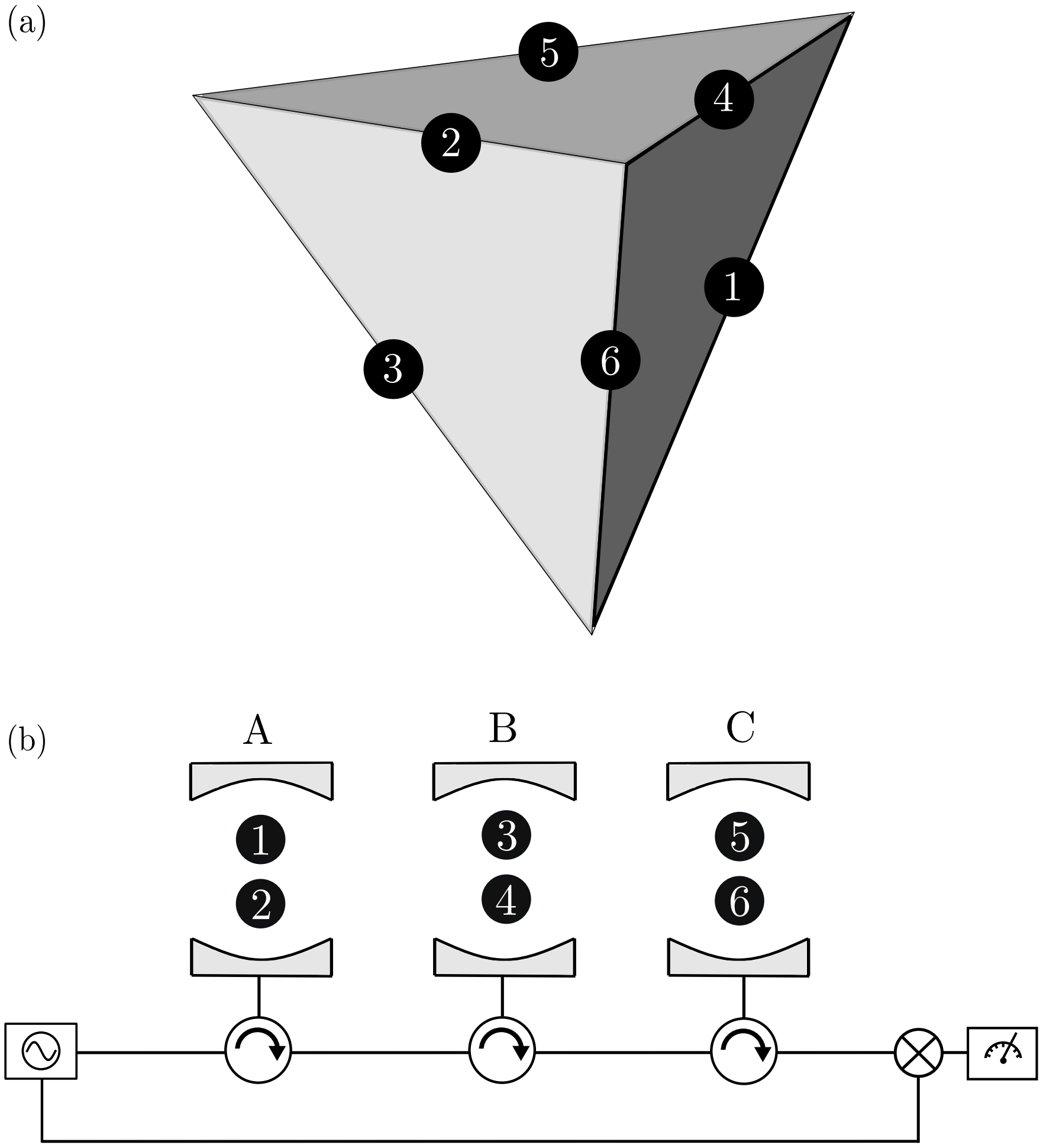

Figure 5.5: (a) Six qubits can be associated with the six edges of a tetrahedron. Each of the stabilizers in Eq. (5.20) is a product of three Pauli-$Z$ or Pauli-$X$ operators acting on qubits that neighbor a common face ($Z$) or vertex ($X$). (b) A potential real-space setup for applications discussed in the main text: Each node $(A,B,C)$ contains a pair of qubits whose associated tetrahedron edges have no face nor vertex in common. Every parity check required for preparation of $|\mathrm{T}\rangle$ [Eq. (5.20)] involves only one qubit at each node. For nodes $(A,B,C)$ implemented via the cavity-QED realization shown in Fig. 5.1, the three qubits not being measured during a given parity check can be made inactive by detuning them from their respective cavities.

The tetrahedron state spans the codespace of a (trivial) CSS code admitting a set of stabilizer generators that can be associated with the faces and vertices of a tetrahedron [Fig. 5.5(a)]:

$$\begin{aligned} S_1 &= Z_1 Z_3 Z_5, & S_4 &= X_1 X_3 X_6, \\ S_2 &= Z_1 Z_4 Z_6, & S_5 &= X_1 X_4 X_5, \\ S_3 &= Z_2 Z_4 Z_5, & S_6 &= X_2 X_3 X_5. \end{aligned} \tag{5.20}$$

It may be verified that the stabilizers associated with the remaining face and vertex are simply given by $S_1 S_2 S_3$ and $S_4 S_5 S_6$.

The stabilizer eigenvector relations $S_i |\mathrm{T}\rangle = |\mathrm{T}\rangle$, $i = 1, 2, \ldots, 6$, admit one common eigenstate, the tetrahedron state:

$$|\mathrm{T}\rangle = \frac{1}{2} \sum_{\beta=\Phi^\pm,\Psi^\pm} |\beta\rangle_{12} |\beta\rangle_{34} |\beta\rangle_{56}, \tag{5.21}$$

where $|\Phi^\pm\rangle = (|00\rangle \pm |11\rangle)/\sqrt{2}$ and $|\Psi^\pm\rangle = (|01\rangle \pm |10\rangle)/\sqrt{2}$. If node $A$ houses qubits 1 and 2, node $B$ qubits 3 and 4, and node $C$ qubits 5 and 6 [Fig. 5.5(b)], then the outcomes of Bell measurements at nodes $A$-$C$ are uniformly distributed across Bell states but perfectly correlated across nodes. A tetrahedron state can therefore provide Alice, Bob, and Charlie with two bits of shared randomness, which could be used to build a shared private key for cryptography.

With the qubits partitioned in the manner described above, every stabilizer $S_i$ in Eq. (5.20) involves one qubit at each node. Preparation of $|\mathrm{T}\rangle$ can be achieved in two steps, starting from a random initial state. In the first step, all $S_i$ are measured using flying-cat parity checks. At this point, each stabilizer measurement returns a random outcome, $\sigma_i = \pm 1$. In the second step, the $-1$ outcomes are treated like an error syndrome and the appropriate correction operator (identified in the manner described below) is applied, transforming the post-measurement state into the target state $|\mathrm{T}\rangle$.

Since $|\mathrm{T}\rangle$ [Eq. (5.21)] is the non-degenerate ground state of $-\sum_i S_i$, the code defined by $S_i$ is trivial, reflecting the trivial genus-0 topology of the tetrahedron. Despite the absence of logical qubits (which, in a toric-code construction [58], requires that $-\sum_i S_i$ have degenerate ground states), the code nonetheless serves as a quantum memory for $|\mathrm{T}\rangle$: Given full syndrome information, consisting of the six eigenvalues $\sigma_i$ of $S_i$, any number of $X$ and $Z$ errors can be corrected using a simple syndrome-decoding algorithm. The procedure is identical for $X$ and $Z$ errors, so without loss of generality, we give the decoding procedure for $X$ errors:

1. Measure the $Z$-basis stabilizers $S_{1-3}$. If the syndrome is $(1,1,1)$, then no correction is required. All other errors can be divided into two classes:
2. Class I errors: If syndrome $j$ in Table 5.2 is obtained, then apply an $X$ gate to qubit $j$.
3. Class II errors: If the syndrome consists entirely of $-1$ outcomes, then apply $X$ gates to *any* pair $(1,2)$, $(3,4)$, or $(5,6)$ of qubits having no face nor vertex in common.

Any number of $X$ errors can be corrected in this way: Single-qubit errors are reversed, while multi-qubit $X$ errors *corrected as single-qubit $X$ errors* are transformed either into $S_{4-6}$ or into products thereof, all of which act trivially on $|\mathrm{T}\rangle$. As an example, consider the scenario where qubits 1 and 3 have undergone bitflip errors. The syndrome $(\sigma_1, \sigma_2, \sigma_3) = (1, -1, 1)$ dictates that an $X$ gate be applied to qubit 6 (Table 5.2).

| $j$ | $\sigma_1$ | $\sigma_2$ | $\sigma_3$ | $\sigma_4$ | $\sigma_5$ | $\sigma_6$ |
|---|---|---|---|---|---|---|
| 1 | $-1$ | $-1$ | $+1$ | $-1$ | $-1$ | $+1$ |
| 2 | $+1$ | $+1$ | $-1$ | $+1$ | $+1$ | $-1$ |
| 3 | $-1$ | $+1$ | $+1$ | $-1$ | $+1$ | $-1$ |
| 4 | $+1$ | $-1$ | $-1$ | $+1$ | $-1$ | $+1$ |
| 5 | $-1$ | $+1$ | $-1$ | $+1$ | $-1$ | $-1$ |
| 6 | $+1$ | $-1$ | $+1$ | $-1$ | $+1$ | $+1$ |

Table 5.2: Syndrome data for decoding Class I errors.

This operation transforms the state $X_1X_3\,|\mathrm{T}\rangle$ into $S_4\,|\mathrm{T}\rangle = |\mathrm{T}\rangle$. For an example of a Class II error, consider the scenario where both qubits 1 and 2 have undergone $X$ errors. In this case, $(\sigma_1,\sigma_2,\sigma_3) = (-1,-1,-1)$, requiring that $X$ gates be applied to any pair $(1,2)$, $(3,4)$, or $(5,6)$ of qubits. If $(1,2)$ is selected, then the errors are reversed. If $(3,4)$ is selected instead, then the correction operator $X_3X_4$ transforms $X_1X_2\,|\mathrm{T}\rangle$ into $S_5S_6\,|\mathrm{T}\rangle = |\mathrm{T}\rangle$.

Multipartite entanglement of $|\mathrm{T}\rangle$ can be detected by measuring the entanglement witness [59, 60]

$$\mathscr{W} = \alpha_{|\mathrm{T}\rangle}\mathbb{1} - |\mathrm{T}\rangle\langle\mathrm{T}|, \tag{5.22}$$

where here, $\alpha_{|\mathrm{T}\rangle}$ is the largest Schmidt coefficient across bipartitions of $|\mathrm{T}\rangle$. This construction is designed to have a non-negative expectation value $\langle\mathscr{W}\rangle \geq 0$ for all biseparable states. A negative expectation value therefore signals detection of genuine multipartite entanglement.

For states (like $|\mathrm{T}\rangle$) having a description in terms of a set of CSS stabilizer generators, it is known that the maximal Schmidt coefficient across bipartitions is $\alpha_{|\mathrm{T}\rangle} = 1/2$, and that the witness $\mathscr{W}$ can be expressed in terms of the CSS stabilizer generators [61]:

$$\mathscr{W} = \frac{3}{2}\mathbb{1} - \prod_{i=1}^{3}\frac{\mathbb{1}+S_i}{2} - \prod_{i=4}^{6}\frac{\mathbb{1}+S_i}{2}. \tag{5.23}$$

The expectation value $\langle\mathscr{W}\rangle$ can be estimated using two measurement settings only: one where all qubits are measured in the $Z$ basis, and one where all qubits are measured in the $X$ basis. The fidelity $F = \langle\mathrm{T}|\rho|\mathrm{T}\rangle = 1/2 - \langle\mathscr{W}\rangle$ of some noisy state $\rho$ relative to $|\mathrm{T}\rangle$ can therefore be estimated using single-qubit measurements, independent of the parity checks that were used to create the state. This provides an avenue for benchmarking the quality of the weight-3 flying-cat parity checks used to prepare $|\mathrm{T}\rangle$.

When a complete set of parity checks is performed starting from some initial state, a set of six eigenvalues $\{\sigma_i\}$ is found. In the absence of measurement errors or errors due to photon loss, the operation required to transform the initial state into $|\mathrm{T}\rangle$ can be correctly identified based on these eigenvalues. A measurement error, however, will lead to the wrong operation being applied, resulting in the creation of a state orthogonal to $|\mathrm{T}\rangle$. Errors on physical qubits can also be introduced by photon loss during a parity check. This loss-induced backaction takes the form of stochastic $Z$ errors ($X$ errors) during measurement of $Z$-basis ($X$-basis) parity checks. Such errors could be corrected in subsequent error-correction cycles using the decoding procedure explained above. However, any errors introduced by the *last* set of parity checks (prior to any measurements occurring at nodes *A*-*C*) remain uncorrected. If, for instance, this last round of checks consists of $Z$-basis parity checks, then $|\mathrm{T}\rangle$ may be affected by uncorrected $Z$ errors, which also produce states orthogonal to $|\mathrm{T}\rangle$. To leading order in $p_{1,2}$, and neglecting corrections $O(p_{1,2}p_{\mathrm{M}})$, we then have

$$\langle \mathscr{W} \rangle \simeq -\frac{1}{2} + [1-(1-p_{\mathrm{M}})^6] + 3(p_1+p_2). \tag{5.24}$$

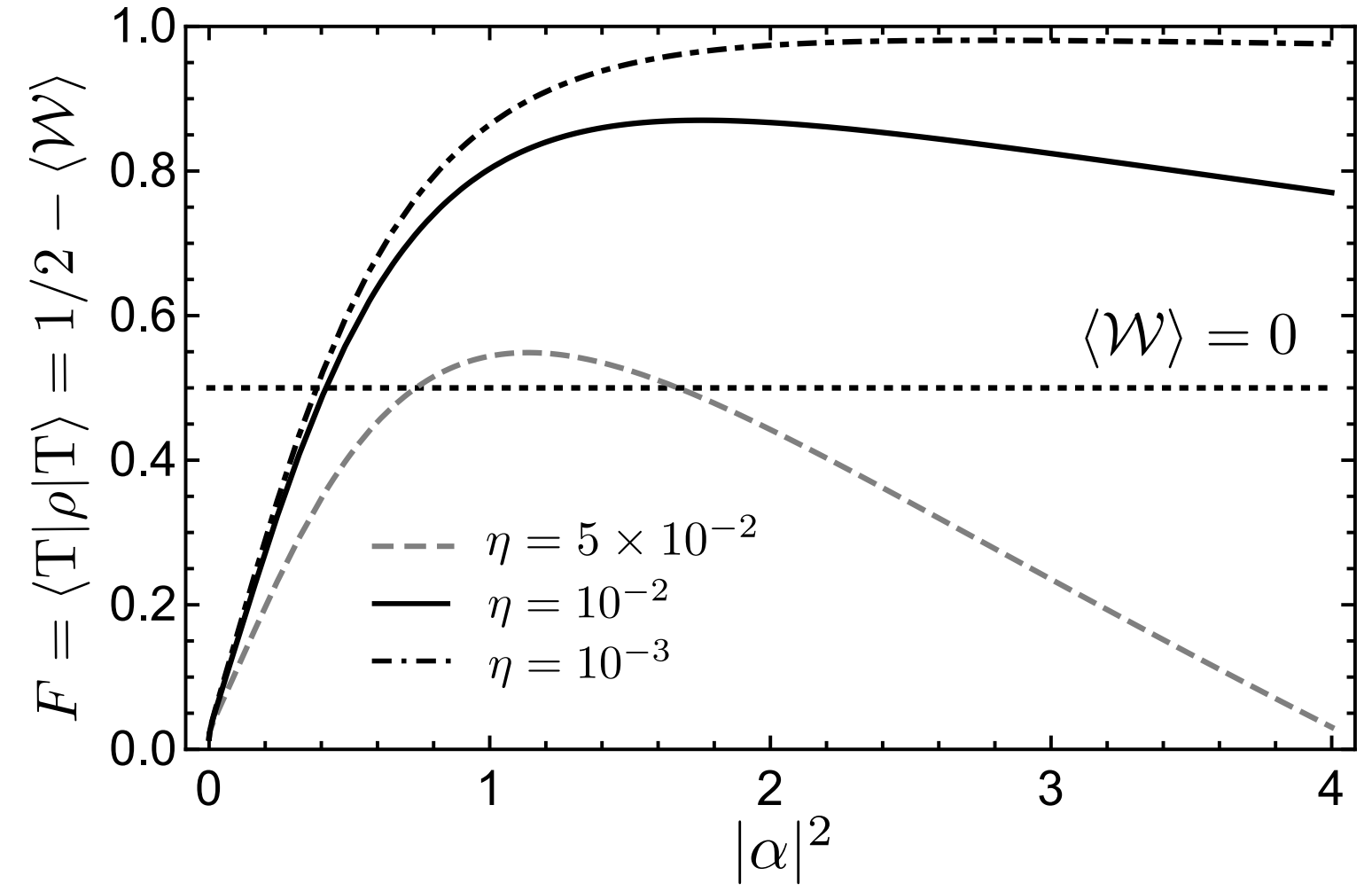

Figure 5.6: The preparation fidelity $F = \langle \mathrm{T}|\rho|\mathrm{T}\rangle$ accounting for measurement errors and errors due to photon loss.

Here, the term $[1-(1-p_{\mathrm{M}})^6]$ gives the probability of making at least one measurement error across the six stabilizer measurements, while the term $3(p_1+p_2)$ gives the probability of photon loss introducing an error during the last set of parity checks (consisting of either three $X$-basis or three $Z$-basis checks). Notably, in the limit $\alpha \to 0$, we have $\langle \mathscr{W} \rangle = 1/2 - 1/2^6$, giving $F = 1/2^6$. This is the fidelity expected from randomly guessing the six stabilizer outcomes and applying a correction based on these guesses.

#### 5.4.2.1 Controlled quantum teleportation

The tetrahedron state can be used as a resource state for three-party controlled quantum teleportation. Controlled quantum teleportation of a single-qubit state using a three-qubit GHZ state was introduced in Ref. [62], and since then, many protocols extending the scheme to multiqubit states using various resource states have been proposed [63–68]. The premise underlying this protocol is that Alice and Bob should be able to teleport quantum states with unit fidelity only with Charlie's cooperation. Here, the fidelity of a teleportation protocol is given by the fidelity of the state reconstructed by Bob relative to Alice's original state. In Ref. [69], Li and Ghose introduced the notion of control power to quantify Charlie's influence over the fidelity of the teleportation protocol. The control power is equal to $1-\bar{F}$ for an average teleportation fidelity $\bar{F}$. For teleportation of a two-qubit state, Charlie's control power must be $1-\bar{F} \geq 3/5$ in order to prevent Alice and Bob from benefiting from their shared entanglement [70]. Not all entangled states meet this criterion, but we now show that the state $|\mathrm{T}\rangle$ does.

We imagine that Alice, Bob, and Charlie share one copy of $|\mathrm{T}\rangle$, with each party holding two qubits according to the partitioning shown in Fig. 5.5(b). Alice has two additional qubits $A_1, A_2$ in state $|\varphi\rangle = a|\Phi^+\rangle + b|\Psi^+\rangle + c|\Psi^-\rangle + d|\Phi^-\rangle$, which she wants to teleport to Bob. She performs Bell measurements on pairs of qubits $(A_1, 1)$ and $(A_2, 2)$ and transmits the results to Bob, who applies the correction operators given in Table 5.3. Once these correction operators have been applied, Bob and Charlie share one of four

| Alice$_1$ | Bob | Alice$_2$ | Bob | Charlie | Bob |
|---|---|---|---|---|---|
| $\lvert\Phi^+\rangle$ | | $\lvert\Phi^+\rangle$ | | $\lvert 0\rangle\lvert +\rangle$ | |
| $\lvert\Psi^+\rangle$ | $X_3$ | $\lvert\Psi^+\rangle$ | $X_4$ | $\lvert 0\rangle\lvert -\rangle$ | $Z_3Z_4$ |
| $\lvert\Phi^-\rangle$ | $Z_3$ | $\lvert\Phi^-\rangle$ | $Z_4$ | $\lvert 1\rangle\lvert +\rangle$ | $X_3X_4$ |
| $\lvert\Psi^-\rangle$ | $Z_3X_3$ | $\lvert\Psi^-\rangle$ | $Z_4X_4$ | $\lvert 1\rangle\lvert -\rangle$ | $Z_3Z_4X_3X_4$ |

Table 5.3: Correction operators for a two-qubit controlled quantum teleportation scheme using $|\mathrm{T}\rangle$ as the entangled resource state.

possible states:

$$\begin{aligned}
&a\left|\Phi^+,\Phi^+\right\rangle + b\left|\Psi^+,\Psi^+\right\rangle + c\left|\Psi^-,\Psi^-\right\rangle + d\left|\Phi^-,\Phi^-\right\rangle,\\
&a\left|\Phi^+,\Psi^+\right\rangle + b\left|\Psi^+,\Phi^+\right\rangle + c\left|\Psi^-,\Phi^-\right\rangle + d\left|\Phi^-,\Psi^-\right\rangle,\\
&a\left|\Phi^+,\Psi^-\right\rangle + b\left|\Psi^+,\Phi^-\right\rangle + c\left|\Psi^-,\Phi^+\right\rangle + d\left|\Phi^-,\Psi^+\right\rangle,\\
&a\left|\Phi^+,\Phi^-\right\rangle + b\left|\Psi^+,\Psi^-\right\rangle + c\left|\Psi^-,\Psi^+\right\rangle + d\left|\Phi^-,\Phi^+\right\rangle.
\end{aligned} \tag{5.25}$$

If Charlie wished to complete the teleportation, he would measure one of his qubits in the $Z$ basis and the other in the $X$ basis. He would then send the measurement outcomes to Bob, who would conditionally apply further corrections to his qubits according to Table 5.3. When Bob is given both Alice's and Charlie's measurement outcomes, the fidelity of the teleportation protocol is unity. With only Alice's information, however, Bob's qubits are in a mixed state $\rho_{\mathrm{B}}$ obtained by tracing over Charlie's qubits. The fidelity $\langle\varphi|\rho_{\mathrm{B}}|\varphi\rangle$ of the teleportation protocol for state $|\varphi\rangle$ can then be averaged over the Haar measure $d\varphi$ to obtain the average fidelity

$$\bar{F} = \int d\varphi\, \langle\varphi|\rho_{\mathrm{B}}|\varphi\rangle = \frac{2}{5}, \tag{5.26}$$

corresponding to a control power of $1-\bar{F} = 3/5$. The average teleportation fidelity that Alice and Bob achieve without Charlie's help is therefore only equal to the fidelity of the best classical strategy [71], in which Alice performs an optimal measurement of her two-qubit state and sends the result to Bob, who then attempts to reconstruct Alice's state on the basis of this classical information. The maximal fidelity for this classical strategy is given by the optimal fidelity for estimation of a quantum state, equal to $2/5$ for a two-qubit system [72–74]. Charlie therefore has the power to prevent Alice and Bob from deriving any advantage (over classical strategies) from their shared entanglement.

We note that the notion of control power given above only has meaning when the three parties share a pure state. Noise can, however, be incorporated under appropriate conditions [75]. Whether the noisy tetrahedron state resulting from a sequence of flying-cat parity checks with finite measurement errors and finite photon loss provides sufficient control power is a question for future study, beyond the scope of the present work.

## 5.5 Feasibility

In this section, we consider the feasibility of implementing flying-cat parity checks in a state-of-the-art circuit QED architecture [76], using currently realized experimental parameters. The basic elements of the parity check consist of: (i) an entangling operation realizing the qubit-state-dependent phase shift of Eq. (5.1), and (ii) propagation of the microwave-frequency light pulse through superconducting transmission lines and circulators [Fig. 5.5(b)]. We begin by discussing error sources related to (i), namely internal cavity losses and errors due to a finite pulse duration, and then errors due to qubit decoherence.

### 5.5.1 Entangling operation

To assess the quality of the entangling operation, here we focus on a circuit-QED setup consisting of a single-sided microwave cavity with a mode at frequency $\omega_c$, coupled to a transmission line with coupling rate $\kappa_0$. In addition, the cavity is coupled via a transverse (Rabi) coupling $g$ to a two-level system encoding a qubit with level splitting $\omega_q$. As in Fig. 5.1, we consider a regime where an incident microwave pulse prepared in a coherent state $|\alpha\rangle$ acquires a qubit-state-dependent phase shift upon reflection [$\alpha \to (-1)^s\alpha$ for $s = 0, 1$]. However, in contrast to the setup in Fig. 5.1, which relies on strong, resonant coupling and the existence of a third level, here we focus on an alternative approach due to its compatibility with qubits having no nearby third level, e.g., electron-spin qubits [77, 78]. A similar analysis could also be carried out for the setup in Fig. 5.1, leading to slightly different conditions.

In the dispersive regime of cavity QED ($|\delta| \gg |g|$, with detuning $\delta = \omega_q - \omega_c$), the cavity and qubit are described approximately by the effective Hamiltonian [79] (setting $\hbar = 1$ here and throughout):

$$H = \omega_c a^\dagger a + \frac{1}{2}\left(\omega_q + \chi\right)\sigma_z + \chi a^\dagger a \sigma_z. \tag{5.27}$$

Here, $\chi = g^2/\delta$ is the dispersive shift, $a$ annihilates a photon in the cavity, and $\sigma_z = |0\rangle\langle 0| - |1\rangle\langle 1|$. In the absence of bit-flips, the Pauli operator $\sigma_z$ is preserved and can be replaced by $\sigma_z \mapsto (-1)^s$ for qubit state $s$. Under the standard (Born-Markov) approximation for a wide-bandwidth transmission line, the cavity field evolves according to the quantum Langevin equation

$$\dot{a}(t) = i[H, a(t)] - \frac{1}{2}\left(\kappa_0 + \kappa_{\text{int}}\right) a(t) - \sqrt{\kappa_0}\, r_{\text{in}}(t), \tag{5.28}$$

where $r_{\text{in}}(t)$ characterizes the complex amplitude of the incoming microwave pulse, and where $\kappa_{\text{int}}$ gives the rate of internal cavity losses. From Eq. (5.28) and the input-output relation [80], $r_{\text{out}}(t) = \int \frac{d\omega}{2\pi} e^{-i\omega t} r_{\text{out}}(\omega) = r_{\text{in}}(t) + \sqrt{\kappa_0}\,\langle a(t)\rangle$, we then find the qubit-state dependent reflection coefficient $R_s(\omega) = r_{\text{out}}(\omega)/r_{\text{in}}(\omega)$:

$$R_s(\omega) = \frac{2i[\omega - \omega_c - (-1)^s\chi] + \kappa_0 - \kappa_{\text{int}}}{2i[\omega - \omega_c - (-1)^s\chi] - \kappa_0 - \kappa_{\text{int}}}. \tag{5.29}$$

For an ideal narrow-bandwidth input pulse with $\omega \simeq \omega_c$, and for negligible internal losses, $\kappa_{\text{int}} \simeq 0$, a special choice of dispersive shift $|\chi| = \kappa_0/2$ leads to a phase shift $\alpha \mapsto R_s(\omega)\alpha \simeq (-1)^s i\alpha$ [39, 42]. The operation of Eq. (5.1) can then be recovered with a change in phase reference, $\alpha \mapsto -i\alpha$. Deviations from the idealized assumptions above (e.g., finite-bandwidth pulses and finite internal losses $\kappa_{\text{int}} \neq 0$) will generally lead to an imperfect entangling operation.

As a measure of quality for the entangling operation, we use the infidelity $\varepsilon$ of the final joint state $\rho_{\text{imperfect}}$ of the qubit and microwave pulse obtained in the presence of imperfections, relative to the ideal final state $|\psi_{\text{ideal}}\rangle = (|0,\alpha\rangle + |1,-\alpha\rangle)/\sqrt{2}$ for a qubit initially prepared in $|+\rangle$:

$$\varepsilon = 1 - \langle\psi_{\text{ideal}}|\rho_{\text{imperfect}}|\psi_{\text{ideal}}\rangle \approx \varepsilon_{\text{qubit}} + \varepsilon_{\text{reflect}}. \tag{5.30}$$

Here, we have divided $\varepsilon$ into a contribution $\varepsilon_{\text{qubit}}$ due to qubit decoherence (neglecting other error sources) and a contribution $\varepsilon_{\text{reflect}}$ due to the combined effects of internal cavity losses and finite pulse bandwidth (neglecting decoherence).

To quantify the effects of both finite $\kappa_{\text{int}}$ and finite bandwidth $\sim \tau^{-1}$ (for a pulse with duration $\tau$), we assume that the spatiotemporal mode supporting state $|\alpha\rangle$ has waveform $u(t)$, normalized according to $\int dt|u(t)|^2 = 1$ [34]. Before the reflection, frequency mode $\omega$ of the transmission line is in a coherent state having amplitude $\alpha u(\omega)$, where here, $u(\omega)$ is the Fourier transform of $u(t)$. For an ideal entangling operation, the waveform would acquire a well-defined $\pm\pi/2$ phase shift, $u(\omega) \mapsto \pm iu(\omega)$, conditioned on the qubit state, in which case we could simply write $|\alpha\rangle \mapsto |\pm i\alpha\rangle$. However, in practice, $\alpha u(\omega) \mapsto \alpha R_s(\omega)u(\omega)$. Evaluating the overlap of the ideal output state with the state resulting from a finite $\kappa_{\text{int}}$ and finite $\tau$ gives

$$\varepsilon_{\text{reflect}} = 1 - \frac{1}{2}\sum_{s=0,1} e^{-\alpha^2 \int \frac{d\omega}{2\pi}|u(\omega)|^2|(-1)^s i - R_s(\omega)|^2}. \tag{5.31}$$

For a Gaussian waveform with width $\tau$, corresponding in the frequency domain to $|u(\omega)|^2 = 2\sqrt{\pi}\tau e^{-(\omega-\omega_c)^2\tau^2}$, the integrand in Eq. (5.31) has finite weight only for $|\omega - \omega_c| \lesssim 1/\tau$. Expanding $R_s(\omega)$ to leading nontrivial order in both $\kappa_{\text{int}}/|\chi|$ and $|\omega - \omega_c|/|\chi| \lesssim 1/(\tau|\chi|)$ then gives

$$|(-1)^s i - R_s(\omega)|^2 \simeq \frac{1}{4}\left(\frac{\kappa_{\text{int}}}{\chi}\right)^2 + \left(\frac{\omega - \omega_c}{\chi}\right)^2. \tag{5.32}$$

Inserting this result into Eq. (5.31), performing the Gaussian integral, then expanding to leading nontrivial order in $\kappa_{\text{int}}/|\chi|$ and $1/(\tau|\chi|)$ gives

$$\varepsilon_{\text{reflect}} \simeq \alpha^2\left(\frac{1}{2\tau^2|\chi|^2} + \frac{\kappa_{\text{int}}^2}{4|\chi|^2}\right). \tag{5.33}$$

Taking a large dispersive shift therefore has two benefits: first, reducing the effects of internal cavity loss $\sim \kappa_{\text{int}}$, and second, allowing for a faster error-correction cycle $\sim \tau$ while maintaining a small error, $\varepsilon_{\text{reflect}}$.

The setup described above has been realized experimentally in, e.g., Ref. [42] for a transmon qubit coupled to a microwave cavity. For the values realized in Ref. [42] ($\kappa_{\text{int}}/2\pi = 0.22$ MHz, $\chi/2\pi = -1.05$ MHz, and $\tau = 500$ ns), and assuming $\alpha = 1$, Eq. (5.33) gives a contribution to the infidelity $\simeq 0.046$ from the term $\propto 1/\tau^2$ and a contribution $\simeq 0.004$ arising from internal losses. The estimate above therefore suggests $\varepsilon_{\text{reflect}} \approx 0.05$, dominated by the finite-bandwidth pulses, although it should be noted that the pulse used in Ref. [42] was square, rather than Gaussian. Doubling the value of the dispersive shift (so that $\chi/2\pi = -2.10$ MHz) would reduce the error $\propto \tau^{-2}$ to 0.01 for the same value of $\tau = 500$ ns, comparable to the $O(1\%)$ threshold of the subsystem surface code [43, 50]. A much smaller value of $\kappa_{\text{int}}/2\pi \sim 100\,\text{Hz}$ also remains well within the state of the art [81]. Therefore, we do not expect $\varepsilon_{\text{reflect}}$ to be a limiting error source for near-term implementations.

The expression for the reflection coefficient $R_s(\omega)$ [Eq. (5.29)] neglects the effects of qubit decoherence, which will also degrade the quality of the entangling operation. To guarantee a low probability of qubit decoherence on the timescale $\tau$ of the pulse, we require that $\min\{T_1, T_2^*\} \gg \tau$, where here, $T_1$ is the timescale for energy relaxation and $T_2^*$ is the inhomogeneous dephasing time due to random fluctuations in the qubit splitting from shot to shot. For dephasing caused by a Markovian (short-correlation-time) environment, the homogeneous dephasing time $T_2$ is equal to $T_2^*$, and the decay of coherences is exponential with timescale $T_2 = T_2^*$. In Ref. [42], decoherence was limited by pure dephasing on a time scale $T_2 = T_2^* = 6\ \mu$s, while the pulse had a duration of $\tau = 500$ ns. To leading order in $\tau/T_2^*$, a quick estimate for the infidelity due to decoherence is then given by

$$\varepsilon_{\text{qubit}} \simeq \tau/T_2^*. \tag{5.34}$$

Inserting the values for $\tau$ and $T_2^*$ from Ref. [42] then gives $\varepsilon_{\text{qubit}} \simeq 0.08$, comparable to the infidelity of 0.11 attributed to the effects of decoherence, based on a more sophisticated error model [42]. Reducing the pulse duration therefore provides a clear avenue towards reducing errors due to decoherence. However, to maintain the same small value of $\varepsilon_{\text{reflect}}$, a reduction in $\tau$ must be accompanied by a proportional increase in $|\chi|$, as discussed above.

### 5.5.2 Transmission losses

For the setup illustrated in Fig. 5.5(b), sources of photon loss will likely include losses $\eta_{\mathrm{trans}}$ incurred in transit, together with losses $\eta_{\mathrm{circ}}$ due to chiral elements like circulators, giving

$$\eta = \eta_{\mathrm{trans}} + \eta_{\mathrm{circ}}. \tag{5.35}$$

Many superconducting interconnects are made of NbTi, for which typical attenuation rates are 5 dB/km, corresponding to losses of 68%/km [82]. The architecture of Ref. [82] instead uses pure Al to achieve losses of 0.15 dB/km (3.4%/km), comparable to the values reached for fiber-optic cables carrying telecom-frequency light. Hence, for a meter-scale experiment, $\eta_{\mathrm{trans}}$ can likely be made negligible relative to other sources of loss.

In microwave platforms, chiral elements like circulators often lead to higher losses than those incurred during free propagation, due in part to the challenges of on-chip integration. In Ref. [6], for instance, circulators led to insertion losses estimated at $\eta_{\mathrm{circ}} = 0.13$. If the dominant source of error is circulator loss, a factor-of-two improvement would be sufficient to demonstrate multipartite entanglement in a tetrahedron state using the layout of Fig. 5.5(b), since $\langle \mathscr{W} \rangle < 0$ can be achieved for the entanglement witness $\mathscr{W}$ [Eq. (5.23)] provided $\eta \lesssim 0.05$ [cf. Fig. 5.6]. Significant effort has gone towards the development of microwave circulators compatible with on-chip integration [83, 84], as this technology has already been identified as a key ingredient for scaling up superconducting processors. As a final comment, the circulators shown in Fig. 5.5(b) are not strictly necessary. They could be replaced by fast dynamical switches that control the path taken by the pulse.

### 5.5.3 Summary: Parametric constraints

For the circuit QED architecture considered above, where the entangling operation of Eq. (5.1) is realized in the dispersive regime, increasing the strength of the dispersive coupling provides a straightforward avenue towards increasing the fidelity of the entangling operation. Decoherence times are another important consideration. For $\min\{T_1, T_2^*\} = 10\,\mu s$, and for the same value of $\kappa_{\mathrm{int}}/2\pi = 0.22$ MHz realized in Ref. [42], a possible combination of $\tau$ and $\chi$ giving $\varepsilon \approx \varepsilon_{\mathrm{qubit}} + \varepsilon_{\mathrm{reflect}} \approx 0.01$ is $\tau = 100$ ns and $|\chi|/2\pi = 10$ MHz. Dispersive shifts as large as 10 MHz have been realized experimentally for transmon qubits [85].

Assuming the layout of Fig. 5.5(b), preparing a tetrahedron state requires six stabilizer measurements, each involving three entangling operations, for a total of eighteen entangling operations. Accounting for errors in the entangling operations, an estimate for the maximum achievable preparation fidelity of the tetrahedron state is therefore $F \leq F_{\mathrm{max}} \simeq 1 - 18\varepsilon$. Since a fidelity $F > 1/2$ is required to demonstrate multipartite entanglement [Fig. 5.6], an ideal scenario is one where all error sources associated with the entangling operation (e.g. qubit dephasing, internal cavity losses, and finite-bandwidth effects) can be controlled to the extent that $18\varepsilon \ll 1/2$ ($\varepsilon \ll 0.03$). If this condition is met, then multipartite entanglement can be demonstrated with a photon loss rate as high as $\eta \approx 0.05$ [Fig. 5.6]. As discussed above, this would require only a factor $\sim 2$ improvement in the circulator insertion losses reported in Ref. [6].

## 5.6 Conclusion

High-quality flying-cat parity checks could provide the ability to perform error correction in distributed architectures, where native qubit-qubit interactions are not typically available across nodes. However, operations involving propagating quasimodes of the electromagnetic field will introduce photon loss as a source of error on code qubits.

In this work, we have analyzed the trade-off between measurement errors and errors due to photon loss in the context of weight-3 flying-cat parity checks. We have shown how the total impact of these error sources could be minimized as a function of the coherent-state amplitude. In addition, we have verified that the horizontal hook errors due to photon loss remain gauge-equivalent to single-qubit errors for gauge operators of the form $X^{\otimes 3}$, $Z^{\otimes 3}$, so that flying-cat parity checks preserve this key advantage of the subsystem surface code. In the context of entanglement distribution, we also described how flying-cat parity checks can be used to generate three-qubit GHZ states, as well as a six-qubit entangled "tetrahedron" state that can be used for controlled quantum teleportation of an arbitrary two-qubit state. Another result of this paper is the conditional post-measurement state $\rho_x$ of the data qubits following a phase-sensitive measurement of the electromagnetic field, incorporating both measurement errors and photon-loss-induced errors. This object can be used to find the conditional probabilities for error patterns on the code qubits, shot-by-shot for each measured value of $x$. In a quantum error correcting code, this probability could be passed to a decoding scheme that takes full advantage of this soft information [86–88]. Finally, we have shown that high-quality flying-cat parity checks are feasible with current or near-future technology.

Flying-cat parity checks rely on sources and detectors for (classical) coherent states of light, rather than the single photons used to link nodes in several other quantum-network implementations. These parity checks therefore do not require single-photon sources or detectors, although they do require a high-quality stable source of coherent light and low-noise phase-sensitive measurements. We have only considered the simplest encoding (a single-mode input coherent state). In future work, it may be interesting to consider whether a more elaborate encoding could be used to simultaneously detect photon loss and parity using, e.g., a cat-code variant [89]. Additionally, rapid qubit gates and dynamical decoupling sequences performed to preserve code qubits coupled to cavities will generically lead to bursts of light that can carry away information [90]. Whether dynamical decoupling can be made compatible with flying-cat parity checks is an open question.

The flexible qubit connectivity enabled by the use of propagating light pulses could also provide an avenue towards realizing the nonlocal interactions required by certain quantum low-density parity-check (LDPC) codes [91–95]. Quantum LDPC codes have shown promise for reducing the number of physical qubits required to encode a logical qubit compared to leading candidates like the surface code.

**Acknowledgments**—We thank K. Wang for useful discussions. We also acknowledge funding from the Natural Sciences and Engineering Research Council of Canada (NSERC) and from the Fonds de Recherche du Québec–Nature et technologies (FRQNT).

# Preface to Chapter 6

The previous chapter discussed a possible strategy for performing parity checks of distant qubits using pulses of radiation. Such parity checks could be realized without the need for long-range gates between stationary qubits, as would be required for parity checks based on the use of ancilla qubits. However, although these pulses of radiation can be used to encode parity information in the exact same manner as an ancilla, they do *not* mediate entangling gates. Long-range gates would enable a wider range of operations on distant qubits, going beyond the parity measurements discussed in the previous chapter.

Over the past few years, there appears to have been a shift in paradigm away from designing and building chips ("quantum processing units", or QPUs) with ever-increasing numbers of qubits. Instead, focus has shifted towards designing modular quantum computers consisting of several spatially separated, semi-independent clusters of qubits integrated into a single device. From a fabrication and engineering point of view, modular architectures would allow faulty chips to be replaced more easily and could offer significant advantages in overcoming the input-output wiring bottlenecks currently blocking the road to true scalability. At the time of writing, classical integration of QPUs has been demonstrated by IBM [1], and simulation strategies based on classically combining the outcomes of modest-sized circuits have allowed the dynamics of certain systems to be simulated with a smaller number of qubits than might naively be expected [2]. In the long run, however, quantum integration enabled by long-range, intermodular gates would allow larger quantum circuits to be run across several QPUs.

Most protocols for implementing two-qubit gates in a circuit-QED setting assume common coupling of both qubits to a single, spectrally isolated cavity mode. Cross-resonance gates and resonator-induced phase gates can both be operated with such a setup. Distant qubits could still be linked by long waveguides acting as cavities, but as the length of a waveguide increases, the corresponding decrease in its free spectral range makes it more difficult to spectrally address a single mode. In this final chapter, we give protocols for performing entangling gates between distant qubits. These gates are mediated by real photons and are therefore compatible with a free spectral range of zero. Although both protocols are based on the well-known "pitch and catch" strategy for Bell-state generation, which typically requires that the "catcher" qubit begin in its ground state, the gates presented in this chapter can be applied to an arbitrary two-qubit initial state. The intent of the research presented here is to expand the toolbox of quantum operations that can be performed in modular architectures.

# 6

# Inter-module entangling gates

*This chapter is the integral text from*

**Protocols for inter-module two-qubit gates mediated by time-bin encoded photons**
Z. M. McIntyre and W. A. Coish

As quantum devices scale to larger numbers of qubits, entangling gates between distant stationary qubits will help provide flexible, long-range connectivity in modular architectures. In this work, we present protocols for implementing long-range two-qubit gates mediated by either Fock-state or time-bin qubits—photonic encodings that are both compatible with the coplanar waveguide resonators commonly used in circuit quantum electrodynamics (QED). These protocols become deterministic in the limit of vanishing photon loss. Additionally, photon loss can be heralded, signaling a failed two-qubit gate attempt. We model the loss of a time-bin qubit to a dielectric environment consisting of an ensemble of two-level systems (TLSs), which are believed to be the dominant mechanism for dielectric loss in circuit QED architectures. The backaction (on the stationary qubits) associated with the loss of the time-bin qubit is strongly suppressed in a non-Markovian regime where the temporal separation of the time bins is short compared to the dielectric environment's correlation time. This result suggests strategies based on a combination of materials-fabrication and time-bin-qubit optimization for ensuring that the loss of a time-bin qubit is not only heralded, but also approximately backaction-free.

## 6.1 Introduction

A promising approach to large-scale quantum information processing involves coupling stationary qubits to microwave photons. The stationary qubits could be superconducting qubits [1] or spins housed in semiconductor heterostructures [2]. As these systems scale up, the complexities associated with wiring and control will likely motivate the adoption of modular architectures where distributed quantum processing units (QPUs) are combined into a single device via classical processing or quantum interconnects [3]. In the near term, techniques like circuit cutting [4, 5] and entanglement forging [6] can be used to simulate the output of larger quantum circuits by classically combining the outputs of smaller circuits run sequentially on a single QPU or in parallel across separate QPUs. Along similar lines, the statistics of long-range entanglement can be replicated using virtual gates [7–9], which simulate two-qubit gates from a quasiprobability decomposition of single-qubit rotations and projective measurements. The costs of simulating non-local virtual gates can be reduced by allowing for the exchange of classical information between QPUs [10], as was recently demonstrated experimentally [11].

Strategies like circuit cutting and entanglement forging are best suited to computations where the structure of the problem motivates a partition into weakly interacting subsystems, each of which can be simulated individually before correlating the outcomes on a classical computer. Such a structure can be exploited to simplify many relevant problems in quantum computational chemistry [12], such as finding the ground states of molecular Hamiltonians [6]. However, entanglement between QPUs would enable a larger class of problems to be tackled in the long term [3].

In a circuit-QED or spin-circuit QED architecture, entanglement between modules housing distinct QPUs can be established via photonic interconnects capable of transmitting quantum information over distances far exceeding the centimeter-scale wavelength of microwave radiation. Cryogenic meter-scale interconnects, combined with a protocol for applying gates across the interconnects, would form the architectural requirements for so-called $l$-type modularity [3]. Significant effort has gone towards developing strategies for quantum-state transfer and Bell-state generation between qubits connected by such interconnects [13–22]. Bell pairs generated using these strategies could be stored and consumed on-demand as part of a gate-teleportation protocol [23, 24], enabling two-qubit gates between distant qubits. However, a more qubit-efficient approach may be to apply long-range gates directly to the code qubits themselves, bypassing the need for ancillary Bell pairs, Bell-pair quantum memory, possible entanglement-purification steps, and the additional gates required by the teleportation protocol itself. Long-range gates could also be used for performing parity checks of distant qubits [25, 26], implementing transversal encoded two-qubit gates [27], reducing the impact of cosmic-ray events on encoded information [28], or as a means of establishing the non-planar connectivity required for certain quantum error correcting codes (quantum low-density parity check codes) [29–32].

Two-qubit gates between atom-qubits in separate optical cavities have been proposed [33] and realized experimentally [34]. These schemes leverage selection rules to entangle a qubit encoded in a multilevel system with a photonic polarization qubit [35, 36], which is then used to mediate the two-qubit gate. Such selection rules exist in many systems featuring optical transitions, including atoms [35] and excitons [37]. Polarization-dependent atomic transitions have also been considered as a way of implementing two-qubit gates between photonic qubits themselves [38, 39]. Although polarization qubits can readily be transmitted through either free space or optical fibers, their integration into circuit QED architectures is not commonly considered due to the fixed polarization of the electromagnetic field housed in the coplanar waveguide resonators commonly used to manipulate, measure, and couple qubits [1, 40].

Here, we describe fixed-polarization protocols for performing two-qubit gates across quantum-photonic interconnects. In contrast to entangling gates like the cross-resonance [41, 42] and resonator-induced-phase gates [43, 44], which require common coupling of the two qubits to a single spectrally isolated standing-wave mode, the approaches presented here couple the stationary qubits to propagating quasimodes of the interconnect—“flying” qubits. The photonic degree-of-freedom required for flying qubits is provided by either a Fock-state encoding (whose basis states are defined by the presence or absence of a photon in the interconnect) or a time-bin encoding (whose basis states are the single-photon Fock states of a pair of orthogonal spatiotemporal modes of the interconnect). Time-bin encodings have been considered for linear-optical quantum computing [45], quantum key distribution [46], Bell-state generation [17, 47, 48], and quantum-state transfer [17]. The two-qubit gate protocols presented here become deterministic (they succeed with unit probability) in the limit of vanishing photon loss. The loss of a photon is a heralded error; hence, rather than contribute to a gate infidelity, errors due to photon loss are instead converted into a reduced probability of successfully executing the gate.

In the most common (Markovian) models of photon loss, absorption of a photon by a dielectric medium occurs at a single instant in time. This absorption time is an indelible trace left in the environment that distinguishes between the two computational basis states of a photonic time-bin qubit immediately before it is absorbed. For the time-bin-qubit mediated gate described in this work, absorption of the intermediary photon in a way that distinguishes between the time-bin states leads to an error (backaction) on the stationary qubits. However, in a more general (non-Markovian) model of dielectric loss, there is room for a photon to be absorbed without introducing backaction. Standard models of photon loss do not account for details of the photon's spatiotemporal mode, but we fully account for this in the present work. Moreover, we model the loss of a time-bin qubit to a bath of two-level systems (TLSs), accounting for a structured spectral density of the TLS environment in order to determine the backaction of photon loss on the stationary qubits involved in the gate. Such TLSs are believed to be responsible for the majority of dielectric loss in circuit QED platforms [49–57]. We calculate the distinguishability of the TLS bath states resulting from the absorption of a photon from either the early or the late time bin. This distinguishability controls the rate of dephasing errors on stationary qubits resulting from the loss of the time-bin qubit. We show that the distinguishability of the states of the TLS ensemble (conditioned on the time-bin state absorbed) can be tuned by modifying either the TLS spectral density or the time-bin separation. This result suggests strategies for improving the performance of the gate protocols presented here in the context of a quantum error-correcting code, where the rate of errors on stationary code qubits should be kept as low as possible.

The structure of the rest of this article is as follows: In Sec. 6.2, we review how cavity-assisted Raman transitions can be used to deterministically emit and absorb single photons. In Sec. 6.3, we present two protocols for applying a controlled-Z (CZ) gate to two distantly separated qubits: The first is appropriate for qubits separated by a distance exceeding the size of the photonic wavepacket, while the second alleviates this restriction at the cost of introducing an ancilla qudit used for measurement of the photonic qubit. In Sec. 6.4, we present the results of a numerical simulation of the gate infidelity accounting for decoherence of the stationary qubits. We discuss the backaction associated with the loss of a time-bin qubit in Sec. 6.5. We offer concluding remarks in Sec. 6.6.

## 6.2 Pitching photons

The emission and subsequent reabsorption of single photons has long been recognized as a useful resource for quantum state transfer and entanglement distribution in quantum networks [58]. In this section, we review how a cavity-assisted Raman transition can be used to controllably emit a single photon into a designated spatiotemporal quasimode of a transmission line coupled to the cavity.

As is well known from the study of atomic ensembles with $\Lambda$-type level structures [59–61], one way to emit and absorb single photons is via cavity-assisted Raman transitions. This process can be described by a Hamiltonian of the form

$$H[\Omega(t)] = i\Omega(t)\,|g,1\rangle\langle f,0| + \text{h.c.}, \tag{6.1}$$

where $|0\rangle$ and $|1\rangle$ are zero- and one-photon Fock states of the cavity. The paradigmatic setup for realizing an interaction of this form [Eq. (6.1)] is one where a cavity mode is coupled off-resonantly with detuning $\Delta$ and strength $g_0$ to the $|g\rangle \leftrightarrow |e\rangle$ transition of a $\Lambda$ system having states $|g\rangle$, $|e\rangle$, and $|f\rangle$, while the atom's $|f\rangle \leftrightarrow |e\rangle$ transition is driven off-resonantly by a classical drive with envelope $\lambda_{\rm d}(t)$. Adiabatically eliminating the $|e\rangle$ state then yields an interaction of the form $H[\Omega(t)]$ with $\Omega(t) = g_0\lambda_{\rm d}(t)/\Delta$ [59], where we have here assumed degeneracy of the $|f\rangle$ and $|g\rangle$ levels. In the dispersive regime of cavity QED (where the difference between the qubit and cavity frequencies far exceeds the strength of the transverse qubit-cavity coupling $g_0$), an interaction of the form of Eq. (6.1) can be realized for superconducting transmon qubits by coherently driving the cavity [62] or qubit [13, 17, 20, 63, 64] at the frequency of the dressed $|f,0\rangle \leftrightarrow |g,1\rangle$ transition. In this case, $\Omega(t) = g_0\lambda_{\rm d}(t)\alpha/(\sqrt{2}\Delta(\Delta+\alpha))$, where $\lambda_{\rm d}(t)$ again denotes the drive envelope and $\alpha$ is the transmon anharmonicity [64].

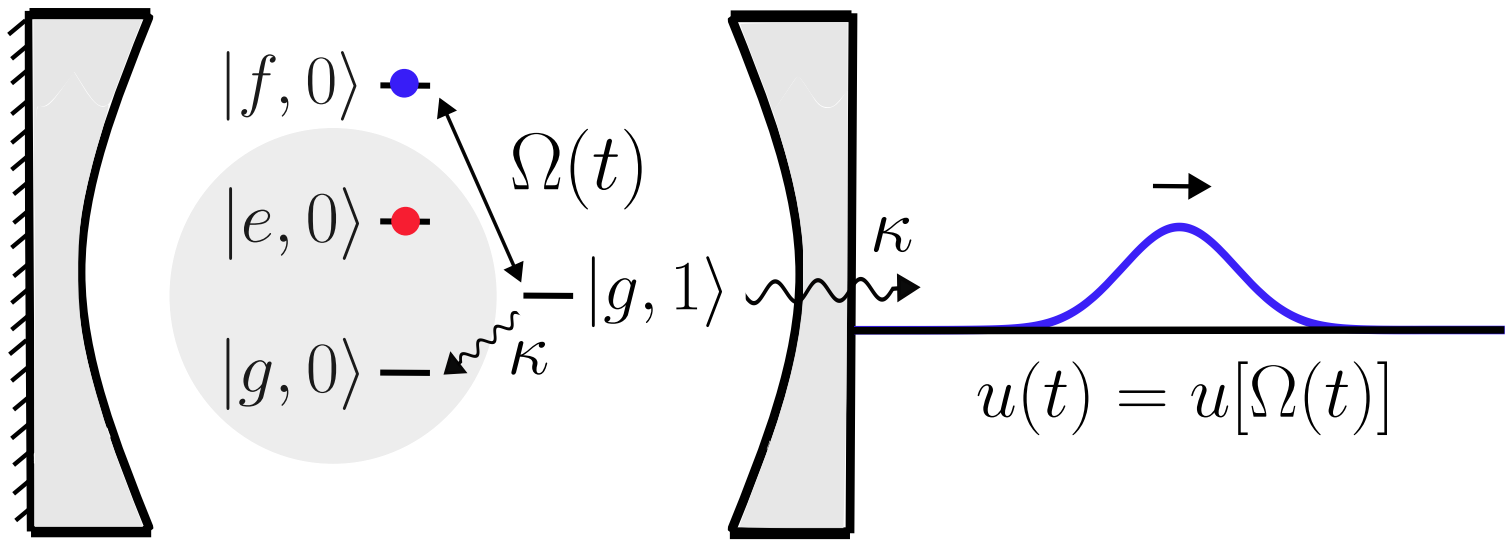

Figure 6.1: A cavity-assisted Raman process with tunable strength $\Omega(t)$ can be used to emit a photon into a chosen quasimode $u$ of a transmission-line resonator coupled to the cavity.

The interaction $H[\Omega(t)]$ can be used to emit a photon, conditioned on the atom being in state $|f\rangle$, into some quasimode $u$ [with a waveform $u(t)$ normalized according to $\int dt\, |u(t)|^2 = 1$] of a transmission-line resonator coupled to the cavity (Fig. 6.1). The resulting single-photon state can be expressed as $|1_u\rangle = r_u^\dagger |\text{vac}\rangle$, where $|\text{vac}\rangle$ is the vacuum state of the transmission line and

$$r_u^\dagger = \int dt\, u(t) r_{\rm out}^\dagger(t). \tag{6.2}$$

Here, $r_{\rm out}(t) = (2\pi)^{-1}\int d\omega e^{-i\omega t} r_\omega$ is an output-field operator written in terms of annihilation operators $r_\omega$ acting on frequency modes of the transmission line and satisfying $[r_\omega, r_{\omega'}^\dagger] = \delta(\omega-\omega')$ [65]. Photon emission conditioned on the atomic state $|f\rangle$ can be realized in superposition to map $(\alpha|e\rangle + \beta|f\rangle)\otimes|\text{vac}\rangle \to \alpha|e,\text{vac}\rangle + \beta|g,1_u\rangle$.

To relate the populated quasimode $u$ to the envelope $\Omega(t) = \Omega_{\rm e}(t)$ required for emitting a photon into mode $u$, we consider evolution described by the master equation

$$\dot{\rho} = -i[H,\rho] + \kappa\mathscr{D}[a]\rho, \tag{6.3}$$

where $H = H[\Omega_{\rm e}(t)]$, $\kappa$ is the rate of cavity decay into the transmission line, and $a$ is an annihilation operator that removes one photon from the cavity mode: $a|1\rangle = |0\rangle$. The damping superoperator $\mathscr{D}[a]$ acts according to $\mathscr{D}[a]\rho = a\rho a^\dagger - \{a^\dagger a, \rho\}/2$.

For $\kappa \gg |\Omega(t)|$, the waveform $u(t)$ of the emitted photon can be related to the output field via the input-output relation (see Ref. [66] and the supplement of Ref. [67] for details): $\langle r_{\rm out}\rangle_t = \sqrt{\kappa}\langle a\rangle_t = u(t)\langle\tau_{gf}\rangle_0$, where $\tau_{gf} = |g\rangle\langle f|$, and where the average is defined as $\langle\mathscr{O}\rangle_t = \text{Tr}\{\mathscr{O}\rho(t)\}$ for operator $\mathscr{O}$. With the density matrix $\rho(t)$ written as $\rho(t) = \sum_{\alpha,\beta}\rho_{\alpha,\beta}(t)|\alpha\rangle\langle\beta|$ for $|\alpha\rangle \in \{|g,0\rangle, |g,1\rangle, |f,0\rangle\}$, we therefore have

$$u(t) = \sqrt{\kappa}\frac{\langle a\rangle_t}{\langle\tau_{gf}\rangle_0} = \sqrt{\kappa}\frac{\rho_{g1,g0}(t)}{\rho_{f0,g0}(0)}. \tag{6.4}$$

To determine the dynamics of the relevant matrix element $\rho_{g1,g0}(t)$, we consider the following set of coupled differential equations obtained from Eq. (6.3):

$$\begin{aligned}\dot{\rho}_{g1,g0}(t) &= \Omega_{\rm e}(t)\rho_{f0,g0}(t) - \frac{\kappa}{2}\rho_{g1,g0}(t),\\ \dot{\rho}_{f0,g0}(t) &= -\Omega_{\rm e}^*(t)\rho_{g1,g0}(t).\end{aligned} \tag{6.5}$$

A simple expression for $\rho_{g1,g0}(t)$ can be derived in the regime where $\Omega_{\rm e}(t)$ varies slowly on a timescale $\kappa^{-1}$ and $\kappa$ far exceeds the strength of the driving, $\kappa \gg |\Omega_{\rm e}(t)|$. In this regime, the dynamics of the cavity field $\langle a\rangle_t = \rho_{g1,g0}(t)$ follows that of the instantaneous (equal time) atomic coherence $\langle \tau_{gf}\rangle_t = \rho_{f0,g0}(t)$, and we can adiabatically eliminate the cavity field by approximating $\dot{\rho}_{g1,g0} \simeq 0$. Solving the above system of equations then gives [cf. Eq. (6.4)]

$$u(t) \simeq \frac{2\Omega_{\rm e}(t)}{\sqrt{\kappa}} e^{-\frac{2}{\kappa}\int_{-\infty}^{t} ds\, |\Omega_{\rm e}(s)|^2}. \tag{6.6}$$

Under the assumption that $\Omega_{\rm e}(t)$ varies slowly on a timescale $\kappa^{-1}$, this expression [Eq. (6.6)] indicates that $u(t)$ varies on a timescale $\sim \kappa/|\Omega_{\rm e}|^2$, which self-consistently exceeds the timescale $\kappa^{-1}$ provided $\kappa \gg |\Omega_{\rm e}(t)|$ [as initially assumed when solving the system of coupled differential equations in Eq. (6.5)].

In order to express $\Omega_{\rm e}(t)$ in terms of $u(t)$, we note that $|u(t)|^2 \simeq -\frac{d}{dt} e^{-\frac{4}{\kappa}\int_{-\infty}^{t} ds\, |\Omega_{\rm e}(s)|^2}$ [cf. Eq. (6.6)], which can be integrated to obtain the relation

$$e^{-\frac{4}{\kappa}\int_{-\infty}^{t} ds\, |\Omega_{\rm e}(s)|^2} \simeq 1 - \int_{-\infty}^{t} ds\, |u(s)|^2. \tag{6.7}$$

Using Eq. (6.7) in conjunction with Eq. (6.6) allows us to solve for $\Omega_{\rm e}(t)$ in terms of $u(t)$, giving [60, 61]

$$\Omega_{\rm e}(t) \simeq \frac{\sqrt{\kappa}}{2}\frac{u(t)}{\sqrt{\int_t^{\infty} ds\, |u(s)|^2}}, \quad |\Omega_{\rm e}(t)| \ll \kappa. \tag{6.8}$$

Here, we have made use of the normalization condition $\int dt\, |u(t)|^2 = 1$ to rewrite the range of the integral in the denominator. *Absorption* of an incident photon in mode $u(t)$ can be realized by shaping the envelope for the absorption pulse $\Omega(t) = \Omega_{\rm a}(t)$ so that its time reverse $\Omega_{\rm a}^*(-t)$ would lead to the emission of a photon into the time-reversed quasimode having waveform $u^*(-t)$ [60]. The pulse required for absorption of a photon with waveform $u(t)$ into a cavity having linewidth $\kappa$ can then be found from Eq. (6.8) and is given by

$$\Omega_{\rm a}(t) \simeq \frac{\sqrt{\kappa}}{2}\frac{u(t)}{\sqrt{\int_{-\infty}^{t} ds\, |u(s)|^2}}, \quad |\Omega_{\rm a}(t)| \ll \kappa. \tag{6.9}$$

The combination of complex conjugation and mapping of $t \to -t$ in the waveform $u^*(-t)$ of the time-reversed quasimode can be understood in a plane-wave basis by noting that $r_\omega^\dagger$ [cf. Eq. (6.2)] populates the plane-wave mode $\langle x|r_\omega^\dagger|\text{vac}\rangle \propto e^{-i\omega(t-x/v)}$ (assuming a linear dispersion and propagation at speed $v$). Reversing the propagation direction of this plane wave can be accomplished through complex conjugating and a mapping of $t \to -t$, giving $e^{-i\omega(t+x/v)}$. This can be generalized from plane waves to quasimodes: For a photon in the quasimode $u$, $\langle x|r_u^\dagger|\text{vac}\rangle \propto \int d\omega\, u(\omega) e^{-i\omega(t-x/v)}$, where $u(\omega) = \int dt\, e^{i\omega t} u(t)$ is the Fourier transform of the waveform $u(t)$. In this case, complex conjugation and a mapping of $t \to -t$ leads to an expression involving $u^*(\omega)$, corresponding to the Fourier transform of $u^*(-t)$—the waveform of the time-reversed mode.

In the next section, we describe two protocols for performing CZ gates between distantly separated stationary qubits. Both protocols require the emission and subsequent re-absorption of a photonic qubit, encoded in either a Fock basis or in a time-bin basis. In the notation established above, a photon encoding a time-bin qubit has a waveform $u_{\mathrm{E}}(t)$ if the photon is emitted at the early time, or $u_{\mathrm{L}}(t) = u_{\mathrm{E}}(t-\tau_{\mathrm{sep}})$ if it is emitted at the late time. Here, $\tau_{\mathrm{sep}}$ is a temporal offset chosen to be sufficiently large for the two time-bin states to be treated as orthogonal $[\int dt\, u_{\mathrm{E}}^*(t) u_{\mathrm{L}}(t) = 0]$.

## 6.3 CZ gates mediated by Fock-state or time-bin qubits

We consider a setup where two stationary qubits (described by two-level subspaces of three-level qudits $\mathrm{Q}_1$ and $\mathrm{Q}_2$) are each coupled to their own cavity. The cavities are themselves coupled by a transmission line, allowing the radiation emitted from one cavity to be channeled into the other. In cases where there is no ambiguity, we also denote by $\mathrm{Q}_i$ the qubit encoded in the three-level system $\mathrm{Q}_i$.

A summary of this section is as follows: In Sec. 6.3.1, we first give steps for realizing an entangling operation between a stationary, cavity-coupled qubit and a photonic qubit encoded in either a Fock-state or time-bin basis. This operation is used as a primitive in constructing the long-range gates presented in Secs. 6.3.2 and 6.3.3. The two CZ-gate protocols differ mainly in whether or not they require a measurement of the flying qubit, which, as we later explain, could be realized by mapping the photonic-qubit basis states onto the states of an ancilla. The "measurement-free" protocol, described in Sec. 6.3.2, does not require measurement of the flying qubit but does require a minimum spatial separation of $\mathrm{Q}_1$ and $\mathrm{Q}_2$, set by the size of the photonic wavepacket. The "ancilla-assisted" protocol, described in Sec. 6.3.3, alleviates this requirement by introducing a measurement of the flying qubit. Both protocols take advantage of the multilevel structure of $\Lambda$-type emitters to herald the loss of the flying qubit. The ancilla-assisted CZ-gate protocol could be mediated by a Fock-state encoded photon in principle, but we choose to present the protocol with a time-bin encoding for the single photon. This is due to the possibility—explored in Sec. 6.5—of combining this gate protocol with engineering of the transmission line's dielectric environment to achieve a situation where, despite the orthogonality of the time-bin basis states, the loss of the time-bin qubit is not only heralded but also approximately backaction-free.

### 6.3.1 A CZ gate between flying and stationary qubits

A crucial ingredient for realizing the long-range gate protocol of Sec. 6.3.2 is an entangling operation of the form

$$U = e^{i\pi|1_u,g\rangle\langle 1_u,g|}. \tag{6.10}$$

This CZ gate acts on a stationary qubit having basis states $|e\rangle$ and $|g\rangle$ and a Fock-state encoded photonic qubit having basis states $|1_u\rangle$ and $|\mathrm{vac}\rangle$. The gate protocol presented in Sec. 6.3.3, below, relies on the same operation for a time-bin encoded photon:

$$U = e^{i\pi|\mathrm{L},g\rangle\langle \mathrm{L},g|}. \tag{6.11}$$

In this case, the photon encodes a time-bin qubit having basis states $|\mathrm{E}\rangle$ and $|\mathrm{L}\rangle$, where $|\mathrm{E}\rangle$ ($|\mathrm{L}\rangle$) is a transmission-line state containing one photon in the spatiotemporal mode with waveform $u_{\mathrm{E}}(t)$ [$u_{\mathrm{L}}(t)$]. There are several ways these CZ gates could be realized using phase shifts of the photonic waveform conditioned on the state of a stationary qubit.

Given a three-level system having states $|g\rangle, |e\rangle, |f\rangle$, a qubit-conditioned phase shift can be realized by resonantly coupling the system's $|e\rangle \leftrightarrow |f\rangle$ transition to a single-sided cavity [Fig. 6.2(a)]. This cavity-qubit interaction leads to hybridization of the bare qubit and cavity states into states of the form $|\pm,1\rangle \propto |f,0\rangle \pm |e,1\rangle$. These states have eigenenergies $\omega_0 \pm g_0$, where here $\omega_0$ is the bare cavity frequency and $g_0$ is the strength of the cavity-qubit coupling. For a qubit in state $|g\rangle$, an incoming photon resonant with the bare cavity frequency will undergo a $\pi$ phase shift provided the photon bandwidth is narrow relative to the cavity linewidth $\kappa$ [68–70]. (This bandwidth condition can be achieved if the photon wavepacket duration $\tau$ satisfies $\tau^{-1} \ll \kappa$.) However, for a qubit in state $|e\rangle$, the cavity density of states at $\omega_0$ is significantly suppressed by the cavity-qubit hybridization provided $g_0 \gg \kappa$. An incoming photon will therefore be reflected *without* a $\pi$ phase shift for a qubit in state $|e\rangle$ [Fig. 6.2(a)]. For a Fock-state encoded photonic qubit, this $\pi$ phase shift implemented conditioned on the stationary qubit being in $|g\rangle$ provides a direct realization of the CZ gate given in Eq. (6.10). Similarly, provided only a photon in state $|\mathrm{L}\rangle$ is allowed to interact with the stationary qubit, the same mechanism could be used to realize a CZ gate between a stationary qubit and a time-bin qubit [Eq. (6.11)]. The time-bin conditioned interaction with the stationary qubit could be realized using a fast switch [73, 74] or tunable coupling to the transmission line [21].

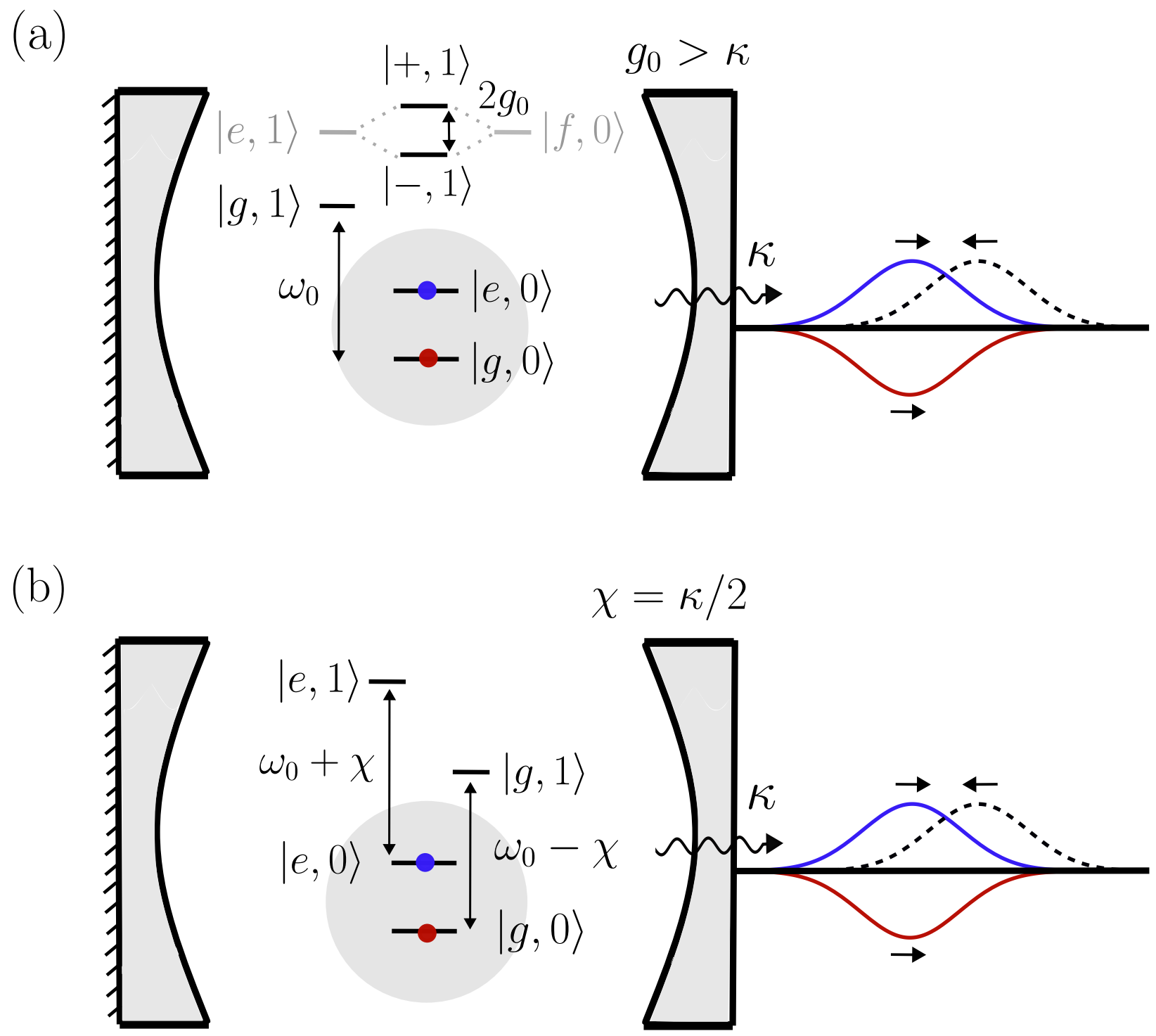

Figure 6.2: Two ways of realizing qubit-state-conditioned phase shifts on the photonic waveform: (a) Strong resonant coupling of the $|e\rangle$ and $|f\rangle$ states suppresses the cavity density-of-states at the bare cavity frequency $\omega_0$ for a qubit in state $|e\rangle$, while for a qubit in state $|g\rangle$, it remains peaked at $\omega_0$. For an incoming photon that is resonant with $\omega_0$ and narrow compared to $\kappa$, the result is a $\pi$ phase shift incurred upon reflection, conditioned on state $|g\rangle$. This mechanism has been demonstrated experimentally in Refs. [68–70]. (b) Dispersive coupling with strength $|\chi| = \kappa/2$ can be used to realize a $\pm\pi/2$ phase shift conditioned on the qubit state, where the sign depends on the sign of $\chi$. This operation is locally equivalent to the $\pi$ phase shift described in (a) and has been demonstrated experimentally in Refs. [71, 72].

We give an alternate mechanism for realizing the operation $U$ with a time-bin qubit [Eq. (6.11)]: For a stationary qubit which is instead *dispersively* coupled to a cavity, an entangling operation locally equivalent to $U$ could be realized by dynamically switching the cavity susceptibility so that it differs for a photon in state $|\mathrm{E}\rangle$ versus $|\mathrm{L}\rangle$ [Fig. 6.2(b)]. In this scenario, the cavity-qubit coupling can be modeled by a time-dependent dispersive interaction of the form

$$H(t) = \chi(t)\sigma_z a^\dagger a, \tag{6.12}$$

where

$$\chi(t) = \frac{\kappa}{2}\Theta(t - T/2). \tag{6.13}$$

Here, $\sigma_z$ is a Pauli-Z operator acting on the qubit, $a$ is an annihilation operator that removes one photon from the cavity, and $\Theta$ is a Heaviside step function with $T$ chosen so that $\Theta(t-T/2) = 0$ [$\Theta(t-T/2) = 1$] for an early (late) photon. With this choice of $\chi(t)$, and for a photon waveform $u_{\mathrm{E,L}}(t)$ having a bandwidth narrow compared to $\kappa$ ($\tau^{-1} \ll \kappa$), a photon in state $|\mathrm{L}\rangle$ will ideally undergo a $+\pi/2$ ($-\pi/2$) phase shift when the stationary qubit is in state $|g\rangle$ ($|e\rangle$), while a photon in state $|\mathrm{E}\rangle$ will undergo a $\pi$ phase shift independent of the state of the stationary qubit. In terms of elementary gates and up to a global phase of $\pi$, this operation can be described as the combined action of (i) the required CZ gate $U$ [Eq. (6.11)] between the time-bin qubit and the stationary qubit and (ii) an $S^\dagger$ gate on the time-bin qubit, where in the basis $\{|E\rangle, |L\rangle\}$, $S$ has matrix elements 1 and $i$ along the diagonal.

Since the quasimode waveforms $u_{\mathrm{E,L}}(t)$ are temporally well separated (to ensure orthogonality), the stepwise behavior of $\chi(t)$ [Eq. (6.13)] could be smoothed out in practice provided the timescale for turning $\chi$ on or off is short compared to the time-bin separation. For circuit QED systems, tuning of the dispersive shift from zero to non-zero values can be achieved on nanosecond timescales using SQUID-based tunable couplers [75, 76].

### 6.3.2 Measurement-free gate

The first long-range gate we present does not require measurement of the photonic qubit, but does require that the size of the photonic wavepacket [set by the spatial extent of its waveform $u(t)$] be less than the distance separating the two qubits $\mathrm{Q}_1$ and $\mathrm{Q}_2$ [Fig. 6.3(a)]. We begin by summarizing the steps involved in performing the gate [Fig. 6.3(b)], before giving details of how each can be realized in practice. We assume an initial state of the two stationary qubits and transmission line of the form

$$|\Psi_0\rangle\,|\mathrm{vac}\rangle = (\alpha\,|gg\rangle + \beta\,|ge\rangle + \gamma|eg\rangle + \delta\,|ee\rangle)\,|\mathrm{vac}\rangle\,, \tag{6.14}$$

where $|s_1 s_2\rangle = |s_1\rangle \otimes |s_2\rangle$ denotes $\mathrm{Q}_i$ in state $|s_i\rangle$, $|\mathrm{vac}\rangle$ is the vacuum state of the transmission line, and where $\alpha, \beta, \gamma, \delta$ are coefficients satisfying the normalization condition $[|\alpha|^2 + |\beta|^2 + |\gamma|^2 + |\delta|^2]^{1/2} = 1$. The gate begins by entangling $\mathrm{Q}_1$ with a Fock-state encoded photonic qubit (having basis states $|1_u\rangle$ and

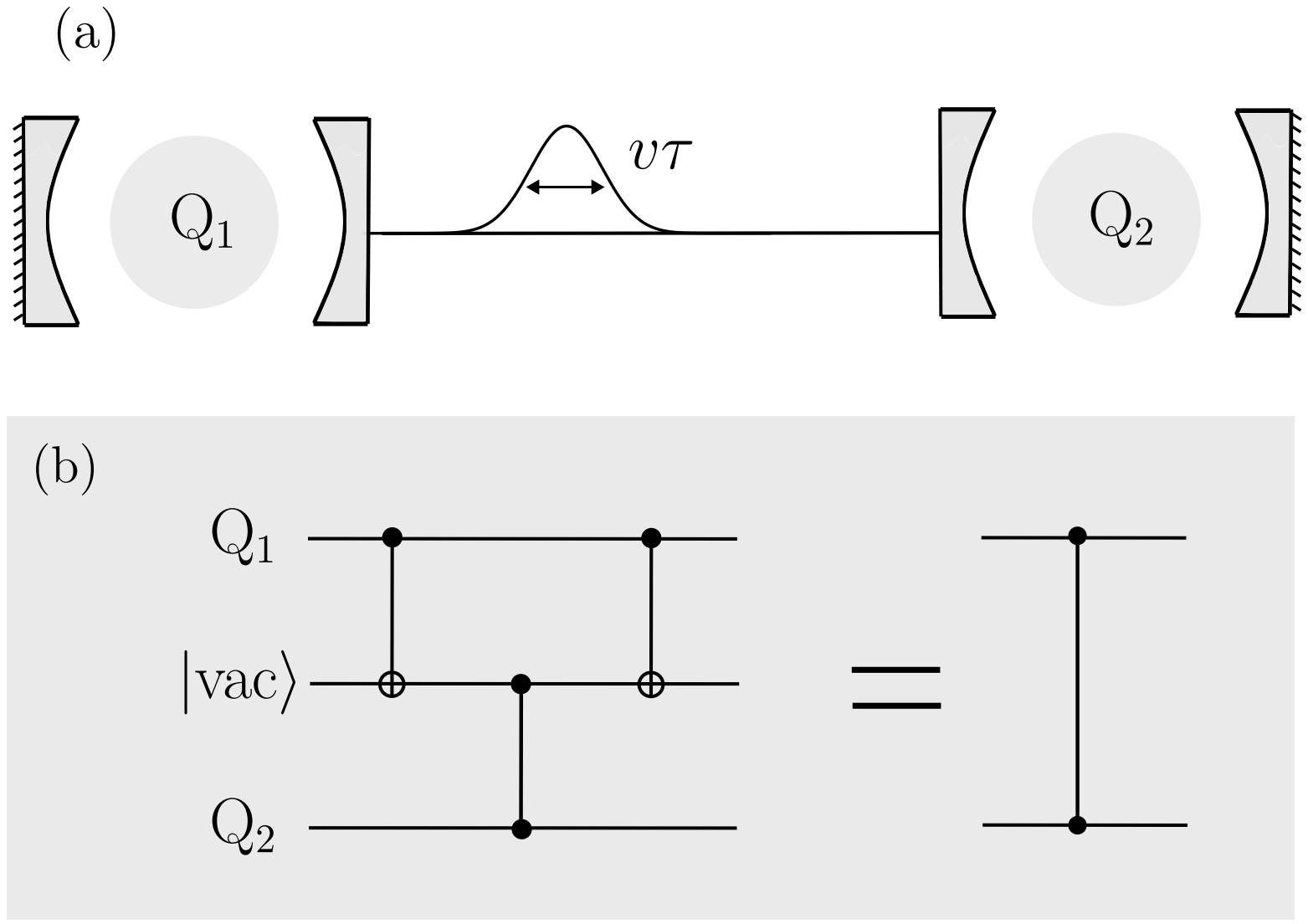

Figure 6.3: (a) Two qubits $Q_1$ and $Q_2$, each coupled to their own cavity mode, are connected by a long transmission line. A CZ gate between $Q_1$ and $Q_2$ can be realized by sending a Fock-state qubit, encoded in the presence or absence of a photon having a waveform of duration $\tau$, from $Q_1$ to $Q_2$ and back. The spatial extent of the photonic waveform is set by $v\tau$, where $v$ is the speed of light in the transmission line. (b) Circuit for applying a CZ gate to $Q_1$ and $Q_2$ using controlled-NOT (CNOT) operations with an auxiliary degree of freedom. These CNOT gates schematically represent conditional photon emission and absorption [cf. Eq. (6.15)].

$|\text{vac}\rangle$) via the mapping

$$\begin{aligned} |e\rangle\,|\text{vac}\rangle &\to |e\rangle\,|\text{vac}\rangle\,, \\ |g\rangle\,|\text{vac}\rangle &\to |g\rangle\,|1_u\rangle\,. \end{aligned} \tag{6.15}$$

The photon is then sent to $Q_2$, where the photonic qubit and $Q_2$ undergo a CZ gate described by $U$ [Eq.(6.10)]. The photon is then sent back to and re-absorbed by $Q_1$, reversing the mapping of Eq. (6.15). As indicated via the circuit identity shown in Fig. 6.3(b), these steps realize a CZ gate between $Q_1$ and $Q_2$.

Having given this overview, we now provide a detailed sequence of steps for realizing the measurement-free gate. Starting from Eq. (6.14), the mapping of Eq. (6.15) can be realized by first applying to $Q_1$ a $\pi_{fe}$ pulse followed by a $\pi_{eg}$ pulse followed by another $\pi_{fe}$ pulse. Here, $\pi_{ab}$ denotes a $\pi$ pulse that exchanges amplitude between levels $|a\rangle$ and $|b\rangle$. These $\pi$ pulses applied to $Q_1$ produce the state

$$(\alpha\,|fg\rangle + \beta\,|fe\rangle + \gamma|eg\rangle + \delta\,|ee\rangle)\,|\text{vac}\rangle\,. \tag{6.16}$$

At this point, a photon can be emitted into the quasimode $u$ (thereby populating the state $|1_u\rangle$) via a cavity-assisted Raman transition, conditioned on $Q_1$ having a nonzero amplitude for the state $|f\rangle$. This completes the mapping of Eq. (6.15). The photon is sent to $Q_2$, after which the photonic qubit and $Q_2$ undergo a CZ gate [Eq. (6.11)]. Following this CZ gate, the state of $Q_1$, $Q_2$, and the transmission line is given by

$$(-\alpha\,|gg\rangle + \beta\,|ge\rangle)\,|1_u\rangle + (\gamma|eg\rangle + \delta\,|ee\rangle)\,|\text{vac}\rangle\,. \tag{6.17}$$

Next, the photon is sent back to and re-absorbed by $Q_1$, sending $|g\rangle\,|1_u\rangle \to |f\rangle\,|\text{vac}\rangle$ and disentangling the state of $Q_1$ and $Q_2$ from the state of the transmission line. Applying a $\pi_{eg}$ pulse, followed by a $\pi_{fe}$ pulse, followed by another $\pi_{eg}$ pulse to $Q_1$ will then yield the final state

$$(-\alpha\,|gg\rangle + \beta\,|ge\rangle + \gamma|eg\rangle + \delta\,|ee\rangle)\,|\text{vac}\rangle\,, \tag{6.18}$$

consistent with the application of a CZ gate $e^{i\pi|gg\rangle\langle gg|}$ to the initial state of $Q_1$ and $Q_2$. If the photon was lost in transit, then the drive used to reabsorb the photon will act trivially [since the Hamiltonian in Eq. (6.1) does not have a finite matrix element linking $|g,0\rangle$ and $|f,0\rangle$]. The final three $\pi$ pulses used to produce Eq. (6.18) will then lead to a finite amplitude of having $Q_1$ in state $|f\rangle$ conditioned on a photon having been emitted. This enables heralding of photon loss through a binary measurement described by the POVM $\{|f\rangle\langle f|, 1-|f\rangle\langle f|\}$ [77].

Since the photonic qubit is emitted and ultimately re-absorbed by the same qubit ($Q_1$), the drives used to emit and absorb the photon cannot overlap in time. The spatial extent of the quasimode $u$ must therefore be comparable to or shorter than the distance separating $Q_1$ and $Q_2$. As an example, this requirement sets a minimum distance of approximately $v\tau \simeq 0.1$ m for a $\tau \simeq 1$-ns photonic pulse traveling at a speed of $v \simeq 0.5c$, where $c$ is the speed of light in vacuum. A more precise estimate for $v$ is [21] $v = c/\sqrt{(1+\varepsilon_r)/2} \approx 0.43c$, assuming a dielectric constant $\varepsilon_r = 10$ (for sapphire). The minimum distance $v\tau$ must in turn be balanced against the requirement that $\tau > \kappa^{-1}$, which, as explained in Sec. 6.3.1, is required for realizing the operation $U$ [Eq. (6.10)]. For $\kappa/2\pi = 50$ MHz [21], for instance, a pulse duration $\tau > 3.2$ ns would be required, corresponding to a pulse length $> 0.41$ m (assuming the value for $v$ given above).

In the next section, we present a second protocol for applying a CZ gate to $Q_1$ and $Q_2$. This second protocol makes use of an ancilla that can be used to reabsorb the photon and therefore does not require the photon emission and absorption to be temporally separated. The protocol is thus compatible with qubits that are separated by a distance smaller than the spatial extent of the photon waveform.

### 6.3.3 Ancilla-assisted gate

In this subsection, we present a second protocol for applying a CZ gate to $Q_1$ and $Q_2$ (Fig. 6.4). In contrast to the measurement-free protocol presented in Sec. 6.3.2, the photonic qubit mediating the gate presented here only travels from $Q_1$ to $Q_2$, rather than there and back. After interacting with $Q_2$, the photonic qubit is measured with the help of an ancilla, $Q_3$. This measurement projects $Q_1$ and $Q_2$ into a state consistent with the application of a CZ gate, $e^{i\pi|gg\rangle\langle gg|}$, up to a potential single-qubit rotation (conditioned on the measurement outcome). Although this gate could be mediated by a Fock-state encoded photonic qubit in principle, we consider a time-bin encoding as it offers the additional possibility of "erasing" the backaction associated with photon loss—a possibility explored in Sec. 6.5. The time-bin qubit mediating the CZ gate presented in this section plays a role analogous to that of the photonic polarization qubit in the proposal of Ref. [33], where a similar protocol for performing a long-range CZ gate was suggested. In the protocol of Ref. [33], a photon encoding a polarization qubit is successively reflected off two distant cavities containing qubits, becoming entangled with both qubits in the process. A final measurement of the photon in an appropriate polarization basis then completes the gate. This protocol has been experimentally implemented between trapped-atom stationary qubits with a polarization-qubit intermediary [34].

Much like the measurement-free gate of Sec. 6.3.2, the ancilla-assisted gate is realized by entangling $Q_1$ with a photonic qubit, here taken to be a time-bin qubit:

$$\begin{aligned} |e\rangle\,|\mathrm{vac}\rangle &\to |e\rangle\,|\mathrm{E}\rangle\,, \\ |g\rangle\,|\mathrm{vac}\rangle &\to |g\rangle\,|\mathrm{L}\rangle\,. \end{aligned} \tag{6.19}$$

The time-bin qubit can then be entangled with $Q_2$ using the unitary operation $U$ [cf. Eq. (6.11)]. The gate is completed by measuring the state of the time-bin qubit in the $X$ basis $|\mathrm{E}\rangle \pm |\mathrm{L}\rangle$. In an optical implementation, the required time-bin Hadamard could be realized using delays and switches [78, 79], or by mapping the time-bin qubit to a polarization degree-of-freedom and applying the Hadamard to the polarization qubit [80]. In the event that a time-bin Hadamard is unavailable or impractical, as would likely be the case in a circuit QED setup, the required $X$-basis measurement could be performed by mapping the state of the time-bin qubit onto the state of an ancillary three-level system $Q_3$ initialized in its ground state [Fig. 6.4(a)] using the

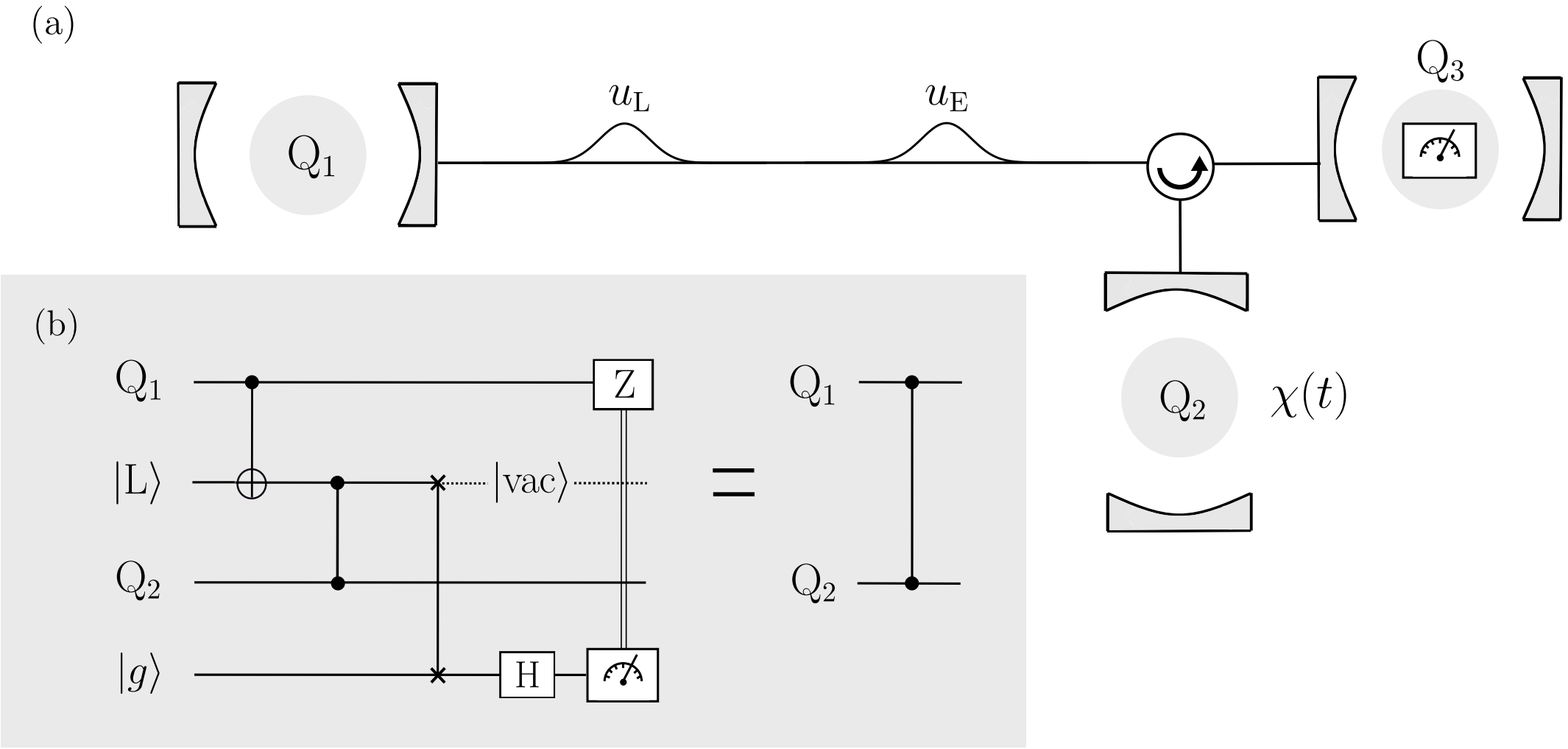

Figure 6.4: (a) Schematic of the imagined setup for the ancilla-assisted CZ gate: A two-qubit gate can be applied to cavity-coupled qubits $\mathrm{Q}_1$ and $\mathrm{Q}_2$ by emitting and re-absorbing a time-bin qubit. Qubit $\mathrm{Q}_1$ plays the role of the quantum emitter that pitches the time-bin qubit, which is subsequently re-absorbed by the ancilla $\mathrm{Q}_3$. While in transit, the photon encoding the time-bin qubit imparts a time-bin conditioned phase to $\mathrm{Q}_2$, enabled by a dispersive shift $\chi(t)$ that can be toggled on during the interaction with either the early ($u_\mathrm{E}$) or late ($u_L$) time-bin waveforms [Eq. (6.13)]. (b) Quantum-circuit description of the suggested protocol for applying a CZ gate to $\mathrm{Q}_1$ and $\mathrm{Q}_2$ using a time-bin intermediary, with the ancilla $\mathrm{Q}_3$ initially prepared in its ground state $|g\rangle$. The entangling operation given in Eq. (6.19) is represented schematically as an initial preparation of the time-bin qubit in state $|\mathrm{L}\rangle$, followed by a CNOT gate acting on $\mathrm{Q}_1$ and the time-bin qubit.

mapping [17]

$$\begin{aligned} |g\rangle\,|\mathrm{E}\rangle &\to |e\rangle\,|\mathrm{vac}\rangle\,, \\ |g\rangle\,|\mathrm{L}\rangle &\to |g\rangle\,|\mathrm{vac}\rangle\,, \\ |g\rangle\,|\mathrm{vac}\rangle &\to |f\rangle\,|\mathrm{vac}\rangle\,. \end{aligned} \tag{6.20}$$

The mappings of Eqs. (6.19) and (6.20) have been demonstrated experimentally with superconducting transmon qubits for the purpose of Bell-state generation [17] and can be realized using $\pi$ pulses together with cavity-assisted Raman transitions. The $X$-basis measurement of the time-bin qubit can then be realized by applying a Hadamard to $\mathrm{Q}_3$ (mapping $|e\rangle \to |+\rangle$ and $|g\rangle \to |-\rangle$) and measuring $\mathrm{Q}_3$ in the basis $\{|g\rangle\,, |e\rangle\,, |f\rangle\}$. In the absence of errors, a measurement of $|g\rangle$ or $|e\rangle$ heralds success: If the measurement outcome is $|e\rangle$, then the post-measurement state of $\mathrm{Q}_1$ and $\mathrm{Q}_2$ corresponds to the state obtained by applying the CZ gate $U_\mathrm{CZ} = e^{i\pi|gg\rangle\langle gg|}$ to the initial state of $\mathrm{Q}_1$ and $\mathrm{Q}_2$ [Eq. (6.14)]. If the outcome is $|g\rangle$, then the applied operation can be transformed to $U_\mathrm{CZ}$ by applying a $Z$ gate to $\mathrm{Q}_1$ [Fig. 6.4(b)]. We quantify the backaction resulting from photon loss, heralded by a measurement of $\mathrm{Q}_3$ in state $|f\rangle$, in Sec. 6.5.

The method considered here for coherently mapping the state of a stationary qubit onto a time-bin degree-of-freedom [Eq. (6.19)] requires the availability of an auxiliary level or degree-of-freedom beyond the two states $|e\rangle$ and $|g\rangle$ spanning the computational subspace. Although we here assume the availability of an auxiliary level $|f\rangle$ (Fig. 6.1), the required auxiliary system could alternatively consist of, e.g., a long-lived nuclear spin as described in Ref. [81].

As with the measurement-free gate, we now give a more detailed sequence of steps for realizing the ancilla-assisted gate protocol: Starting again from the initial state given in Eq. (6.14), the amplitude for the state $|g\rangle$ ($|e\rangle$) of $\mathrm{Q}_1$ is first transferred to $|e\rangle$ ($|f\rangle$) using a $\pi_{fe}$ pulse followed by a $\pi_{eg}$ pulse. Using a cavity-assisted Raman transition, a photon can be transferred into the cavity coupled to $\mathrm{Q}_1$, conditioned on $\mathrm{Q}_1$ starting in the state $|f\rangle$. This photon will leak into the transmission line and populate the state $|\mathrm{E}\rangle$, leaving the cavity in its vacuum state independent of the state of $Q_1$. The state of $Q_1$, $Q_2$, $Q_3$ (initialized in $|g\rangle$) and the transmission line is then

$$(\alpha\,|eg\rangle + \beta\,|ee\rangle)\,|g\rangle\,|\text{vac}\rangle + (\gamma|gg\rangle + \delta\,|ge\rangle)\,|g\rangle\,|\text{E}\rangle\,. \tag{6.21}$$

The basis ordering for the stationary qubits in Eq. (6.21) is $|Q_1,Q_2\rangle\,|Q_3\rangle$. We assume a switch [73, 74], a tunable coupler [21], or a time-dependent dispersive shift $\chi(t)$ [Fig. 6.4(b)] is used to ensure that a photon in state $|\text{E}\rangle$ does not interact with $Q_2$. This action results in an $|\text{E}\rangle$-conditioned identity operation on $Q_2$ [cf. Eq. (6.11)], up to a phase in the case of $\chi(t)$ (see Sec. 6.3.1). A photon in state $|\text{E}\rangle$ is then re-absorbed by $Q_3$, giving

$$(\alpha\,|eg\rangle + \beta\,|ee\rangle)\,|g\rangle\,|\text{vac}\rangle + (\gamma|gg\rangle + \delta\,|ge\rangle)\,|f\rangle\,|\text{vac}\rangle\,. \tag{6.22}$$

Next, a $\pi_{fe}$ pulse is applied to $Q_1$, followed by a $\pi_{eg}$ pulse. A $\pi_{fe}$ pulse is concurrently applied to $Q_3$, so that the state of $Q_3$ is $|e\rangle$ conditioned on the photon having been in state $|\text{E}\rangle$ [cf. Eq. (6.20)]. Driving the $|f,0\rangle \leftrightarrow |g,1\rangle$ transition of $Q_1$ [Eq. (6.1)] will again transfer a photon into the transmission line, conditioned on $Q_1$ being in state $|f\rangle$. However, a photon emitted at this stage will instead populate the state $|\text{L}\rangle$, leading to the state

$$(\alpha\,|gg\rangle + \beta\,|ge\rangle)\,|g\rangle\,|\text{L}\rangle + (\gamma|eg\rangle + \delta\,|ee\rangle)\,|e\rangle\,|\text{vac}\rangle\,. \tag{6.23}$$

During the time that $|\text{L}\rangle$ is populated, the switch, tunable coupler, or dispersive shift $\chi(t)$ is instead configured so that a photon in state $|\text{L}\rangle$ *does* interact with $Q_2$, picking up a phase shift of $\pi$ conditioned on $Q_2$ being in state $|g\rangle$ [cf. Eq. (6.11)]. This interaction sends $\alpha \to -\alpha$ in Eq. (6.23), above.

Finally, a photon in state $|\text{L}\rangle$ is re-absorbed by $Q_3$, at which point the transmission line (in state $|\text{vac}\rangle$) becomes disentangled from $Q_1$, $Q_2$, and $Q_3$, whose state is now given by

$$(-\alpha\,|gg\rangle + \beta\,|ge\rangle)\,|f\rangle + (\gamma|eg\rangle + \delta\,|ee\rangle)\,|e\rangle\,. \tag{6.24}$$

A $\pi$-pulse sequence consisting of back-to-back $\pi_{eg}$, $\pi_{fe}$, and $\pi_{eg}$ pulses applied to $Q_3$ can be used to transfer population in state $|f\rangle$ of $Q_3$ to $|g\rangle$. The final state of $Q_3$ is therefore $|g\rangle$ ($|e\rangle$) conditioned on the time-bin qubit having been in state $|\text{L}\rangle$ ($|\text{E}\rangle$) [Eq. (6.20)]:

$$(-\alpha\,|gg\rangle + \beta\,|ge\rangle)\,|g\rangle + (\gamma|eg\rangle + \delta\,|ee\rangle)\,|e\rangle\,. \tag{6.25}$$

If the photon was lost in transit, then the drives [leading to $\Omega(t)$] applied to $Q_3$ as part of the time-bin-qubit reabsorption both act trivially [since there is no matrix element linking $|g,0\rangle$ and $|f,0\rangle$ in Eq. (6.1)]. The ancilla $Q_3$ therefore remains in state $|g\rangle$ until the final three $\pi$ pulses, which ultimately leave $Q_3$ in state $|f\rangle$ [Eq. (6.20)].

As explained above, an $X$-basis measurement of the time-bin qubit can then be realized by applying a Hadamard to $Q_3$ and measuring $Q_3$ in the computational basis [Fig. 6.4(b)]. Photon loss is heralded by a measurement of $Q_3$ in state $|f\rangle$.

## 6.4 Gate infidelity

In this section, we report estimates for the gate infidelity

$$\varepsilon = 1 - F \tag{6.26}$$

of the ancilla-assisted gate, post-selected on not having lost the photon encoding the time-bin qubit (a.k.a. post-selected on a measurement of $Q_3$ in $|g\rangle$ or $|e\rangle$). Here,

$$F = \int d\psi\,\langle\psi|U_{\text{CZ}}^{\dagger}\mathscr{M}(|\psi\rangle\langle\psi|)U_{\text{CZ}}|\psi\rangle \tag{6.27}$$

is the post-selected gate fidelity, where $d\psi$ is the two-qubit Haar measure, $U_{\mathrm{CZ}} = e^{i\pi|gg\rangle\langle gg|}$ is a unitary describing the action of an ideal CZ gate, and $\mathscr{M}(|\psi\rangle\langle\psi|)$ is the two-qubit state obtained in the presence of errors in the CZ-gate protocol, modeled in the manner we now describe. We focus on the ancilla-assisted gate for this analysis due to its compatibility with photon wavepackets longer than the distance separating the qubits. This compatibility yields better flexibility for optimizing the duration $\tau$ of the photonic waveform given fixed values of the qubit and cavity parameters.

Given some initial state $|\psi\rangle \otimes |g\rangle$ of $Q_1$, $Q_2$, and $Q_3$, we numerically evaluate $\mathscr{M}(|\psi\rangle\langle\psi|)$ by solving a master equation

$$\dot{\rho} = -i[H_{\mathrm{tot}}(t), \rho] + \sum_{j=0} \mathscr{D}[L_j]\rho \equiv \mathscr{L}_{\mathrm{tot}}\rho \tag{6.28}$$

for the joint state $\rho(T)$ of $Q_i$ $(i = 1,2,3)$ prior to the final $X$-basis measurement of $Q_3$ at time $T$. Here, $\mathscr{D}[L]\rho = L\rho L^\dagger - \frac{1}{2}\{L^\dagger L, \rho\}$ is the usual damping superoperator. The state $\mathscr{M}(|\psi\rangle\langle\psi|)$ of $Q_1$ and $Q_2$ entering Eq. (6.27) corresponds to the state obtained following the measurement of $Q_3$, including the $Z$ gate applied to $Q_1$ conditioned on a measurement of $Q_3$ in state $|g\rangle$ [Fig. 6.4(b)]. The Hamiltonian $H_{\mathrm{tot}}(t)$ includes $H[\Omega(t)]$ [cf. Eq. (6.1)] and $\pi$ pulses (here treated as ideal and instantaneous) applied to $Q_1$ and $Q_3$ as part of the time-bin-qubit emission and absorption, as well as a time-dependent dispersive coupling $\chi(t)$ to $Q_2$ [Eq. (6.13)]. The unidirectional propagation of radiation from one cavity to the next is modeled via the SLH formalism [82]. The specific terms entering $H_{\mathrm{tot}}(t)$ are defined in the Appendix.

We denote by $\tau_{\alpha\beta,i}$ the operator $\tau_{\alpha\beta} = |\alpha\rangle\langle\beta|$ acting on $Q_i$. To account for qubit decoherence, we include damping processes generated by $L_j \in \{\sqrt{\Gamma}\tau_{eg,i}, \sqrt{\Gamma/2}(\tau_{ee,i} - \tau_{gg,i})\}_{i=1,2,3} \cup \{\sqrt{\Gamma}\tau_{fe,i}, \sqrt{\Gamma/2}(\tau_{ff,i} - \tau_{ee,i})\}_{i=1,3}$. These generators describe relaxation of the $|f\rangle$ states of $Q_1$ and $Q_3$ into the qubits' respective $|e\rangle$ states, as well as relaxation $|e\rangle \to |g\rangle$ for each of $Q_i$, $i = 1,2,3$. (The $|f\rangle$ level of $Q_2$ is not populated at any point.) Qubit dephasing is assumed to be relaxation-limited, leading to a factor-of-2 difference between the rates associated with relaxation and dephasing. For simplicity, we take all relaxation rates to be equal. The linewidths $\kappa_i$ of the cavities coupled to $Q_i$ $(i = 1,2,3)$ are taken to be equal as well: $\kappa_i = \kappa$.

The measurement of $Q_3$ is treated as ideal in simulation. We provide an estimate of the gate infidelity due to measurement errors at the end of this section. Evaluating the gate infidelity $\varepsilon = 1 - F$ [cf. Eq. (6.27)] in the manner described above yields the results shown in Fig. 6.5 for $\kappa/2\pi = 50$ MHz. The quality of the "slowly varying" approximation made in deriving the drive envelopes $\Omega_{\mathrm{e,a}}(t)$ [Eqs. (6.8)-(6.9)] is expected to increase with $\kappa\tau$ as the dynamics of the cavity mode coupled to $Q_1$ ($Q_3$) become more instantaneously related to those of $Q_1$ ($Q_3$). As described in Sec. 6.3.1, the quality of the CZ gate between $Q_2$ and the time-bin qubit will also increase with increasing $\kappa\tau$ since realizing a well-defined phase $\phi$ on a photonic waveform (sending, e.g., $|\mathrm{E}\rangle \to e^{i\phi}|\mathrm{E}\rangle$) requires that the bandwidth $\sim \tau^{-1}$ of the waveform be narrow compared to the cavity linewidth $\kappa$. For a Gaussian $u(t)$ with standard deviation $\tau$, and in the absence of other sources of error, we expect these finite-bandwidth effects to control the gate infidelity ($\varepsilon = \varepsilon_{\kappa\tau}$) according to [26]

$$\varepsilon_{\kappa\tau} = \frac{A}{(\kappa\tau)^2}, \quad \kappa\tau \gg 1, \quad \Gamma = T_1^{-1} \to 0, \tag{6.29}$$

where $A$ is a numerical prefactor related to the slope of the $T_1 \to \infty$ line in Fig. 6.5(a). For a finite $T_1$, however, increasing $\tau$ results in a longer gate duration and consequently an increased contribution $\varepsilon_{T_1}$ to the gate infidelity due to qubit decoherence, which we expect to scale as

$$\varepsilon_{T_1} = B\frac{\tau}{T_1}, \quad \tau < T_1. \tag{6.30}$$

Here, $B$ is again a numerical prefactor. The competition between $\varepsilon_{\kappa\tau}$ and $\varepsilon_{T_1}$ leads to a minimum in $\varepsilon(\kappa\tau)$ (Fig. 6.5), implying the existence of an optimal $\tau = \tau_{\mathrm{opt}}$. This minimum occurs at the value of $\tau$ for which $d(\varepsilon_{\kappa\tau}+\varepsilon_{T_1})/d\tau = 0$, leading to a predicted scaling for $\tau_{\text{opt}}$ given by

$$\tau_{\text{opt}} = \frac{K}{\kappa}(\kappa T_1)^{\xi}, \quad \xi = 1/3, \tag{6.31}$$

where $K = (2A/B)^{1/3}$. The minimum $\varepsilon(\kappa\tau_{\text{opt}}) = \inf_{\tau}\varepsilon(\kappa\tau) = (\varepsilon_{\kappa\tau}+\varepsilon_{T_1})|_{\tau_{\text{opt}}}$ is therefore predicted to scale according to

$$\varepsilon(\kappa\tau_{\text{opt}}) = D(\kappa T_1)^{\zeta}, \quad \zeta = -2/3, \tag{6.32}$$

where $D = 3(AB^2/2)^{1/3}$. The predicted scaling exponents $\xi$ and $\zeta$ [Eqs. (6.31) and (6.32)] resulting from this simple analysis are relatively consistent with the numerical results shown in Fig. 6.5(b), where the slopes of the lines of best fit (corresponding to the scaling exponents) are

$$\begin{aligned} \xi_{\text{est}} &= 0.341 \pm 0.004, \\ \zeta_{\text{est}} &= -0.6628 \pm 0.0008, \end{aligned} \tag{6.33}$$

and where the prefactors $K$ and $D$ extracted from the same fits are $K = 0.89 \pm 0.04$ and $D = 30.3 \pm 0.3$. The deviation from the predicted scaling exponents of $\xi = 1/3$ and $\zeta = -2/3$ can most likely be attributed to a more sophisticated interplay between finite-bandwidth effects and qubit decoherence than what was assumed in deriving Eqs. (6.31) and (6.32).

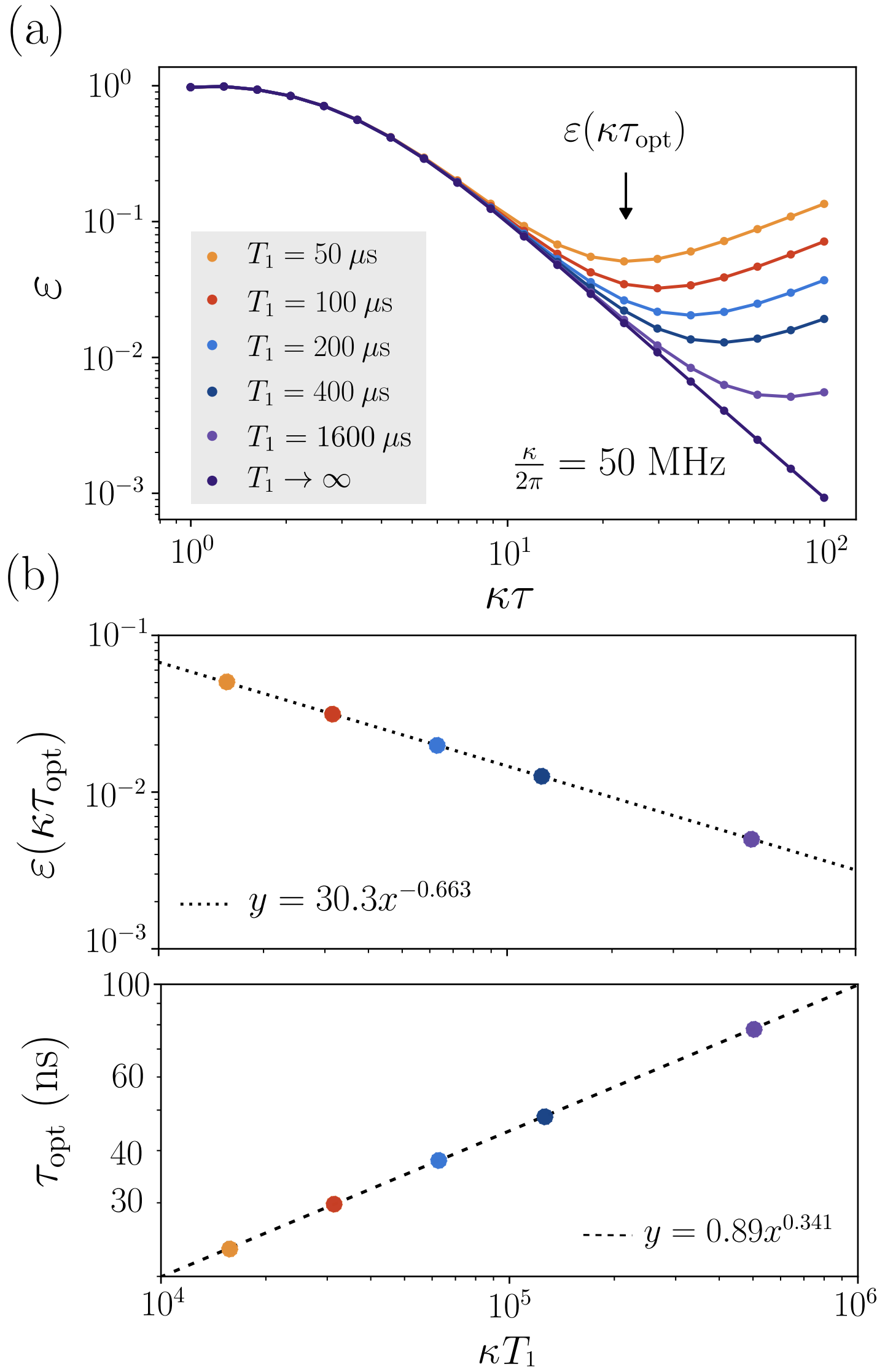

Figure 6.5: (a) The gate infidelity $\varepsilon = 1 - F$ evaluated as a function of $\tau$ for a fixed value of $\kappa/2\pi = 50$ MHz and a total gate duration $T = 16\tau$. In the absence of qubit decoherence ($T_1 \to \infty$), the gate infidelity is proportional to $(\kappa\tau)^{-2}$ for $\kappa\tau \gtrsim 10$, while for a finite $T_1$, competition between the errors due to the finite pulse bandwidth and the errors due to decoherence leads to a minimum in $\varepsilon(\kappa\tau)$, and consequently to an optimal time-bin duration $\tau_{\text{opt}}$. Points indicate values of $\kappa\tau$ used in simulation, while solid lines are intended to provide a guide to the eye. (b) Top panel: The minimum error $\varepsilon(\kappa\tau_{\text{opt}})$ as a function of $T_1$ for the same values of $T_1$ as considered in (a). The dotted line is a least-squares fit whose slope $m = -0.663$ gives a numerical estimate of the scaling exponent $\zeta$ in Eq. (6.32). Bottom panel: The optimal time-bin duration $\tau_{\text{opt}}$, also as a function of $T_1$ for the same values of $T_1$ as shown in (a). The slope $m = 0.341$ of the line of best fit gives a numerical estimate of the scaling exponent $\xi$.

The simulation results displayed in Fig. 6.5 were obtained under the assumption that the measurement of $Q_3$ was ideal. Measurement errors occurring with probability $p_{\text{m}}$ will instead lead to a fidelity

$$F(p_{\text{m}}) = (1-p_{\text{m}})F(0) + p_{\text{m}}F', \tag{6.34}$$

where an expression for $F = F(0)$ was given in Eq. (6.27), and where $F'$ is the gate fidelity obtained in the presence of a measurement error. Such a measurement error simply leads to the wrong correction operator being applied: a $Z$ gate is applied to $Q_1$ when the measurement outcome is $|e\rangle$, or not applied when the outcome is $|g\rangle$. In the absence of other sources of error (e.g. qubit decoherence), we therefore have the following simple expression for the gate fidelity conditioned on a measurement error:

$$F' = \int d\psi \langle\psi|U_{\text{CZ}}^{\dagger} Z_1 U_{\text{CZ}}|\psi\rangle = 0. \tag{6.35}$$

In the presence of measurement errors, the gate fidelity is therefore given by

$$F(p_{\text{m}}) = (1-p_{\text{m}})F(0). \tag{6.36}$$

A finite $p_{\text{m}}$ can therefore be accounted for through a simple rescaling of the $\varepsilon$ axis of Fig. 6.5 by a factor of $1-p_{\text{m}}$.

## 6.5 Time-bin qubit erasure

Dielectric loss due to two-level systems (TLSs) has been identified as a key limitation in current solid-state quantum-computing platforms [49–57]. These TLSs may interact with electric fields via a charge dipole and thus limit the lifetimes of superconducting qubits and microwave resonators. In this section, we estimate the size of the errors introduced due to the loss of a time-bin encoded photonic qubit during the ancilla-assisted gate. Photon loss can be heralded by measuring $Q_3$ as described in Sec. 6.3.3. We identify a regime where the loss of the photon encoding the time-bin qubit is approximately backaction-free, leading to a reduction in the occurrence of photon-loss-related errors on stationary qubits. The avenues given here for reducing backaction-induced errors are only relevant to the ancilla-assisted gate: Since the photonic qubit mediating the measurement-free gate of Sec. 6.3.2 travels from $Q_1$ to $Q_2$ *and back*, simply heralding the loss of the photon would not provide any information about whether the photon was lost before or after interacting with $Q_2$. By contrast, in the case of the ancilla-assisted gate, we can reasonably assume that a photon lost in transit is lost before reaching $Q_2$.

In order to quantify the backaction caused by photon loss during the ancilla-assisted gate, we model the interaction of the time-bin encoded photon with an ensemble of TLSs located along the transmission line. We take the initial state of $Q_1$, $Q_2$, the time-bin qubit, and the TLS bath to be of the form $|\Psi\rangle\,|\mathscr{E}_0\rangle$, where $|\mathscr{E}_0\rangle$ is the initial state of the TLS bath, and where $|\Psi\rangle$ is the state resulting from the mapping of Eq. (6.19) applied to an initial state $|\Psi_0\rangle\,|\mathrm{vac}\rangle$ [Eq. (6.14)]. We can express $|\Psi\rangle$ as a state of the form

$$|\Psi\rangle = \sum_{\lambda=\mathrm{E},\mathrm{L}} \sqrt{p_\lambda}\,|\Psi_\lambda\rangle\,|\lambda\rangle\,, \tag{6.37}$$

where $p_\mathrm{E} = |\gamma|^2 + |\delta|^2$ and $p_\mathrm{L} = |\alpha|^2 + |\beta|^2$ are the probabilities of emitting the photon in the early (E) and late (L) time bins, respectively, and where $|\Psi_\lambda\rangle = \langle\lambda|\Psi\rangle/\sqrt{p_\lambda}$.

For both the early and late time bins ($\lambda = \mathrm{E},\mathrm{L}$), we assume that the photon is lost with probability $q$ while travelling between $Q_1$ and $Q_2$:

$$|\lambda,\mathscr{E}_0\rangle \to \sqrt{1-q}\,|\lambda,\mathscr{E}_0\rangle + \sqrt{q}\,|\mathrm{vac},\mathscr{E}_\lambda\rangle\,. \tag{6.38}$$

In Eq. (6.38), $|\mathscr{E}_\lambda\rangle$ is the state of the environment conditioned on having lost a photon from time-bin $\lambda$.

As explained in Sec. 6.3.3, the loss of the time-bin qubit is heralded by a measurement of $Q_3$ in state $|f\rangle$. The post-measurement state $\rho_f$ of $Q_1$ and $Q_2$, conditioned on such a measurement outcome $|f\rangle$, is obtained by tracing over the state of the environment and is given by

$$\rho_f = \sum_\lambda p_\lambda\,|\Psi_\lambda\rangle\langle\Psi_\lambda| + \sqrt{p_\mathrm{E}p_\mathrm{L}}\,(C\,|\Psi_\mathrm{L}\rangle\langle\Psi_\mathrm{E}| + \mathrm{h.c.})\,, \tag{6.39}$$

where the coherence factor $C$ depends on the overlap

$$C = \langle\mathscr{E}_\mathrm{E}|\mathscr{E}_\mathrm{L}\rangle = |C|e^{i\varphi}, \quad \varphi = \arg C. \tag{6.40}$$

In the ideal case, $C = 1$ and the original two-qubit state $|\Psi_0\rangle\langle\Psi_0| = \rho_f$ remains pure despite the loss of the photon: $\mathrm{Tr}\,\rho_f^2 = 2|C|^2 - 1 = 1$. In general, however, it may be the case that $C \neq 1$. For $C \neq 1$, the distinguishability of the environmental states $|\mathscr{E}_\lambda\rangle$ [cf. Eq. (6.40)] leads to a loss of coherence in an effective two-dimensional subspace spanned by $|\Psi_\lambda\rangle$. Since $|\Psi_\lambda\rangle$ are product states with $Q_1$ in state $|e\rangle$ ($|g\rangle$) for $|\Psi_\mathrm{E}\rangle$ ($|\Psi_\mathrm{L}\rangle$), the state $\rho_f$ with $C \neq 1$ can be expressed as the output of a single-qubit dephasing channel applied to the original state $|\Psi_0\rangle$ [Eq. (6.14)]:

$$e^{-i\frac{\varphi}{2}Z_1}\rho_f e^{i\frac{\varphi}{2}Z_1} = (1-\eta)\,|\Psi_0\rangle\langle\Psi_0| + \eta Z_1\,|\Psi_0\rangle\langle\Psi_0|\,Z_1, \tag{6.41}$$

where $\eta = (1-|C|)/2$. In the event that $C$ has a phase ($\varphi \neq 0$) which is known, it can be compensated deterministically through a rotation of $Q_1$ as shown above. The backaction of photon loss therefore amounts to a phase flip on $Q_1$ occurring with probability $\eta$. Such a phase flip could be detected and corrected via an error-correcting code prior to re-attempting the failed gate.

To estimate the size of $C$, which sets the strength of the dephasing channel [Eq. (6.41)], we model the absorption of the photon by a dielectric medium. We assume the dielectric is composed of many two-level systems (TLSs) located at positions $x_j$ along the transmission line housing the time-bin qubit, and that these TLSs couple to photons in the waveguide with a linear interaction. We therefore take the total lab-frame

Hamiltonian $H$ of the transmission line and TLS environment to be of the form

$$H = H_0 + H_{\text{int}}, \tag{6.42}$$

where

$$H_0 = \sum_k \omega_k r_k^\dagger r_k + \sum_j \omega_j \left|\uparrow_j\right\rangle\!\left\langle\uparrow_j\right| \tag{6.43}$$

generates the decoupled evolution of the transmission-line modes $r_k$ (having frequency $\omega_k = v|k|$) and TLS environment. Here, the annihilation operators $r_k$ satisfy $[r_k, r_{k'}^\dagger] = \delta_{k,k'}$, and $\left|\uparrow_j\right\rangle\!\left\langle\uparrow_j\right|$ is a projector onto the excited state of TLS $j$ with frequency splitting $\omega_j$. The Hamiltonian $H_{\text{int}}$ describing the interaction between the transmission-line modes and TLS environment is of the form

$$H_{\text{int}} = \sqrt{v\tau}\sum_j g_j \left[\psi(x_j) + \psi^\dagger(x_j)\right]\sigma_j^x, \tag{6.44}$$

where $\sigma_j^x = \sigma_j^+ + \sigma_j^-$ is a Pauli-X operator for TLS $j$ with $\sigma_j^- = \left|\downarrow_j\right\rangle\!\left\langle\uparrow_j\right| = (\sigma_j^+)^\dagger$. The operator $\psi(x)$ in Eq. (6.44) is a field operator for the transmission line, given by $\psi(x) = L^{-1/2}\sum_k e^{ikx} r_k$ for a transmission line of length $L$, while $\sqrt{v\tau}$ is a prefactor with dimensions of $[\text{length}]^{1/2}$ required by dimensional considerations. For this choice, $g_j$ is the single-photon coupling strength between TLS $j$ and a photon in quasimode $u$ having an effective mode volume $\propto v\tau$. The only essential restrictions on Eq. (6.44) are that the coupling is linear in the (electric or magnetic) field of the transmission line and that the interaction between the field and TLS is short-ranged and local at each site $x_j$. The form given in Eq. (6.44) is consistent, e.g., with electric-dipole coupling, in which case $g_j$ would be proportional to the transition dipole of TLS $j$, but this form could equally well describe coupling between the magnetic field in the transmission line and the magnetic dipoles of environmental spins. In Eq. (6.44), we have neglected longitudinal coupling terms $\sigma_j^z$, which are also generically present but average out in a rotating-wave approximation [83].

In an interaction picture defined with respect to $H_0$, $\tilde{H}(t) = U_0 H U_0^\dagger - iU_0^\dagger \dot{U}_0 = \tilde{H}_{\text{int}}(t) + \text{counter-rot.}$ with

$$\tilde{H}_{\text{int}}(t) = \sqrt{\frac{v\tau}{L}}\sum_{j,k} g_j \left[e^{i(kx_j - \omega_k t)} e^{i\omega_j t} r_k \sigma_j^+ + \text{h.c.}\right], \tag{6.45}$$

where we have neglected counter-rotating terms in a rotating-wave approximation valid for $|g_j| \ll |\omega_j|$. For the purpose of the calculation presented in this section, it will be helpful to introduce an orthonormal set $\{u'\}$ of quasimodes associated with bosonic operators $r_{u'}$ satisfying the usual commutation relation $[r_{u'}, r_{u''}^\dagger] = \delta_{u',u''}$. For $u' = u$, the operator $r_u$ annihilates the quasimode $u$ introduced in Eq. (6.2). We can write $\psi(x)$ in terms of these operators as $\psi(x) = \sum_{u'} \phi_{u'}(x) r_{u'}$, where $\phi_{u'}(x) = L^{-1/2}\sum_k e^{ikx}\phi_{u'k}$ for $\phi_{u'k} = \langle k|u'\rangle$. Using the relation

$$r_k = \sum_{u'} \phi_{u'k} r_{u'}, \tag{6.46}$$

we can then rewrite Eq. (6.45) as

$$\tilde{H}_{\text{int}}(t) = \sqrt{v\tau}\sum_{j,u'} g_j \left[\phi_{u'}(x_j - vt) e^{i\omega_j t} r_{u'} \sigma_j^+ + \text{h.c.}\right], \tag{6.47}$$

where we have used the fact that $\phi_{u'}(x_j - vt) = L^{-1/2}\sum_k e^{i(kx_j - \omega_k t)}\phi_{u'k}$ with $\omega_k = v|k|$.

To calculate the coherence factor $C$, we consider the evolution of the initial state $|1_u, \mathscr{E}_0\rangle$, where $|1_u\rangle = r_u^\dagger|\text{vac}\rangle$ [cf. Eq. (6.2)] and $|\mathscr{E}_0\rangle = |\downarrow\downarrow\cdots\downarrow\rangle$. When the evolution is dominated by a single action of $\tilde{H}_{\text{int}}(t)$

(Born approximation) the state will approximately take the form

$$|1_u(t)\rangle \simeq \alpha_0(t)\,|1_u,\mathscr{E}_0\rangle + \sum_j \alpha_j(t)\,|\mathrm{vac},j\rangle\,, \tag{6.48}$$

where $\alpha_j(t)$ are coefficients and the state $|j\rangle$ is given by $|j\rangle = \sigma_j^+\,|\mathscr{E}_0\rangle = |\downarrow\cdots\uparrow_j\cdots\downarrow\rangle$. Knowing how a photon decays out of a specific quasimode $u$ will allow us to calculate the post-measurement states $|\mathscr{E}_\lambda\rangle$ of the TLS bath conditioned on a measurement of $Q_3$ in state $|f\rangle$. Recall that the state $|\mathscr{E}_\lambda\rangle$ is obtained due to decay out of the quasimode $u = u_\lambda$ defining the time-bin qubit [cf. Eq. (6.38)].

We calculate the leading-order transition amplitude $\alpha_j(t)$ associated with the loss of a photon initially in quasimode $u$ due to absorption by TLS $j$, $\alpha_j(t) = \langle \mathrm{vac},j|\,\tilde{U}(t)\,|1_u,\mathscr{E}_0\rangle$, where

$$\tilde{U}(t) = \mathscr{T}e^{-i\int_{-\infty}^{t}dt'\tilde{H}_{\mathrm{int}}(t')} \simeq 1 - i\int_{-\infty}^{t} dt'\tilde{H}_{\mathrm{int}}(t'). \tag{6.49}$$

This gives

$$\alpha_j(t) \simeq -i\sqrt{v\tau}g_j\int_{-\infty}^{t} dt'\,\phi_u(x_j - vt')e^{i\omega_j t'}. \tag{6.50}$$

Since the wavepacket $\phi_u(x-vt)$ propagates without dispersion, the quasimode $u$ can equivalently be described by a function of time at a single point in space, given by the waveform $u(t) = \sqrt{v}\phi_u(-vt)e^{i\omega_0 t}$, where the factor $\sqrt{v}$ is required to satisfy the normalization condition $\int dx\,|\phi_u(x)|^2 = \int dt\,|u(t)|^2 = 1$, and where the phase factor $e^{i\omega_0 t}$ results from the relationship between the lab-frame wavepacket $\phi_u(x-vt)$ and an interaction-picture waveform $u(t)$. Inserting this relationship into Eq. (6.50) gives

$$\alpha_j(t) \simeq -ig_j\sqrt{\tau}\int_{-\infty}^{t} dt'\,u(t' - x_j/v)e^{i\delta\omega_j t'}, \tag{6.51}$$

where we have introduced the detuning $\delta\omega_j$ between TLS $j$ and the central frequency $\omega_0$ of the photon:

$$\delta\omega_j = \omega_j - \omega_0. \tag{6.52}$$

For times satisfying $t - x_j/v \gg \tau$, the integral in Eq. (6.51) no longer picks up any weight since the propagating single-photon pulse is no longer interacting with the TLS at position $x_j$. We can therefore extend the upper integration limit in Eq. (6.51) to $+\infty$, yielding the steady-state amplitudes

$$\alpha_j(t) \simeq \bar{\alpha}_j = -ig_j\sqrt{\tau}\,e^{i\delta\omega_j x_j/v}u(\delta\omega_j). \tag{6.53}$$

Note that $u(\omega)$ has dimensions of $[\mathrm{time}]^{1/2}$, following from the normalization condition $(2\pi)^{-1}\int d\omega\,|u(\omega)|^2 = 1$. The quantity $|\bar{\alpha}_j|^2$ gives the excited-state population of TLS $j$ resulting from its interaction with a photon in quasimode $u$. Rather intuitively, this population is proportional to the spectral weight $\propto |u(\delta\omega_j)|^2$ of the photon at the detuning $\delta\omega_j$ of TLS $j$ from the central frequency of the photon.

To relate this result back to the map of Eq. (6.38), we substitute Eq. (6.53) into Eq. (6.48) for times $t$ satisfying $t - x_j/v \gg \tau$, giving

$$|1_{u_\lambda}(t)\rangle \simeq \sqrt{1-q}\,|\lambda,\mathscr{E}_0\rangle + \sqrt{q}\,|\mathrm{vac}\rangle\sum_j \frac{\bar{\alpha}_{j,\lambda}}{\sqrt{q}}\,|j\rangle\,, \tag{6.54}$$

where $\bar{\alpha}_{j,\lambda}$ is given by Eq. (6.53) with $u(\omega) = u_\lambda(\omega)$. Under the assumption that $u_{\mathrm{L}}(t-\tau_{\mathrm{sep}}) = u_{\mathrm{E}}(t) \equiv u(t)$, we have the simple relation $u_{\mathrm{L}}(\omega) = e^{-i\omega\tau_{\mathrm{sep}}}u(\omega)$ between the frequency-domain waveform of a photon occupying $|\mathrm{E}\rangle$ or $|\mathrm{L}\rangle$. In terms of quantities characterizing the quasimode and TLS's, the probability $q$ of losing the photon is given by

$$q = \sum_j |\bar{\alpha}_j|^2 = \tau \int \frac{d\omega}{2\pi} J(\omega) |u(\omega)|^2, \tag{6.55}$$

where $J(\omega) = 2\pi \sum_j |g_j|^2 \delta(\omega - \delta\omega_j)$ is the spectral density of the TLS bath. In terms of the amplitudes $\bar{\alpha}_{j,\lambda}$, we can therefore write the environmental states $|\mathscr{E}_\lambda\rangle$ conditioned on having lost a photon from time bin $\lambda = \mathrm{E}, \mathrm{L}$ as

$$|\mathscr{E}_\lambda\rangle = \frac{1}{\sqrt{q}} \sum_j \bar{\alpha}_{j,\lambda} |j\rangle . \tag{6.56}$$

Using Eq. (6.56), we can then re-express the coherence factor $C$ as

$$C = \frac{\int d\omega\, J(\omega) |u(\omega)|^2 e^{-i\omega\tau_{\mathrm{sep}}}}{\int d\omega\, J(\omega) |u(\omega)|^2}. \tag{6.57}$$

As expected, the loss of qubit coherence ($C \neq 1$) results from the temporal separation $\tau_{\mathrm{sep}}$ of the quasimodes defining the early and late time-bin states. However, it is interesting to note that the orthogonality of the time bin states $|\mathrm{E}\rangle$ and $|\mathrm{L}\rangle$ does not imply orthogonality of the environmental states $|\mathscr{E}_{\mathrm{E}}\rangle, |\mathscr{E}_{\mathrm{L}}\rangle$. This in turn implies that there are conditions under which photon loss can have zero backaction on the stationary qubits. In other words, for an appropriate environment spectral density, time-bin waveform, and time-bin separation, the time-bin information can be erased with the loss of the time-bin photon.

One regime where photon loss *does* lead to significant backaction is the regime where the spectral density $J(\omega)$ is approximately flat over the support of a Gaussian $u(\omega)$ having standard deviation $\tau$, for which

$$|u(\omega)|^2 = 2\sqrt{\pi}\tau e^{-\tau^2\omega^2}. \tag{6.58}$$

In that case, setting $J(\omega) = J_0$ in Eq. (6.57) gives

$$C = e^{-\left(\frac{\tau_{\mathrm{sep}}}{2\tau}\right)^2}, \quad J(\omega) = J_0, \tag{6.59}$$

which in turn gives $\eta \approx 1/2$ [cf. Eq. (6.41)] since $\tau_{\mathrm{sep}} > \tau$ is needed to ensure orthogonality of the time-bin states. The loss of a photon into an environment having a bandwidth that is wide compared to $\tau_{\mathrm{sep}}^{-1}$ therefore gives rise to significant dephasing of the stationary qubits [Eq. (6.41)].

The form of $C$ [Eq. (6.57)] suggests that if $J(\omega)$ is instead narrow in frequency compared to the photon bandwidth, then it — rather than $u(\omega)$ — will determine the qualitative features of $C$. For simplicity, we consider a Gaussian spectral density,

$$J(\omega) = \sqrt{2\pi} \frac{g^2}{\Lambda} e^{-\frac{(\omega - \delta\bar{\omega})^2}{2\Lambda^2}}, \tag{6.60}$$

where $\Lambda$ sets the bandwidth of the environment, $\delta\bar{\omega}$ gives the average of $\delta\omega_j = \omega_j - \omega_0$, and where $g^2 = \sum_j |g_j|^2$. For the same Gaussian $u(\omega)$ as in Eq. (6.58), this choice of $J(\omega)$ [Eq. (6.60)] gives

$$q = 2\sqrt{\pi}(g\tau)^2 \frac{e^{-\frac{(\delta\bar{\omega}\tau)^2}{1+2\Lambda^2\tau^2}}}{\sqrt{1+2\Lambda^2\tau^2}}, \tag{6.61}$$

$$C = e^{-i\delta\bar{\omega}\tau_{\mathrm{sep}}} e^{-\frac{(\Lambda\tau_{\mathrm{sep}})^2}{2+4\Lambda^2\tau^2}}, \tag{6.62}$$

where the phase $\varphi = \arg C$ depending on $\delta\bar{\omega}$ could be corrected via a single-qubit rotation provided $\delta\bar{\omega}$ is known [cf. Eq. (6.41)]. As expected, the photon-loss probability $q$ decreases as the average $\bar{\omega}$ of the TLS frequency distribution moves further off resonance from the central frequency $\omega_0$ of the photon. Notably, however, there is no corresponding decrease in $C$ with $\delta\bar{\omega}$ since $C$ is post-selected on having already lost the photon. In the wide-bandwidth regime $\Lambda \gg \tau_{\mathrm{sep}}^{-1}$, we recover the exponential suppression of $C$ in the ratio of timescales $(\tau_{\mathrm{sep}}/\tau)^2 \gtrsim 1$ already given in Eq. (6.59). In the opposite, narrow-bandwidth regime $\Lambda \ll \tau_{\mathrm{sep}}^{-1}$, we instead find that

$$|C| = 1 - \frac{1}{2}(\Lambda\tau_{\mathrm{sep}})^2 + O((\Lambda\tau_{\mathrm{sep}})^4), \tag{6.63}$$

yielding dephasing errors occurring with probability $\eta \approx (\Lambda\tau_{\mathrm{sep}})^2/4 \ll 1$. In this regime, the backaction due to photon loss, as quantified by the rate of dephasing errors, is strongly suppressed relative to the wide-bandwidth (Markovian) case. This is a consequence of the TLS bath having a correlation time $\tau_{\mathrm{c}} \sim \Lambda^{-1}$ that is long compared to the time-bin separation $\tau_{\mathrm{sep}}$.

For a time-bin separation of $\tau_{\mathrm{sep}} \approx 100$ ns, consistent with the optimal values of $\tau = \tau_{\mathrm{opt}}$ shown in Fig. 6.5(c), we have $\Lambda\tau_{\mathrm{sep}} = 1$ for $\Lambda/(2\pi) \approx 1.6$ MHz, corresponding to a bath correlation time of $\tau_{\mathrm{c}} \approx 0.1$ $\mu$s. For this choice of $\tau_{\mathrm{sep}}$, a correlation time $\gg 0.1$ $\mu$s would therefore be required to ensure that the loss of the time-bin qubit is approximately backaction-free. Realistic values for the correlation time will depend on the nature of the coupling $g_j$ between the quasimode field $\psi(x)$ and the environmental TLSs, the amount of inhomogeneous broadening experienced by the TLSs, and the lifetime of the TLSs. Charge-like two-level systems with lifetimes on the order of milliseconds have been measured [55, 56]. Although the spectral characteristics of charge-like TLSs would likely be difficult to control, naturally occurring narrow-band mechanisms associated with, e.g., well-defined transitions between hyperfine-split spin sublevels [51] in surface hydrogen may eventually become the limiting source of photon loss as materials-fabrication techniques continue to yield improvements in the densities of charge-like TLSs in materials commonly used for circuit QED. Given a certain dielectric material, a simple measurement exists to determine whether it enables backaction-free loss of time-bin encoded photons: After entangling a time-bin qubit with a stationary qubit according to Eq. (6.19), the photon is sent directly to a single-photon detector. If no click is registered (signaling the loss of the photon), then the qubit is measured in the $X$ or $Y$ basis to yield information about the real and imaginary parts of $C = \langle \mathscr{E}_{\mathrm{E}} | \mathscr{E}_{\mathrm{L}} \rangle$.

## 6.6 Conclusion

In this work, we have presented two possible strategies for realizing entangling gates between qubits connected by meter-scale photonic interconnects. Such interconnects, equipped with a two-qubit entangling gate protocol, would provide the quantum links required for so-called $l$-type modular architectures [3]. Both CZ gate protocols presented here are compatible with a transmission-line free spectral range of zero and could be used to implement gates over arbitrarily long distances in principle. The gates are deterministic with heralded photon loss, and notably, they do not rely on the use of a polarization degree-of-freedom for the photonic qubit encoding. This ensures compatibility with the coplanar waveguide resonators commonly used in circuit QED architectures. The measurement-free gate requires that the distance between the two qubits exceed the size of the photonic wavepacket and does not require measurement of the photonic qubit. The ancilla-assisted gate lifts this requirement on the minimum qubit separation at the cost of introducing a measurement of the photonic (time-bin) qubit, which could be realized straightforwardly with the help of an ancilla.

We have also analyzed the dephasing channel associated with the loss of a time-bin qubit, which cannot be treated using existing loss models since the details of the spatiotemporal modes defining the time-bin qubit must be taken into account. We show that in spite of the orthogonality of the time-bin states, there are conditions under which the loss of the photon causes only minimal dephasing of the stationary qubits. Whether or not these conditions are met depends on the bandwidth (and consequently, correlation time) of the dielectric environment that absorbs the photon, relative to the temporal separation of the time-bin states. This result highlights the role that materials fabrication could potentially play in reducing errors on stationary qubits connected by photonic time-bin qubits. An investigation of whether current dielectric materials could be made to satisfy the requirements for backaction-free loss of time-bin encoded photons would provide an interesting line of future inquiry, beyond the scope of the present work.

**Acknowledgments**—We acknowledge funding from the Natural Sciences and Engineering Research Council of Canada (NSERC) and from the Fonds de Recherche du Québec–Nature et technologies (FRQNT).

# Appendix to Chapter 6

## E1 Master-equation simulation

In this appendix, we provide details concerning the master-equation simulation of the gate infidelity $\varepsilon$ [Eq. (6.27)] presented in Sec. 6.4. For simplicity, we implement the simulation using Gaussian time-bin waveforms having a width set by $\tau$:

$$u(t,t_{\text{peak}}) = \frac{e^{-\frac{(t-t_{\text{peak}})^2}{2\tau^2}}}{\pi^{1/4}\sqrt{\tau}}. \tag{E1}$$

We also take the total duration of the gate to be fixed at $T = 16\tau$.

Given some initial state $|\psi\rangle$ of $Q_1$ and $Q_2$, we evaluate the joint final state $\rho_\psi(T)$ of $Q_1$, $Q_2$, and $Q_3$ (prepared in the ground state) by solving the master equation given in Eq. (6.28). The formal solution of Eq. (6.28) can be written as

$$\rho_\psi(T) = \underbrace{\mathscr{T} e^{-i\int_0^T dt\, \mathscr{L}_{\text{tot}}(t)}}_{\equiv \Lambda_{\text{tot}}(T)} \rho_\psi(0), \tag{E2}$$

where $\rho_\psi(0) = \pi_{fg,1}\,|\psi\rangle\langle\psi|\,\pi_{fg,1} \otimes |g\rangle\langle g|$, $\mathscr{T}$ is a time-ordering operator, and $\pi_{ab,i}$ is a $\pi$ pulse between levels $|a\rangle$ and $|b\rangle$ of qubit $Q_i$. Concretely, the simulation proceeds by evaluating $\Lambda_{\text{tot}}(T)$ as

$$\Lambda_{\text{tot}}(T) = \Lambda(T, T/2) L_\pi \Lambda(T/2, 0). \tag{E3}$$

Here, $L_\pi$ describes the action of (ideal) $\pi_{fe,1}$, $\pi_{eg,1}$, and $\pi_{fe,3}$ pulses applied to $Q_1$ and $Q_3$ as part of the time-bin photon emission and absorption processes. The superoperator $\Lambda$ describing evolution before and after the $\pi$ pulses is given by

$$\Lambda(t_f, t_i) = \mathscr{T} e^{-i\int_{t_i}^{t_f} dt\, \mathscr{L}(t)}, \tag{E4}$$

where $\mathscr{L}(t)$ generates evolution according to

$$\dot{\rho} = -i[H_{\text{tot}}(t), \rho] + \sum_{j=0} \mathscr{D}[L_j]\rho \equiv -i\mathscr{L}\rho. \tag{E5}$$

Here, the damping superoperator $\mathscr{D}[L]$ acts according to $\mathscr{D}[L]\rho = L\rho L^\dagger - \frac{1}{2}\{L^\dagger L, \rho\}$, and the sum over $L_j$ describes the contribution of various dissipative processes (explained in the main text). We model the directed routing of photons via the SLH formalism [82], which eliminates the need to explicitly simulate the field propagating between cavities. This simplification can be justified provided the propagation time is short compared to the timescale set by the cavity linewidth $\kappa_i$ [82]. Following the SLH approach, we therefore include the dissipator

$$L_0 = \sum_{j=1}^{3} \sqrt{\kappa_j} a_j \tag{E6}$$

in the sum over $\mathscr{D}[L_j]$, together with the following bilinear terms $\propto \sqrt{\kappa_i \kappa_j}$ in the Hamiltonian $H(t)$ governing the coherent evolution of the qubits and cavities:

$$H_{\rm tot}(t) = \sum_{j=1}^{3} H_j(t) + \frac{i}{2}\sum_{\substack{i=1\\ j>i}} (\sqrt{\kappa_i\kappa_j} a_i^\dagger a_j - \text{h.c.}). \tag{E7}$$

The particular combination of coherent terms $\propto \sqrt{\kappa_i\kappa_j}$ and dissipator $\mathscr{D}[L_0]$ ensures a master-equation description that is consistent with the unidirectional propagation of photons [82]. We have also assumed in writing Eq. (E7) that all cavities are resonant.

In addition to the terms $\propto \sqrt{\kappa_i\kappa_j}$, the Hamiltonian $H(t)$ also includes terms $H_i(t)$ describing cavity-qubit coupling for qubits $Q_i$ ($i = 1,2,3$). The Hamiltonians $H_1(t)$ and $H_3(t)$ describe the emission (by $Q_1$) and subsequent re-absorption (by $Q_3$) of a photon with waveform $u(t,t_{\rm peak})$, conditioned on $Q_1$ being in the $|f\rangle$ state:

$$H_j(t) = i\Omega_j(t)\left(\tau_{fg,j}a_j^\dagger - \text{h.c.}\right), \quad j = 1,3. \tag{E8}$$

Here, $\tau_{ab,j}$ denotes the pseudospin operator $\tau_{ab} = |a\rangle\langle b|$ for qubit $Q_j$, and $a_j$ is a bosonic annihilation operator that removes one photon from the cavity mode coupled to $Q_j$. The drives $\Omega_j$ are defined piecewise according to

$$\Omega_j(t) = \sum_{\alpha=\rm E,L} \Omega_{\alpha,j}(t)\Pi_\alpha(t), \tag{E9}$$

where $\Pi_{\rm E}(t) = \Theta(t)\Theta(T/2-t)$ is equal to 1 for $t \in (0,T/2)$ and 0 otherwise. Similarly, $\Pi_{\rm L}(t) = \Theta(t-T/2)\Theta(T-t)$ is equal to 1 for $t \in (T/2,T)$ and 0 otherwise. The drive $\Omega_{\alpha,1}(t)$ acting on $Q_1$ is given by Eq. (6.8) with $u(t) = u_\alpha(t)$, while the drive $\Omega_{\alpha,3}(t)$ acting on $Q_3$ is given by Eq. (6.9) with $u(t) = u_\alpha(t)$. In both cases, $u_{\rm E}(t)$ and $u_{\rm L}(t)$ are given by Eq. (E1) with $t_{\rm peak} = T/4$ and $t_{\rm peak} = 3T/4$, respectively. Finally, the Hamiltonian $H(t)$ includes a term describing the dispersive coupling of $Q_2$ to its cavity, given by

$$H_2(t) = \chi(t)\left(\tau_{ee,2} - \tau_{gg,2}\right)a_2^\dagger a_2, \tag{E10}$$

where $\chi(t)$ is given by Eq. (6.13).

Given $\rho_\psi(T)$, we apply an $S$ gate to $Q_1$ and measure $Q_3$ in the $X$ basis (Fig. 6.4b). The $S$ gate ensures that in the absence of errors, the final state of $Q_1$ and $Q_2$ is given by $U_{\rm CZ}|\psi\rangle$ with $U_{\rm CZ} = e^{i\pi|gg\rangle\langle gg|}$. The need for such an $S$ gate is a consequence of the mechanism [Eq. (E10)] used to realize an entangling gate between the time-bin qubit and $Q_2$. As discussed in Sec. 6.3.1, the mechanism of Eq. (E10) produces an operation equal to the product of $U = e^{i\pi|{\rm L},g\rangle\langle {\rm L},g|}$ [Eq. (6.11)] and an $S^\dagger$ gate on the time-bin qubit. Since the state of $Q_1$ is correlated with that of the time-bin qubit via the mapping of Eq. (6.19), the $S^\dagger$ gate on the time-bin qubit can be canceled by applying an $S$ gate to $Q_1$. Such a gate would not be required in practice (it can be tracked in software) but is included in Eq. (E11), below, to allow for comparison to an ideal operation of the form $U_{\rm CZ} = e^{i\pi|gg\rangle\langle gg|}$.

The post-measurement state $\rho_{\psi,\pm}(T)$ of $Q_1$ and $Q_2$, conditioned on a measurement of $Q_3$ in state $|\pm\rangle \propto |e\rangle \pm |g\rangle$, is then given by

$$\rho_{\psi,\pm} = \frac{{\rm Tr}_3\{|\pm\rangle\langle\pm|S\rho_\psi(T)S^\dagger\}}{P_\pm}, \tag{E11}$$

where ${\rm Tr}_3$ denotes a partial trace over the state of $Q_3$ and $P_\pm = {\rm Tr}\{|\pm\rangle\langle\pm|\rho_\psi(T)\}$ gives the probability of obtaining a measurement outcome $|\pm\rangle$. Conditioned on a measurement of $Q_3$ in state $|-\rangle$, a $Z$ gate is applied to $Q_1$ [Fig. 6.4(b)]. Both the measurement and the $Z$ gate are treated as ideal. This procedure yields the state $\mathscr{M}(|\psi\rangle\langle\psi|)$ entering the expression for $F$ [Eq. (6.27)]:

$$\mathscr{M}(|\psi\rangle\langle\psi|) = \rho_{\psi,+} = Z_1\rho_{\psi,-}Z_1. \tag{E12}$$

# Conclusion

This thesis has considered various aspects of quantum information processing with cavity quantum electrodynamics (QED), beginning with spectroscopic characterization of the qubit's environment and moving towards the long-term goal of scalable, modular quantum-computing architectures equipped with protocols for manipulating and correcting the states of distant qubits. After introducing the basics of cavity QED in Chapter 1, we saw in Chapter 2 how spectroscopic measurements of the field leaking out of a cavity could be used to reconstruct the echo envelope of a qubit undergoing a dynamical decoupling sequence. This echo envelope has a phase arising from the non-commutation of the bath operators coupling to the qubit—a genuine quantum effect whose signature could be detected by monitoring both quadratures of the cavity output field. Since properties of the bath spectral density could be revealed via measurements of the output field—and probed at different frequencies by modifying the filter function characterizing the dynamical decoupling sequence—there remains the possibility that in the future, this approach to noise spectroscopy could be combined with pulse-shaping or optimal-control strategies designed to isolate the qubit from the dominant decoherence mechanisms due to its environment.

In Chapter 3, we introduced a novel quantum-optical effect wherein a parametrically modulated longitudinal coupling could be used to generate entanglement between the state of a cavity-coupled qubit and the path taken by a multiphoton coherent-state wavepacket. This effect is based on destructive interference between the incoming coherent-state wavepacket and the qubit-state-conditioned displacement of the cavity mode. In future work, it would be interesting to understand whether such an effect could be harnessed for qubit readout. As discussed in the introductory chapter of this thesis, the standard approach for determining the state of a qubit via measurements of a cavity involves measuring the phase shift experienced by a weak coherent probe tone upon reflection from the cavity. We can view the action of the probe tone as encoding the state of the qubit into the state of a *single* mode housing the reflected field. Through an appropriate modulation of a longitudinal coupling strength, the qubit-state-conditioned destructive interference underpinning the which-path entangler instead encodes information about the qubit state into the state of *two* orthogonal modes. It may be the case that correlated measurements of these orthogonal modes could lead to improved assignment fidelities relative to uncorrelated single-mode measurements. The standard calculations of signal-to-noise ratio used to analyze, e.g., dispersive readout are not obviously suited to answering this question, so the question of how to construct a metric that captures any potential benefits of correlated two-mode measurements for qubit readout remains to be answered. More fundamental than the question of qubit readout is the question of how to realize the which-path entangler across different qubit platforms. The implementation discussed in the appendix to Chapter 3 involves a flopping-mode spin qubit, and notably, we showed that by operating the flopping-mode qubit with a special relationship between the Zeeman gradient and the double-dot detuning, a parametrically modulated longitudinal coupling of the form $\tilde{g}_1(t)\,|1\rangle\langle 1|\,(a-a^\dagger)$ could be realized by modulating the tunnel coupling. Although the literature on modulated longitudinal coupling almost universally treats the coupling as $\propto \sigma_z$, there is no reason to assume that the modulation of any given parameter (e.g., the tunnel coupling) will affect the two qubit states symmetrically. This issue was foreshadowed in the introductory chapter of this thesis, but to reiterate: A *time-independent* longitudinal cavity-qubit coupling can generically be expressed in the form $(g_z\sigma_z + g_0)(a - a^\dagger)$, where the term $\propto g_0$ is proportional to the identity on the qubit. In this case, the qubit-state-independent term can be eliminated from the Hamiltonian by transforming to a displaced frame and re-defining the qubit states. This is done implicitly throughout the literature and is justified in the static case. However, in the presence of a modulation designed to induce a time dependence of $g_z \to g_z(t)$, the displacement transformation used to eliminate the qubit-state-independent term must be treated carefully, as the same modulation could also induce a time dependence of $g_0 \to g_0(t)$. It would therefore be interesting and useful to understand under what circumstances modulating a certain parameter, like the tunnel coupling in the case of the flopping-mode qubit, would give rise to the asymmetric longitudinal coupling $\tilde{g}_1(t)\,|1\rangle\langle 1|\,(a - a^\dagger)$ required for realizing the which-path entangler across other qubit platforms involving, e.g., superconducting qubits or Rydberg atoms.

We also discussed in Chapter 3 a method for converting qubit–which-path entanglement into entanglement between distant stationary qubits (Bell states). This protocol involved combining the reflected and transmitted fields at a 50:50 beamsplitter in order to re-encode the which-path degree-of-freedom into a parity degree-of-freedom—a cat qubit whose basis states $|\alpha\rangle \pm |-\alpha\rangle$ are characterized by their photon-number parity. The resulting entanglement between the stationary qubit and the flying-cat qubit could be converted into a stationary-qubit Bell state through a photon-number-conditioned operation on a second stationary qubit, followed by a homodyne measurement of the cat qubit. In Chapter 5, we considered the use of these same photon-number-conditioned operations for realizing stabilizer measurements (parity checks) of distant stationary qubits using pulses of classical radiation in place of the ancilla qubits typically considered as parity meters in the context of quantum error correction. We showed that in the case of the subsystem surface code, photon loss during these parity checks does not lead to correlated (horizontal hook) errors on the encoded information: The two-qubit errors resulting from photon loss are equivalent to single-qubit errors, modulo a gauge operator. This simple observation highlights the potential of these ancilla-qubit-free parity checks operated in combination with this particular choice of error-correcting code. That being said, both the Bell states generated in the manner of Chapter 3 and the parity checks implemented in the manner of Chapter 5 retain a level of susceptibility to photon loss: In the case of the Bell states, the two-qubit concurrence decreases exponentially in the probability of losing a single photon, while in the case of the parity checks, the loss of a photon leads to an error on the code qubits. Although these errors could be dealt with in principle through either entanglement distillation protocols (for Bell states) or through the error-correcting code itself (for parity checks), an interesting avenue of future study would seek to mitigate the impact of photon loss on either or both of these protocols by replacing the cat qubits with a higher-order cat-code.

In Chapter 4, we considered the problem of Heisenberg-limited quantum interferometry, focusing on the measurement protocol used to extract information about the interferometric phase $\phi$. In general, there are two main considerations to be borne in mind when evaluating a measurement protocol: First, there is the question of how close it comes to saturating the quantum Cramér-Rao bound in the asymptotic limit, i.e., the question of how close it comes to optimality. Second, there are the technical requirements of the measurement scheme itself. The theoretical literature on quantum metrology has focused predominantly on photon counting as a strategy for realizing optimal measurements. By contrast, we showed in Chapter 4 that homodyne detection can be made optimal for estimation of the phase difference $\phi$, provided knowledge of the average phase—acquired beforehand in a separate experiment, or inferred from a symmetry of the problem—is incorporated into the local oscillator settings. To our knowledge, this is the first optimal homodyning scheme to appear in the literature, which is notable since homodyne detection is significantly less technologically demanding than number-resolving measurements. Our results hold for two input states: the qubit–which-path states introduced in Chapter 3, as well as entangled coherent states (ECSs) given by two-mode superpositions of the form $|\alpha, 0\rangle + |0, \alpha\rangle$. A natural extension of the results presented in Chapter 4 would be to consider other input states commonly used for quantum metrology, and to see whether optimal homodyning schemes could be constructed for those states as well. Although we focused chiefly on the measurement protocol used to estimate $\phi$, state preparation is another major technical challenge that must be overcome if one wishes to exploit the precision enhancements associated with the use of non-classical resource states. The preparation of an ECS, for instance, involves mixing a coherent state $|\alpha/\sqrt{2}\rangle$ and a Schrödinger cat state $|\alpha/\sqrt{2}\rangle + |-\alpha/\sqrt{2}\rangle$ on a 50:50 beamsplitter. Exploiting the Heisenberg-limited scaling of an ECS therefore requires a supply of large-amplitude cat states. Such large-amplitude cat states could be created through serial single-photon subtractions from the squeezed vacuum. However, seeing as single-photon subtraction is typically realized probabilistically (by passing the squeezed vacuum through a weakly reflective beamsplitter and conditioning on detector clicks in one of the output modes), the creation of large-amplitude cat states is usually associated with a massive probabilistic overhead, making the use of ECSs for quantum interferometry somewhat inconvenient even notwithstanding the question of which measurement protocol to use. Incidentally, there exists in addition to the question of how to construct an optimal measurement given a certain state (our focus in this thesis) the question of which state is best suited to a particular task given certain experimental constraints. The second of these questions has already been answered by Lang and Caves [1] for an interferometer powered by a coherent state at one input, and subject to a constraint on the total number of photons passing through the interferometer: They found that the optimal state to inject into the second input port is squeezed vacuum, in the sense that this mix of $|\text{coherent state}\rangle \otimes |\text{squeezed vacuum}\rangle$ maximizes the *quantum* Fisher information associated with estimation of a difference phase $\phi$. It is *also* known that a low-amplitude ECS can be approximated by mixing a coherent state with squeezed vacuum having an optimized squeezing parameter [2]. In view of this information, we now take the liberty of applying some transitive reasoning: Since squeezed vacuum is typically much easier to generate than Schrödinger cat states, the obvious question is to what extent our homodyning protocol—known to be optimal for ECS—is also optimal for the experimentally much less challenging input state $|\text{coherent state}\rangle \otimes |\text{squeezed vacuum}\rangle$, which can be made to approximate an ECS in a certain regime. The combination of an experimentally simple state preparation together with an experimentally simple optimal measurement scheme would go a long way toward helping to realize the benefits of quantum interferometry for real-life sensing applications.

This discussion of ECSs as challenging to produce highlights another potential application of the which-path entangler as a deterministic source of ECSs: Whereas one typically imagines generating the cat state required for an ECS via repeated, probabilistic single-photon subtractions from the squeezed vacuum, as explained in the preceding paragraph, the mechanism underpinning the which-path entangler of Chapter 3 could be used to produce a state of the form $|\uparrow,0,\alpha\rangle + |\downarrow,\alpha,0\rangle$ *deterministically*. A measurement of the qubit in the basis $|\uparrow\rangle \pm |\downarrow\rangle$ will then produce an ECS. This ECS can be understood as a photonic Bell state with qubits having basis stats $|0\rangle$ and $|\alpha\rangle$. This then suggests the possibility of lowering the resources required for realizing an all-photonic quantum repeater [3]. In the seminal scheme of Ref. [3], stations housing quantum emitters are used to produce *repeater graph states* consisting of highly entangled photonic cluster states, different parts of which are sent to different repeater nodes. At these repeater nodes, Bell measurements are used to project photons from different stations into Bell states, thereby growing entanglement throughout the network of repeater stations. The motivation for using repeater graph states is twofold: The first reason is to combat photon loss, which requires a certain level of redundancy in the entangled photonic states being distributed across the network. The second reason is related to the success probability of the aforementioned Bell measurements, which is capped at 50% for single photons and linear optics even in the absence of imperfections. For photonic qubits encoded in coherent states, however, the ability to discriminate all four Bell states, though imperfect due to the nonorthogonality of finite-amplitude coherent states, can be made arbitrarily close to 1 by using larger coherent-state amplitudes [4]. Although photon loss would remain a concern, a deterministic source of ECSs could potentially help reduce certain intrinsic sources of probabilism associated with linear-optical Bell measurements, with obvious applications for linear-optical quantum computing as well.

Finally, in Chapter 6, we turned our attention to the question of how to realize two-qubit gates between distant qubits. Such gates would enable distinct quantum chips to operate as a single device while helping to alleviate the input-output wiring bottlenecks that would almost certainly be encountered in scaling up a single-chip layout. Whereas two-qubit gates in circuit QED are typically mediated by a single spectrally isolated cavity mode coupled to both qubits, the protocols we describe could operate with a free spectral range of zero and are therefore compatible with very distantly separated qubits. This would in turn provide significant flexibility in device layout, especially since the protocols described in Chapter 6 are compatible with microwave-to-optical transduction, at least in principle. The gate protocols also come with a heralding scheme that could be used to flag instances where the loss of the photonic qubit has led to failure of the gate. We have investigated the backaction associated with the loss of a time-bin qubit to a dielectric environment, which cannot be treated using standard models of photon loss since these models do not typically consider the time at which the photon was lost. We showed that despite the orthogonality of the time-bin basis states, the loss of a time-bin qubit to a non-Markovian environment (having a correlation time that is long compared to the time-bin separation) leads to significantly less backaction than would be experienced for loss to a Markovian environment. This suggests possible strategies for tailoring either the dielectric environment or time-bin parameters to engineer a situation where the loss of a time-bin qubit is almost backaction-free.